%% file: QFormExcitations.tex
\documentclass[10pt]{article}
\pdfoutput=1
\usepackage{mathStyle}
\usepackage{amsmath,amssymb,amsthm,latexsym,bbm,calc,wasysym,mathtools,empheq, bbold}
\usepackage{mathrsfs}
\usepackage[english]{babel}
\usepackage{relsize}
\usepackage{graphicx,color}
\usepackage[dvipsnames]{xcolor}
\usepackage[makeroom]{cancel}

\usepackage{tikz}
\usetikzlibrary{shapes.geometric}
\usetikzlibrary{arrows,matrix,calc,scopes,decorations.markings, scopes}
\usetikzlibrary{arrows.meta}

\usepackage{xspace}
\usepackage{pifont,dsfont}
\usepackage{marvosym}
\usepackage{slashed}
\usepackage{subfigure}
\usepackage{sidecap}
\usepackage{stmaryrd}
\usepackage{empheq}

\usepackage[export]{adjustbox}
\mathtoolsset{showonlyrefs=false}
\usepackage{placeins}
\usepackage{enumitem}

\usepackage{amsthm}
\newtheoremstyle{mydef}%
	{0.9em} 
	{0.7em}
	{\hangindent=2em}
	{1.8em}
	{\scshape}
	{.}
	{.5em}
	{}%
\theoremstyle{mydef}
\newtheorem{definition}{Definition}
\numberwithin{definition}{section}

\theoremstyle{mydef}

\numberwithin{lemma}{section}

\theoremstyle{mydef}

\numberwithin{theorem}{section}

\theoremstyle{mydef}
\newtheorem{example}{Example}
\numberwithin{example}{section}

\theoremstyle{mydef}
\newtheorem{convention}{Convention}
\numberwithin{convention}{section}

\theoremstyle{mydef}
\newtheorem{property}{Property}
\numberwithin{property}{section}

\newcommand{\la}{\langle}
\newcommand{\ra}{\rangle}
\newcommand{\q}{\quad}
\newcommand{\nn}{\nonumber}
\newcommand{\rU}{{\rm U}}

\newcommand{\sss}{\scriptstyle}

\newcommand{\smilo}{\! \smile \!}
\newcommand{\zo}{\text{\small $0$}}
\newcommand{\snum}[1]{\text{\small $#1$}}
\newcommand{\unit}{\mathbbm{1}}
\newcommand{\cM}{\mathcal{M}}
\newcommand{\smo}{\,{\sss \otimes}\,}
\newcommand{\dr}{\,{\! \text{\tiny dR}}\,}
\newcommand{\ec}{\text{\footnotesize $Q$}}
\newcommand{\mc}{\text{\footnotesize $M$}}

\newcommand\myeq{\mathrel{\overset{d}{=}}}

\hypersetup{%
	bookmarks=true,         
	bookmarksopen=true,
	unicode=true,           
	pdftoolbar=true,        
	pdfmenubar=true,        
	pdffitwindow=false,     
	pdfstartview={FitH},    
	pdftitle={On 2-form gauge models of topological phases},
	pdfauthor={},
	pdfsubject={},          
	pdfcreator={},          
	pdfproducer={pdfLaTeX}, 
	pdfkeywords={2-form} {topological phases} {gauge theory},
	pdfnewwindow=true,      
	colorlinks,
	citecolor=BrickRed,
	filecolor=BrickRed,
	linkcolor=BlueViolet,
	urlcolor=BrickRed,
	linktocpage=true
}

\include{tikzsb}

\title{\boldmath On 2-form gauge models of topological phases}

\author[\Square,\pentagon]{Clement Delcamp,}

\affiliation[\Square]{Max-Planck-Institut f{\"u}r Quantenoptik \\ Hans-Kopfermann-Str. 1, 85748 Garching, Germany}
\affiliation[\pentagon]{Perimeter Institute for Theoretical Physics\\ 31 Caroline Street North, Waterloo, Ontario  N2L 2Y5, Canada}

\author[\hexagon]{Apoorv Tiwari}
\affiliation[\hexagon]{Department of Physics, University of Zurich \\ Winterthurerstrasse 190, 8057 Zurich, Switzerland}

\emailAdd{clement.delcamp@mpq.mpg.de}
\emailAdd{t.apoorv@gmail.com}

\abstract{\\~\\ 
    We explore various aspects of 2-form topological gauge theories in (3+1)d. These theories can be constructed as sigma models with target space the second classifying space $B^2G$ of the symmetry group $G$, and they are classified by cohomology classes of $B^2G$. Discrete topological gauge theories can typically be embedded into continuous quantum field theories. In the 2-form case, the continuous theory is shown to be a strict 2-group gauge theory. This embedding is studied by carefully constructing the space of $q$-form connections using the technology of Deligne-Beilinson cohomology. The same techniques can then be used to study more general models built from Postnikov towers. For finite symmetry groups, 2-form topological theories have a natural lattice interpretation, which we use to construct a lattice Hamiltonian model in (3+1)d that is exactly solvable. This construction relies on the introduction of a cohomology, dubbed 2-form cohomology, of algebraic cocycles that are identified with the simplicial cocycles of $B^2G$ as provided by the so-called $W$-construction of Eilenberg-MacLane spaces. We show algebraically and geometrically how a 2-form 4-cocycle reduces to the associator and the braiding isomorphisms of a premodular category of  $G$-graded vector spaces. This is used to show the correspondence between our 2-form gauge model and the Walker-Wang model.}

\begin{document} 
	\vspace*{-2em}
	\maketitle
	\flushbottom
	\newpage

\section{Introduction}

\bigskip\bigskip\noindent Over the last several decades, \emph{quantum field theories} have emerged as the central language in which
modern theoretical physics is formulated. For instance, \emph{quantum phases of matter} may succinctly be defined as equivalence classes of quantum field theories, and a given quantum model is a concrete realization of a phase.  \emph{Topological} quantum field theories (TQFTs) form a subclass of quantum field theories that are particularly tractable. Indeed, topological theories are much simpler than conventional theories as they associate finite dimensional Hilbert spaces to codimension-one submanifolds and have trivial Hamiltonian evolution. From a mathematical point of view, TQFTs can usually be reformulated algebraically in terms of finite sets of data. Such a reformulation, which bears a strong category theoretical flavor, was initially pioneered by Atiyah in \cite{Atiyah:1989vu} who defined a TQFT as a \emph{symmetric monoidal functor} from a certain category of \emph{bordisms} to the category of finite dimensional vector spaces.\footnote{
	For example, a ($d$+1)-dimensional TQFT $\mathcal{Z}$ is a symmetric monoidal functor that assigns to every \emph{oriented closed} $d$-manifold $\cM$ a vector space $\mathcal{Z}[\cM]$ over the field $k$ and to every bordism $\mathcal{B}: \cM_1 \to \cM_2$ between two oriented closed $d$-manifolds a linear map of vector spaces $\mathcal{Z}[\mathcal{B}]:\mathcal{Z}[\cM_1] \to \mathcal{Z}[\cM_2]$, together with the following isomorphisms
	\begin{equation}
	\nn
	\mathcal{Z}[\varnothing] \simeq k \q , \q \mathcal{Z}[\cM_1 \sqcup \cM_2] \simeq \mathcal{Z}[\cM_1] \otimes \mathcal{Z}[\cM_2] \; .
	\end{equation}
	This data is subject to some coherence relations that ensure the topological nature of the theory. Moreover, it can be readily generalized to accommodate manifolds with additional structure such as spin structure or framing by suitably replacing the category of oriented bordisms.} 
This proposal was further developed by Baez and Dolan in \cite{Baez:1995xq} who suggested that \emph{higher category theory} was the correct framework to capture the local structure inherent to quantum theory. More precisely, they proposed that a ($d$+1)-dimensional \emph{fully extended} TQFT, which is capable of capturing locality all the way down to points, should be understood as a ($d$+1)-functor between a higher ($d$+1)-category of bordisms\footnote{It is a category of extended bordisms whose objects are points, 1-morphisms are 1-bordisms between disjoint union of points, 2-morphisms are bordisms between 1-bordisms, and so on and so forth.} and a higher \emph{symmetric monoidal} ($d$+1)-category. This came to be known as the \emph{cobordism hypothesis} \cite{lurie2009higher, Lurie:2009keu, Freed:2012hx}.  These mathematical definitions that are motivated by topological invariance on the one hand and locality on the other hand  severely constrain the structure of TQFTs, and can therefore be used as a classifying tool for topological theories in a given spacetime dimension.

It is believed that at long wavelengths \emph{gapped} phases of matter, i.e. phases that have a spectral gap above the ground state that persists in the thermodynamics limit, are described by equivalence classes of topological quantum field theories.\footnote{Nevertheless, it is not completely clear whether there is a bijection between physically realizable gapped phase of matter and TQFTs. The subtle relation between TQFTs and gapped phases was carefully studied in \cite{Gaiotto:2017zba} for theories displaying a global symmetry.} Therefore, the above mentioned mathematical constraints turn out to have  profound physical consequences and serve as an organizational tool for the space of gapped phases of matter. Furthermore, given a TQFT describing deep \emph{infrared} physics, it is often possible to construct an exactly solvable model in terms of a \emph{lattice Hamiltonian projector}. The model may then be deformed away from its exactly solvable projector in order study dynamical properties within the corresponding phase. This is one of the reasons why understanding topological theories and building the corresponding exactly solvable models is a worthwhile endeavor. 

Naturally, the map from the space of \emph{ultraviolet} models to the space of TQFTs is surjective. Since quantum models are understood in terms of correlation functions of the observables that they furnish, going from the ultraviolet to the topological infrared is performed by a map that only retains the topological part of the correlation functions. As a matter of fact, it is a defining feature of topological theories to be blind to operators that are irrelevant under the renormalization group. Therefore, perturbing a TQFT away from its deep infrared fixed point, while maintaining its gap, may be thought of as going towards the ultraviolet regime.

\bigskip \noindent There is a particular class of fully extended TQFTs, known as \emph{Dijkgraaf-Witten} theories \cite{dijkgraaf1990topological}, that are mathematically well-defined in all dimensions. These theories are constructed from finite groups and have a topological gauge theory interpretation. Given a ($d$+1)-manifold $\cM$ and a finite group $G$,  they depend on a single datum, namely a \emph{cohomology class} $[\omega]\in H^{d+1}(BG,\mathbb R/\mathbb Z)$ where $BG$ is the \emph{classifying space} of the group $G$, which has the property that its only non-vanishing homotopy group is the fundamental group and it equals the group $G$ itself. Dijkgraaf-Witten theories can be cast in two equivalent ways: $(i)$ as topological sigma models whose target space is $BG$ and the sum in the partition function being performed over homotopy classes of maps from the spacetime manifold $\cM$ to $BG$, $(ii)$ as topological lattice gauge theories defined on a triangulation of the spacetime manifold together with a \emph{$G$-coloring}, i.e. an assignment of group elements in $G$ to every 1-simplex of the triangulation that satisfies compatibility conditions. 
Although, the first approach $(i)$ is more mathematically succinct, the latter point of view  $(ii)$ has the advantage of being more physically transparent, i.e the fields, observables and gauge transformations can be more explicitly defined and studied. This happens to be very useful when studying for instance the excitations of the theory and their properties. 

The equivalence between the two aforementioned approaches  is conceptually straightforward and yet slightly subtle:
The topological action for the sigma model approach is provided by integrating the pullback of the cohomology class $[\omega]$ onto the manifold $\cM$, while in the lattice gauge theory picture, the topological action is provided by evaluating the cocycle on each $G$-colored ($d$+1)-simplices of the triangulation. But this relies implicitly on the fact that for discrete groups the cohomology $H^{d+1}(BG,\mathbb R/\mathbb Z)$ as an algebraic description. More precisely, it uses the fact there is an equivalence between the cohomology $H^{d+1}(BG,\mathbb R/\mathbb Z)$ of \emph{simplicial} cocycles of $BG$ and the cohomology $H^{d+1}(G,\mathbb R/\mathbb Z)$ of \emph{algebraic} group cocycles of $G$. Instead of representing ($d$+1)-cochains as simplices, they are then defined as functions from $G^{d+1}$ to $\mathbb R/\mathbb Z$, and the coboundary operator is modified accordingly. This second approach in terms of group cohomology is naturally the one used in order to construct exactly solvable models that are lattice Hamiltonian realizations of Dijkgraaf-Witten theories \cite{Hu:2012wx, Wan:2014woa}. It turns out that a similar correspondence can also be established for topological theories that have a \emph{higher gauge} theory interpretation. It is however not as straightforward as we explain at length in the present manuscript. 

It is possible to define different sigma models that generalize the Dijkgraaf-Witten construction by choosing different target spaces. The most natural generalization is obtained by replacing the classifying space $BG$ of the discrete group $G$ by the $q$-th classifying space $B^qG$.\footnote{Since the partition sum is built by summing over homotopy classes of maps to $B^qG$, we really mean $B^qG$ up to homotopy equivalence here.} The $q$-th classifying space $B^qG$ is an example of \emph{Eilenberg-MacLane space} $K(G,q)$ which has the property that only its $q$-th homotopy group is non-vanishing and equals the group $G$ itself, i.e. $\pi_{n}(K(G,q))=\delta_{q,n}G$ \cite{eilenberg1953groups, eilenberg1954groups}.\footnote{The classifying space $BG$ is thus an example of Eilenberg-MacLane space $K(G,1)$.} Interestingly, the same way Dijkgraaf-Witten theories have a lattice gauge theory interpretation, a topological sigma model whose target space is an Eilenberg-MacLane space $K(G,q)$ can be interpreted as a $q$-form topological lattice gauge theory, i.e. a theory that contains ($q$$-$1)-dimensional symmetry operators instead of point-like ones. Theories displaying a ($q$$-$1)-form gauge invariance have a gauge field that is locally described by a $q$-form. A further generalization involves building topological sigma models whose target spaces are provided by \emph{Postnikov towers}. A Postnikov tower is a topological space constructed as a sequence of \emph{fibrations} of simpler topological spaces. In particular, in this manuscript we will be interested in Postnikov towers which are built as fibrations of Eilenberg-MacLane spaces. In analogy to Dijkgraaf-Witten theories, these may be understood as topological \emph{higher group} gauge theories that contain several gauge fields. More specifically, for every Eilenberg-MacLane space $K(G,q)$ contained in the Postnikov tower, the gauge theory will include a corresponding $q$-form gauge field. In the lattice gauge theory picture, a $q$-form gauge field is defined by coloring the $q$-simplices of the triangulation with elements of the group $G$ that satisfy some consistency criteria in the form of cocycle conditions. The precise form of these cocycle conditions is obtained from the data that goes into building the Postnikov tower. The corresponding gauge transformations are built from the same data. These different generalizations are presented in sec.~\ref{sec:Sigma}.

\bigskip \noindent Throughout this manuscript, we focus most of our attention on (3+1)d topological sigma models with the second classifying space $B^2G$ as the target space where $G$ is a finite abelian group, or equivalently discrete (3+1)d 2-form topological lattice gauge theories. As explained above, such higher form gauge theories arise naturally from a mathematical point of view. But they also happen to be physically motivated. For instance, it is known that \emph{Yang-Mills theory} is confining and the gauge bosons are gapped at long wavelengths, and it was argued in \cite{kapustin2014topological} that the infrared physics of the confining phase is captured by a non-trivial 2-form topological gauge theory. The gauge group of this 2-form gauge theory is the magnetic gauge group that survives in the infra red \cite{Aharony:2013hda, gaiotto2015generalized}. These 2-form gauge theories have also appeared in various other contexts in the literature \cite{Pfeiffer:2003je, Freed:2009qp, Walker:2011mda, vonKeyserlingk:2012zn, gaiotto2017theta,delcamp2018gauge, Benini:2018reh, Hsin:2018vcg, Wan:2018bns, Wan:2018zql, Guo:2018vij, Wan:2018djl, Wan:2019oyr}.  One particular reason for the interest in such TQFTs resides in the fact that they host a topologically ordered surface.

Given a finite abelian group $G$, 2-form topological theories are classified by a single datum, namely a cohomology class $[\omega] \in H^{4}(B^2G,\mathbb R/\mathbb Z)$.
It was shown by Eilenberg and MacLane in a series of seminal papers  \cite{eilenberg1953groups, eilenberg1954groups} that the cohomology group $H^{4}(B^2G,\mathbb R/\mathbb Z)$ is isomorphic to the group of (possibly degenerate) $\mathbb R/\mathbb Z$-valued quadratic functions on $G$. This result allows for an explicit expression of the topological action in terms of a quadratic form and a quadratic operation known as the \emph{Pontrjagin square} on $H^{2}(\mathcal M,G)$ that is the space of fields of the 2-form theory \cite{kapustin2014anomalies, gaiotto2015generalized}. Moreover, the \emph{topological order} living at the surface  can be described in terms of a categorical structure whose input data is the same as the one labeling the bulk theory, namely a finite abelian group and a quadratic form. If the quadratic form is degenerate, then the topological order is non-trivial.\footnote{We define non-trivial topological orders as the ones that have long-range entanglement, non-trivial ground state degeneracy that depends on the topology and fractionalized excitations.} Furthermore, abelian Chern-Simons theories are labeled by precisely the same data. As a matter of fact, it was shown in \cite{Freed:2009qp} that the 2-form theory is precisely the \emph{anomaly theory} for the \emph{framing anomaly} within the abelian Chern-Simons theory. Therefore, we may interpret abelian Chern-simons as a \emph{framed topological} quantum field theory or as a TQFT along with the corresponding (3+1)d 2-form topological gauge theory. 

Besides topological gauge theories, there exist other TQFTs which have been extensively studied. For instance, in (2+1)d it is possible to define a topological theory from any \emph{modular tensor category} using the \emph{Turaev-Viro} construction \cite{Turaev:1994xb, Turaev:1992hq, Barrett:1993ab} and the corresponding Hamiltonian realization is provided by the \emph{Levin-Wen models} \cite{Levin:2004mi}. Similarly, in (3+1)d it is possible to define a topological theory for any \emph{premodular tensor category}\footnote{By \emph{premodular} category we mean a braided fusion category. A premodular category is then modular if its $S$-matrix is non-degenerate.} using the \emph{Crane-Yetter} construction \cite{Crane:1993cm, Crane:1993if, Crane:1994ji} and the corresponding Hamiltonian realization is provided by the \emph{Walker-Wang} models \cite{Walker:2011mda}. But, when the input data of the premodular category is a finite abelian group and a quadratic form, the Walker-Wang model provides a Hamiltonian realization of a 2-form gauge theory that describes the topological order mentioned above. 

\bigskip 
\noindent
\emph{Our study pursues two complementary approaches:} The first one relies on a formulation of 2-form gauge theories in the continuum. 
Indeed, it is often possible to embed discrete gauge theories, especially the ones built from abelian groups, into continuous gauge theories. This embedding, if possible, is such that partition function of the discrete gauge theory and the one of  the continuous theory are equal. A well-known example of such a procedure is the embedding of a $\mathbb Z_{n}$-gauge theory in ($d$+1)-dimensions into a BF theory with a $\rU(1)$-connection 1-form $A$ and a $\rU(1)$-dynamical field ($d$$-$1)-form $B$.\footnote{The action of the continuous BF theory reads $\mathcal{S}=2\pi in \int B \wedge d_{\dr}A$ where $d_{\dr}$ is the usual exterior derivative on forms so that $d_{\dr} A$ is the curvature $2$-form.} A special example of this scenario is the embedding of the \emph{toric code} model, i.e. a $\mathbb Z_2$-gauge theory, into a $\rU(1)$ BF theory. Similarly, discrete 2-form gauge theories may also be embedded into continuous $\rU(1)$ gauge theories. But in this case the gauge structure is not the usual one. Indeed, gauge connections are now locally described by some number of $1$-form and $2$-form fields that do not transform independently under $0$-form and $1$-form gauge transformations. In sec.~\ref{sec:DB}, we study such gauge bundles in detail and show that they form so-called \emph{strict 2-group bundles} \cite{baez2004higher, Baez:2003fs,Baez:2004in, Delcamp:2017pcw}. We do so by carefully constructing the configuration space of $q$-form $\rU(1)$ connections using the technology of \emph{Deligne-Beilinson cohomology} \cite{deligne1971theorie, Thuillier:2015vma, Mathieu:2015mda} and then building the configuration space of strict 2-group bundles by taking a certain twisted product of 1-form and 2-form gauge bundles. Although the continuous formulation thus obtained gives access to powerful tools  familiar to quantum field theories, it is sometimes more convenient to work in the discrete within the Hamiltonian formalism. This takes us to our second approach.

Our second approach involves defining a 2-form gauge model Hamiltonian realization directly in terms of a cocycle in $H^4(B^2G, \mathbb R / \mathbb Z)$. More precisely, the model is defined in terms of a cocycle in a cohomology that is the algebraic analogue of $H^4(B^2G, \mathbb R / \mathbb Z)$, i.e. a cohomology of algebraic cocycles on $G$ that is in one-to-one correspondence with the cohomology of simplicial cocycles on $B^2G$. We dubbed this cohomology of algebraic cochains \emph{2-form cohomology} and its definition relies on the so-called \emph{$W$-construction} of Eilenberg-MacLane spaces $K(G,2)$. After reviewing basic facts regarding Eilenberg-MacLane spaces as well as the general $W$-construction in sec.~\ref{sec:EM}, we define precisely this 2-form cohomology in sec.~\ref{sec:2formco}. The 2-form Hamiltonian model is finally constructed in sec.~\ref{sec:Ham2Forms}. Using solely the cocycle conditions, it is possible to show explicitly how a 2-form 4-cocycle can be reduced to a group 3-cocycle $\alpha$ and a group 2-cochain $R$ that satisfies the so-called \emph{hexagon equations}. Together, $\alpha$ and $R$ define an \emph{associator} and a \emph{braiding}, respectively, which are precisely the isomoprhisms entering the definition of a certain premoludar category, namely the premodular category of \emph{$G$-graded vector spaces}. As a matter of fact, it can even be shown that the set of equivalence classes of pairs $(\alpha, R)$ is isomorphic to the cohomology $H^4(B^2G, \mathbb R / \mathbb Z)$.

The algebraic correspondence mentioned above between a pair $(\alpha,R)$ of associator and braiding on one side, and a 2-form 4-cocycle on the other, can also be displayed graphically:  In the lattice Hamiltonian picture, the 2-form cocycle arises as the amplitude of local unitary transformations performed on fixed point ground states. In (3+1)d, these local unitary transformations are expressed in terms of 2--3 and 1--4 \emph{Pachner moves} \cite{PACHNER1991129}. But we show in sec.~\ref{sec:Ham2Forms} how these moves reduce to the moves defined in the context of the Walker-Wang model whose amplitudes are provided by the associator and the braiding isomorphisms. This algebraic and geometric correspondence can then be used to show explicitly how our Hamiltonian model is related to the Walker-Wang model for the category of $G$-graded vector spaces. This is the purpose of sec.~~\ref{sec:WW}. Most interestingly, we can display how the \emph{ad hoc} splitting into three-valent vertices required for the definition of the Walker-Wang Hamiltonian is now directly encoded in the definition of the 2-form cocycle itself. This makes the definition of our model more compact and more systematic.

\bigskip \bigskip

\begin{center}
	\textbf{Organization of the paper}
\end{center}
\noindent
In sec.~\ref{sec:Sigma}, we first review the definition of the Dijkgraaf-Witten model both as a sigma model and as a lattice gauge theory. We then present a generalization obtained by choosing the target space to be the $q$-th classifying space of a discrete abelian group. We review known material about sigma models whose target spaces are provided by the second classifying space of a finite abelian group and review their classification. In sec.~\ref{sec:DB}, we introduce Deligne-Beilinson cohomology and show that the $q$-th Deligne-Beilinson cohomology group is isomorphic to the space of gauge inequivalent $q$-form $\rU(1)$ connections. This can also be used in order to construct strict 2-group connections that naturally appear when trying to embed theories based on finite abelian groups into continuous toric gauge theories. We then move on to the study of the lattice realization of a 2-form topological gauge theory. In sec.~\ref{sec:EM}, we review the theory of Eilenberg-MacLane spaces as well as their so-called $W$-construction. We use the $W$-construction in sec.~\ref{sec:2formco} to define the 2-form cohomology. The lattice Hamiltonian of the (3+1)d 2-form model is defined in sec.~\ref{sec:Ham2Forms} and the excitations yielded by the Hamiltonian are briefly discussed. Finally, in sec.~\ref{sec:WW} our lattice model is compared to the Walker-Wang model for the category of $G$-graded vector spaces. The paper also contains a couple of appendices. In particular, App.~\ref{sec:app_quant} provides further detail regarding the quantization and the invertibility of 2-form theories, while in app.~\ref{sec:app_DB} we propose explicit expressions of $q$-form topological actions using the language of Deligne-Beilinson cohomology.

\medskip \noindent \emph{Sections \ref{sec:Sigma}--\ref{sec:DB} and sections \ref{sec:EM}--\ref{sec:WW} offer two different perspectives on the study of 2-form topological gauge theories. These two parts are complementary and almost self-contained. If the reader is mainly interested in the Hamiltonian lattice realization of 2-form gauge theories, it is therefore possible to jump directly to sec.~\ref{sec:EM}.}
\newpage

\section{Topological gauge theories as topological sigma models \label{sec:Sigma}}
\setlength{\fboxsep}{2\fboxsep}

In this section, we introduce different topological theories as sigma models. We also explain how these can be formulated as lattice (higher) gauge models. This lattice interpretation will be at the heart of the study carried out in sec.~\ref{sec:EM} onwards.

\subsection{Dijkgraaf-Witten theory}
Dijkgraaf and Witten defined in \cite{dijkgraaf1990topological} a \emph{topological gauge theory} for a finite group $G$ in general spacetime dimension ($d$+1).\footnote{Although their paper only discusses (2+1)d, generalization to any dimension is very straightforward.} They showed that different topological $G$-gauge theories were classified by a single datum, namely a \emph{cohomology class}
\begin{align}
	[\omega]\in H^{d+1}(BG,\mathbb R/\mathbb Z)
\end{align}
where $BG$ is the \emph{classifying space} of the group $G$ that has the distinguished property that its only non-vanishing homotopy group is the fundamental group $\pi_1(G)$, and $\pi_1(G)$ equals the group $G$ itself. The gauge theory is built as a sigma model with the \emph{target space} being $BG$. The partition sum is performed over homotopy classes of maps $[\gamma]: \cM \to BG$ where $\cM$ is an oriented ($d$+1)-manifold. To each map $\gamma$, we associate a topological action that is the integral over $\cM$ of the pull-back $\gamma^\star \omega \in  H^{d+1}(\cM,\mathbb R/\mathbb Z)$ of $\omega $. The partition function takes a simple form
\begin{align}
	\mathcal Z^{BG}_{\omega}[\cM]=\frac{1}{|G|^{b_0}}\sum_{[\gamma]: \cM \to BG}e^{2\pi i \la \gamma^{\star} \omega, [\cM]\ra }
	\label{eq:DW_pf}
\end{align}
where $b_0$ is the $0$-th Betti number, $[\cM] \in H_{d+1}(\cM,\mathbb{Z})$ the fundamental homology cycle of $\cM$ and $\la {\sss \bullet}, {\sss \bullet } \ra$ the canonical pairing defined as $\la \gamma^{\star} \omega, [\cM]\ra = \int_{\cM} \gamma^{\star}\omega$. Since the only non-vanishing homotopy group of $BG$ is its fundamental group, homotopy classes of maps from  $\cM$ to $BG$ are homomorphisms $\text{Hom}(\pi_1(\mathcal M),\pi_1(BG)=G)/\sim$ where the equivalence relation $\sim$ is generated by null homotopic maps. The partition sum can therefore be rewritten 
\begin{align}
	\mathcal Z^{BG}_{\omega}[\cM]=\frac{1}{|G|^{b_0}}\sum_{A\in \text{Hom}(\pi_1(\cM),G)/\sim}e^{2\pi i\la \omega(A), [\cM]\ra}
	\label{eq:DW_pf3}
\end{align}
where $A$ is a representative in a homotopy class $[\gamma]$ and $\omega(A)$ the evaluation of $\gamma^{*}\omega$ on $A$. When the group $G$ is abelian the partition sum is over a cohomology group which is the natural abelianization of the homotopy group. In other words, maps $\gamma$ become $G$-valued 1-cocycles and the null homotopic maps are $G$-valued 1-coboundaries (written as $d\phi$) so that the configuration space of the sigma model is $H^{1}(\cM,G)$. 

\bigskip \noindent Alternatively, \eqref{eq:DW_pf3} can be recast as a lattice gauge theory. In order to do so, let us endow $\cM$ with a triangulation $\triangle$. Thanks to the path-connectedness of $BG$, one can smoothly deform maps $\gamma$ so that the space of paths in $BG$ that is $G$ up to homotopy can be mapped to the 1-simplices of $\triangle$. The contractible paths are then mapped to the identity group element. In practice, this means that we assign to every $1$-simplex $(xy)\subset \triangle$ a group element $g_{xy}$ such that for every 2-simplex $(xyz)$ whose boundary is associated with a contractible path, the flatness condition (or 1-cocycle condition) $\la dg, (xyz)\ra \equiv g_{yz}\cdot g_{xz}^{-1} \cdot g_{xy} =\mathbb{1}  $ is imposed. This is merely the statement that a flat $G$-connection can have non-trivial holonomies along non-contractible closed paths only. Non-trivial group elements are thus assigned to non-contractible cycles of $\cM$ so that each assignment is an element of $\text{Hom}(\pi_1(\cM),G)$. We refer to such an assignment of group elements as a \emph{$G$-coloring} and we denote by ${\rm Col}(\cM,G)$ the set of $G$-colorings. The Dijkgraaf-Witten partition function then reads 
\begin{equation}
	\mathcal{Z}_{\omega}^{BG}[\cM] = \frac{1}{|G|^{|\triangle^0|}} \sum_{g \in {\rm Col}(\cM,G)}\prod_{\triangle^{d+1}}e^{2 \pi i \mathcal{S}_\omega[g,\triangle^{d+1}]}
	\label{eq:DW_pf2}
\end{equation}
where $|\triangle^0|$ is the number of $0$-simplices. The topological action is provided by $\mathcal{S}_\omega[g,\triangle^{d+1}] := \epsilon(\triangle^{d+1})\la \omega(g), \triangle^{d+1} \ra$ such that $\epsilon(\triangle^{d+1}) = \pm 1$ is determined by the orientation of the ($d$+1)-simplex and $\la \omega(g), \triangle^{d+1}\ra$ is the evaluation of the cocycle $\omega$ on the $G$-colored simplex $\triangle^{d+1}$.

\medskip \noindent So there are multiple constructions of the Dijkgraaf-Witten partition: $(i)$ As a topological sigma model with target space the classifying space $BG$ which gives the formulation \eqref{eq:DW_pf}. $(ii)$ Upon noticing that the homotopy classes of maps satisfy $[\cM,BG]\simeq H^{1}(\cM,G)$, one obtains \eqref{eq:DW_pf3}. $(iii)$ After endowing the space-time manifold $\cM$ with a triangulation, a lattice construction can be obtained which leads to \eqref{eq:DW_pf2}. The relation between $g\in {\rm Col}(\cM,G)$ and $A\in \text{Hom}(\pi_1(\cM),G)/\sim$ is that $A$ corresponds to an equivalence class of $g$'s where the equivalence relations are gauge transformations.     

\subsection{Generalized topological gauge theories}
The compact expression \eqref{eq:DW_pf} for the Dijkgraaf-Witten partition function can be readily generalized to the scenario where $BG$ is replaced by some other space $X$. For several different choices of $X$, the space of homotopy classes of maps $[\cM,X]$ is isomorphic to a generalized cohomology group on $\cM$. One then may study topological sigma models, with the space $X$ as the target space, that provides generalizations of conventional topological gauge theories. Similar to the Dijkgraaf-Witten partition functions above \eqref{eq:DW_pf}--\eqref{eq:DW_pf2}, such generalized gauge theories can also be built as lattice (higher) gauge theories on triangulated space-time manifolds. We first describe the construction of these generalized gauge theories as sigma models and then as topological lattice theories.

A topological sigma model can be constructed by generalizing the Dijkgraaf-Witten partition function as follows:
\begin{align}
	\mathcal{Z}_\omega^X[\cM]=\frac{1}{\mathcal{N}_X}\sum_{[\gamma]\in \pi_0[\text{Map}(\cM,X)]} e^{2 \pi i \la \gamma^{\star} \omega, [\cM] \ra} \; ,
\end{align}
where $\omega \in C^{d+1}(X,\mathbb R/\mathbb Z)$ is a 
($d$+1)-cochain, $\mathcal M$ is a compact oriented ($d$+1)-manifold, $[\mathcal{M}] \in H_{d+1}(\mathcal{M}, \mathbb{Z})$ its fundamental homology cycle and $\mathcal N_{X}$ is a normalization constant that depends on the manifold and the choice of target space $X$. The sum in the partition function is over homotopy classes $[\gamma]$ of maps $\gamma$ from $\cM$ to $X$.

Naturally, the choice of ($d$+1)-cochain $\omega$ is constrained: Given an oriented ($d$+2)-bordism $\mathcal{W}: \cM_{1}\sqcup \cM_2\to \cM_3$, it is required that \cite{dijkgraaf1990topological, Freed:1991bn} 
\begin{align}
	0=&\;\langle \gamma^{\star}\omega,[\cM_1] \rangle+\langle \gamma^{\star}\omega,[\cM_2] \rangle-\langle \gamma^{\star}\omega,[\cM_3] \rangle \nn \\
	=&\; \langle \gamma^{\star}\omega,[\partial \mathcal{W}] \rangle \nn \\
	=&\; \langle \gamma^{\star}d \omega ,[\mathcal{W}] \rangle
	\label{bordCond}
\end{align}
where $d: C^{d}(X,\mathbb R/\mathbb Z)\to C^{d+1}(X,\mathbb R/\mathbb Z)$ is the coboundary operator on the space of cochains. Condition \eqref{bordCond} is required to hold for every bordism $\mathcal{W}$ which implies that $\omega$ must be a cocycle in $Z^{d+1}(X,\mathbb R/\mathbb Z)$. When $\cM$ is closed, modifying the cocycle $\omega$ by a coboundary $d \phi$ where $\phi \in C^{d}(X,\mathbb R/\mathbb Z)$ has clearly no effect. However, when $\cM$ is an open manifold, this alters the action by a boundary term that can be absorbed into a $\rU(1)$ phase upon quantization of the theory. Correspondingly, the redefined Hilbert space preserves amplitudes and as such describes the same theory. Putting everything together, we obtain that distinct topological sigma models are classified by cohomology classes $[\omega]\in H^{d+1}(X,\mathbb R/\mathbb Z)$.

Let us now consider several examples of sigma models that correspond to different choices of target space $X$:

\begin{example}[\emph{$X$ is the $q$-th classifying space $B^qG$ of a finite abelian group $G$}] 
	This example is the immediate generalization of the Dijkgraaf-Witten theory obtained by considering the target space $X$ to be the so-called $q$-th classifying space $B^qG$ of a finite abelian group $G$ for $q>1$.\footnote{Explicit constructions of classifying spaces are provided in sec.~\ref{sec:EM}.} This space satisfies the defining property $\pi_n(B^qG)=\delta_{n,q} G$. Similarly to the above construction, one can build models which have a \emph{higher-form topological gauge theory} interpretation by providing a cohomology class $[\omega]\in H^{d+1}(B^qG,\mathbb R/\mathbb Z)$. The partition sum looks almost identical to the Dijkgraaf-Witten partition function:
	\begin{align}
		\mathcal Z^{B^qG}_{\omega}[\cM]=\frac{1}{|G|^{b_{0 \to q-1}}}\sum_{[\gamma]: \cM \to B^qG}e^{2\pi i \la \gamma^\star , [\cM]\ra}
		\label{eq:higher_DW_pf}
	\end{align}
	where  $b_{0 \to q-1} := \sum_{i=0}^{q-1} b_{q-i}(\cM)(-1)^{i}$ and $b_i(\cM)$ is the $i$-th Betti number of the manifold $\cM$. Like the classifying space $BG$, the $q$-th classifying space $B^qG$ can be constructed as a simplicial complex so that the simplicial map $\gamma: \cM \to B^qG$ is furnished by a $G$-valued $q$-cocycle $A$. The null homotopic maps can be extended to a cone above $\cM$ and take the form $d\phi\in B^{q}(\cM,G)$. These null homotopies represent the ($q$$-$1)-form gauge transformations in the $q$-form gauge theory, i.e $A\sim A+ d\phi$. Hence the homotopy classes of maps are isomorphic to the cohomology classes $[\cM,B^qG]\simeq H^{q}(\cM,G)$.
\end{example}
\noindent
In the following sections, we almost exclusively restrict ourselves to the study of the sigma models whose target space are provided by the second classifying space $B^2G$ of a finite abelian group $G$. Such models have a natural interpretation in terms of 2-form gauge theories. Nevertheless, before going into the details of these theories, it is enlightening to sketch out some further generalizations.

\begin{example}[\emph{$X$ is a two-stage Postnikov tower}] 
	Following the theory of Postnikov towers \cite{hatcher2002algebraic}, let us denote the $q_1$-th classifying space of a finite abelian group $G_1$ by $E_1 := B^{q_1}G_1$. As described in the previous example, a topological ($q_1$-form) gauge theory can be built wherein the local fields are cocycles $A_{1}\in Z^{q_1}(\cM,G_1)$ which represent maps from $\gamma :\cM\to E_1  $. Furthermore, the null homotopic maps are captured by coboundaries $d\phi_1\in B^{q_1}(\cM,G_1)$. Since the partition sum is over homotopy classes of maps, we must identify $A_1\sim A_1+d\phi_1$ which we recognize as the ($q_1$$-$1)-form gauge invariance so that gauge inequivalent configurations are isomorphic to $H^{q_1}(\cM,G)$. The target space $E_1$ is referred to as a one-stage Postnikov tower. Things get more interesting if we consider a 2-stage Postnikov tower $E_2= E_1 \rtimes_{\alpha_2} B^{q_2}G_2$ where $[\alpha_2]\in H^{q_2+1}(E_1,G_2)$ such that $E_2$ fits in the exact sequence
	\begin{align}
		0\to B^{q_2}G_2 \to E_2 \to E_1 \to 0
	\end{align}
	whose \emph{extension class} is $[\alpha_2]$. A map from $\cM$ to $E_2$ is furnished by a tuple of local data $\mathbb A_2$ defined as $\mathbb A_2=\left\{(A_1,A_{2})\in C^{q_1}(\cM,G_1)\times C^{q_2}(\cM,G_2)\right\}$. Furthermore, it is required that $\mathbb A_{2}$ is in the kernel of a differential operator denoted by $D_{E_2}$, i.e. $D_{E_2}\mathbb{A}_2 = 0$, such that
	\begin{align}
		D_{E_2}\; :\; C^{q_1}(\cM,G_1)\times C^{q_2}(\cM,G_2) &\, \longrightarrow \, C^{q_1+1}(\cM,G_1)\times C^{q_2+1}(\cM,G_2) \nn \\
		(A_1,A_2)  = \mathbb{A}_2 & \, \longmapsto \, D_{E_2}  \mathbb{A}_2 := \left( dA_{1}, dA_2-\alpha_2(A_1) \right) \; .
	\end{align}
	In other words, $A_{1}$ and $A_2$ satisfy some cocycle conditions \emph{twisted} by the extension class $[\alpha_2]$. Similarly, a null homotopy is provided by the image of an operator $D^{\flat}_{E_2}$ that acts on a tuple $
	\mathbb{\Phi}_{2}=\left\{(\phi_1,\phi_2)\in C^{q_1-1}(\cM,G_1)\times C^{q_2-1}(\cM,G_2)\right\}$ via
	\begin{align}
		D^{\flat}_{E_2}\; : \; C^{q_1-1}(\cM,G_1)\times C^{q_2-1}(\cM,G_2) &\, \longrightarrow \, C^{q_1} (\cM,G_1)\times C^{q_2}(\cM,G_2) \nn \\
		(\phi_1, \phi_2) = \mathbb{\Phi}_2 &\, \longmapsto \, D^{\flat}_{E_2}\mathbb{\Phi}_2 := \left(d\phi_1, d\phi_2 + \zeta_{2}(A_1,\phi_1)\right)
	\end{align}
	where $\zeta_2(A_1,\phi_1)$ is a descendant of $\alpha_2$ satisfying
	\begin{align}
		\alpha_2(A_1+d\phi_1)-\alpha_2(A_1)=d\zeta_2(A_1,\phi_1) \; .
	\end{align}
	We can easily check that $D_{E_2}\circ D_{E_2}^{\flat}=0$ so that one can define a cohomology $H_{E_2}^{\vec{q}}(\cM):= \text{ker}D_{E_2} / \text{im}D^{\flat}_{E_2}$ where $\vec{q}=(q_1,q_2)$. Homotopy classes of maps $[\gamma]:\cM\to E_2$ are in one-to-one correspondence with the equivalence classes of the cohomology we just defined. Given a class $[\omega]\in H^{d+1}(E_2,\mathbb R/\mathbb Z)$, we can thus define a topological gauge theory whose partition function reads
	\begin{align}
		\mathcal{Z}^{E_2}_\omega[\cM]=\frac{1}{|G_1|^{b_{0 \to q_1-1}}|G_2|^{b_{0 \to q_2-1}}}\sum_{[\mathbb A_2]\in H^{\vec{q}}_{E_2}(\cM)}e^{2\pi i\la\omega(\mathbb A_2),[\cM] \ra} \; .
	\end{align}
\end{example}
\noindent
In the case where $q_1=1$ and $q_2=2$, the previous construction reduces to a \emph{(weak) 2-group bundle} which has been recently studied in several papers, see for instance \cite{Kapustin:2013uxa, Cordova:2018cvg, delcamp2018gauge, Benini:2018reh}. Topological gauge models built from 2-group connections can be found in \cite{Kapustin:2013uxa, Bullivant:2016clk, Cui:2016bmd, Bullivant:2017sjz, delcamp2018gauge, 2018arXiv180809394Z}. This construction can be even further generalized to so-called $k$-stage Postnikov towers (see app.~\ref{app:Postnikov}).

\subsection{Topological lattice (higher) gauge theories\label{sec:LGT}}
In order to build a lattice (higher) gauge theory which corresponds to a certain topological sigma model described above, we can proceed as follows: Let the target space of the sigma model be $X$ and let us endow the space-time manifold $\cM$ with a triangulation $\triangle$. For each non-vanishing homotopy group $\pi_{q_i}(X)= G_i$, we introduce a $G_i$-valued $q_i$-cochain on $\cM$. Locally, this amounts to labeling the $q_i$-simplices of the triangulation with elements in $G_i$. Furthermore, we introduce constraints on the labelings of the different simplices that are analogous to the cocycle conditions satisfied by the data representing a homotopy class of a map from $\cM$ to $X$. Labelings satisfying such constraints are referred to as $X$-\emph{colorings} of the triangulation $\cM$ and the set of all colorings is denoted by ${\rm Col}(\cM,X)$.\footnote{Actually this set has a monoidal structure which makes it a group or a generalization thereof.} Denoting a given coloring by $ g \in {\rm Col}(\cM,X)$, the partition function takes the form
\begin{equation}
	\mathcal{Z}_{\omega}^{X}[\cM] = \frac{1}{\mathcal N_{X}^{\triangle}} \sum_{g \in {\rm Col}(\cM,X)}\prod_{\triangle^{d+1}}e^{2 \pi i \mathcal{S}_\omega[g,\triangle^{d+1}]}
	\label{eq:lattice}
\end{equation}
where $\mathcal N_{X}^{\triangle}$ is a normalization constant and $\mathcal{S}_{\omega}[{g},\triangle^{d+1}]$ is the topological action whose value depends on the local data ${g}$ as well as a representative of the class $[\omega]\in H^{d+1}(X,\mathbb R/\mathbb Z)$.

\begin{example}[\emph{$q$-form lattice gauge theories}]
	Let us construct the lattice realization of a $q$-form topological gauge theory that corresponds to a topological sigma model with target space $B^qG$. 
	Flat $q$-form connections (dubbed flat $G_{[q]}$-connections) can have non-trivial $q$-holonomies along non contractible closed $q$-paths only. Therefore, a flat $G_{[q]}$-connection can be defined as a homomorphism from the $q$-th homotopy group $\pi_q(\cM)$ to $G$. Locally, this means that a flat $G_{[q]}$-connection is fully characterized by a $q$-cochain valued in $G$ satisfying $dg = \zo$, with $\zo \in G$ the unit element. In practice, we assign to every $q$-simplex $\triangle^q = (v_0 \ldots v_q) \subset \triangle$ a group element $g_{v_0 \ldots v_q} = \la g,(v_0 \ldots v_q) \ra$ such that for every ($q$+1)-simplex $\triangle^{q+1} = (v_0 \ldots v_{q+1}) \subset \triangle$, we impose the \emph{$q$-flatness condition}
	\begin{equation}
		\la dg, (v_0 \ldots v_{q+1})\ra = \sum_{i =0}^q(-1)^{i+1} g_{v_0 \ldots \hat{v}_i \ldots v_{q+1}} = \zo 
	\end{equation}
	where the notation $\hat{{\sss \bullet}}$ indicates that the corresponding vertex is omitted from the list. 
	Such a labeling is referred to as a $G_{[q]}$-coloring and the set of $G_{[q]}$-colorings is denoted by ${\rm Col}(\cM,G_{[q]})$. Note that a ($q$$-$1)-form gauge transformation is defined as a gauge parameter $\phi$ which acts on such colorings as 
	\begin{equation}
		\phi \triangleright g_{v_0 \ldots v_q} = g_{v_0 \ldots v_q} + \la d\phi , (v_0 \ldots v_q)\ra  \; .
	\end{equation}
	The topological action is provided by pulling back a class representative in a cohomology class $[\omega]\in H^{d+1}(G_{[q]},\mathbb R/\mathbb Z) \equiv H^{d+1}(B^qG,\mathbb R/\mathbb Z)$ and evaluating it on a choice of $G_{[q]}$-coloring $g \in {\rm{Col}}(\cM,G_{[q]})$. The partition function finally looks like 
	\begin{align}
		\boxed{
			\mathcal{Z}_{\omega}^{G_{[q]}}[\cM] =\frac{1}{|G|^{|\triangle^{0\to q-1}|}} \sum_{g \in {\rm{Col}}(\cM,G_{[q]})}\prod_{\triangle^{d+1}}e^{2 \pi i \mathcal{S}_\omega[g,\triangle^{d+1}]} }
		\label{eq:latticenform}
	\end{align} 
	where $|\triangle^{0 \to q-1}| := \sum_{i=0}^{q-1} |\triangle^{q-i}(\cM)|(-1)^{i}$ such that $|\triangle^i(\cM)|$ is the number of $i$-simplices in the triangulation $\triangle$ of $\cM$. 
\end{example}
\noindent
In the following sections, we focus our attention on 2-form topological gauge theories and their lattice realization as defined in the previous example. In particular, in sec.~\ref{sec:EM}, we will carefully build the cohomology group $H^{d+1}(B^qG,\mathbb R/\mathbb Z)$ for the case $q=2$ so as to provide a more explicit expression for \eqref{eq:latticenform} which can be used to construct a lattice Hamiltonian realization of this topological theory. As before, this lattice construction can be readily generalized to sigma models whose target space is provided by a Postnikov tower (see app.~\ref{app:Postnikov}).

\subsection{2-form topological action}
Let us explore in more detail topological sigma models that have a 2-form gauge theory interpretation. In particular, we wish to emphasize the role played by the classification of the relevant cohomology group in terms of quadratic forms. 

\bigskip \noindent
We explained above how 2-form topological gauge theories for a finite \emph{abelian} group $G$ can be built as topological sigma models with target space the second classifying space $X=B^2G$ of $G$. Homotopy classes of maps $[M,B^2G]$ can be labeled by $B\in H^{2}(\cM,G)$ and the homotopies of these maps are gauge transformations $B \sim B + d \phi$ where $\phi \in C^{1}(\cM,G)$. The partition function is provided by \eqref{eq:higher_DW_pf} which we repeat below:
\begin{align}
	\mathcal Z^{B^2G}_{\omega}[\cM]=\frac{1}{|G|^{b_1(\cM)-b_0(\cM)}}\sum_{B\in H^{2}(\cM,G)}e^{2\pi i \la \omega(B), \cM\ra} \; .
	\label{eq:2form_DW_pf}
\end{align}
Restricting to (3+1)d, the topological actions are classified by $[\omega] \in H^{4}(B^2G,\mathbb R/\mathbb Z)$. But, since $H_{3}(B^2G,\mathbb Z)=0$, we may write
\begin{align}
	H^{4}(B^2G,\mathbb R/\mathbb Z)=&\; \text{Hom}\left(H_{4}(B^2G),\mathbb R/\mathbb Z\right) \nonumber \\
	=&\;\text{Hom}\left(\Gamma(G),\mathbb R/\mathbb Z\right) \nonumber
\end{align}
where $\Gamma(G) \simeq H_{4}(B^2G,\mathbb Z)$ is known as the \emph{universal quadratic group} for $G$ \cite{eilenberg1953groups, eilenberg1954groups}. Before stating the defining property of $\Gamma(G)$, let us first recall the definition of a quadratic form:
\begin{definition}[\emph{Quadratic form}]
	A \emph{quadratic form} on a finite abelian group $G$ valued in $\mathbb R/\mathbb Z$ is a function $q:G \to \mathbb R/\mathbb Z$ such that $q(g) = q(-g)$ and 
	\begin{equation*}
		b:(g,h)\mapsto q(g)+ q(h)-q(g+h)
	\end{equation*}
	is bilinear, i.e. $b(g_1+g_2,h) = b(g_1,h)+b(g_2,h)$, $\forall g_1,g_2,h \in G$. 
\end{definition}

\noindent
Conversely, any lattice with a symmetric bilinear form $b$ defines a quadratic form via $q(x):=\frac{1}{2}b(x,x)$. Furthermore, it can be checked that the value of $b$ and $q$ on the generators of $G$ completely determine these forms. 

The universal quadratic group $\Gamma(G)$ is uniquely defined by the property that any quadratic function $q:G\to \mathbb R/\mathbb Z$ may be written as the composition ${q}=\widetilde{q}\circ \gamma$ where $\gamma: G\to \Gamma(G)$ and $\widetilde{q} \in \text{Hom}(\Gamma(G), \mathbb R/\mathbb Z)$. For instance, the universal quadratic group of $\mathbb{Z}_n$ is $\Gamma(\mathbb Z_{n})=\mathbb Z_{n}$ or $ \mathbb{Z}_{2n}$ for $n$ an odd integer or an even integer, respectively. The universal quadratic group of any finite abelian group of the form $G=\oplus_{I}\mathbb Z_{n_I} $ is then
\begin{align}
	\Gamma(G)=\Big[\bigoplus_{I}\Gamma(\mathbb Z_{n_I})\Big]\oplus \Big[ \bigoplus_{I<J}\mathbb Z_{\text{gcd}(n_I,n_J)} \Big] 
\end{align}
where ${\rm gcd}(n_I,n_J)$ is the \emph{greatest common divisor} of $n_I$ and $n_J$.
It was shown by Eilenberg and MacLane \cite{eilenberg1954groups} that the cohomology group $H^{4}(B^2G,\mathbb R/\mathbb Z)$ is isomorphic to the group of quadratic functions. Following the above discussion, the topological action in \eqref{eq:2form_DW_pf} can thus be defined as the composition of a canonical quadratic operation $\mathfrak P: H^{2}(\cM,G)\to H^{4}(\cM,\Gamma(G))$ known as the \emph{Pontrjagin square}, with a homomorphism $\widetilde{q}$ from $\Gamma(G)$ to $\mathbb R/\mathbb Z$, i.e. $\omega(B) \equiv \widetilde{q}_{*}\mathfrak P(B) \in H^{4}(\cM,\mathbb R/\mathbb Z)$ (see app.~\ref{sec:app_Pontr} and \cite{Freed:2009qp, kapustin2014topological, gaiotto2015generalized} for more details). 
The form of the topological action $\omega(B) \equiv \widetilde{q}_{*}\mathfrak P(B) \in H^{4}(\cM,\mathbb R/\mathbb Z)$ naturally depends on a choice of homomorphism $\widetilde{q}\in \text{Hom}(\Gamma(G),\mathbb R/\mathbb Z)$. Since the universal quadratic group for $\mathbb Z_n$ depends on whether $n$ is even or odd, without loss of generality let us write our gauge group $G=\bigoplus_{I}\mathbb Z_{n_I}$ such that $n_{I}$ is even if $I \leq K$ and odd for $I>K$. An element of $\mathbb{a}\in \Gamma(G)$  takes the form $\mathbb{a}\equiv \left\{a_{I},a_{IJ}\right\}$ where 
\begin{align}
	a_{I}\in&\;\left\{
	\begin{array}{@{}ll@{}}
	\left\{0,\dots, 2n_{I}-1\right\}, & \text{if}\ I\leq K \\
	\left\{ 0, \dots ,n_{I}-1 \right\}, & \text{if} \ I>K 
	\end{array}\right. \nonumber \\
	a_{IJ}\in&\; \left\{0,\dots , \text{gcd}(n_I,n_J)-1\right\} \; .
\end{align}
Similarly, a homomorphism $\widetilde{q}\in \text{Hom}(\Gamma(G),\mathbb R/\mathbb Z) \simeq \Gamma(G)$ is prescribed by $\left\{p_{I},p_{IJ}\right\}$
\begin{align}
	p_{I}\in&\;\left\{
	\begin{array}{@{}ll@{}}
	\left\{0,\dots, 2n_{I}-1\right\}, & \text{if}\ I\leq K \\
	\left\{ 0, \dots ,n_{I}-1 \right\}, & \text{if} \ I>K 
	\end{array}\right. \nonumber \\
	p_{IJ}\in&\; \left\{0,\dots , \text{gcd}(n_I,n_J)-1\right\} 
\end{align}
via the map 
\begin{align}
	\widetilde{q}(\mathbb{a})=\sum_{I\leq K}\frac{p_{I}a_I}{2n_I} + \sum_{I>K}\frac{p_I a_{I}}{n_I} + \sum_{I<J} \frac{p_{IJ} a_{IJ}}{\text{gcd}(n_I,n_J)} \; .
\end{align}
Then, for a field configuration $B^{I}\in Z^{2}(\cM,\mathbb Z_{n_I})$ the action takes the form 
\begin{align}
	\mathcal{S}_p[{B^I},\cM]=&\; 2\pi i\int_{\cM}\widetilde{q}_{*}\mathfrak P(\sum_{I}B^{I}) \nonumber \\
	=&\;\sum_{I\leq K} \frac{2\pi ip_{I}}{2n_{I}} \int_{\cM} \mathfrak P(B^{I}) + 
	\sum_{I>K} \frac{2\pi ip_{I}}{n_{I}} \int_{\cM} \mathfrak P(B^{I}) + \sum_{I < J} \frac{2\pi i p_{IJ}}{\text{gcd}(n_I,n_J)}\int_{\cM} B^{I} \smilo B^{J} \; .
	\label{eq:2form_action}
\end{align}
It can be checked that $\mathfrak P(B^{I}+d\lambda^I)- \mathfrak P(B^{I})\myeq 0 \; (\text{mod} \ n_{I} \ \text{or} \ 2n_I)$ when $n_{I}$ is odd or even, respectively. There is also a 2-form global symmetry $B^{I} \mapsto B^{I} +\beta^{I}$ where $\beta^{I} \in Z^{2}(\cM,\mathbb Z_{n_{I}})$ and $\text{Sq}^{2}(\beta^{I})=0$.\footnote{$\text{Sq}^2: H^{2}(\cM,\mathbb Z_{n_I})\to H^{4}(\cM,\mathbb Z_{n_I})$ as $[\beta^{I}]\to [\beta^{I}] \smilo [\beta^{I}]$. Therefore we need to impose that $[\beta^I] \smilo [\beta^I] =0\in H^{4}(\cM,\mathbb Z_{n_I}) $ so that the action is invariant under the global symmetry transformation.}
The partition function for the above topological gauge theory was computed in \cite{Freed:2009qp, gaiotto2015generalized} for the case where $\cM$ has vanishing torsion in all its homology groups.\footnote{When $\cM$ has non-vanishing torsion, the space of $\mathbb Z_n$-bundles $H^{2}(\cM,\mathbb Z_{n})$ fits within the exact sequence \cite{freed2007heisenberg}
	\begin{align}
		0\to H^{2}(\cM,\mathbb Z)\otimes \mathbb Z_n \to H^{2}(\cM,\mathbb Z_{n}) \to \text{Tor}\left(H^{3}(\cM,\mathbb Z)\right)\otimes \mathbb Z_{n}\to 0 
	\end{align}
	where $\text{Tor}$ refers to the torsion subgroup. If we consider the simpler case of isomorphism classes of 1-form $\mathbb Z_{n}$-bundles that fit in the sequence
	\begin{align}
		0\to H^{1}(\cM,\mathbb Z)\otimes \mathbb Z_n \to H^{1}(\cM,\mathbb Z_{n}) \xrightarrow{\mathfrak b} \text{Tor}\left(H^{2}(\cM,\mathbb Z)\right)\otimes \mathbb Z_{n}\to 0
	\end{align}
	where $\mathfrak b$ is the \emph{Bockstein map}, we may evaluate the topological action $\frac{1}{n}\int_{\cM}A\smilo dA$ on a manifold with non-vanishing torsion such as the lens space $L(n,1)$ for which $H_{1}(L(n,1),\mathbb Z)=\mathbb Z_{n}$. Then, one has
	\begin{align}
		\frac{1}{n}\int_{L(n,1)}A\smilo d_{\dr} A=&\; \frac{1}{n}\int_{\text{p.d}([d_{\dr}A] \in H^2(\cM,\mathbb Z))}A  
		= \frac{\ell}{n}\int_{[L]\in H_1(\cM,\mathbb Z)}A= \frac{\ell^2}{n}
	\end{align}
	where we have used the fact that the generator of the first homology group is the \emph{Poincar\'e dual} (p.d) to the generator of the integer cohomology $H^{2}(\cM,\mathbb Z)$. We assume the configuration where $\text{p.d}[dA]=\ell [L]$ where $\ell\in [0,n-1]\cap \mathbb Z$. It is not clear to us what the equivalent statement for higher cup products is. Such a duality would be needed to compute $\int_{\cM}B\smilo_{\! 1} d_{\dr} B$ on some general manifold.} 
In that case, we may write 
\begin{equation}	
	B^{I}=\sum_{a=1}^{b_2(\cM)}\frac{b^{I}_{a}h_{a}}{n_I} 
\end{equation}
where $h_{a}$ is a basis element in $H^{2}(\cM,\mathbb Z)$. The topological action evaluates to 
\begin{align}
	\mathcal{S}_p[\vec{b}^I,\cM]=&\; \sum_{I\leq K}\frac{\pi i p_I (b^I)^{\top}\, \mathbb I \, b^{I}}{n_I}  + \sum_{I>K}\frac{2\pi i p_I (b^I)^{\top}\, \mathbb I \, b^{I}}{n_I}  
	+\sum_{I<J}\frac{2\pi i p_{IJ} (b^I)^{\top} \, \mathbb I \, b^J}{\text{gcd}(n_I,n_J)} 
\end{align}
where $\vec{b}^{I}\in (\mathbb Z/2n_I\mathbb Z)^{b_{2}(\cM)}$ when $n_I$ is even and $\vec{b}^{I}\in (\mathbb Z/n_I\mathbb Z)^{b_{2}(\cM)} $ when $n_I$ is odd. The object $\mathbb{I}$ defined as $(\mathbb I)_{ab}=\int_{\cM}h_a\smilo h_{b} $ is the intersection pairing in $H^{2}(\cM,\mathbb Z)$. The partition function finally reads
\begin{align}
	\mathcal Z_p^{B^2G}[\cM]=\frac{1}{(\prod_{I}n_I)^{b_{1}(\cM)-b_0(\cM))}}\sum_{\vec{b}^{I}}e^{\mathcal{S}_p[\vec{b}^I, \cM]}
	\label{pf_BcupB} \; .
\end{align}
 The topological theories defined so far are all constructed from finite groups. As such, these theories are naturally defined in the discrete on a lattice. However, it is often desirable to have a continuous formulation of a theory. Such a formulation, if it exists, may give one access to powerful and sometimes familiar tools of quantum field theory. It turns out that topological gauge theories for finite \emph{abelian} groups can be naturally embedded into continuous toric gauge theories. The simplest example of this statement is the $\mathbb Z_{2}$ topological gauge theory in ($d$+1)-dimensions, or equivalently the $G=\mathbb Z_2$ Dijkgraaf-Witten theory with a trivial cohomology class in $H^{d+1}(BG,\mathbb R/\mathbb  Z)$.\footnote{In $2+1$ dimensions, this is described by the familiar toric code Hamiltonian \cite{Kitaev:1997wr}.} The continuous topological gauge theory that embeds $\mathbb Z_{2}$ gauge theory is the BF theory described by the action
\begin{align}
	\mathcal{S}[A,B,\cM]=4\pi i\int_{\cM}B \wedge d_{\dr} A
\end{align} 
where $A$ is a 1-form $\rU(1)$ gauge field, $B$ is a ($d$$-$1)-form $\rU(1)$ gauge field, and $d_{\dr}$ is the usual exterior derivative on differential forms. One obtains the $\mathbb Z_{2}$ gauge theory by simply integrating over $B$ in the path integral. Indeed, integrating over the globally defined field configurations imposes that $d_{\dr}A=0$, i.e $A$ is a locally flat $\rU(1)$ connection while summing over the topological sectors (monopole configurations) of $B$  imposes that the holonomies of $A$ are $\mathbb Z_{2}$ quantized. This makes $A$ a $\mathbb Z_{2}$ gauge field and reduces the BF theory to a cohomologically trivial $\mathbb Z_2$ gauge theory. 

Such formulations of (3+1)-dimensional Dijkgraaf-Witten theories in terms of (muli-component coupled) BF theories have been studied at length in recent years \cite{Wang:2014pma, Tiwari:2016zru, Tiwari:2017wqf, putrov2016braiding, wang2018tunneling}. Next we discuss embedding the above finite gauge theory into a continuous topological gauge theory built from toric $\rU(1)$ 1-form and 2-form gauge fields. See for example \cite{kapustin2014coupling, gaiotto2015generalized} for earlier works studying this theory. For the above parameters $\left\{p_{I}, p_{IJ}\right\}$, the continuous action takes the form 
\begin{align}
	\label{eq:action_BwedgeB}
	&\mathcal{S}_p[A^{I}, B^{I},\cM] \\ \nn
	& \q \!\! =
	2\pi i \!\!
	\int_{\cM} \!\! \Big(n_I\delta_{IJ}B^{I}\wedge d_{\dr}\! A^{J} +\sum_{I\leq K}\frac{p_{I}n_{I}}{2} B^{I} \wedge B^{I} + 	\sum_{I>K}p_{I}n_IB^{I}\wedge B^{I} + \sum_{I<J}p_{IJ}\text{lcm}(n_I,n_J) B^{I}\wedge B^{J}\Big) 
\end{align}
where $\text{lcm}(n_I,n_J)$ is the \emph{lowest common multiple} of $n_I$ and $n_J$. The partition function evaluated for \eqref{eq:action_BwedgeB} matches with \eqref{pf_BcupB}. This can be shown quite explicitly, at least for manifolds with vanishing torsion: Integrating over $A^{I}$ enforces $B^{I}$ to be flat with holonomies on closed non-contractible surfaces restricted to integer multiples of $1 / N_I$. In other words, $B^{I}\in \text{Hom}(H_{2}(\cM,\mathbb Z),Z_{N_I})$ which is simply a flat 2-form $\mathbb Z_{N_I}$-bundle. But this continuous formulation of the 2-form gauge theory has an interesting gauge structure due to the presence of the cohomological twist. The conserved charges (or Gau{\ss} operators) that generate the gauge transformations take the form
\begin{alignat}{2}
	\mathcal Q_{B^I}& = 2\pi n_1\Big(d_{\dr} A^{I}+p_I B^{I}+ \sum_{J}\frac{p_{IJ}\text{lcm}(n_I,n_J)}{n_J} B^J\Big) \; , & \q & \;\;  I \leq K \nn \\ \nn
	\mathcal Q_{B^I}& = 2\pi n_1\Big(d_{\dr} A^{I}+2p_I B^{I}+ \sum_{J}\frac{p_{IJ}\text{lcm}(n_I,n_J)}{n_J} B^J\Big) \; , & \q & \;\; I > K \\
		\mathcal Q_{A^I}& = 2\pi n_{I}d_{\dr} B^{I}.
		\label{eq:bb_gauss}
\end{alignat}
These charges generate the non-standard $\rU(1)$ 0-form and 1-form gauge transformations 
\begin{alignat}{2}
	A^{I} &\to A^{I}+d_{\dr}\lambda^I-p_I\theta^I-\sum_{J}\frac{{p}_{IJ}\text{lcm}(n_{I},n_{J})}{n_{I}}\theta^J \; , & \q & \;\; I\leq K \nn \\
	A^{I}&\to A^{I}+d_{\dr}\lambda^I-2p_I\theta^I-\sum_{J}\frac{{p}_{IJ}\text{lcm}(n_{I},n_{J})}{n_{I}}\theta^J
	\; , & \q & \;\;  I >K \nn \\
	B^{I}& \to B^{I} +d_{\dr}\theta^{I} 
	\label{eq:strict_gt}
\end{alignat}
where $\lambda^{I}$ are circle-valued scalars and $\theta^{I}$ are 1-form fields. Both these gauge transformations have quantized periods, i.e $d \lambda^{I} \in \Omega^{1}_{\mathbb Z}(\cM)$ and $d\theta^{I}\in \Omega_{\mathbb Z}^2(\cM)$.\footnote{We use the notation $\Omega_{\mathbb Z}^{p}(\cM)$ to denote the space of $q$-forms with integer periods on any $p$-cycle $\mathfrak{L}^{(p)}\in Z_{p}(\cM,\mathbb Z)$ i.e for some $\xi\in \Omega^{p}_{\mathbb Z}(\cM)$, $\oint_{\mathfrak{L}^{(p)}}\xi \in  \mathbb Z$.} Hence we see that embedding the discrete 2-form theory into a continuous theory indeed has a non-trivial effect on the gauge structure. The 1-form and 2-form fields no longer transform independently under gauge transformations. This is due to the fact that although the canonical commutation relations of the theory \eqref{eq:action_BwedgeB} are the usual BF type-commutation relations, the charge operators are modified and consequently the gauge transformations are modified as well. We may write the constraints \eqref{eq:bb_gauss} as $d A^{I}+\left[t(B)\right]^{I}=0$ where $t\in \text{Hom}(\rU(1)^N,\rU(1)^N)\simeq {\rm GL}(N,\mathbb Z)$ is parametrized by $\left\{p_{I}, p_{IJ}\right\}$ and $N$ is the number of flavor fields ($I=1,\dots, N$). Putting all this together we realize that \eqref{eq:action_BwedgeB} actually describes a gauge theory built from a \emph{strict 2-group} rather than ordinary groups. A strict 2-group $\mathcal G$ is built from four pieces of data $\mathcal G=\left\{G,H,t,\triangleright \right\}$ where $G,H$ are groups ($H$ is necessarily abelian), $t\in \text{Hom}(H,G)$ and $\triangleright: G\to \text{Aut}(H)$. The gauge transformations of a strict toric 2-group have exactly the form \eqref{eq:strict_gt}. Hence we realize a non-trivial fact that the partition functions for topological gauge theories, one a toric strict 2-group theory and the other a finite 2-form theory are dual to one another.

Numerous properties of the topological action \eqref{eq:action_BwedgeB} are reviewed in app.~\ref{sec:app_quant}. In the next section, we study in detail the configuration space of 1-form and 2-form $\rU(1)$ gauge theories before returning to the discussion of strict 2-group bundles in sec.~\ref{sec:strict} that embeds a finite group bundle and also encodes the non-trivial cohomological twist.

\section{Deligne-Beilinson cohomology and higher gauge theory\label{sec:DB}}

In this section, we describe the configuration space of \emph{twisted} 2-form gauge theory for a finite abelian group $G$. As described above such 2-form gauge theories can be embedded into $\rU(1)$ gauge theories that involve both 1-form and 2-form $\rU(1)$ gauge fields. However, these different fields transform under gauge transformations in an unconventional way. In order to have a better understanding of this formulation, it is necessary to have a systematic understanding of the configuration space of gauge inequivalent configurations. Here we present such an understanding using the technology of \emph{Deligne-Beilinson} (DB) cohomology \cite{deligne1971theorie}. An alternative approach is provided by Cheeger-Simons differential cohomology \cite{cheeger1985differential, hopkins2002quadratic, freed2007heisenberg} that may be employed to systematize the configuration space of $q$-form $\rU(1)$ gauge theory. The two approaches of DB cohomology and Cheeger-Simons differential cohomology are equivalent \cite{simons2008axiomatic} however in this work we stick to the former. In order to be self-consistent we begin by assembling the necessary ingredients to describe $q$-form $\rU(1)$ connections using DB cohomology \cite{Thuillier:2015vma, Mathieu:2015mda}.

\subsection{Preliminaries and definitions}
Let us briefly revisit the physical understanding of a 1-form $\rU(1)$ connection. Locally a 1-form connection $A$ is simply a 1-form field. There is an equivalence relation related to gauge transformations which are redundancies of the physical description. These gauge transformations act as $A\to A+ d\lambda$ where $d\lambda\in \Omega^{1}_{\mathbb Z}(\cM)$. Hence the gauge invariant information is encoded in holonomies
\begin{align}
	\text{hol}_A(\mathfrak{L}^{(1)}):=\oint_{\mathfrak{L}^{(1)}}A \q (\text{mod}  \ \mathbb Z) \; ,
\end{align}
or equivalently in Wilson operators $W^{\ec}(\mathfrak{L}^{(1)}):= \exp\big\{2\pi i\ec \oint_{\mathfrak{L}^{(1)}}A\big\}$ where $\mathfrak{L}^{(1)}$ is a 1-cycle on $\cM$. Furthermore, for topologically non-trivial bundles, i.e those with non-vanishing \emph{Chern number}, there is no globally defined 1-form connection. Instead, one has to work with a field strength $F \in \Omega^{2}_{ \mathbb Z}(\cM)$. On contractible patches, the field strength and holonomies agree via
\begin{align}
	\text{hol}_{A}(\mathfrak{L}^{(1)})=\int_{\partial^{-1}\mathfrak{L}^{(1)}}F  \q (\text{mod} \ \mathbb Z)
\end{align}
where $\partial^{-1}\mathfrak{L}^{(1)}$ is a surface that bounds $\mathfrak{L}^{(1)}$. We shall now see that all this data fits neatly together into the Deligne-Beilinson cohomology group.  In order to do so, we need first to introduce the basic notions of  oriented open cover, \v{C}ech-de Rham bicomplex and polyhedral decomposition:

\begin{definition}[\emph{Oriented and ordered open cover}] Let $\cM$ be a closed smooth and oriented manifold defined with an open cover $\mathcal U=\{U_{i}\}_{i \in \mathcal I}$ such that $\bigcup_{i \in \mathcal I}U_{i}= \cM$. We denote overlaps of sets as 
	\begin{align}
		U_{i_0i_{1}}=&\; U_{i_{0}}\cap U_{i_{1}} \nn \\
		U_{i_0i_{1}i_{2}}=&\; U_{i_{0}}\cap U_{i_{1}}\cap U_{i_2} 
		\nn \\
		\vdots&\; \nn \\ 
		U_{i_0i_{1}i_{2}\ldots i_p}=&\; U_{i_{0}}\cap U_{i_{1}}\cap U_{i_2}\cdots \cap U_{i_p} \; .
	\end{align}
	The index of $U_{i_0i_{1}i_{2}\ldots i_p}$ is referred to as the \emph{\v{C}ech index} of this intersection and $p\in \mathbb Z$ as the \emph{\v{C}ech degree}. We only consider overlaps whose indices are ordered i.e $i_0<i_1 <\cdots < i_p$ and refer to $\mathcal U$ as an ordered cover of $\cM$. Let the collection of all non-vanishing overlaps of ordered ($p$+1)-open sets be denoted by $\mathbb U_p$. Since $\cM$ is compact, the cardinality of $\mathbb U_p$ and of $\mathcal U$ is finite.
\end{definition}
\noindent We denote by $\Omega^{r}(\mathbb U_{p})$ the space of de Rham $r$-forms assigned to all elements in $\mathbb U_{p}$ and $\mu^{r}_{p}\in \Omega^r(\mathbb U_p)$ a generic element. The quantity $n_{p}\in \text{Map}(\mathbb U_{p},\mathbb Z)=: \Omega^{-1}(\mathbb U_{p})$ denotes an assignment of integers to all elements of $\mathbb U_{p}$. One can define two independent differential operators that act on $\mu^{r}_{p}$, namely the de Rham differential $d_{\dr}^{r}$ and the \v{C}ech differential $d_{p}$ 
\begin{align}
	d_{\dr}^{r}:&\; \Omega^{r}(\mathbb U_{p})\to \Omega^{r+1}(\mathbb U_{p}) \nn \\
	d_p:&\; \Omega^{r}(\mathbb U_{p})\to \Omega^{r}(\mathbb U_{p+1}) 
\end{align}
that satisfy the properties $d_{\dr}^{r+1}\circ d_{\dr}^r=0$ and $d_{p+1}\circ d_{p}=0$. The action of $d_{\dr}^{r}$ is simply given by the exterior derivative that acts locally on each open set, while the \v{C}ech differential acts as
\begin{align}
	\left(d_{p}\mu^{r}_{p}\right)_{i_0i_{1}\dots i_{p+1}}=\sum_{j=0}^{p+1}(-1)^{j}\left(\mu^{r}_{p}\right)_{i_{0}\dots\hat{i}_{j}\dots i_{p+1}}  \in \Omega^{r}(\mathbb U_{p+1}) \; .
\end{align}
The \emph{\v{C}ech-de-Rham bicomplex} is a bicomplex of cochains $\Omega^{r}(\mathbb U_p)$ labeled by two indices $r$ and $p$ which are the de Rham and \v{C}ech degrees, respectively. The maps between cochains are provided by $d_{\dr}^{r}$ and $d_{p}$ as described above. Furthermore, we define a completion of the de Rham complex via the differential $d_{\dr}^{-1}: \Omega^{-1}(\mathbb U_{p})\to \Omega^{0}(\mathbb U_{p})$  where $d_{\dr}^{-1}$ is simply the injection of integers into the space of (constant) functions.

\medskip \noindent Let $Z_{p}(\cM,\mathbb Z)$ denote the space of oriented $p$-cycles in $\cM$. In order to integrate $p$-cochains on $\cM$ over $p$-cycles, we need to introduce the notion of polyhedral decomposition:
\begin{figure}[t]
	\centering
	\begin{tikzpicture}
	\node[inner sep=0pt] (a) at (0,0)
	{\includegraphics[scale=1.1]{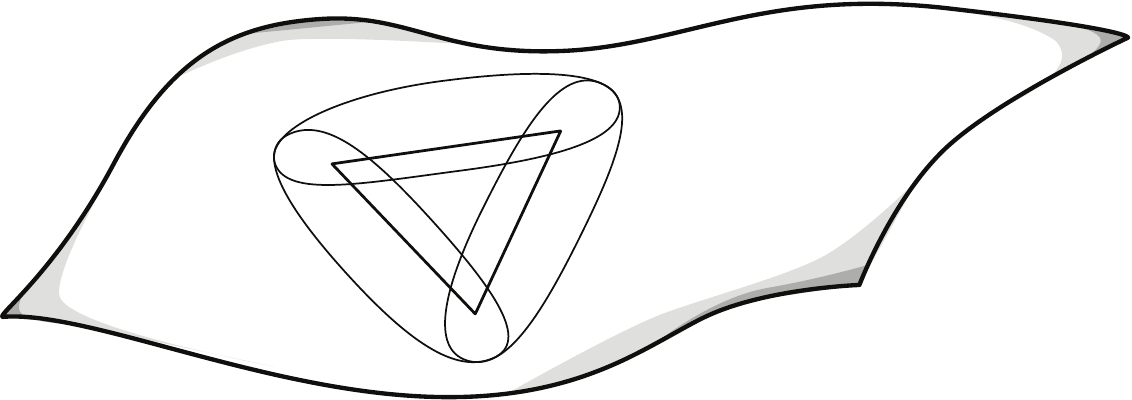}};
	\node[above] at (-1,-1.85) {\small{$\mathfrak l^{(0)}_{12}$}};
	\node[above] at (0.1,0.65) {\small{$\mathfrak l^{(0)}_{23}$}};
	\node[above] at (-2.9,0.15) {\small{$\mathfrak l^{(0)}_{13}$}};
	\node[above] at (-1,-2.25) {\small{$U_{12}$}};
	\node[above] at (0.85,1) {\small{$U_{23}$}};
	\node[above] at (-3.6,0.3) {\small{$U_{13}$}};
	\node[above] at (-2.1,-0.95) {\small{$\mathfrak l^{(1)}_{1}$}};
	\node[above] at (-0.2,-0.6) {\small{$\mathfrak l^{(1)}_{2}$}};
	\node[above] at (-1.4,0.5) {\small{$\mathfrak l^{(1)}_{3}$}};
	\node[above] at (-2.8,-1.1) {\small{$U_{1}$}};
	\node[above] at (0.33,-0.9) {\small{$U_{2}$}};
	\node[above] at (-1.5,1.3) {\small{$U_{3}$}};
	\end{tikzpicture}
	\caption{ Polyhedral decomposition of a 1-cycle $\mathfrak L^{(1)}=\Sigma_{i=1}^{3}\mathfrak l^{(1)}_i$ subordinate to a choice of open cover $\mathcal U=\bigcup_{i=1,2,3}U_{i}$. The 1-chains $\mathfrak l^{(1)}_i\in U_{i}$ and 0-chains $\mathfrak l^{(0)}_{ij}\in U_{ij}$. 
	}
	\label{fig:poly_1_cycle}
\end{figure}
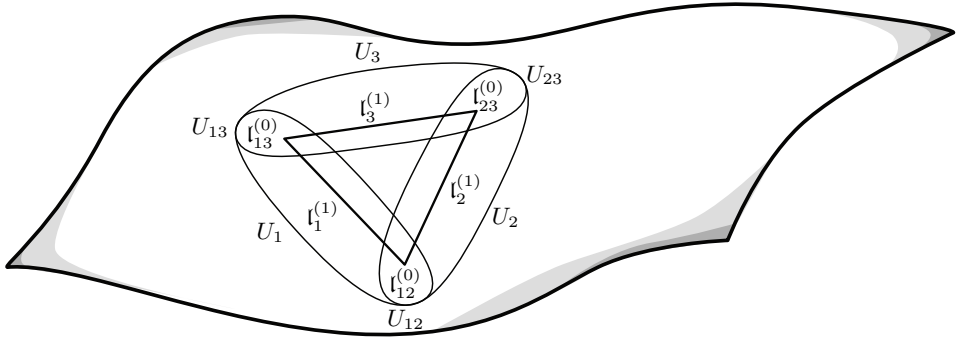
\begin{definition}[\emph{Polyhedral decomposition}] 
	Let $\mathfrak L^{(p)}$ be a $p$-cycle, then a polyhedral decomposition of $\mathfrak L^{(p)}$ subordinate to a given open cover is given by decomposing $\mathfrak L^{(p)}=\sum_{i_0}\mathfrak l^{(p)}_{i_0}$ such that $\mathfrak L^{(p)}_{i_0}\subset U_{i_0}$. We define a boundary map $\partial$ whose action reads
	\begin{align}
		\partial \mathfrak l^{(p)}_{i_0}=\sum_{i_1} \mathfrak l^{(p-1)}_{i_{1}i_0}- \mathfrak l^{(p-1)}_{i_{0}i_1}
	\end{align}
	where $\mathfrak l^{(p-1)}_{i_{0}i_1}\subset U_{i_0i_1}$. The boundary operator further acts as
	\begin{align}
		\partial \mathfrak l^{(p-1)}_{i_0i_1}=\sum_{i_2}\left[\mathfrak l^{(p-2)}_{i_2i_0i_1} -\mathfrak l^{(p-2)}_{i_0i_2i_1} +\mathfrak l^{(p-2)}_{i_0i_1i_2}\right]
	\end{align}
	where $\mathfrak l^{(p-2)}_{i_0i_1i_2}\subset U_{i_0i_1i_2}$. This process is iterative and after $k$ iterations, we obtain
	\begin{align}
		\partial \mathfrak l^{(p-k)}_{i_0i_{1}\dots i_{k}}=\sum_{j=1}^{k-1}\mathfrak l^{(p-k-1)}_{i_{0}i_1\dots i_{j-1}i_j i_{j+1}\dots i_k}+\mathfrak l^{(p-k-1)}_{i_{k+1}i_0i_1\dots i_k}+ \mathfrak l^{(p-k-1)}_{i_0i_1\dots i_ki_{k+1}}
	\end{align}
	where as before $\mathfrak l^{(p-k-1)}_{i_0i_{1}\dots i_{k+1}}\subset U_{i_0i_{1}\dots i_{k+1}}$. Note that some of the entries in this sum vanish (e.g. $\mathfrak l^{(p-1)}_{i_1i_0}$, $\mathfrak l^{(p-2)}_{i_0i_2i_1}$) since we only consider an \emph{ordered} cover. 
\end{definition}
\noindent
In the following, we work with four-manifolds, therefore we do not need to iterate this procedure defined above more than four times. It is important to note that it is always possible to find a good open cover with respect to which a given $p$-cycle admits a polyhedral decomposition. Let us consider a few simple examples to illustrate the previous definition:
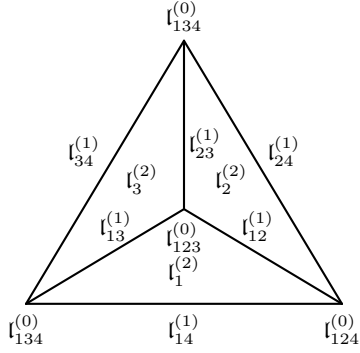
\begin{figure}[bt]
	\centering
	\begin{tikzpicture}[scale=0.7,baseline=0em, line width=0.8pt]
	\coordinate (a) at (0,-0.2);
	\draw[] (-3,-2) -- (a);
	\draw[] (3,-2) -- (a);
	\draw[] (0,3) -- (a);
	\draw[] (-3,-2) -- (3,-2);
	\draw[] (-3,-2) -- (0,3);
	\draw[] (3,-2) -- (0,3);
	\node[below] at (-3,-2) {\small{$\mathfrak{l}^{(0)}_{134}$}};
	\node[below] at (3,-2) {\small{$\mathfrak{l}^{(0)}_{124}$}};
	\node[above] at (0,3) {\small{$\mathfrak{l}^{(0)}_{134}$}};
	\node[below, yshift=-0.1em] at (0,-0.2) {\small{$\mathfrak{l}^{(0)}_{123}$}};
	\node[below] at (0,-2) {\small{$\mathfrak{l}^{(1)}_{14}$}};
	\node[right] at (1.4,0.9) {\small{$\mathfrak{l}^{(1)}_{24}$}};
	\node[left] at (-1.4,0.9) {\small{$\mathfrak{l}^{(1)}_{34}$}};
	\node[right, xshift=-0.2em, yshift=0.1em] at (0,1) {\small{$\mathfrak{l}^{(1)}_{23}$}};
	\node[below] at (0,-0.9) {\small{$\mathfrak{l}^{(2)}_{1}$}};
	\node[above] at (1.4,-1) {\small{$\mathfrak{l}^{(1)}_{12}$}};
	\node[above] at (-1.3,-1) {\small{$\mathfrak{l}^{(1)}_{13}$}};
	\node[above] at (0.9,-0.1) {\small{$\mathfrak{l}^{(2)}_{2}$}};
	\node[above] at (-0.8,-0.1) {\small{$\mathfrak{l}^{(2)}_{3}$}};
	\end{tikzpicture}
	\caption{ Polyhedral decomposition of a 2-cycle $\mathfrak L^{(2)}=\Sigma_{i=1}^{4}\mathfrak l^{(1)}_i$ subordinate to a choice of open cover $\mathcal U=\bigcup_{i=1,2,3,4}U_{i}$. The open cover has not been illustrated in the figure above to avoid clutter but it is such that the 2-chains $\mathfrak l^{(1)}_{i}\in U_{i}$, the 1-chains $\mathfrak l^{(2)}_{ij}\in U_{ij}$ and the 0-chains $\mathfrak l^{(0)}_{ijk}\in U_{ijk}$. 
	}
	\label{fig:poly_2_cycle}
\end{figure}
\begin{example}
	Let $\mathfrak L^{(1)}\in Z_{1}(\cM,\mathbb Z)$ be a given 1-cycle as shown in fig.~\ref{fig:poly_1_cycle}. The polyhedral decomposition of $\mathfrak L^{(1)}$ can be fixed for a given open cover $\mathcal U=\{ U_{i} \}_i$. We write $\mathfrak L^{(1)}=\mathfrak l^{(1)}_1+\mathfrak l^{(1)}_2+\mathfrak l^{(1)}_3$ where $\mathfrak l^{(1)}_i \in U_{i}$. The boundary operator acts as
	\begin{align}
		\partial \mathfrak L^{(1)}=&\; \partial \mathfrak l^{(1)}_{1} +\partial \mathfrak l^{(1)}_{2}+ \partial \mathfrak l^{(1)}_{3 } \nn \\
		=&\; (\mathfrak l^{(0)}_{21}-\mathfrak l^{(0)}_{12}+\mathfrak l^{(0)}_{31}-\mathfrak l^{(0)}_{13}) + (\mathfrak l^{(0)}_{12}-\mathfrak l^{(0)}_{21}+\mathfrak l^{(0)}_{32}-\mathfrak l^{(0)}_{23}) + (\mathfrak l^{(0)}_{23}-\mathfrak l^{(0)}_{32} +\mathfrak l^{(0)}_{13}-\mathfrak l^{(0)}_{31}) \nn \\
		=&\; (-\mathfrak l^{(0)}_{12}-\mathfrak l^{(0)}_{13}) + (\mathfrak l^{(0)}_{12}-\mathfrak l^{(0)}_{23})+(\mathfrak l^{(0)}_{23}+\mathfrak l^{(0)}_{13}) \nn \\
		=&\; 0 \; .
	\end{align} 
	In the third equality, we used the fact that we are working with an ordered cover, therefore 0-chains of the form $\mathfrak l^{(0)}_{ij}$ where $j<i$ vanish.
\end{example}
\begin{example}
	Let $\mathfrak L^{(2)}\in Z_{2}(\cM,\mathbb Z)$ be a 2-cycle whose polyhedral decomposition is illustrated in fig.~\ref{fig:poly_2_cycle}. Then $\mathfrak L^{(2)}=\sum_{i=1}^{4}\mathfrak l^{(2)}_{i}$ and $\partial \mathfrak L^{(2)}=\sum_{i=1}^{4}\partial \mathfrak l^{(2)}_{i}$, where for instance
	\begin{align}
		\partial \mathfrak l^{(2)}_{1}=&\;(\mathfrak l^{(1)}_{21}-\mathfrak l^{(1)}_{12})+(\mathfrak l^{(1)}_{31}-\mathfrak l^{(1)}_{13})+(\mathfrak l^{(1)}_{41}-\mathfrak l^{(1)}_{14}) \nn \\
		=&\; -\mathfrak l^{(1)}_{12} -\mathfrak l^{(1)}_{13}-\mathfrak l^{(1)}_{14} \; .
	\end{align} 
	It is easy to check that $\partial \mathfrak L^{(2)}=0$ as it should be.
\end{example}
\noindent
We now have all the ingredients to introduce the \v{C}ech-de Rham construction of Deligne-Beilinson (DB) cohomology:  
\begin{definition}[\emph{Deligne-Beilinson cohomology}] We call a DB $q$-cochain a ($q$+2)-tuple of data of the form:
	\begin{align}
		(\mu^{q}_{0}, \mu^{q-1}_{1}, \ldots, \mu^{0}_{q}, n_{q+1} ) \in 
		\Omega^{q}(\mathbb U_{0})\times \Omega^{q-1}(\mathbb U_{1}) \times  \cdots \times \Omega^{0}(\mathbb U_{q}) \times \Omega^{-1}(\mathbb U_{q+1}) 
	\end{align}
	and denote the space of DB $q$-cochains by $C^{q}_{\rm DB}(\cM,\mathbb Z)$. We define two  differential operators $D_{(q-1,q)}: C^{q-1}(\cM,\mathbb Z)\to C^{q}(\cM,\mathbb Z)$ and $D_{(q,q)}: C^{q}(\cM,\mathbb Z)\to C^{q+1}(\cM,\mathbb Z)$ via
	\begin{align}
		D_{(q-1,q)}:=&\;(d_0 + d_{\dr}^{q-1})-(d_{1}+d_{\dr}^{q-2})+ \dots + (-1)^{q}(d_{q}+d_{\dr}^{-1}) \nn \\
		=&\; \sum_{i=0}^{q}(-1)^{i}(d_{i}+d_{\dr}^{q-1-i}) \\
		D_{(q,q)}=&\;(d_0+0)-(d_{1}+d_{\dr}^{q-1})+\dots + (-1)^{q+1}(d_{q+1}+d_{\dr}^{-1}) \nn \\
		=&\; d_{0} +\sum_{i=1}^{q+1}(-1)^{i}(d_{i}+d_{\dr}^{q-i}) 
	\end{align}
	where the first index in the subscript is meant to denote the degree of DB cochain that the given codifferential operator acts on, and the second index denotes the maximum de Rham degree in the image of the given operator. It can easily be checked that $D_{(q,q)}\circ D_{(q-1,q)}=0$. A DB $q$-cocycle is defined as a DB $q$-cochain in the kernel of the operator $D_{(q,q)}$, while a DB $q$-coboundary is a $q$-cochain in the image of $D_{(q-1,q)}$. We may then define the $q$-th Deligne-Beilinson cohomolgy as the following quotient 
	\begin{align}
		H^{q}_{\rm DB}(\cM,\mathbb Z)=\frac{\text{ker}(D_{(q,q)})}{ \text{im}(D_{(q-1,q)})} \; .
	\end{align}
\end{definition}
\noindent 	The DB cohomology as defined above has degree one lower than corresponding differential cohomology defined for example in \cite{ hopkins2002quadratic, freed2007heisenberg}. Here, we follow the conventions of \cite{Thuillier:2015vma, Mathieu:2015mda}.

\subsection{Configuration space for $q$-form $\rU(1)$ connections}

We defined above the Deligne-Beilinson cohomology of cochains on a \v{C}ech-de-Rham bicomplex. We will now use this technology in order to define the configuration space of $q$-form connections. Below, we illustrate this construction with a couple of examples of $\rU(1)$ connections at low form degree and check that they are indeed described by DB cohomology classes. But, before getting to this we provide some intuition about why this somewhat intricately defined cohomology group is isomorphic to the space of gauge inequivalent configurations of $\rU(1)$ fields.

 A $q$-form $\rU(1)$ connection is usually defined by specifying $q$-forms on open sets. However, for topologically non-trivial bundles, it is not possible to describe a connection via a globally defined $q$-form, in which case one works with a covering of open sets with representatives of the connection defined locally as $q$-forms on each of the open sets. On overlaps of open sets these $q$-forms need to be glued together via ($q$$-$1)-form gauge transformations. The ($q$$-$1)-form gauge transformation fields in turn are only defined on double overlaps of open sets and not globally. A gluing condition needs to be provided for them on triple overlaps via a ($q$$-$2)-form gauge field. This process continues iteratively until a specification of integers on ($q$+1)-overlaps of open sets and finally the consistency condition for this specification requires that the oriented sum of these integers must vanish on the corresponding overlap of $q+2$ open sets. All this data defined on open sets as well as overlaps of open sets at various degrees can be succinctly described as a DB $q$-cochain. Furthermore, the various gluing conditions are nothing but the statement that the DB cochain must actually be a DB cocycle. Finally, there are some redundancies in this description that can very naturally be understood as the image of a DB codifferential operator acting on the space of DB ($q$$-$1)-cochains. Upon modding out by this redundancy, what we obtain are the isomorphism classes of gauge inequivalent $q$-form $\rU(1)$ fields on $\cM$ but defined as such this is nothing but the $q$-th DB cohomology group. We illustrate this idea through a few simple examples. Let us first consider the case of 1-form connections: 
\begin{example}[\emph{1-form $\rU(1)$ connections}]
	DB 1-cochains are defined by the data $\mathbb A\equiv (\mu^{1}_{0},\mu^{0}_{1},n^{\mathbb A}_{2})$ where $\mu_0^1$ are 1-forms defined on local contractible patches, $\mu_1^0$ are functions defined on overlaps of open sets and $n_2^{\mathbb A}$ are integers defined on double overlaps. As described above, this is precisely the data one requires to build a connection for a 1-form $\rU(1)$ bundle. All this data can be glued together by imposing that $D_{(1,1)}\mathbb A=0$. This cocycle condition implies
	\begin{align}
		(d_0 \mu^{1}_0)_{i_0i_1}\equiv&\;(\mu^{1}_0)_{i_1}-(\mu^{1}_0)_{i_0}=(d_{\dr}^0 \mu^0_{1})_{i_0i_1} \nn \\
		(d_1 \mu^{0}_1)_{i_0i_1i_2}\equiv &\; (\mu_1^{0})_{i_1i_2}-(\mu_1^{0})_{i_0i_2}+(\mu_1^{0})_{i_0i_1}=(d_{\dr}^{-1} n^{\mathbb A}_{2})_{i_0i_1i_2} \nn \\
		(d_2 n^{\mathbb A}_2)_{i_0i_1i_2i_3}\equiv&\; (n_{2}^{\mathbb A})_{i_{1}i_2i_3}-(n_{2}^{\mathbb A})_{i_{0}i_2i_3}+(n_{2}^{\mathbb A})_{i_{0}i_1i_3}-(n_{2}^{\mathbb A})_{i_{0}i_1i_2}=0 \; .
		\label{eq:1form_gluing_rel}
	\end{align} 
	It remains to quotient by the redundancies which physically correspond to 0-form gauge transformations and mathematically correspond to DB 1-coboundaries. Given $\Lambda\equiv (\lambda^{0}_{0},m^{\mathbb A}_{1})\in C^{0}_{\rm DB}(\cM,\mathbb Z)$, we need to impose $\mathbb A \sim \mathbb A + D_{(0,1)}\Lambda$. Explicitly, it reads
	\begin{align}
		(\mu^{1}_{0})_{i_0}\sim&\; (\mu^{1}_{0}+d_{\dr}^0\lambda^{0}_0)_{i_0} \nn \\
		(\mu^{0}_{1})_{i_0i_1}\sim&\; (\mu^{0}_{1} + d_0\lambda^{0}_0-d_{\dr}^{-1}m^{\mathbb A}_1)_{i_1i_2} \nn \\
		(n^{\mathbb A}_2)_{i_0i_1i_2} \sim &\; (n^{\mathbb A}_2-d_1 m^{\mathbb A}_1)_{i_0i_1i_2}
	\end{align}
	which are nothing but 0-form $\rU(1)$ gauge transformations. For completeness, we can check that
	\begin{align}
		D_{(1,1)}\circ D_{(0,1)}=&\;\left[d_0 - (d_1+d_{\dr}^0)+(d_2+d_{\dr}^{-1})\right]\circ \left[(d_0 +d_{\dr}^0)-(d_1+d_{\dr}^{-1})\right] \nn \\ 
		=&\; d_{2}(d_{\dr}^0-d_{\dr}^{-1}) \nn \\
		=&\;0
	\end{align}
	where the third line follows from the fact that $C^{1}_{\rm DB}(\cM,\mathbb Z)\subset \text{ker}(d_2)$. It is well-known that the field strength of a $\rU(1)$ connection is quantized to have integer periods. This can be readily checked: Since $d_{\dr}^{1}\mu_{0,i_0}^{1}-d_{\dr}^1\mu^{1}_{0,i_1}=d_{\dr}^1\circ d^{0}\mu^0_{1,i_0i_1}=0$, we can use $d_{\dr}^1\mu_{0}^{1}$ as local representative of the field strength. Let the field strength corresponding to a connection $\mathbb A$ be denoted by $F_{\mathbb A}$. Then on an open set $U_i$ we may write the local representative of the field strength as $(F_{\mathbb A})_{i_0}:= d_{\dr}^{1}\mu^{1}_{0,i_0}$. Given a 2-cycle $\mathfrak L^{(2)}$ together with a polyhedral decomposition, we obtain 
	\begin{align}
		\oint_{\mathfrak L^{(2)}}F_{\mathbb A}=&\;\sum_{i_0}\int_{\mathfrak l^{(2)}_{i_0}}(d_{\dr}^1\mu_{0}^{1})_{i_0}
		=\sum_{i_0}\int_{\partial \mathfrak l^{(2)}_{i_0}}(\mu_{0}^{1})_{i_0}
		\nn \\
		=&\; \sum_{i_0,i_1}\int_{\mathfrak l^{(1)}_{i_0i_1}}(d_0\mu_{0}^{1})_{i_0i_1}=\sum_{i_0,i_1}\int_{\mathfrak l^{(1)}_{i_0i_1}}(d_{\dr}^0\mu_{1}^{0})_{i_0i_1}  \nn \\
		=&\; \sum_{i_0,i_1,i_2}\int_{\mathfrak l^{(0)}_{i_0i_1i_2}}(d_1\mu_1^0)_{i_0i_1i_2} =\sum_{i_0,i_1,i_2}\int_{\mathfrak l^{(0)}_{i_0i_1i_2}}(d_{\dr}^{-1}n_2^{\mathbb A})_{i_0i_1i_2}
		\nn \\
		=&\; \sum_{i_0,i_1,i_2}(d_{\dr}^{-1}n_2^{\mathbb A})\Big |_{\mathfrak l^{(0)}_{i_0i_1i_2}} \in \mathbb Z 
	\end{align} 
	which is obviously the expected quantization of field strength. Note finally that given a 1-cycle $\mathfrak L^{(1)}$ together with a polyhedral decomposition, the holonomy of $\mathbb A$ along $\mathfrak L^{(1)}$ takes the form
	\begin{align}
		W^{\ec}(\mathfrak L^{(1)}):=&\; \exp\bigg\{2\pi i \ec \oint_{\mathfrak L^{(1)}}\mathbb A \bigg\} \nn \\
		=&\; \exp\bigg\{2\pi i \ec \bigg(\sum_{i_0}\int_{\mathfrak l^{(1)}_{i_0}}(\mu^{1}_{0})_{i_0}
		-\sum_{i_0,i_1}(\mu_{1}^{0})\Big|_{\mathfrak{l}^{(0)}_{i_0i_1}}
		\bigg) 
		\bigg\} \; , 
		\label{Wilson_line}
	\end{align}
	which is invariant under (0-form) gauge transformations.
\end{example}
\noindent
Following exactly the same steps, we define 2-form connections:
\begin{example}[\emph{2-form $\rU(1)$ connections}]
	Deligne-Beilisnon 2-cochains are defined by the data $\mathbb B\equiv (\nu^{2}_0,\nu^{1}_{1},\nu^{0}_2, n^{\mathbb B}_3)$. Similar to the case of 1-form connections, this is precisely the data one needs to construct/describe a 2-form $\rU(1)$ connection in the most general case. However, in order to glue all this data together correctly we need to impose that $\mathbb B$ is in the kernel of $D_{(2,2)}$. Writing $D_{(2,2)}\mathbb B=0$ explicitly, we get
	\begin{align}
		(d_0 \nu^{2}_0)_{i_0i_1}\equiv&\; ( \nu^{2}_0)_{i_1}-( \nu^{2}_0)_{i_0} =(d_{\dr}^1 \nu^1_{1})_{i_0i_1} \nn \\
		(d_1 \nu^{1}_1)_{i_0i_1i_2}\equiv&\; (\nu^{1}_1)_{i_1i_2}-(\nu^{1}_1)_{i_0i_2}+(\nu^{1}_1)_{i_0i_1}=-(d_{\dr}^{0} \nu^{0}_{2})_{i_0i_1i_2} \nn \\
		(d_2 \nu^{0}_2)_{i_0i_1i_2i_3}\equiv &\; 
		(\nu^0_2)_{i_1i_2i_3}-
		(\nu^0_2)_{i_0i_2i_3}
		+(\nu^0_2)_{i_0i_1i_3}
		-(\nu^0_2)_{i_0i_1i_2}
		=(d_{\dr}^{-1} n^{\mathbb B}_{3})_{i_0i_1i_2i_3}\nn \\
		(d_3 n^{\mathbb B}_3)_{i_0i_1i_2i_3i_4}\equiv&\; \sum_{j=0}^{4}(-1)^{j}(n_3^{\mathbb B})_{i_0\dots \hat{i}_j \dots i_4}= 0 \; .
	\end{align}
	It remains to quotient by 1-form gauge transformations which in the context of the DB construction implies modding out by coboundaries in the image of $D_{(1,2)}$. Given $\Theta\equiv (\theta^{1}_0,\theta^{0}_1,m^{\mathbb B}_2)\in C^{1}_{\rm DB}(\cM,\mathbb Z)$, we need to impose $\mathbb B \sim \mathbb B + D_{(1,2)}\Theta$. Explicitly, it reads
	\begin{align}
		(\nu^{2}_{0})_{i_0}\sim&\; (\nu^{2}_{0}+d_{\dr}^1\theta^{1}_0)_{i_0} \nn \\
		(\nu^{1}_{1})_{i_0i_1}\sim&\; (\nu^{1}_{1} + d_0\theta^{1}_0-d_{\dr}^{0}\theta^0_1)_{i_0i_1} \nn \\
		(\nu^{0}_{2})_{i_0i_1i_2}\sim&\; (\nu^{0}_{2} - d_1\theta^{0}_1+d_{\dr}^{-1}m^{\mathbb B}_2)_{i_0i_1i_2} \nn \\
		(n^{\mathbb B}_3)_{i_0i_1i_2i_3} \sim &\; (n^{\mathbb B}_3-d_2 m^{\mathbb B}_2)_{i_0i_1i_2i_3} \; .
		\label{2_form_gauge}
	\end{align}
	We could check explicitly that $D_{(2,2)}\circ D_{(1,2)}=d_{3}(-d_{\dr}^{1}+d_{\dr}^0-d_{\dr}^{-1})=0$ using the fact that $C^{1}_{\rm DB}(
	\cM,\mathbb Z)\subset \text{ker}(d_{3})$. Similar to 1-form connections, the field strength of a 2-form $\rU(1)$ connection satisfies a generalized \emph{Dirac quantization condition} which means that the monopole charge is integer quantized, i.e.  $\oint_{\cM}F_{\mathbb B}\in \mathbb Z $. This can be demonstrated explicitly using $(d^{2}\nu_{0}^{2})_{i_0}$ as a local representative of $F_{\mathbb B}$ on an open set $U_i$. Given a 3-cycle $\mathfrak L^{(3)}$ together with a polyhedral decomposition, we obtain indeed
	\begin{align}
		\oint_{\mathfrak L^{(3)}}F_{\mathbb B}=&\;\sum_{i_0}\int_{\mathfrak l^{(3)}_{i_0}}(d_{\dr}^2\nu_{0}^{2})_{i_0}
		=\sum_{i_0}\int_{\partial \mathfrak l^{(3)}_{i_0}}(\nu_{0}^{2})_{i_0}
		\nn \\
		=&\; \sum_{i_0,i_1}\int_{\mathfrak l^{(2)}_{i_0i_1}}(d_0\nu_{0}^{2})_{i_0i_1}=\sum_{i_0,i_1}\int_{\mathfrak l^{(2)}_{i_0i_1}}(d_{\dr}^1\nu_{1}^{1})_{i_0i_1}  \nn \\
		=&\; \sum_{i_0,i_1,i_2}\int_{\mathfrak l^{(1)}_{i_0i_1i_2}}(d_1\nu_1^1)_{i_0i_1i_2} =\sum_{i_0,i_1,i_2}\int_{\mathfrak l^{(1)}_{i_0i_1i_2}}(d_{\dr}^{0}\nu_2^{0})_{i_0i_1i_2}
		\nn \\
		=&\; \sum_{i_0,i_1,i_2,i_3}\int_{\mathfrak l^{(0)}_{i_0i_1i_2i_3}}(d_2\nu_2^0)_{i_0i_1i_2i_3} =\sum_{i_0,i_1,i_2,i_3}\int_{\mathfrak l^{(0)}_{i_0i_1i_2i_3}}(d_{\dr}^{-1}n_3^{\mathbb B})_{i_0i_1i_2i_3}
		\nn \\
		=&\; \sum_{i_0,i_1,i_2,i_3}(d_{\dr}^{-1}n_3^{\mathbb B})\Big |_{\mathfrak l^{(0)}_{i_0i_1i_2i_3}} \in \mathbb Z \; .
	\end{align}
	Note finally that given a 2-cycle $\mathfrak L^{(2)}$ together with a polyhedral decomposition, the (2-)holonomy of $\mathbb B$ along $\mathfrak L^{(2)}$ takes the gauge invariant form
	\begin{align}
		U^{\mc}(\mathfrak L^{(2)}):=&\; \exp\bigg\{2\pi i \mc \oint_{\mathfrak L^{(2)}}\mathbb B\bigg\} \nn \\
		=&\; \exp\bigg\{2\pi i \mc \bigg(\sum_{i_0}\int_{\mathfrak l^{(2)}_{i_0}}(\nu^{2}_{0})_{i_0}
		-\sum_{i_0,i_1}\int_{\mathfrak l^{(1)}_{i_0i_1}}(\nu_{1}^{1})_{i_0i_1}+ \sum_{i_0,i_1,i_2}\nu_{2}^{0}\Big |_{\mathfrak l^{(0)}_{i_0i_1i_2}}
		\bigg) 
		\bigg\} \; .
		\label{Wilson_surface}
	\end{align}
\end{example}
\noindent
So the space of $q$-form connections is equivalent to the space of equivalence classes in the $q$-th DB cohomology $H^{q}_{\rm DB}(\cM,\mathbb Z)$, as illustrated above for the $q=1,2$ cases. We say a connection $\mathbb A^{(q)}\in H^{q}_{\rm DB}(\cM,\mathbb Z)$ is \emph{flat} if it lies in the kernel of the $D_{(q,q+1)}$ operator and thus we have the following isomorphism:
\begin{align*}
	\boxed{
	\big\{\text{\,Equivalence classes of flat $q$-form $\rU(1)$ connections on $\cM$} \,\big\} \simeq H^{q}_{\rm DB}(\cM,\mathbb Z)\cap \text{ker}(D_{(q,q+1)}) \; . }
\end{align*}
This follows from the fact that a $\rU(1)$ $q$-form connection $\mathbb A^{(q)}=(\mu^{q}_0,\mu^{q-1}_1,\ldots, \mu^{0}_{q},n_{q+1}) \in H^{q}_{\rm DB}(\cM,\mathbb Z)$ needs to satisfy a single extra constraint in order to be in the kernel of $D_{(q,q+1)}$ that is
\begin{align}
	d^{q}\mu^{q}_1=0 \; .
\end{align} 
Hence the curvature of the $q$-form connection vanishes locally on each open set. 

\subsection{Strict 2-group connections}\label{sec:strict}

Having described the space of gauge inequivalent configurations of higher form $\rU(1)$ gauge theories in terms of Deligne-Beilinson cohomology, in this subsection we explore a scenario where the group bundle is a non-trivial product of bundles corresponding to 1-form $\rU(1)$ connections and 2-form $\rU(1)$ connections. Here by non-trivial product we mean that locally the data required on open sets, overlaps of open sets and so on is identical to that of a direct sum of some number of 1-form connections and 2-form connections. However, the gluing relations which were previously related to certain DB cocycle conditions are \emph{twisted} in a way that we make precise below. Also, the redundancies or gauge transformations which were related to DB coboundaries are altered accordingly. This is the relevant situation when discussing the embedding of a finite group 2-form gauge theory into a \emph{toric gauge theory}. That particular field theory \eqref{eq:action_BwedgeB} is the motivation for this subsection. By constructing the Gau{\ss} operators and the gauge transformations within this theory, we inferred that these transformations correspond to those of a toric strict 2-group bundle. Below we first briefly describe strict 2-groups and then carefully construct the corresponding strict 2-group bundles.  

\medskip \noindent A \emph{strict toric 2-group} \cite{baez2004higher, Baez:2003fs,Delcamp:2017pcw} is defined by four pieces of data, namely $\mathcal G=\left\{\rU(1)^{P},\rU(1)^{Q},t,\triangleright \right\}$ where 
\begin{align}
	t:&\;\rU(1)^{P}\to\rU(1)^{Q} \nonumber \\ 
	\triangleright:&\;\rU(1)^{Q}\to \text{Aut}(\rU(1)^{P}) \; .
\end{align}
This data needs to satisfy some consistency conditions which ensure that $t$ and $\triangleright$ interact well with one another.\footnote{These consistency relations make $\mathcal G$ equivalent to a \emph{crossed module}. For details please see \cite{baez2004higher} and references therein.} The consistency relations for some $\mathbf{A}\in \rU(1)^{Q}$ and ${\mathbf B}\in \rU(1)^{P}$ are $t({\mathbf A}\triangleright {\mathbf B})=t({\mathbf B})$ and $t({\mathbf B})\triangleright {\mathbf B}'={\mathbf B}'$. In the following, we choose $\triangleright = \text{id}$. A homomorphism $t$ may be written as
\begin{align}
\left[t({\mathbf B})\right]_{I}= \prod_{J=1}^{P}h_{J}^{p_{IJ}}
\end{align}
where $I\in 1,\dots ,P $, $J\in 1,\dots, Q$ and ${\mathbf B}\equiv (h_{1},\dots, h_{P})\in \rU(1)^{P}$. In order to build a $\mathcal G$-bundle, we require local data which corresponds to $P$ 2-form $\rU(1)$ connections and $Q$ 1-form $\rU(1)$ connections. Therefore, the local fields are $Q$ DB 1-cochains $\mathbb A^{I}$ and $P$ DB 2-cochains $\mathbb B^{J}$: 
\begin{align}
	\mathbb A^{I}=&\;(\mu^{1,I}_0,\mu^{0,I}_1,n^{\mathbb A,I}_2) \nonumber \\
	\mathbb B^{J}=&\;(\nu^{2,J}_0,\nu^{1,J}_{1},\nu^{0,J}_2,n^{\mathbb B,J}_3) \; .
\end{align}
Henceforth, in order to keep the notation light, we specialize to the case $P=Q=1$ which can be readily generalized to $P,Q\in \mathbb Z$.  Although the local data corresponds to a direct sum of an ordinary 1-form and 2-form $\rU(1)$ gauge theory, the gluing (cocycle) conditions and gauge transformations are twisted by the homomorphism $t\in \text{Hom}(\rU(1),\rU(1)) \simeq \mathbb Z$. We note that since $\rU(1)\simeq \mathbb R/\mathbb Z$ fits in the canonical exact sequence $0\to \mathbb Z\to \mathbb R \to \mathbb R/\mathbb Z$, the homomorphism lifts to $t\in \text{Hom}(\mathbb R,\mathbb R)$ and $t\in \text{Hom}(\mathbb Z,\mathbb Z)$.\footnote{We use `$t$' for the lifted homomorphisms as well in order to keep the notation light.} Hence the homomorphism acts on all the local data of the \v{C}ech-de Rham bicomplex. This is an essential ingredient in writing consistent gluing relations.  

The space of strict 2-group $\mathcal G=\{\rU(1),\rU(1),t, {\rm id} \}$-connections on $\cM$ is spanned by tuples of DB cochains $(\mathbb A,\mathbb B) \in C^{1}_{\rm DB}(\cM,\mathbb Z)\times C^{2}_{\rm DB}(\cM,\mathbb Z)$ satisfying the conditions
\begin{align}
	\nn
	D_{(2,2)}\mathbb B & = (d_0\nu_0^{2}-d_{\dr}^{1}\nu_1^1)_{i_0i_1} 
	+ (-d_1 \nu_1^{1}+d_{\dr}^{0}\nu_2^{0})_{i_0i_1i_2} + (d_2 \nu_{2}^{0}-d_{\dr}^{-1}n_3^{\mathbb B})_{i_0i_1i_2i_3} + (-d_3 n_3^{\mathbb B})_{i_0i_1i_2i_3i_4} \\ &= 0 
	\\ \nn 
	D_{(1,1)}^{t(\mathbb B)}\mathbb A &=   \left(d_{0}\mu^1_0 -d_{\dr}^0 \mu_1^{0}+t(\nu_1^1)\right)_{i_0i_1} 
	\left(-d_{1}\mu_{1}^0 +d_{\dr}^{-1}n_2^{\mathbb A}+t(\nu_2^0)\right)_{i_0i_1i_2} + \left( d_2 n_2^{\mathbb A} +t(n_{3}^{\mathbb B}) \right)_{i_0i_1i_2i_3} \\
	& = 0  \; .
	\label{eq:strict_gluing}
\end{align}
Let us look at the above gluing conditions a bit more closely. For example the 1-form connection $\mathbb A$ involves an assignment of $(\mu_{0}^{1})_{i}$ on open sets $U_i$. On the overlap $U_{i_0i_1}$ of two open sets $U_{i_0}$ and $U_{i_1}$ the local 1-form representatives are glued together by imposing
\begin{align}
	(\mu_0^{1})_{i_1}-(\mu_0^{1})_{i_0}=d_{\dr}^{0}(\mu_1^{0})_{i_0i_1}-t(\nu_1^1)_{i_0i_1} \; .
\end{align}
Hence the gluing condition for the 1-form connection has been altered by the presence of the 2-form connection. Similarly, the gluing conditions on overlaps of all degrees are modified. In other words we need to impose that all the parenthesis in \eqref{eq:strict_gluing} vanish independently. Furthermore, this data is defined up to the following gauge transformations
\begin{align}
	\mathbb A\sim&\;  \mathbb A + D^{t(\Theta)}_{(0,1)}\Lambda \,=:\,\mathbb A + D_{(0,1)}\Lambda -t(\Theta) \nn \\
	\mathbb B \sim&\; \mathbb B + D_{(1,2)}\Theta 
	\label{strict_gauge}
\end{align}
where $\Lambda \in C^{1}_{\rm DB}(\cM,\mathbb Z)$ and $\Theta\in C_{\rm DB}^{2}(\cM,\mathbb Z)$. Note that \eqref{strict_gauge} is nothing but \eqref{eq:strict_gt} written more precisely in terms of the Deligne-Beilinson data. More explicitly, in terms of the local data, the former equivalence reads
\begin{align}
	(\mu^{1}_{0})_{i_0}\sim&\; (\mu^{1}_{0}+d_{\dr}^0\lambda^{0}_0-t(\theta_{0}^1))_{i_0} \nn \\
	(\mu^{0}_{1})_{i_0i_1}\sim&\; (\mu^{0}_{1} + d_0\lambda^{0}_0-d_{\dr}^{-1}m^{\mathbb A}_1-t(\theta_{1}^0))_{i_0i_1} \nn \\
	(n^{\mathbb A}_2)_{i_0i_1i_2} \sim &\; (n^{\mathbb A}_2-d_1 m^{\mathbb A}_1-t(m^{\mathbb B}_2))_{i_0i_1i_2} \; ,
	\label{eq:strict_gt_DB}
\end{align}
while the gauge transformations for $\mathbb B$ are the same as those for ordinary 2-form $\rU(1)$ connections \eqref{2_form_gauge}. We can readily check that $D^{t}_{(1,1)}\circ D^{t}_{(0,1)}=0$ so that one may define an \emph{affine cohomology theory}. The space of gauge inequivalent configurations of a strict 2-group $\mathcal G$ are isomorphic to this cohomology space that we denote by $H^{2,1}_{\mathcal G}(\cM)$. 

\begin{definition}
	The affine cohomology group $H^{2,1}_{\mathcal G}(\cM)$ is defined as the group of cohomology classes equivalent to isomorphism classes of gauge configurations of a toric strict 2-group gauge theory for the strict 2-group $\mathcal G$. $H^{2,1}_{\mathcal G}(\cM)$ are spanned by tuples of DB cochains $(\mathbb A,\mathbb B) \in C^{1}_{\text{DB}}(\cM,\mathbb Z)\times C^{2}_{\text{DB}}(\cM,\mathbb Z)$ that satisfy the condition \eqref{eq:strict_gluing} modulo those that are of the form $(D_{(0,1)}^{t(\Theta)}\Lambda, D_{(1,2)}\Theta)$ where $(\Lambda, \Theta)\in C^{0}_{\text{DB}}(\cM,\mathbb Z)\times C^{1}_{\text{DB}}(\cM,\mathbb Z)$. 
\end{definition}

\bigskip \noindent
Having defined $H^{2,1}_{\mathcal G}(\cM)$, we then consider the subspace of flat connections. This will be important in what follows as it is the configuration space of topological $\mathcal G$-gauge theories. 
\begin{definition}
	The space of flat strict 2-group $\mathcal G$-connections on $\cM$ is the set of tuples of DB-cochains $(\mathbb A,\mathbb B)$ that satisfy the conditions
	\begin{align}
		D_{(1,2)}\mathbb A+t(\mathbb B)=&\;0 \nn \\
		D_{(2,3)}\mathbb B=&\; 0 
	\end{align} 
	which, in terms of the local data, translates into
	\begin{align}
		\nn
		D_{(2,3)}\mathbb B&=\; (d_{\dr}^2\nu_0^{2})_{i_0}+(d_0\nu_0^{2}-d_{\dr}^{1}\nu_1^1)_{i_0i_1} + (-d_1 \nu_1^{1}+d_{\dr}^{0}\nu_2^{0})_{i_0i_1i_2} \\
		\nn &\phantom{=}+ (d_2 \nu_{2}^{0}-d_{\dr}^{-1}n_3^{\mathbb B})_{i_0i_1i_2i_3} + (-d_3 n_3^{\mathbb B})_{i_0i_1i_2i_3i_4}\\ &= 0 \\ \nn
		D_{(1,2)}\mathbb A +t(\mathbb B)&= \left(d_{\dr}^1\mu_0^{1}+t(\nu_0^{2})\right)_{i_0} + \left(d_{0}\mu^1_0 -d_{\dr}^0 \mu_1^{0}+t(\nu_1^1)\right)_{i_0i_1} \nn \\
		&\phantom{=}+ \left(-d_{1}\mu_{1}^0 +d_{\dr}^{-1}n_2^{\mathbb A}+t(\nu_2^0)\right)_{i_0i_1i_2} + \left( d_2 n_2^{\mathbb A} +t(n_{3}^{\mathbb B}) \right)_{i_0i_1i_2i_3} \\ &= 0 \; .
	\end{align}
\end{definition}
\noindent It is easy to check that the flatness condition is preserved under the gauge  transformations \eqref{strict_gauge}. Indeed,
\begin{align}
	D_{(1,2)}\mathbb A+t(\mathbb B)\rightarrow&\; D_{(1,2)}\mathbb A +t(\mathbb B)+D_{(1,2)}\circ D_{(0,1)}\Lambda \\ \nn =&\; D_{(1,2)}\mathbb A+t(\mathbb B)
\end{align}
where we made use of the fact that $D_{(1,2)}\circ D_{(0,1)}=(d^1_{\dr}+ D_{(1,1)})\circ D_{(0,1)} 	= d_{\dr}^{1}\circ D_{(0,1)} =  0$ that follows from $\text{im}(D_{(0,1)}) \cap \mu^{1}(\cM) \subset \text{im}(d_{\dr}^{0})$. 

We now want to compute the integral of the curvature of the 2-group connection and reading off whether it satisfies any quantization conditions. First of all, we can immediately infer that since the gauge transformations of $\mathbb B$ are unaltered compared to the case of the 2-form gauge theory previously studied, the quantization condition also remains unaltered, i.e
\begin{align}
	\oint_{\mathfrak L^{(3)}}F_{\mathbb B}\in \mathbb Z
\end{align}
where  $\mathfrak L^{(3)}\in Z_{3}(\cM,\mathbb Z)$. The situation is different as far as the curvature $F_{\mathbb A}$ is concerned. Let us first try to construct a local representative of $F_{\mathbb A}$. The simplest possibility is $d_{\dr}^{1}\mu_{0}^{1}$. Doing so, we realize that
\begin{align}
	d_{\dr}^{1}(\mu_{0}^{1})_{i_0}-d_{\dr}^{1}(\mu_{0}^{1})_{i_0}=-d_{\dr}^{1}\left(t(\nu_1^1)\right)_{i_0i_1}
\end{align}
so that $(F_{\mathbb A})_{i_0}:=\big(d_{\dr}^1\mu_{0}^{1}+t(\nu_0^{2})\big)_{i_0}$ can serve as a local representative since $(F_{\mathbb A})_{i_0}-(F_{\mathbb A})_{i_1}=0$. Using this representative, we may integrate the curvature over a closed 2-cycle $\mathfrak L^{(2)}$ in $\cM$
\begin{align}
	\oint_{\mathfrak L^{(2)}}F_{\mathbb A}=&\;\sum_{i_0}\int_{\mathfrak l^{(2)}_{i_0}}\left(d_{\dr}^1\mu_{0}^{1}+t(\nu_0^{2})\right)_{i_0} \nn = \sum_{i_0,i_1}\int_{\mathfrak l^{(1)}_{i_0}}\left(d_0\mu_{0}^{1}\right)_{i_0i_1}+\int_{\mathfrak l^{(2)}_{i_0}}t(\nu_0^2)_{i_0} \nn \\
	=&\; \sum_{i_0,i_1}\int_{\mathfrak l^{(1)}_{i_0i_1}}\left(d_{\dr}^0\mu_{1}^{0}-t(\nu_1^1)\right)_{i_0i_1}+\int_{\mathfrak l^{(2)}_{i_0}}t(\nu_0^2)_{i_0} \nn \\
	=&\; \sum_{i_0,i_1,i_2}\int_{\mathfrak l^{(0)}_{i_0i_1i_2}}\left(d_1\mu_{1}^{0}\right)_{i_0i_1i_2} +\int_{\mathfrak l^{(2)}_{i_0}}t(\nu_0^2)_{i_0} 
	-\sum_{i_0,i_1}\int_{\mathfrak l^{(1)}_{i_0i_1}}\left(t(\nu_1^1)\right)_{i_0i_1}
	\nn \\
	=&\; \sum_{i_0,i_1,i_2} d_{\dr}^{-1}n_{2}^{\mathbb A}\Big|_{\mathfrak l^{(0)}_{i_0i_1i_2}} +\int_{\mathfrak l^{(2)}_{i_0}}t(\nu_0^2)_{i_0} 
	-\sum_{i_0,i_1}\int_{\mathfrak l^{(1)}_{i_0i_1}}t(\nu_1^1)_{i_0i_1}
	+\sum_{i_0,i_1,i_2}\nu_2^0\Big|_{\mathfrak l^{(0)}_{i_0i_1i_2}} \nn \\
	\in&\; \mathbb Z + \oint_{\mathfrak L^{(2)}}\mathbb B \; .
\end{align}
Hence the field strength of a strict 2-group connection is not quantized but rather, as expected, the quantization is shifted by the holonomy of $\mathbb B$.

\medskip \noindent
Since 2-group connections comprise 1-form and 2-form gauge fields, we expect the gauge invariant operators to be Wilson lines as well as Wilson surfaces. The gauge transformations for the connection $\mathbb B$ are the same as the ones entering the definition of a 2-form connection so that the surface operators are the same as the ones defined in \eqref{Wilson_surface}, i.e.
\begin{align}
	U^{\mc}(\mathfrak L^{(2)})=\exp \bigg\{2\pi i \mc \oint_{\mathfrak L^{(2)}} \mathbb B \bigg\} \; .
	\label{eq:Strict_wilson_surf}
\end{align} 
The line operators are a bit more subtle since the naive guess \eqref{Wilson_line} is not gauge invariant. Furthermore, a Wilson line can only be defined for homologically trivial 1-cycles in order to be (2-group) gauge invariant. Instead, the gauge invariant operator takes the form
\begin{align}
	W^{\ec}(\mathfrak L^{(1)},\partial^{-1}\mathfrak L^{(1)}):=\exp \bigg\{2\pi i \ec \oint_{\mathfrak L^{(1)}}\mathbb A +2\pi i \ec \int_{\partial^{-1} \mathfrak L^{(1)}}t(\mathbb B)\bigg\}
	\label{eq:strict_wilson_line}
\end{align}
where $\partial^{-1}\mathfrak L^{(1)}$ is a 2-chain whose boundary is $\mathfrak L^{(1)}$. The corresponding polyhedral decomposition can be obtained by attaching a single disc-like region to the 1-cycle $\mathfrak L^{(1)}$. Let us first focus on the l.h.s term of \eqref{eq:strict_wilson_line} whose integrand only depends on $\mathbb A$. As mentioned earlier, the integral of $\mathbb{A}$ over $\mathfrak L^{(1)}$ is not invariant under gauge transformations by itself due to the modified gauge structure. Indeed, under gauge transformations one has
\begin{align}
	\exp \bigg\{ 2\pi i \ec\oint_{\mathfrak L^{(1)}} \mathbb A \bigg\}
	\to&\; \exp \bigg\{ 2\pi i \ec \oint_{\mathfrak L^{(1)}}\mathbb A  - 2\pi i \ec \sum_{i_0}\int_{\mathfrak l^{(1)}_{i_0}}t \big( (\theta^{1}_0)_{i_0}\big)+\sum_{i_0,i_1}t \big(\theta_{1}^{0}\big)\Big|_{\mathfrak l^{(0)}_{i_0i_1}} \bigg\} \nn \\
	=&\; \exp\bigg\{2\pi i \ec \oint_{\mathfrak L^{(1)}}\mathbb A- t(\Theta)\bigg\} \; .
\end{align}
The piece of data on the r.h.s that depends on $\mathbb B$ requires a bit more care. We attach a disc-like region to $\mathfrak L^{(1)}$ and introduce an open set labeled by $U_{{\underline{0}}}$ with the convention that 
$\underline{0}<i_0$ for all $i_0$. By introducing this open set, every open set $U_{i_0}$ in $\mathfrak L^{(1)}$ becomes an overlap of two open sets $U_{\underline{0}i_0}$ in $\partial^{-1}\mathfrak L^{(1)} $ and in turn every overlap of two open sets $U_{i_0i_1}$ in $\mathfrak L^{(1)}$ becomes an overlap of three open sets $U_{\underline{0}i_0i_1}$ in $\partial^{-1}\mathfrak L^{(1)}$. We may now integrate $\mathbb B\in C^{2}_{\rm DB}(\cM,\mathbb Z)$ over $\partial^{-1}\mathfrak L^{(1)}$ and write how it is modified under gauge transformations
\begin{align}
	\exp\bigg\{2\pi i \ec \oint_{\partial^{-1}\mathfrak L^{(1)}}t(\mathbb B)\bigg\}		
	=&\; \exp\bigg\{2\pi i \ec t\bigg(\int_{\mathfrak l^{(2)}_{\underline{0}}}(\nu^{2}_{0})_{\underline{0}}
	-\sum_{\underline{0},i_1}\int_{\mathfrak l^{(1)}_{\underline{0}i_1}}(\nu_{1}^{1})_{\underline{0}i_1}+ \sum_{\underline{0},i_1,i_2}\nu_{2}^{0}(\mathfrak l^{(0)}_{\underline{0}i_1i_2})
	\bigg) 
	\bigg\} \nn \\
	\to &\; \exp\bigg\{ 2\pi i \ec \int_{\partial^{-1}\mathfrak L^{(1)}}\big(t(\mathbb B) + t(D_{(1,2)}\Theta) \big) \bigg\} \nn \\
	=&\; \exp\bigg\{2\pi i \ec \bigg(\int_{\partial^{-1}\mathfrak L^{(1)}} t(\mathbb B) + \oint_{\mathfrak L^{(1)}}t(\Theta) \bigg) \bigg\} \; .
\end{align}
This confirms that \eqref{eq:Strict_wilson_surf} and \eqref{eq:strict_wilson_line} are the gauge invariant operators for a strict 2-group toric gauge theory. To conclude, we have shown above that the gauge transformations for the continuous topological gauge theory \eqref{eq:action_BwedgeB} correspond to a strict toric 2-group bundle. Furthermore, such a bundle can be defined rigorously using methods based on Deligne-Beilinson cohomology. Above, we constructed such a bundle, studied the quantization conditions for its topological sectors and constructed gauge invariant functions (operators in the quantum theory) in terms of local data.

Using the same technology, it is possible to write down rigorous actions for higher-form topological phases in terms of Deligne-Beilinson cocycles. Some explicit examples are provided in app.~\ref{sec:app_DB}. Note that this construction can also be adapted in order to describe flat connections for \emph{weak} 2-group bundles and more generally for models built from Postnikov towers.

\vfill{}
\noindent
\emph{In the previous sections, we explored properties of higher-form topological theories with a special emphasis on 2-form topological gauge theories. More specifically, we explained how these models could be defined as sigma models with target space the classifying spaces $B^qG$ of a finite group $G$. Since these theories are defined in terms of finite groups, they are naturally defined in the discrete. However, it is also possible to embed them in the continuum. We explained that in the case of 2-form gauge theories, the corresponding continuous theory is based on a strict 2-group. We defined the space of gauge inequivalent $q$-form $\rU(1)$ connections as the space of $q$-th Deligne-Beilinson cohomology classes. We then used this result in order to provide a rigorous definition of the strict 2-group continuous theory.}

\emph{
We shall now investigate the same theories but on the other end of the spectrum, namely in terms of Hamiltonian lattice realizations. More specifically, we are going to define the lattice Hamiltonian realization of a 2-form topological theory for a finite abelian group. To do so, we need a more explicit definition of the cohomology group $H^4(B^2G,\mathbb{R} / \mathbb{Z})$, which in turn requires a better understanding of the space $B^2G$. But since the second classifying space $B^2G$ is an example of Eilenberg-MacLane space $K(G,2)$, we shall present the general theory of Eilenberg-MacLane spaces defined as abelian simplicial groups and present their so-called $W$-construction. This can in turn be used to define the so-called 2-form cohomology that is identified with the cohomology of the classifying space $B^2G$ as provided by this W-construction. All the results presented in the following sections will then follow from the properties of the 2-form cohomology. Interestingly, the classification of the cohomology group $H^4(B^2G,\mathbb{R} / \mathbb{Z})$ in terms of quadratic forms also plays a very important role on the lattice.} 

\newpage
\section{Eilenberg-MacLane spaces \label{sec:EM}}

This section provides some review material for algebraic topology in order to motivate the construction of the 2-form cohomology presented in the next section. More specifically, we review the algebraic structure of Eilenberg-MacLane spaces that are defined as:

\begin{definition}[\emph{Eilenberg-MacLane space}]
	Let $ q \in \mathbb{N}$ and $G$ a group (abelian if $q \geq 2$), then an \emph{Eilenberg-MacLane space} $K(G,q)$ is a connected topological space such that $\pi_q(K(G,q)) \simeq G$ and $\pi_n(K(G,q)) \simeq 0$ if $n \neq q$, where $\pi_n$ denotes the $n$-th homotopy group.
\end{definition}
\noindent
Eilenberg-MacLane spaces satisfy the following fundamental property:
\begin{property}
	Eilenberg-MacLane spaces are unique up to homotopy equivalence. 
\end{property}
\noindent
Therefore, we will often abusively refer to any Eilenberg-MacLane space $K(G,q)$ as $K(G,q)$ and, in particular, we identify thereafter the $q$-th classifying space $B^q(G)$, which is a space $K(G,q)$, with $K(G,q)$.
There exists different constructions of Eilenberg-MacLane spaces \cite{eilenberg1943relations, eilenberg1950relations, eilenberg1953groups, eilenberg1954groups, may1992simplicial, may1999concise, hatcher2002algebraic}. In this paper, we define them as \emph{simplicial abelian groups} and we focus specifically on the so-called \emph{$W$-construction}. This is the formulation we will use in sec.~\ref{sec:2formco} in order to define the 2-form cohomology.

\subsection{Abelian simplicial groups}
Let us first present the general definition of an abelian simplicial group and then illustrate it by constructing the space $K(G,1)$, with $G$ a finite group. An abelian simplicial group can be succinctly defined as a \emph{simplicial object} in the category of abelian groups \cite{may1992simplicial}. Nevertheless, we provide below a more explicit definition. Let us first introduce the notion of simplicial set:

\begin{definition}[\emph{Simplicial set}]
		A \emph{simplicial set} $X$ is a collection $\{X_n\}_{n \in \mathbb{N}}$ of sets, together with homomorphisms 
		\begin{alignat}{2}
			\partial_i =\partial_i^{(n)} &: X_n \to X_{n-1} \; , & \q & i= 0,\ldots, n \; , \; n > 0 \; ,\\
			\eta_i = \eta_i^{(n)} &: X_n \to X_{n+1} \; ,  & \q & i= 0,\ldots, n \; , \; n > 0 \; ,
		\end{alignat}
		subject to the identities
		\begin{alignat}{2}
			\label{eq:boundProp}
			\partial_i \partial_j &= \partial_{j-1}\partial_i \; ,& \q &{\rm if}\;\; i < j \; ,\\
			\eta_i \eta_j &= \eta_{j+1}\eta_i \; ,& \q &{\rm if}\;\; i \le j \; ,\\
			\partial_i \eta_j &= \eta_{j-1}\partial_i \; ,& \q &{\rm if}\;\; i < j \; ,  \\
			\partial_i \eta_i &= \partial_{i+1}\eta_i = {\rm id} \; ,& \q & \;  \\
			\partial_i \eta_j &= \eta_{j}\partial_{i-1} \; ,& \q &{\rm if}\;\; i > j+1 \; .								
		\end{alignat}
\end{definition}
\noindent
The maps $\partial_i$ and $\eta_i$ are referred to as \emph{face} and \emph{degeneracy} operators, respectively. The elements of $X_n$ are usually referred to as \emph{$n$-simplices} and $i,j$ in the equations above label the faces of these simplices. 
Given a simplicial set $X$, we define the \emph{boundary} map $\partial = \partial^{(n)} : X_n \to X_{n-1}$ as $\partial^{(0)} \equiv 0$ for $n=0$, and for $n > 0$ as
\begin{equation}
	\label{boundary}
	\partial^{(n)} = \partial_0 - \partial_1 + \cdots + (-1)^n \partial_n \; .
\end{equation}
From the identity \eqref{eq:boundProp} follows the usual rule $\partial \circ \partial \equiv 0$.
The simplest example of simplicial set is provided by the \emph{standard $n$-simplex}:
\begin{example}[\emph{Standard $n$-simplex}]
	Let us first define an $n$-\emph{simplex} $\triangle^n$ as the smallest convex set in $\mathbb{R}^n$ containing $n+1$ points denoted by $v_0, \ldots, v_n$ such that they do not lie in a hyperplane of dimension less than $n$. The points $v_i$ are $0$-simplices and are identified with the \emph{vertices} of the $n$-simplex. In the following, we denote such $n$-simplex by $(v_0 \ldots v_n)$. Furthermore, the vertices are endowed with an ordering which induces an orientation of the edges $(v_i\,v_j)$, $i<j$, according to increasing subscripts. We then define a \emph{face} of an $n$-simplex $\triangle^n$ as a subsimplex defined by the vertices which form a subset of $\{v_0, \ldots, v_n\}$. The $i$-th face of the $n$-simplex can be defined as the image of the map $\partial_i$ such that
	\begin{equation}
	\partial_i(v_0 \, \ldots \, v_n) = (v_0 \, \ldots \, \hat{v}_i \, \ldots \, v_n) := (v_0 \, \ldots \, v_{i-1} \, v_{i+1} \, \ldots \, v_n) \; ,
	\end{equation}
	where the notation $\hat{{\sss \bullet}}$ indicates that the corresponding vertex is omitted from the list. 
	The oriented boundary of an $n$-simplex is then obtained as the image of the operator $\partial$ defined according to \eqref{boundary} as
	\begin{equation}
	\partial^{(n)} (v_0 \, \ldots \, v_n) := \sum_{i=0}^n (-1)^i (v_0 \, \ldots \, \hat{v}_i \, \ldots \, v_n) \; .
	\end{equation}
	Furthermore, the $i$-th degenerate simplex of an $n$-simplex is obtained as the image of the map $\eta_i$ defined as
	\begin{equation}
	\eta_i(v_0 \,\ldots \,v_n) := (v_0 \,\ldots \,v_{i-1} \,v_i \,v_i \,v_{i+1} \, \ldots \,v_q) \; .
	\end{equation}
	The set of $n$-tuples $(v_0 \,\ldots \,v_n)$ together with the face and degeneracy maps introduced above naturally form a simplicial set that is referred to as the \emph{standard $n$-simplex}. 
\end{example} 
\noindent
We can now straightforwardly define a simplicial group:
\begin{definition}[\emph{Simplicial group}]
	A \emph{simplicial group} is a simplicial set $X$ such that each $X_n$ is a group and the degeneracy and face operators are homomorphisms between them. If all the $X_n$ are abelian, then $X$ is an \emph{abelian simplicial group}.
\end{definition}
\noindent
Given a simplicial group $X$, since the face and degeneracy maps are group homomorphisms, the boundary map $\partial$ defined as in \eqref{boundary} is also a homomorphism. Therefore, together with the property $\partial \circ \partial \equiv 0$, the simplicial group $X$ defines a chain complex with chain groups $\{X_n\}_{n \in \mathbb{N}}$ \cite{eilenberg1943relations}. This last remark is the main reason why the study of Eilenberg-MacLane spaces, which are examples of abelian simplicial groups, is relevant to group cohomology and its generalizations.

\subsection{Classifying space $BG$}
Let us now illustrate the concepts introduced above with the construction of the classifying space $BG \equiv B(G)$ of a finite group $G$, which is an Eilenberg-MacLane space $K(G,1)$. We follow an admittedly minimal (but hopefully pedagogical) approach to define such classifying space, however this is enough for the purpose at hand. More details can be found in \cite{eilenberg1943relations, eilenberg1950relations, eilenberg1953groups, eilenberg1954groups, may1992simplicial, may1999concise, hatcher2002algebraic}.

The construction of the classifying space $B(G)$ mimics the construction of the standard $n$-simplex such that the $n$-simplices are now abstract simplices whose vertices are labeled by group variables:

\begin{definition}[\emph{Classifying space}]
	Let $G$ be a finite group and $E(G)$ the simplicial set such that $E(G)_n=G^{n+1}$. The $n$-simplices of $E(G)$ are therefore identified with the ordered ($n$+1)-tuples $(g_0,\ldots,g_n)$, with $g_i \in G$. The boundary of an $n$-simplex $(g_0, \ldots, g_n)$ reads
	\begin{equation}
		\partial^{(n)} (g_0, \ldots, g_n) := \sum_{i=0}^n (-1)^i (g_0, \ldots, \hat{g}_i, \ldots, g_n) \; ,
	\end{equation}
	and the $i$-th degenerate simplex of an $n$-simplex reads
	\begin{equation}
		\eta_i (g_0, \ldots, g_n) := (g_0,\ldots,g_{i-1},g_i,g_i,g_{i+1},\ldots,g_q) \; .
	\end{equation}
	The group $G$ has a left action on $E(G)$ by left multiplication such that for all $g \in G$,
	\begin{equation}
		g \triangleright (g_0, \ldots , g_n) = (gg_0, \ldots, gg_n) \; .
	\end{equation}
	The classifying space $B(G)$ of $G$ is finally defined as the quotient space $B(G) = E(G)/G$. The simplicial set structure of $B(G)$ is inherited from the one of $E(G)$. Furthermore, because of the homeomorphism between $B(G \times G)$ and $B(G) \times B(G)$, the classifying space $B(G)$ inherits the multiplication rule on $G$ as the composite $B(G \times G) \simeq B(G) \times B(G) \to B(G)$, so that $B(G)$ is a simplicial group.
\end{definition}

\noindent
By definition, the $n$-simplices of $B(G)$ satisfy the equivalence relation $(g_0, \ldots, g_n) \sim (gg_0, \ldots, gg_n)$ which implicitly identifies all the 0-simplices (or vertices) of $B(G)$ so that it only contains a single $0$-simplex, namely $(g)$. The presentation of $B(G)$ as constructed above is sometimes referred to as the \emph{homogeneous} one as opposed to the \emph{non-homogeneous} one that we will now present.

Let us consider $n$-tuples $[g_1, \ldots , g_n]$ of elements $g_i \in G$. To each such tuple, we associate an $n$-simplex of the simplicial group $B(G)$ as follows
\begin{equation}
		[g_1,\ldots,g_n] \longrightarrow (\mathbbm{1}, g_1, g_1g_2, \ldots, g_1 \cdots g_n)
\end{equation}
where $\mathbbm{1}$ denotes the group identity. 
Conversely, to each $n$-simplex $(g_0, \ldots, g_n)$, we can assign an $n$-tuple according to
\begin{equation}
	(g_0, \ldots, g_n) \longrightarrow [g_0^{-1}g_1, g_1^{-1}g_2, \ldots, g_{n-1}^{-1}g_n]
\end{equation}
which provides a one-to-one correspondence between the $n$-simplices $(g_0,\ldots,g_n)$ and the $n$-tuples $[g_1, \ldots, g_n]$. In the following, we regard each $n$-simplex of $B(G)$ as such an $n$-tuple so that ordered products of $g_i$ variables label the 1-simplices of $B(G)$. It is straightforward to see that the action of the boundary map $\partial$ can now be rewritten 
\begin{equation}
	\label{boundaryBAR}
	\partial^{(n)} [g_1 , \ldots , g_n] = [g_2, \ldots, g_n] + \sum_{i=1}^{n-1}(-1)^i[g_1, \ldots, g_{i-1}, g_ig_{i+1},g_{i+2}, \ldots , g_n] + (-1)^n [g_1 , \ldots , g_{n-1}] \; .
\end{equation}
Since the definition of the classifying space mimics the one of the standard $n$-simplex, it is easy to see how we can represent geometrically the relations presented above by drawing simplices and labeling their edges with group variables and product of group variables (when working with the non-homogeneous presentation). Note that given a 2-simplex, the oriented product of the group variables labeling its boundary 1-simplices is always equal to the identity,\footnote{This follows directly from the definition of the non-homogeneous presentation. Let us consider the $2$-simplex $[g_1,g_2]$. By applying the boundary map \eqref{boundaryBAR}, we obtain $
	\partial^{(2)}[g_1,g_2] = [g_2]-[g_1g_2]+[g_1] 
$,
which informs us that the 1-simplices bounding $[g_1,g_2]$ are labeled by $g_2$, $g_1g_2$ and $g_1$, respectively. Note that the orientation of $[g_1g_2]$ is opposite to the one of $[g_1]$ and $[g_2]$ so that the oriented product is indeed $g_2 \cdot g_2^{-1}g_1^{-1} \cdot g_1 = \mathbbm{1}$.} hence the correspondence with (1-form) flat connections on the lattice. The fact that we can represent $n$-simplices of $B(G)$ graphically will turn out to be very useful in the following when dealing with more complex formulas.

\medskip \noindent 
As alluded earlier, a simplicial group together with a boundary homomorphism satisfying $\partial \circ \partial \equiv 0$ forms a chain complex whose chain groups are given by $G^{n}$ in the case of $B(G)$. Consequently, we can think of an $n$-tuple $[g_1, \ldots, g_n]$ as an $n$-chain which we choose to be valued in a $G$-module $\mathcal{A}$, defining the space of $n$-chains $C_n(B(G),\mathcal{A}) $. We obtain the dual cohomology by defining an $n$-dimensional cochain over $\mathcal{A}$ as a function which associates to each $n$-simplex of the simplicial group an element of $\mathcal{A}$, so that an $n$-cochain can be thought of as a function of $n$ variables on $G$ valued in $\mathcal{A}$, and by dualizing the boundary operator. The resulting cohomology turns out to be identified with the so-called \emph{group cohomology} whose definition is recalled below so that algebraic cocycles on $G$ are equivalent to simplicial cocycles on $B(G)$:

	\begin{definition}[\emph{Group cohomology}]
	Let $G$ be a finite group and $\mathcal{A}$ a $G$-module which is an abelian group. The group $G$ has an action $\triangleright$ on $\mathcal{A}$ which commutes with the multiplication rule of $\mathcal{A}$. We call an \emph{$n$-cochain} a function $\omega_n:G^n \to \mathcal{A}$ and we denote by $C^n(G,\mathcal{A})$ the space of $n$-cochains. We define the \emph{coboundary} operator $d^{(n)}:C^n(G,\mathcal{A}) \to C^{n+1}(G,\mathcal{A})$ via
	\begin{align}
		&d^{(n)}\omega(g_1, \ldots, g_{n+1}) \\[-0.8em] & \q =  g_1 \triangleright \omega(g_2,\ldots,g_{n+1})\omega(g_1,\ldots,g_n)^{(-1)^{n+1}}\prod_{i=1}^n \omega(g_1,\ldots,g_{i-1},g_i g_{i+1},g_{i+2}, \ldots,g_{n+1})^{(-1)^i} \nn
	\end{align}
	where we chose to write the product rule in $\mathcal{A}$ multiplicatively.
	An \emph{$n$-cocycle} is then defined as an $n$-cochain that satisfies
	\begin{equation}
		\label{eq:defCocCond}
		d^{(n)}\omega_n = 1 \; .
	\end{equation}
	We refer to \eqref{eq:defCocCond} as the \emph{group $n$-cocycle condition} and the subgroup of $n$-cocycles is denoted by $Z^n(G,\mathcal{A})$. Given an ($n$$-$1)-cochain $\omega_{n-1}$, we define an \emph{$n$-coboundary} as an $n$-cocycle of the form
	\begin{equation}
		\omega_n = d^{(n-1)}\omega_{n-1} \; .
	\end{equation}
	The subgroup of $n$-coboundaries is denoted by $B^n(G,\mathcal{A})$. We finally construct the $n$-th (group) \emph{cohomology group} as the quotient space of $n$-cocycles defined up to $n$-coboundaries:
	\begin{equation}
		H^n(G,\mathcal{A}) := \frac{Z^n(G,\mathcal{A})}{B^n(G,\mathcal{A})} = \frac{{\rm ker}(d^{(n)})}{{\rm im}(d^{(n-1)})} \; .
	\end{equation} 
\end{definition}

\bigskip
\noindent
It turns out that the construction of the classifying space $B(G)$ as presented above is not confined to finite groups. More precisely it is possible to generalize it so as to assign to any abelian simplicial group $X$ a classifying space $B(X)$. There exist several such generalizations that yield different simplicial groups but whose cohomology groups are isomorphic. The most well-known generalizing procedure is referred to as the \emph{bar construction} \cite{eilenberg1953groups} of a simplicial group. In the case where $X$ is chosen to be a finite abelian group, $B(X)$ is itself an abelian simplicial group so that the procedure can be iterated. More precisely, starting from the canonical simplicial group constructed out of a finite abelian group $G$, and under this bar construction, it is possible to define a $q$-th classifying space $B^q(G)$ of $G$ recursively as
\begin{equation}
	B^0(G) = G \q , \q B^q(G) = B(B^{q-1}(G)) 
\end{equation}
which is an Eilenberg-MacLane space $K(G,q)$, i.e. 
\begin{equation}
	\pi_n(B^q(G)) = \pi_{n-1}(B^{q-1}(G)) = \ldots = \pi_{n-q}(G) = 
	\begin{cases}
	G \;\; \text{if} \;\, q=n \\ \, 0 \;\;\, \text{otherwise} 
	\end{cases}  \; . 
\end{equation}
We do not expose the details of this bar construction here since it does not serve our purpose well. Instead, we will make use of an alternative construction which we now present.

\subsection{$W$-construction \label{sec:Wconstr}}

In this section, we introduce a construction where the Eilenberg-MacLane space $B^{q+1}(G)$ (or $K(G,q+1)$) is obtained recursively from $B^q(G)$ via a uniform process denoted by $W$. This construction is different from the bar construction but they both yield homologically equivalent complexes. It provides us with a specific presentation for the $n$-cycles of the homology group $H_n(B^q(G),\mathcal{A})$ which, after dualization, will be used to define what we will call the $q$-form cohomology group $H^n(G_{[q]},\mathcal{A})$. This cohomology group is identified with the cohomology group $H^n(B^q(G),\mathcal{A})$ the same way as the cohomology of a finite group is identified with the cohomology its classifying space.\footnote{Note that the group does not need to be abelian as long as we are only interested in the (first) classifying space and, a fortiori, the group cohomology.} The reason why this $W$-construction is more relevant than the bar construction is because the corresponding $n$-cochains are naturally defined as functions of a certain number of variables that is equal to the number group variables necessary to define a flat $q$-form connections on an $n$-simplex. More specifically, under this construction a 2-form $n$-cochain depends on $\frac{n(n-1)}{2}$ group variables as expected from the study of 2-form flat connections.\footnote{Recall that given a finite abelian group $G$ and a manifold $\mathcal{M}$ equipped with a triangulation $\triangle$, we defined a flat 2-form connection by assigning to every $2$-simplex $(012) \subset \triangle$, a group element $g_{012} = \la g , (012) \ra$ such that for every 3-simplex $(0123) \subset \triangle$, the cocycle condition $\la dg, (0123)\ra  = g_{123}-g_{023}+g_{013}-g_{012} = 0$ is imposed. It follows that given an $n$-simplex of $\triangle$ and a flat 2-form connection, only $\frac{n(n-1)}{2}$ variables are independent.} 

In general, given a simplicial group $X$, we define a new simplicial group denoted by $W(X)$ via the recursive formulas
\begin{equation}
		W(X)_0 = \{\la \, \ra\} \q , \q W(X)_{n+1} = X_n \otimes W(X)_{n} 
\end{equation}
where $\la \, \ra$ denotes the single element of $W(X)_0$ and $W(X)_n$ the set of $n$-simplices of $W(X)$. It is also possible to define $W(X)_n$ directly without recursion. Indeed, we have
\begin{equation}
	\label{Wq}
	W(X)_n = X_{n-1} \otimes X_{n-2} \otimes \cdots \otimes X_0 \; .
\end{equation}
We denote the elements of an $n$-fold product by 
\begin{equation}
	\la x_{n-1},x_{n-2}, \ldots , x_0\ra = x_{n-1} \otimes x_{n-2} \otimes \cdots \otimes x_0 \otimes \la \, \ra  
\end{equation} 
such that $x_i \in X_i$.
Furthermore, the face and degeneracy operators $W(X)_n$ satisfy the following properties
\begin{align}
	\nn
	\eta_0 \la \, \ra  &= \la \mathbbm{1}_0 \ra \\ \nn
	\partial_i \la x_0 \ra & = \la \, \ra \\ \nn
	\partial_0 \la x_{n-1} , \ldots , x_0 \ra &= \la x_{n-2}, \ldots, x_0\ra   \\ 
		\label{eq:recurBound}
	\partial_{i}\la x_{n-1}, \ldots, x_0 \ra &= \la \partial_{i-1} x_{n-1} ,\partial_{i-2}x_{n-2}, \ldots , \partial_1x_{n-i+1}, x_{n-i-1}\cdot \partial_0x_{n-i}, x_{n-i-2}, \ldots, x_0\ra \\ \nn
	\partial_{n} \la x_{n-1}, \ldots , x_0\ra & = \la \partial_{n-1} x_{n-1}, \partial_{n-2}x_{n-2},\ldots, \partial_1 x_1 \ra 
	\\ \nn
	\eta_{0} \la x_{n-1}, \ldots, x_0\ra &= \la \mathbbm{1}_{n-1}, x_{n-2}, \ldots , x_0\ra\\ \nn
	\eta_{i} \la x_{n-1}, \ldots, x_0\ra &= \la \eta_{i-1}x_{n-1}, \ldots, \eta_0 x_{n-i}, \mathbbm{1}_{n-i},x_{n-i-1}, \ldots, x_0\ra
\end{align}
where $\mathbbm{1}_i$ denotes the identity element of $X_i$.
Regarding $B^0(G) = G$ as a simplicial group where the $n$-simplices are identified with the elements $g_i \in G$, such that $\partial_i g = g = \eta_i g$, then $W(B^0(G)) \simeq B(G)$ is the simplicial group whose $n$-simplices are the $n$-tuples
\begin{equation}
		\la g_{n-1}, \ldots, g_0\ra \equiv [g_{n-1} , \ldots , g_0]
\end{equation}
which matches exactly the non-homogeneous bar construction of the classifying space sketched earlier (this equality is only valid for $BG$). Applying the uniform process $W$ one more time then provides $B^2(G)$. It follows directly from the definition \eqref{Wq} that an element of $(B^2(G))_n$ explicitly depends on $\sum_{i=0}^{n-1} i = \frac{n(n-1)}{2} = \binom{n}{2}$ variables in $G$, as expected. Let us now illustrate these definitions by looking at the first $2,3,4$-simplices of $B^2(G)$:
\begin{example}\label{ex:3simplex}
		Let us first consider a 3-simplex $\triangle^3:=\la \la g_{3}, g_{2} \ra , \la g_1 \ra , \la \, \ra \ra \in (B^2(G))_{3} = BG_{2} \otimes BG_{1} \otimes BG_{0}$. The boundary of this simplex is obtained via $\partial^{(3)}  = \sum_i (-1)^i \partial_i$ using \eqref{eq:recurBound} such that
	\begin{align}
		\partial_0 \triangle^3 &= \la g_1 , \la \, \ra \ra \q , \q
		\partial_1 \triangle^3 = \la g_{2}+g_1 , \la \, \ra \ra \q , \q
		\partial_2 \triangle^3= \la g_{3} + g_{2} , \la \, \ra \ra \q , \q 
		\partial_3 \triangle^3= \la g_{3} , \la \, \ra \ra 
	\end{align}
	where we chose to write the product rule in $G$ additively since the group is abelian. 
	It follows that the boundary of the $3$-simplex $\la \la g_{3}, g_{2} \ra, \la g_1 \ra , \la \, \ra \ra$ reads 
	\begin{align}
		\partial^{(3)} \triangle^3 &=
		\sum_{i=0}^{4}(-1)^i \partial_i  \la \la g_{3},g_{2}\ra, \la g_1\ra , \la \, \ra \ra \\ \nn
		&= \la g_1, \la \, \ra \ra - \la g_{2}+g_1, \la \, \ra \ra + \la g_{3}+g_{2}, \la \, \ra \ra - \la g_{3}, \la \, \ra \ra \; .
	\end{align}
\end{example}
\medskip
\begin{example}\label{ex:4simplex}
	Let us now consider a 4-simplex $\triangle^4:=\la \la g_{6},g_{5},g_{4}\ra, \la g_{3}, g_{2} \ra , \la g_1 \ra , \la \, \ra \ra \in (B^2(G))_{4} = BG_{3} \otimes  BG_{2} \otimes BG_{1} \otimes BG_{0}$. The boundary of this simplex is obtained via $\partial^{(4)}  = \sum_i (-1)^i \partial_i$ such that
	\begin{align}
		\partial_0 \triangle^4&= \la \la g_{3}, g_{2} \ra, \la g_1 \ra , \la \, \ra \ra \\
		\partial_1 \triangle^4&= \la \la g_{5}+g_{3}, g_{4}+g_{2} \ra, \la g_1 \ra , \la \, \ra \ra \\
		\partial_2 \triangle^4&= \la \la g_{6}+g_{5}, g_{4} \ra, \la g_{2} + g_{1} \ra , \la \, \ra \ra \\
		\partial_3 \triangle^4&= \la \la g_{6}, g_{5}+g_{4} \ra, \la g_{3}+g_{2} \ra , \la \, \ra \ra \\
		\partial_4 \triangle^4&= \la \la g_{6}, g_{5} \ra , \la g_{3}\ra , \la \, \ra \ra   \; .
	\end{align} 
	For the sake of clarity, let us develop one of the computations. For instance, we have
	\begin{align}
		\partial_2 \la \la g_{6},g_{5},g_{4}\ra, \la g_{3}, g_{2} \ra , \la g_1 \ra , \la \, \ra \ra &= \la \partial_1 \la g_{6},g_{5},g_{4}\ra, \partial_0 \la g_{3}, g_{2} \ra \cdot  \la g_1 \ra , \la \, \ra \ra \\
		& = \la \la \partial_0 g_{6}+  g_{5},g_{4}\ra, \la g_{2} \ra \cdot \la g_1 \ra , \la \, \ra \ra \\
		&= \la \la g_{6} + g_{5} , g_{4}\ra, \la g_{2}+g_1\ra , \la \, \ra \ra
	\end{align}
	where we used the fact that $\partial_i g = g$.
\end{example}
\medskip
\begin{example}\label{ex:5simplex}
	Let us consider a 5-simplex $\triangle^5:=\la \la g_{10},g_{9},g_{8},g_{7}\ra, \la g_{6},g_{5},g_{4} \ra, \la g_{3},g_{2} \ra, \la g_1 \ra, \la \, \ra \ra \in (B^2(G))_{5} = BG_{4} \otimes \cdots \otimes BG_{0}$. The boundary of this simplex is obtained via $\partial^{(5)}  = \sum_i (-1)^i \partial_i$ such that 
	\begin{align}
		\partial_0 \triangle^5&= \la \la g_{6}, g_{5}, g_{4} \ra, \la g_{3}, g_{2} \ra, \la g_1 \ra , \la \, \ra \ra \\
		\partial_1 \triangle^5&= \la \la g_{9}+ g_{6}, g_{8}+g_{5}, g_{7}+g_{4}\ra, \la g_{3}, g_{2} \ra, \la g_1 \ra , \la \, \ra \ra \\
		\partial_2 \triangle^5&= \la \la g_{10}+g_{9}, g_{8}, g_{7}\ra, \la g_{5}+g_{3}, g_{4}+g_{2} \ra, \la g_{1} \ra , \la \, \ra  \ra \\
		\partial_3 \triangle^5&= \la \la g_{10}, g_{9}+g_{8}, g_{7}\ra ,\la g_{6}+ g_{5},g_{4} \ra, \la g_{2}+g_1 \ra , \la \, \ra \ra \\
		\partial_4 \triangle^5&= \la \la g_{10}, g_{9}, g_{8}+g_{7}\ra,\la g_{6}, g_{5} +g_4\ra , \la g_{3}+g_2\ra , \la \, \ra \ra  \\
		\partial_5 \triangle^5&= \la \la g_{10}, g_{9}, g_{8}\ra, \la g_{6}, g_{5}\ra, \la g_{3}\ra , \la \, \ra \ra \; .
	\end{align} 
\end{example}

\noindent
As for the classifying space $B(G)$, since $\partial$ is a homomorphism of the group structure of $B(G)$ inherited from $G$ and since $\partial \circ \partial \equiv 0$, we can define a homology theory of the simplicial group by considering finite chains valued in a $G$-module $\mathcal{A}$ which we identify with the $n$-simplices. More specifically, we assign for instance to a $4$-simplex $\la \la g_{6},g_{5},g_{4} \ra, \la g_{3},g_{2} \ra, \la g_1 \ra , \la \, \ra \ra$ a 4-chain. Similarly, we can define a cohomology theory by defining an $n$-dimensional cochain over the $G$-module $\mathcal{A}$ as a function which associates to each $n$-simplex of $B^2(G)$ an element of the $G$-module $\mathcal{A}$ and by defining the coboundary map $d$ dual to $\partial$. Using the $W$-construction of $B^2(G)$, this provides $n$-cochains as functions of $\frac{n(n-1)}{2} = \binom{n}{2}$ variables on $G$ valued in $\mathcal{A}$, as required from the definition of flat 2-form connections. We refer to this cohomology as the $2$-form cohomology whose fundamental properties are explored in the next section.

\section{2-form (co)homology \label{sec:2formco}}
In the previous section, we recalled the construction of the classifying space of a finite group $G$ as a simplicial group, and explained how this becomes a chain complex when identifying the $n$-simplices with $n$-chains, from which we can define the cohomology $H^n(BG, \mathcal{A})$ of simplicial cocycles that is equivalent to the cohomology $H^n(G, \mathcal{A})$ of algebraic cocycles. In this section, we use the $W$-construction of the second classifying space $B^2(G)$ of a finite abelian group $G$ in order to define the so-called 2-form cohomology $H^n(G_{[2]},\mathcal{A})$ that is the cohomology of algebraic cocycles identified with the cohomology $H^n(B^2(G),\mathcal{A})$ of simplicial cocycles.

\subsection{Definition and 2-form cocycle conditions}

Let $G$ be a finite abelian group and $\mathcal{A}$ an abelian group that is a $G$-module whose product rule is written multiplicatively. We assume for notational convenience that $G$ has a trivial action on the abelian group $\mathcal{A}$.\footnote{Relaxing this assumption could allow to explore orientation-reversing elements.} We call a 2-form \emph{$n$-cochain} a function $\omega_n:G^{\binom{n}{2}} \to \mathcal{A}$. We denote by $C^n(G_{[2]},\mathcal{A})$ the space of 2-form $n$-cochains. Given a 2-form $n$-cochain $\omega \in C^n(G_{[2]},\mathcal{A})$, we make use of the following notation
\begin{equation}
	\omega(g_{\binom{n}{2}},g_{\binom{n}{2}-1},\ldots | g_{\binom{n-1}{2}},\ldots | \ldots | g_3, g_2 | g_1) \in \mathcal{A}
\end{equation}
where the presence of \emph{bars} $|$ is there to remind of the underlying tensor product structure of the corresponding $n$-simplices according to \eqref{Wq}. We find this notation convenient in order to make the algebraic structure more manifest but this can easily be omitted as well.   

We then define the 2-form \emph{coboundary} operator $d^{(n)}:C^n(G_{[2]},\mathcal{A}) \to C^{n+1}(G_{[2]},\mathcal{A})$ as the canonical dual of the boundary operator $\partial$ on the space of 2-form $n$-chains, where $\partial$ is inherited from the boundary operator of the simplicial group $B^2(G)$ as provided by the $W$-construction. More precisely, we think of a given evaluation of the $n$-cochain $\omega \in C^n(G_{[2]},\mathcal{A})$ as a pairing between $\omega$ and the corresponding $n$-simplex in $B^2(G)$ identified with the relevant $n$-chain, i.e.
\begin{equation} \nn
	\omega(g_{\binom{n}{2}},g_{\binom{n}{2}-1},\ldots | g_{\binom{n-1}{2}},\ldots | \ldots | g_3, g_2 | g_1) =:
	\la \, \omega \,  , \triangle^n \ra 
\end{equation}
with 
\begin{equation}
	\triangle^n :=
	\la \la g_{\binom{n}{2}},g_{\binom{n}{2}-1},\ldots \ra , 
	\la g_{\binom{n-1}{2}},\ldots \ra,  \ldots , \la g_3, g_2 \ra, \la g_1 \ra, \la \, \ra \ra \in (B^2(G))_{n} = BG_{n-1} \otimes \cdots \otimes BG_{0}
\end{equation}
so that the action of the coboundary operator can be defined directly in terms of the boundary operator on the $n$-simplex via Stoke's theorem
\begin{equation}
	\la \, d^{(n)} \omega \, , \, \triangle^{n+1 }\, \ra  = \la \, \omega \, , \,  \partial^{(n+1)} \triangle^{n+1} \, \ra \; .
\end{equation} 
In particular, it follows from ex.~\ref{ex:3simplex}, \ref{ex:4simplex} and $\ref{ex:5simplex}$ the action of the 2-form coboundary operator $d^{(2)}$:

\begin{equation}
	\boxed{
	d^{(2)}\beta(a,b|c):= \big\la d^{(2)}\beta, \la \la a,b \ra, \la c \ra, \la \, \ra \big\ra
	= \big\la \beta, \partial^{(3)}\la \la a,b \ra, \la c \ra, \la \, \ra \big\ra
	= \frac{\beta(c)\, \beta(a+b)}{\beta(b+c)\, \beta(a)} 
	} \; ,
\end{equation}
the action of the 2-form coboundary operator $d^{(3)}$:
\begin{equation}
	\label{eq:2formCoc3}
	\boxed{
	d^{(3)}\alpha(a,b,c|d,e|f) =\frac{\alpha(d,e|f) \, \alpha(a+b,c|e+f) \, \alpha(a,b|d)}{\alpha(b+d,c+e|f) \, \alpha(a,b+c|d+e)} 
	} \; ,
\end{equation}
and the action of the 2-form coboundary operator $d^{(4)}$:
\begin{equation}
	\label{eq:2formCoc4}
	\boxed{
	d^{(4)}\omega(a,b,c,d|e,f,g|h,i|j) 
	= \frac{\omega(e,f,g|h,i|j)\, \omega(a+b,c,d|f+h,g+i|j)\, \omega(a,b,c+d|e,f+g|h+i)}{\omega(b+e,c+f,d+g|h,i|j)\, \omega(a,b+c,d|e+f,g|i+j)\, \omega(a,b,c|e,f|h)} 
	} \; ,
\end{equation}
respectively, where $a,b,\ldots,i,j \in G$. Using the general recursive definition \eqref{eq:recurBound} together with the correspondence spelled out above, we can then find the defining formula of any 2-form coboundary operator $d^{(n)}$. Nevertheless, we will only make use of the previous three formulas in this work. A 2-form \emph{$n$-cocycle} is then defined as a 2-form $n$-cochain that satisfies
\begin{equation}
	d^{(n)}\omega_n = 1 \; .
\end{equation}
The subgroup of 2-form $n$-cocycles is denoted by $Z^n(G_{[2]},\mathcal{A})$. Given a 2-form ($n$$-$1)-cochain $\omega_{n-1}$, we define an \emph{$n$-coboundary} as a 2-form $n$-cocycle of the form
\begin{equation}
	\omega_n = d^{(n-1)}\omega_{n-1} \; .
\end{equation}
The subgroup of 2-form $n$-coboundaries is denoted by $B^q(G_{[2]},\mathcal{A})$. We finally construct the 2-form $q$-th \emph{cohomology group} as the quotient space of 2-form $n$-cocycles defined up to 2-form $n$-coboundaries:
\begin{equation}
	H^n(G_{[2]},\mathcal{A}) := \frac{Z^n(G_{[2]},\mathcal{A})}{B^n(G_{[2]},\mathcal{A})} = \frac{{\rm ker}(d^{(n)})}{{\rm im}(d^{(n-1)})} \; .
\end{equation}

\subsection{Geometric realization\label{sec:geomInter}}
We mentioned above how we can think of a given evaluation of the cocycle $\omega$ as a pairing between $\omega$ and the corresponding $n$-simplex identified with the relevant $n$-chain so that a 2-form $n$-cocycle assigns to each $n$-simplex of $B^2(G)$ an element of the group $\mathcal{A}$. This can be used in order to provide a geometric interpretation to the cocycle conditions presented above.

In order to do so, we need the geometric realization of $B^2(G)$ which is, loosely speaking, obtained by identifying a given $n$-simplex of $B^2(G)$ with a standard $n$-simplex together with a $G_{[2]}$-coloring, i.e. an assignment of group elements $g \in G$ to every 2-simplex such that for every 3-simplex the 2-cocycle constraint $dg=0$ is satisfied. Let us for instance consider the $3$-simplex $\la \la g_{3}, g_{2} \ra, \la g_1 \ra , \la \, \ra \ra \in (B^2(G))_3$. Recall that its boundary reads
\begin{align}
	\partial^{(3)} \la \la g_{3},g_{2}\ra, \la g_1\ra, \la \, \ra \ra &= \la g_1, \la \, \ra \ra - \la g_{2}+g_1, \la \, \ra \ra + \la g_{3}+g_{2}, \la \, \ra \ra - \la g_{3}, \la \, \ra \ra
\end{align}
so that $\la g_1 , \la \, \ra \ra$, $\la g_{20}+g_1 , \la \, \ra \ra$, $\la g_{21}+g_{20} , \la \, \ra \ra$ and $\la g_{21} , \la \, \ra \ra$ are 2-simplices. Let us now think about the 3-simplex $\la \la g_{3}, g_{2} \ra, \la g_1 \ra , \la \, \ra \ra \in (B^2(G))_3$ as a standard 3-simplex $\snum{(0123)}$ so that $\snum{0}$, $\snum{1}$, $\snum{2}$ and $\snum{3}$ label its vertices, together with a $G_{[2]}$-coloring which assigns to every face $(xyz)$ the group element $g_{xyz} = \la g , (xyz)\ra$. From the expression of the boundary of the 3-simplex $\la \la g_{3},g_{2}\ra, \la g_1\ra, \la \, \ra \ra$, we read-off the correspondence
\begin{align} 
	\label{eq:color2Simp}
	g_{123} = g_1 \q , \q  g_{023} = g_{2} + g_1 \q , \q g_{013} = g_{3}+ g_{2} \q , \q g_{012} = g_{3} \; .
\end{align}	
We can check that this coloring automatically satisfies the cocycle constraint $dg=0$ since $\la dg , \snum{(0123)}\ra = g_{123}-g_{023}+g_{013}-g_{012} = g_1-(g_2+g_1)+(g_3+g_2)-g_3=\zo$. So given a 3-simplex of $B^2(G)$, we can assign to it a standard $3$-simplex whose $G_{[2]}$-coloring is provided by \eqref{eq:color2Simp}. Conversely, given a standard 3-simplex $\snum{(0123)}$ with a given $G_{[2]}$-coloring, we assign to it a 3-simplex of $B^2(G)$ which reads $\la \la g_{012},g_{013}-g_{012}\ra,\la g_{123}\ra, \la \, \ra \ra$. Since a 2-form 3-cocycle assigns to a 3-simplex of $B^2(G)$ an element of the group $\mathcal{A}$, we can use the previous correspondence to further assign to the standard $3$-simplex $\snum{(0123)}$ the evaluation
\begin{equation}
	\label{eq:corres3Coc}
	 \la \alpha , \snum{(0123)} \ra 
	\equiv \alpha(g_{012}, g_{013}-g_{012}| g_{123}) \; .
\end{equation}
Note that we use a slightly abusive notation since we treat the standard $3$-simplex $\snum{(0123)}$ as $\la \la g_{012},g_{013}-g_{012}\ra,\la g_{123}\ra, \la \, \ra \ra$ in light of the identification we have just made. The same procedure can be iterated so as to assign to a given standard $n$-simplex together with a $G_{[2]}$-coloring the evaluation of a 2-form $n$-cocycle. For instance, we will make extensive use of the following correspondence between a 2-form 4-cocycle and the standard 4-simplex $\snum{(01234)}$:
\begin{equation}
	\label{eq:pairing4Coc}
	\boxed{
	\la \omega , \snum{(01234)} \ra 
	=\omega(g_{012}, g_{013}-g_{012}, g_{014} - g_{013}| 
	g_{123}, g_{124}-g_{123}| g_{234})} \; .
\end{equation}
Similarly, the evaluation of a 2-form 5-cocycle assigned to the standard 5-simplex $\snum{(012345)}$ reads
\begin{align*}
	 & \la \pi, \snum{(012345)}\ra \\ \nn & \q = \pi(g_{012}, g_{013}-g_{012}, g_{014}-g_{013}, g_{015}-g_{014}|g_{123},g_{124}-g_{123},g_{125}-g_{124}|g_{234},g_{235}-g_{234}|g_{345})
\end{align*}
from which we can easily read-off the structure underlying this construction and therefore `guess' the subsequent formulas.

Let us now use this correspondence in order to provide a geometric interpretation to the cocycle conditions. It turns out that an $n$-cocycle condition is associated with a so-called $n$-dimensional \emph{Pachner move} \cite{PACHNER1991129}. Given a \emph{piecewise linear manifold} $\mathcal{M}$ and its triangulation $\triangle$, a Pachner move replaces $\triangle$ by another triangulation $\triangle'$ associated with a manifold $\mathcal{M}'$ homeomorphic to $\mathcal{M}$. Given two triangulations of a given manifold, it is always possible to relate one to the other by a finite sequence of Pachner moves. In three dimensions, we distinguish two Pachner moves, namely the 2--3 and the 1--4 move denoted by $\mathcal{P}_{2 \mapsto 3}$ and $\mathcal{P}_{1 \mapsto 4}$, respectively. The 1--4 move subdivides a 3-simplex into four 3-simplices by introducing an additional vertex inside, while the 2--3 move decomposes the gluing of two 3-simplices into three 3-simplices. Graphically, this latter move can be represented as
\begin{equation}
	\label{eq:23Pachner}
	\mathcal{P}_{2 \mapsto 3} : \PPONE{1}{1}{-0.3} \longmapsto \PPTWO{1}{1}{-0.3} \; ,
\end{equation}
where the initial 3-simplices $\snum{(0123)}$ and $\snum{(0234)}$ are replaced by $\snum{(0124)}$, $\snum{(1234)}$ and $\snum{(0134)}$. Let $\alpha$ be a 2-form 3-cochain. Using the correspondence \eqref{eq:corres3Coc}, we assign to each one of the five 3-simplices $\triangle^3=(wxyz)$ a term $\la \alpha, (wxyz)\ra$, and we finally obtain the 2-form 3-cocycle condition as
\begin{equation}
	\prod_{i=1}^5 \la \alpha, \triangle^3_i\ra^{\epsilon(\triangle^3_i)} = 1
\end{equation}   
where $\epsilon(\triangle^3)= \pm 1$ is a sign factor which depends on the orientation of each $3$-simplex as determined by the following convention:

\begin{convention}[\emph{Orientation convention of a 3-simplex\label{conv:or3S}}]
	Pick one of the 2-simplices bounding the 3-simplex $\triangle^3$ and look at the remaining vertex through this triangle. If the vertices of the 2-simplex are ordered in a clock-wise fashion, then the orientation is positive, otherwise it is negative. For instance, we have
	\begin{equation}
		\epsilon \bigg( \convONE{0.7}{0}{1}{2} \bigg) = +1 \q , \q
		\epsilon \bigg( \convONE{0.7}{0}{2}{1} \bigg) = -1
	\end{equation}
	where $\otimes$ represent the fourth vertex as seen from behind the triangle.
\end{convention}
\noindent
Putting everything together, the 2-form 3-cocycle condition associated with the $\mathcal{P}_{2 \mapsto 3}$ move \eqref{eq:23Pachner} reads 
\begin{align}	
	&d^{(3)}\alpha(g_{012}, g_{013}-g_{012}, g_{014} - g_{013}| 
	g_{123}, g_{124}-g_{123}| g_{234}) \\
	& \q =\frac{\alpha(g_{123},g_{124}-g_{123}|g_{234}) \, \alpha(g_{013},g_{014}-g_{013}|g_{134}) \, \alpha(g_{012},g_{013}-g_{012}|g_{123})}{\alpha(g_{023},g_{024}-g_{023}|g_{234}) \, \alpha(g_{012},g_{014}-g_{012}|g_{124})} 
\end{align}
which reproduces exactly \eqref{eq:2formCoc3} when choosing $g_{012} = a$, $g_{013}-g_{012} = b$, $g_{014}-g_{013} = c$, $g_{123} = d$, $g_{124}-g_{123} = e$ and $g_{234} = f$. Similarly, we can provide a geometric interpretation for any cocycle condition. In the following, these geometric interpretations will be put to use in order to define our lattice Hamiltonian model.

\subsection{Normalization conditions \label{sec:normCond}}

In the following section, we construct a lattice Hamiltonian model whose input data is a finite abelian group $G$ and a 2-form cohomology class $[\omega] \in H^4(G_{[2]},\rU(1))$. Before doing so, it is convenient to derive some normalization conditions for 2-form cocycles. This will allow us to greatly simplify some formulas but also to provide useful information regarding the algebraic structure of the cocycles. The list of normalization conditions presented here may not be exhaustive but these are all the ones we need for our purpose. 

By definition, $[\omega]\in H^n(G_{[2]},\rU(1))$ is an equivalence class of 2-form $n$-cocycles defined up to 2-form $n$-coboundaries and in order to represent the class $[\omega]$, we can choose any cocycle $\omega \in [\omega]$. The purpose of this section is, given an equivalence class $[\omega]$, to find a representative $\omega \in [\omega]$ which satisfies as many normalization conditions as possible. Since the data entering the definition of our lattice model is a 2-form 4-cocycle, we will focus our attention on the cohomology group $H^4(G_{[2]},\rU(1))$ but the strategy presented here is very general and can be applied to any 2-form cocycle.

\bigskip \noindent
Let $[\omega]$ be a 2-form cohomology class in $H^4(G_{[2]},\rU(1))$ and let $\omega, \omega' \in [\omega]$ be two different representatives of this class. By definition, 4-cocycles within the same cohomology class are equivalent up to 4-coboundaries, and therefore there exists a 2-form 3-cochain $\alpha$ such that
\begin{align}
	\nn
	\omega'(a,b,c|d,e|f) &= \omega(a,b,c|d,e|f)\cdot d^{(3)}\alpha(a,b,c|d,e|f) \\
	\label{eq:upto4Cbdry}
	& = \omega(a,b,c|d,e|f) \cdot
	\frac{\alpha(d,e|f) \, \alpha(a+b,c,|e+f) \, \alpha(a,b|d)}{\alpha(b+d,c+e|f) \, \alpha(a,b+c|d+e)} \; .
\end{align}
In other words, given a representative $\omega \in [\omega] $, we can obtain another representative $\omega' \in [\omega]$ by choosing a 2-form 3-cochain $\alpha \in C^3(G_{[2]},\mathcal{A})$ and apply formula \eqref{eq:upto4Cbdry}. For instance, choosing $\alpha$ such that $\alpha(\zo,\zo|\zo)=\omega^{-1}(\zo,\zo,\zo|\zo,\zo|\zo)$, we obtain a representative $\omega'$ such that $\omega'(\zo,\zo,\zo|\zo,\zo|\zo)=1$. Therefore, it is always possible to choose a 4-cocycle $\omega$ such that $\omega(\zo,\zo,\zo|\zo,\zo|\zo)=1$. This is the simplest normalization condition. We will now apply the same strategy to derive several additional normalization conditions. 

Let us consider equation \eqref{eq:upto4Cbdry} for which $a =b =c =\zo$:
\begin{align}
	\nn
	\omega'(\zo,\zo,\zo|d,e|f) 
	& = \omega(\zo,\zo,\zo|d,e|f) \cdot
	\frac{\alpha(d,e|f) \, \alpha(\zo,\zo,|e+f) \, \alpha(\zo,\zo|d)}{\alpha(d,e|f) \, \alpha(\zo,\zo|d+e)}
	\\ &= \omega(\zo,\zo,\zo|d,e|f) \cdot
	\frac{ \alpha(\zo,\zo,|e+f) \, \alpha(\zo,\zo|d)}{ \alpha(\zo,\zo|d+e)} \; .
\end{align}
By choosing $\alpha \in C^3(G_{[2]}, \rU(1))$ such that $\alpha(\zo,\zo|x) := \omega(\zo,\zo,\zo|\zo,\zo|x)^{-1}$ and using the cocycle condition 
\begin{equation}
	d^{(4)}\omega(\zo,\zo,\zo,\zo|\zo,\zo,\zo|h,i|j) 
	= \frac{\omega(\zo,\zo,\zo|h,i|j)\, \omega(\zo,\zo,\zo|h,i|j)\, \omega(\zo,\zo,\zo|\zo,\zo|h+i)}{\omega(\zo,\zo,\zo|h,i|j)\, \omega(\zo,\zo,\zo|\zo,\zo|i+j)\, \omega(\zo,\zo,\zo|\zo,\zo|h)} = 1
\end{equation}
we obtain that $\omega'(\zo,\zo,\zo|d,e|f)= 1 $ for all $d,e,f \in G$. So we can always choose a 4-cocycle $\omega$ which satisfies $\omega(\zo,\zo,\zo|d,e|f)= 1$. Consider now the following equation:
\begin{align}
	\omega'(\zo,b,c|\zo,\zo|\zo) 
	& = \omega(\zo,b,c|\zo,\zo|\zo) \cdot
	\frac{\alpha(\zo,b|\zo)}{\alpha(\zo,b+c|\zo)} \; .
\end{align}
Choosing $\alpha \in C^3(G_{[2]},\mathcal{A})$ such that $\alpha(\zo,x|\zo) = \omega(\zo,x,-x|\zo,\zo|\zo)^{-1}$ and using the cocycle condition
\begin{equation}
	d^{(4)}\omega(\zo,b,c,- b - c|\zo,\zo,\zo|\zo,\zo|\zo) = \frac{\omega(\zo,b,-b|\zo,\zo|\zo)}{\omega(\zo,b+c,-b-c|\zo,\zo|\zo)\, \omega(\zo,b,c|\zo,\zo|\zo)}=  1
\end{equation}
we find that $\omega'(\zo,b,c|\zo,\zo|\zo) = 1$. Similarly, it is always possible to find an $\omega$ which satisfies $\omega(a,\zo,c|\zo,\zo|\zo) = 1 =  \omega(a,b,\zo|\zo,\zo|\zo)$. 

Let $\omega \in Z^4(G{[2]},\mathcal{A})$ be a 2-form 4-cocycle which fulfills all the normalization conditions derived above. It then follows directly from the cocycle condition that it also satisfies
\begin{align}
	 \omega(a,\zo,c|\zo,e|\zo) = \omega(\zo,b,c|\zo,\zo|f) = \omega(a,\zo,\zo|d,\zo|\zo) =  \omega(a,b,\zo|d,\zo|\zo) = 1 \; .
\end{align}
We can find another representative of the cohomology class $[\omega]$ according to
\begin{equation}
	\omega'(a,\zo,\zo|d,e|\zo) = \omega(a,\zo,\zo|d,e|\zo) \cdot
	\frac{\alpha(a,\zo|e)\, \alpha(a,\zo,|d)}{\alpha(a,\zo|d+e)} \; .
\end{equation}
Choosing $\alpha \in C^3(G_{[2],M})$ such that $\alpha(x,\zo|z) = \omega(x,\zo,\zo|\zo,\zo|z)^{-1}$ and using the cocycle condition $d^{(4)}\omega(a,\zo,\zo,\zo|\zo,\zo,\zo|h,i|\zo) = 1$, we have $\omega'(a,\zo,\zo|d,e|\zo)=1$ for all $a,d,e \in G$. Similarly, considering
\begin{equation}
	\omega'(\zo,b,\zo|d,\zo|f) = \omega(\zo,b,\zo|d,\zo|f) \cdot
	\frac{\alpha(d,\zo|f)\, \alpha(b,\zo,|f)}{\alpha(b+d,\zo|f)}
\end{equation}
and choosing $\alpha(x,\zo|z) = \omega(x,\zo,\zo|\zo,\zo|z)^{-1}$ with the cocycle condition $d^{(4)}\omega(\zo,b,\zo,\zo|e,\zo,\zo|\zo,\zo|j) = 1$, we have $\omega'(\zo,b,\zo|d,\zo|f)=1$ for all $b,d,f \in G$. Let us also remark that the normalization conditions above imply that 
\begin{equation}
	\label{eq:Equalities3C}
	\omega(\zo,b,\zo|f,g|\zo) = \omega(b,f,g|\zo,\zo|\zo)^{-1} \q {\rm and} \q \omega(\zo,\zo,c|\zo,f|i) = \omega(\zo,c,\zo|f,i|\zo)^{-1} \; . 
\end{equation}
These last two identities will turn out to be very useful in the following. Putting everything together, it is always possible to find a 2-form cocycle $\omega \in Z^4(G_{[2]},\mathcal{A})$ which satisfies the following normalization conditions
\begin{empheq}[box=\fbox]{align}
	\label{normCond2}
	\omega(\zo,b,c|\zo,\zo|\zo)  = \omega(a,\zo,c|\zo,\zo|\zo)  = \omega(a,b,\zo|\zo,\zo|\zo) &= 1 = \omega(\zo,\zo,\zo|d,e|f)\\
	\label{normCond3}
	\omega(a,\zo,\zo|d,e|\zo)  = \omega(a,b,\zo|d,\zo|\zo) = \omega(\zo,b,\zo|d,\zo|f) &= 1 =
	\omega(a,\zo,c|\zo,e|\zo)  = \omega(\zo,b,c|\zo,\zo|f)
\end{empheq}
for all $a,b,c,d,e,f \in G$.

\bigskip \noindent 
Let us close this section with a few remarks regarding 2-form 3-cocycles.
Let us consider $[\alpha] \in H^3(G_{[2]},\mathcal{A})$ and let $\alpha \in [\alpha]$ be a representative. By definition $\alpha$ satisfies the 2-form 3-cocycle condition
\begin{equation}
	\alpha(d,e|f) \, \alpha(a+b,c|e+f) \, \alpha(a,b|d) =\alpha(b+d,c+e|f) \, \alpha(a,b+c|d+e)
\end{equation}
from which immediately follows that $\alpha(\zo,\zo|\zo)= 1$. Similarly, by looking at the cocycle conditions $d^{(3)}\alpha(a,\zo,\zo|\zo,\zo|f)=1$ and $d^{(3)}\alpha(\zo,\zo,c|\zo,\zo|\zo)=1$, we obtain that 
\begin{equation}
	\alpha(a,\zo|f) = 1 = \alpha(\zo,c|\zo) 
\end{equation}
for all $a,c,f \in G$, respectively. Let us furthermore consider the following cocycle condition
\begin{equation}
	\label{eq:cobounA}
	d^{(3)}\alpha(a,b,c|\zo,\zo|\zo) =\frac{\alpha(\zo,\zo|\zo)\, \alpha(a+b,c,|\zo)\, \alpha(a,b|\zo)}{\alpha(b,c|\zo)\,\alpha(a,b+c|\zo)} =1 \; .
\end{equation}
Using the fact that $\alpha(\zo,\zo|\zo)=1$, we deduce that $\beta \in Z^2(G,{\rm U}(1))$ where $\beta(a,b)\equiv \alpha^{-1}(a,b|\zo) $. In other words, the 2-form cocycle $\alpha$ evaluated on $\la \la a,b \ra, \la \zo \ra , \la \, \ra\ra$ satisfies the group 2-cocycle condition, i.e.
\begin{equation}
	d^{(2)}\beta(a,b,c)=\frac{\beta(b,c)\beta(a,b+c)}{\beta(a+b,c)\beta(a,b)}=1 \; .
\end{equation} We further deduce from the cocycle condition $d^{(3)}\alpha(\zo,b,\zo|\zo,e|\zo)=1$ that $\alpha(b,e|\zo)= \alpha(\zo,b|e)^{-1}$ which together with $d^{(3)}\alpha(\zo,\zo,c|d,\zo|\zo) = 1$ provides
\begin{equation}
	\label{eq:cobounR}
	\boxed{
		d^{(3)}\alpha(\zo,\zo,c|d,\zo|\zo) = \frac{\alpha(c,d|\zo)}{\alpha(d,c|\zo)} \equiv \frac{\beta(d,c)}{\beta(c,d)}=1 } 
\end{equation}
so that $\beta$ defines a \emph{symmetric} group 2-cocycle. This special 3-coboundary will be important in the following.

\newpage
\section{Hamiltonian realization of 2-form TQFTs\label{sec:Ham2Forms}}
In this section, we use the results of the previous section in order to construct and study an exactly solvable model whose ground state is described by a topological lattice 2-form gauge theory. In the next section, we will study how this model turns out to be related to the Walker-Wang model for abelian braided fusion categories.

\subsection{Fixed point wave functions \label{sec:fixedWaves}}

One can in general define gapped phases of matter in terms of equivalence classes of many body wave functions under \emph{local unitary transformations} \cite{Levin:2004mi, Chen:2010gda}. Given a graph or a lattice, these local transformations can be used so as to implement a wave function \emph{renormalization group flow}. Defining equivalence classes of wave functions under such transformations then boils down to finding so-called \emph{fixed point wave functions}. By definition, these fixed point wave functions are expected to capture the \emph{long-range entanglement} pattern which is the defining feature of intrinsic topological order. 

Levin and Wen introduced in \cite{Levin:2004mi} so-called \emph{string net models} as a way to systematically construct fixed point wave functions in two dimensions. A string net is essentially defined in terms of a graph labeled by objects satisfying compatibility conditions such that each graph with a given consistent labeling represents a state (or many-body wave function). The linear superposition of spatial configurations of string nets define the Hilbert space of the model. Local unitary transformations are defined at the level of the graph and uniquely specify the fixed point wave functions. These fixed point wave functions are then found to be ground states of given Hamiltonians.  

In this section we follow an approach similar to Levin and Wen to construct an exactly solvable model whose fixed point wave functions define the ground states of a lattice Hamiltonian which has a 2-form gauge theory interpretation. Our setup is the following: Let $\mathcal{M}$ be a compact oriented four-manifold and $\Sigma$ a closed three-dimensional hypersurface equipped with a triangulation $\triangle$. Each 2-simplex $\triangle^2 = (xyz) \subset \triangle$ of this triangulation is decorated by a group element $g_{xyz} \in G$ with $G$ a finite abelian group such that the group identity $\zo \in G$ is the vacuum sector. Furthermore, we define compatibility conditions referred to as \emph{branching rules} at every $3$-simplex of the triangulation: Given a 3-simplex $\triangle^3 =(wxyz) \subset \triangle$, the branching rules impose that the oriented product of the super-selection sectors labeling the 2-simplices vanishes, i.e. $g_{xyz} - g_{wyz} + g_{wxz} - g_{wxy} = 0$. Using the differential on cochains and the canonical pairing between simplices and cochains, this can be rewritten: $\forall \, (wxyz) \subset \triangle$, $\la dg, (wxyz)\ra = \zo$. A labeling of the 2-simplices of $\triangle$ such that the branching rules are everywhere satisfied is said to be $consistent$ and we refer to it as a $G_{[2]}$-coloring. The set of $G_{[2]}$-colorings is denoted by ${\rm Col}(\Sigma, G_{[2]})$ and defines a local description of the set of flat 2-form connections.

Since the branching rules effectively reduce to the group multiplication, the input data of our lattice Hamiltonian model is particularly simple, namely a tuple $\{G, [\omega]\}$ where $G$ is a finite abelian group and $[\omega] \in H^4(G_{[2]},\rU(1))$ a class of 2-form 4-cocycles valued in $\rU(1)$. Since the model is defined in terms of a 2-form 4-cocycle up to 4-coboundaries, we can choose whichever representative of the class we want to carry out our calculations: We choose it so as to satisfy the normalization conditions \eqref{normCond2} and \eqref{normCond3} derived in sec.~\ref{sec:normCond}. Furthermore, the fixed point wave functions are defined as the states satisfying the following relations under the corresponding local unitary transformations:
\begin{align}
 	\label{eq:loc23}
	\Bigg| \PPONE{1}{1}{-0.3} \Bigg\rangle &= \sum_{\substack{g_{014},g_{134} \\ g_{124}}} \la \omega , \snum{(01234)}\ra^{\epsilon(01234)} \Bigg| \PPTWO{1}{1}{-0.3} \Bigg\rangle \\
	\label{eq:loc41}
	\Bigg| \PFourOne{1}{1}{-0.5} \Bigg\rangle &=  \la \omega , \snum{(01234)} \ra ^{-\epsilon(01234)} \Bigg| \PFourOne{1}{2}{-0.5} \Bigg\rangle	 
\end{align}
where the $G_{[2]}$-coloring is left implicit. These two equations dictate how a fixed point wave function is modified under a 2--3 Pachner move $\mathcal{P}_{2 \mapsto 3}$ and a 4--1 Pachner move $\mathcal{P}_{4 \mapsto 1}$, respectively. The equations associated with the opposite moves, namely $\mathcal{P}_{3 \mapsto 2}$ and $\mathcal{P}_{1 \mapsto 4}$, are obtained in an obvious way. Both equation \eqref{eq:loc23} and \eqref{eq:loc41} depends on a factor $\la \omega , \snum{(01234)} \ra^{\pm \epsilon(01234)}$ that is the pairing between the 2-form 4-cocycle $\omega$ and the 2-form chain identified with the standard 4-simplex $\snum{(01234)} \subset \triangle$ that is the 4-simplex bounded by the 3-simplices appearing in the corresponding Pachner moves. The sign factor $\epsilon\snum{(01234)}$ depends on the orientation of the 4-simplex. Since the convention for the $\mathcal{P}_{4 \mapsto 1}$ follows from the one of the $\mathcal{P}_{2 \mapsto 3}$, we only explain the latter one in detail:

\begin{convention}[2--3 Pachner move\label{conv:P23}] 
		Pick one of the 3-simplices in the source complex and assume that it is labeled by $(wxyz)$ such that $w < x < y < z$. The remaining vertex of the source complex is labeled by $o$. We denote by $\epsilon(wxyzo)$ the sign factor associated with the corresponding 2--3 Pachner move. First, determine the orientation $\epsilon(wxyz)$ of the 3-simplex $(wxyz)$ according to convention~\ref{conv:or3S}, then consider the list of vertices $\{o,w,x,y,z\}$. If it takes an even number of permutations to bring this list to the ascending ordered one, then $\epsilon(wxyzo) = +\epsilon(wxyz)$, otherwise $\epsilon(wxyzo) = -\epsilon(wxyz)$. The same convention applies to find the orientation of a 4-simplex.
\end{convention}

\noindent
Applying this convention to equation \eqref{eq:loc23}, we find that $\epsilon\snum{(01234)} = \epsilon\snum{(0123)} = -1$.

The constraints \eqref{eq:loc23} and \eqref{eq:loc41} satisfied by the wave functions under local unitary transformations are only valid together with the corresponding consistency conditions. We will show later that these are guaranteed by the fact that $\omega$ is a 2-form 4-cocycle, but before doing so we are going to investigate in more detail these local unitary transformations. First of all, let us present an alternative presentation for these mutations which is closer related to the original formulation of Levin and Wen for 2d string net models as well as the one of the 3d Walker-Wang model which we review in sec.~\ref{sec:WW}. Instead of working with the triangulation $\triangle$, we consider the one-skeleton of its dual polyhedral decomposition denoted by $\Upsilon$ such that the 2-simplices $\triangle^2 \subset \triangle$ are dual to links ${\mathsf l} \subset \Upsilon$ and the 3-simplices $\triangle^3 \subset \triangle$ are dual to nodes $\mathsf{n} \subset \Upsilon$. The branching rules or (compatibility conditions or 2-flatness constraints) are now encoded at every node. Because we now work on the dual graph, it is inconvenient to keep the labeling completely implicit, thus we label each link by its dual 2-simplex. Note, however, that it is not strictly necessary to specify explicitly the orientation of each edge since it can be easily deduced from the definition of the constraint $\la dg, (wxyz) \ra = g_{xyz}-g_{wyz}+g_{wxz}-g_{wxy}=\zo$. In terms of the dual graph $\Upsilon$, equation \eqref{eq:loc23} becomes
\begin{equation}
	\label{eq:loc23Dual}
	\Bigg| \PachnerDualONE{0.3}{1} \Bigg\rangle = \sum_{\substack{g_{014},g_{134} \\ g_{124}}} \la \omega, \snum{(01234)} \ra^{-1} \Bigg| \PachnerDualTWO{0.3}{1} \Bigg\rangle \; ,
\end{equation}
while equation \eqref{eq:loc41} now reads
\begin{equation}
	\label{eq:loc41Dual}
	\Bigg| \PachnerDualTHREE{0.3}{1} \Bigg\rangle =  \la \omega, \snum{(01234)} \ra^{+1} \Bigg| \PachnerDualFOUR{0.3}{1} \Bigg\rangle \; .
\end{equation}
Now is a good time to recall that the pairing between the 2-form 4-cocycle and the standard 4-simplex identified with the corresponding 4-chain (of the simplicial group $B^2(G)$) is explicitly given by
\begin{align}
	\label{eq:corres4Coc} 
	\la \omega, \snum{(01234)} \ra &= \omega(g_{012}, g_{013} - g_{012},g_{014} - g_{013}| g_{123}, g_{124} - g_{123} | g_{234}) \ .
\end{align}
In the following, we are interested in special cases of the equations above which correspond to setting some of the group variables appearing in \eqref{eq:corres4Coc} to the identity. In particular, we study what these special cases reveal about the algebraic structure of 2-form 4-cocycles using arguments similar to the ones presented at the end of sec.~\ref{sec:normCond}. Let us for instance consider the $\mathcal{P}_{3 \mapsto 2}$ move dual to the $\mathcal{P}_{2 \mapsto 3}$ move depicted in \eqref{eq:loc23Dual} but with a different distribution of vertices and such that $g_{013} = g_{134} = g_{234} = g_{012} =  0$:
\begin{equation}
	\nn
	\Bigg| \hspace{1em}\PachnerDualTWO{0.3}{2} \Bigg\rangle = \la \omega, \snum{(01234)} \ra \Bigg| \hspace{1em} \PachnerDualONE{0.3}{2} \Bigg\rangle 
\end{equation}
where the dashed line represents links (or 2-simplices) labeled by the vacuum sector. Furthermore, we set $g_{014} =g_{034}= b$ and $g_{023}= g_{123} = g_{124} = a$ so that
\begin{align} 
	 \la \omega, \snum{(01234)}\ra =  \omega(\zo, \zo, b | a , \zo | \zo) \; .
\end{align}
It should be obvious from this presentation that when setting $g_{013} = g_{134} = g_{234} = g_{012} =  0$, the $\mathcal{P}_{2 \mapsto 3}$ move effectively reduces to a \emph{braiding move}. The 2-form 4-cocycle correspondingly reduces to a group 2-cochain $R:(a,b)\mapsto \omega(\zo,\zo,b|a,\zo|\zo)$ such that
\begin{equation}
	\Bigg| \hspace{1em} \PachnerDualTWO{0.3}{2} \Bigg\rangle
	:= R(g_{023},g_{014})  \Bigg| \hspace{1em}\PachnerDualONE{0.3}{2} \Bigg\rangle
\end{equation}
which can effectively be represented in terms of string diagrams as
\begin{equation}
	\label{eq:effectBraid}
	\Rmove{0.3}{2} \q \overset{R(a,b)}{\longrightarrow} \q
	\Rmove{0.3}{1} \q , \q 
	\Rmove{0.3}{3} \q \overset{R(b,a)^{-1}}{\longrightarrow} \q
	\Rmove{0.3}{1}
\end{equation}
where the r.h.s corresponds to the situation where the left strand undercrosses the right one.
Note that instead of starting from the $\mathcal{P}_{2 \mapsto 3}$ move, we could have considered a special case of the $\mathcal{P}_{4 \mapsto 1}$ move instead and it would have led exactly to the same result. 

\bigskip \noindent
There is another interesting special case of the same $\mathcal{P}_{3 \mapsto 2}$ move that is obtained by setting $g_{124} = g_{234} = g_{134} = g_{123} = \zo$:\footnote{Note that we use yet another distribution of vertices compared to the one in \eqref{eq:loc23Dual}. Nevertheless, it obviously does not matter how we choose such labeling and we could have performed the same analysis with any other. However, we always choose the one which makes the evaluation $\la \omega, (01234)\ra$ as simple as possible. In general, because of the inherent redundancy of the algebraic structure underlying the 2-form 4-cocycle, there are many ways to write the same thing. We tried to choose our examples so as to make the results as manifest as possible.}
\begin{equation}
	\nn
	\Bigg| \PachnerDualTWO{0.3}{3} \hspace{1em} \Bigg\rangle =  \la \omega, \snum{(01234)} \ra^{-1} \Bigg|  \PachnerDualONE{0.3}{3} \hspace{1em} \Bigg\rangle	 \; .
\end{equation}
Let us furthermore set $g_{012} = a $, $g_{023}= b$ and $g_{034} = c$ so that
\begin{align} 
	\la \omega, \snum{(01234)}\ra =  \omega(a, b, c | \zo , \zo | \zo) \; .
\end{align}
It should be obvious from this presentation that when setting $g_{124} = g_{234} = g_{134} = g_{123} = \zo$, the $\mathcal{P}_{2 \mapsto 3}$ move effectively reduces to a 2--2 Pachner move $\mathcal{P}_{2 \mapsto 2}$ if we were to consider the two-dimensional triangulation dual to the three-valent graph defined by the bold edges. Indeed, the 2-form 4-cocycle reduces to a group 3-cochain $\alpha : (a,b,c) \mapsto \omega^{-1}(a,b,c|\zo,\zo|\zo)$ such that
\begin{equation}
	\nn
	\Bigg| \PachnerDualTWO{0.3}{3} \hspace{1em}  \Bigg\rangle
	= \alpha(g_{012},g_{023},g_{034}) \Bigg| \PachnerDualONE{0.3}{3} \hspace{1em} \Bigg\rangle 
\end{equation}
which can effectively be represented in terms of string diagrams as
\begin{equation}
	\label{eq:effectFmove}
	\hexagonMoveONE{0.3}{2} \q \overset{\alpha(a,b,c)}{\longrightarrow} \q  \hexagonMoveFIVE{0.3}{2} \; .
\end{equation}
What we have just shown, from a graphical point of view, is how the local unitary transformation associated with a 2-form 4-cochain reduces to a braiding move or a 2--2 Pachner move. In the following subsection, we study the coherence relations of these transformations which allows us to make this correspondence more precise. 

\subsection{Consistency conditions\label{sec:consistency}}

We presented earlier the equations that fixed wave functions must satisfy under local unitary transformations. However, for these equations to be self-consistent, some coherence relations must be satisfied. We are now going to study these coherence relations with an emphasis on the correspondence put forward at the end of the previous subsection.

\begin{figure}[]
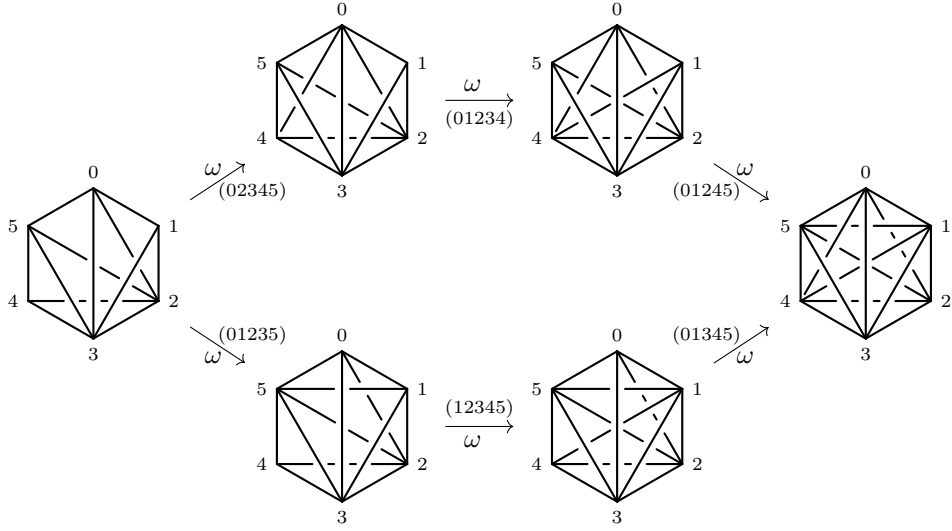

	\center
	\assocTWO{}
	\caption{Consistency condition of the 2--3 Pachner move whose amplitude is given by the 2-form 4-cocycle $\omega \in Z^4(G_{[2]},\rU(1))$. Starting from the union of three 3-simplices, there exist two different successions of 2--3 Pachner moves which lead to the same union of six 3-simplices. On each arrow we indicate the 4-simplex bounded by the five 3-simplices involved in the Pachner move, on which the 2-form cocycle is evaluated.}
	\label{fig:pentagonator}
\end{figure}

Let us consider the union of three 3-simplices. There are two different sequences of  three $\mathcal{P}_{2 \mapsto 3}$ moves which lead to the same complex that is the union of six 3-simplices. This situation is represented in fig.~\ref{fig:pentagonator} where the direction of the arrows is decided according to convention~\ref{conv:P23}. According to equation \eqref{eq:loc23Dual}, the amplitude of the map performing each $\mathcal{P}_{2 \mapsto 3}$ move is given in terms of the 2-form 4-cochain $\omega$ or its inverse. Applying this definition to the two sequences of $\mathcal{P}_{2 \mapsto 3}$ and requiring that any composition of maps yielding the same final configuration must be identified, the coherence relation implies the following equality:
\begin{equation}
	\label{eq:CohRel4Coc}
	\la\omega, \snum{(02345)} \ra \la \omega , \snum{(01245)} \ra \la \omega , \snum{(01234)}\ra = 
	\la\omega ,\snum{(01235)}\ra \la\omega, \snum{(01345)}\ra \la \omega , \snum{(12345)}\ra \; . 
\end{equation}
It turns out that this equality is nothing else than the 2-form 4-cocycle condition and fig.~\ref{fig:pentagonator} its graphical interpretation. The coherence of the local unitary transformation is therefore ensured by the fact that $\omega \in Z^4(G_{[2]},\rU(1))$. To check explicitly that this is indeed the cocycle condition $d^{(4)}\omega(a,b,c,d|e,f,g|h,i|j) = 1$ as written in \eqref{eq:2formCoc4}, we just need to use \eqref{eq:pairing4Coc} and label the face variables as follows: $g_{012} = a$, $g_{013} = a+b$, $g_{014} = a+b+c$, $g_{015} =a+b+c+d$ $g_{123} =e$, $g_{124} = e+f$, $g_{125} = e+f+g$, $g_{234} = h$, $g_{235}= h+i$ and $g_{345} = j$. Note that there is another way to interpret the 2-form 4-cocycle condition following the lines of sec.~\ref{sec:geomInter}. Indeed, considering a 3--3 Pachner move and identifying each one of the oriented six 4-simplices with a 2-form 4-cocycle, we would obtain \eqref{eq:CohRel4Coc} as well. This relies on the fact that there is a canonical way to assign a cyclic sequence of five $\mathcal{P}_{2 \mapsto 2}$ moves to a $\mathcal{P}_{2 \mapsto 3}$ move,  a cyclic sequence of six $\mathcal{P}_{2 \mapsto 3}$ moves to a $\mathcal{P}_{3 \mapsto 3}$ move, and so on and so forth, as exploited in \cite{delcamp2018gauge}. 

In the same way we investigated earlier special cases of the $\mathcal{P}_{2 \mapsto 3}$ move, we will now study special cases of the cocycle condition $d^{(4)}\omega(a,b,c,d|e,f,g|h,i|j) = 1$. In particular, we are interested in the graphical interpretation of these special cases in light of the correspondence between \eqref{eq:CohRel4Coc} and fig.~\ref{fig:pentagonator}.

\bigskip \noindent
Let us consider the 4-coboundary $d^{(4)}\omega(a,b,c,d|\zo,\zo,\zo|\zo,\zo|\zo)$ which yields the cocycle condition
\begin{equation}
	\frac{\omega(\zo,\zo,\zo|\zo,\zo|\zo)\, \omega(a+b,c,d|\zo,\zo|\zo)\, \omega(a,b,c+d|\zo,\zo|\zo)}{\omega(b,c,d|\zo,\zo|\zo)\, \omega(a,b+c,d|\zo,\zo|\zo)\, \omega(a,b,c|\zo,\zo|\zo)} = 1 \; .
\end{equation}
First of all, according to the normalization conditions satisfied by $\omega$, one has $\omega(\zo,\zo,\zo|\zo,\zo|\zo) = 1$.
It then follows straightforwardly that the group 3-cochain $\alpha$ defined as $\alpha : (a,b,c) \mapsto \omega^{-1}(a,b,c|\zo,\zo|\zo)$ is a group 3-cocycle in $H^3(G,{\rm U}(1))$ satisfying the group 3-cocycle condition
\begin{equation}
	\label{eq:Gr3Coc}
	\boxed{
	\alpha(a+b,c,d)\, \alpha(a,b,c+d) = \alpha(b,c,d) \, \alpha(a,b+c,d) \, \alpha(a,b,c) 
	}
\end{equation} 
also referred to as the \emph{pentagon relation} in the context of monoidal category theory.
Together with the relations \eqref{eq:Equalities3C} derived directly from the normalization conditions, this yields the following set of equalities 
\begin{equation}
	\label{eq:3CocRelations}
	\boxed{
	\alpha(a,b,c) = \omega(a,b,c|\zo,\zo|\zo)^{-1} = \omega(\zo,a,\zo|b,c|\zo) = \omega(\zo,\zo,a|\zo,b|c)^{-1}
	}
\end{equation}
which we use several times below. Equation~\eqref{eq:3CocRelations} is a good example of the inherent redundancy underlying the $W$-construction of $B^2(G)$. This redundancy is the main reason why we need to choose carefully our examples and our conventions in order for the results to be manifest and not to be lost in the redundancy of the description. 

\begin{figure}[]
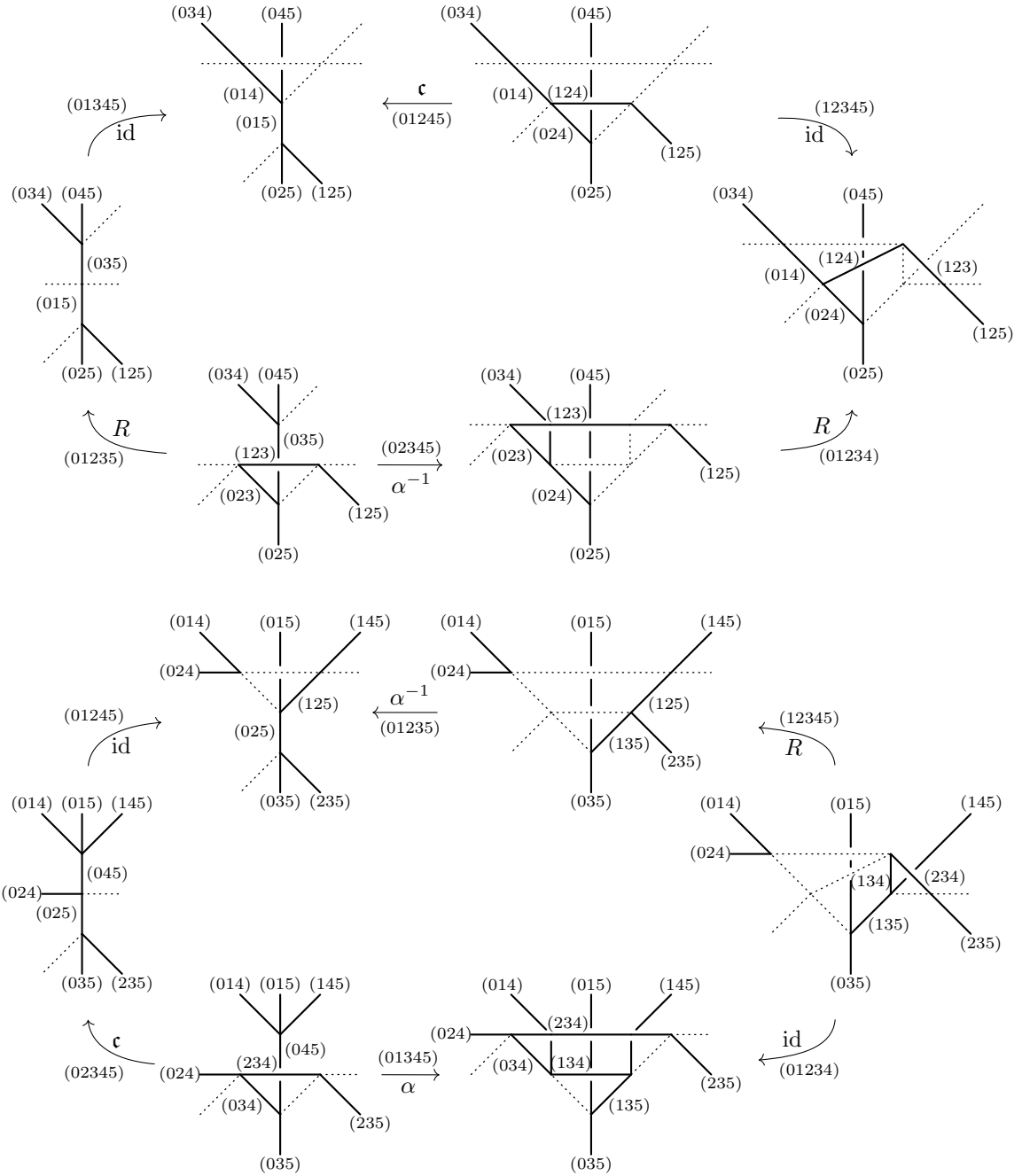

	\center
	\fourCocycleONE{1} 
	\newline
	\fourCocycleTWO{2} 
	\label{fig:InterHalfHex}
	\caption{Graphical depiction of the 2-form cocycle condition $d^{(4)}\omega(0,0,c,d|e,0,0|0,0|0) = 1$ and $d^{(4)}\omega(0,0,c,0|0,0,g|h,0|0) = 1$. The dashed line are labeled by the identity group element $\zo \in G$. Each arrow of the diagram is labeled by a 4-simplex $(abcde)$ such that $\la \omega, (abcde)\ra$ is the evaluation of the 2-form 4-cocycle $\omega$ that is the amplitude of the corresponding 2-3 Pachner move, as well as a symbol $\alpha$, $R$ or ${\rm id}$ depending on whether the 2-3 Pachner move effectively reduces to a 2-2 Pachner move, a braiding move or a trivial move, respectively. Together, these two consistency conditions effectively reduce to a so-called \emph{hexagon} relation.}
\end{figure}

\bigskip \noindent
Let us now consider the 4-coboundary $d^{(4)}\omega(\zo,\zo,c,d|e, \zo, \zo|\zo, \zo| \zo)$ which yields the cocycle condition
\begin{equation}
	\label{eq:firstHexNotSimp}
	 \frac{\omega(e,\zo,\zo|\zo,\zo|\zo)\,\omega(\zo,c,d|\zo,\zo|\zo)\, \omega(\zo,\zo,c+d|e,\zo|\zo)}{\omega(e,c,d|\zo,\zo|\zo) \, \omega(\zo,c,d|e,\zo|\zo) \, \omega(\zo,\zo,c|e,\zo|\zo)}=1 \; .
\end{equation}
Firstly, it follows from the normalization conditions \eqref{normCond2} satisfied by $\omega$ that $\omega(e,\zo,\zo|\zo,\zo|\zo) =1$ and $\omega(\zo,c,d|\zo,\zo|\zo)= 1$. Secondly, we recognize terms which reduce to the group 2-cochain $R$ and the group 3-cocycle $\alpha$. Thirdly, we define a group 3-cochain denoted by $\mathfrak{c}$ as $\mathfrak{c}:(a,b,c) \mapsto \omega(\zo,a,b|c,\zo|\zo)$. Putting everything together, the cocycle condition $\la d^{(4)}\omega, \snum{(012345)}\ra :=  d^{(4)}\omega(\zo,\zo,c,d|e, \zo, \zo|\zo, \zo| \zo) =1 $ reads
\begin{equation}
	\label{eq:firstHexSimp}
	\frac{R(e,c+d) \, \alpha(e,c,d)}{\mathfrak{c}(c,d,e) \, R(e,c)}=1 
\end{equation}
such that $\la \omega , \snum{(02345)}\ra = \alpha(e,c,d)^{-1}$, $\la \omega, \snum{(01245)}\ra =\mathfrak{c}(c,d,e)$, $\la \omega, \snum{(01235)}\ra = R(e,c+d)$ and $\la \omega, \snum{(01234)}\ra = R(e,c)$.
In fig.~\ref{fig:InterHalfHex} (top panel), we provide a graphical interpretation to this 2-form 4-cocycle condition. In order to obtain this figure, we proceed as follows: $(i)$ Reproduce fig.~\ref{fig:pentagonator} but for the one-skeleton of the polyhedral decomposition $\Upsilon$ dual to the triangulation $\triangle$ and for a different (judicious) numbering of the vertices. $(ii)$ Identify the 2-simplex variables from \eqref{eq:firstHexNotSimp} using the correspondence \eqref{eq:pairing4Coc} and draw the dual links with a dashed line when the corresponding labeling vanishes. Focusing on the bold edges appearing in fig.~\ref{fig:InterHalfHex}, we can draw several remarks: Firstly, as expected, the two $\mathcal{P}_{2 \mapsto 3}$ moves associated with the 4-cocycle evaluations normalized to one do not modify the combinatorics of the diagram. Secondly, two $\mathcal{P}_{2 \mapsto 3}$ effectively reduce to a braiding move which is consistent with the fact two $R$-matrices appear in equation \eqref{eq:firstHexSimp}. Thirdly, one $\mathcal{P}_{2 \mapsto 3}$ effectively reduces to a $\mathcal{P}_{2 \mapsto 2}$ which is consistent with the presence of the group 3-cocycle $\alpha$ in equation \eqref{eq:firstHexSimp}. Finally, there is the move corresponding to the term $\mathfrak{c}(c,d,e)$. Putting everything together, the 2-form 4-cocycle condition can effectively be graphically interpreted in terms of string diagrams as

\begin{equation}
	\label{eq:decompC}
\begin{tikzpicture}[scale=1,baseline=0em]
	\matrix[matrix of math nodes,row sep =2em,column sep=2em, ampersand replacement=\&] (m) {
	\hexagonMoveTWO{0.3}{1}  \& \hexagonMoveFOUR{0.3}{1} \& \hexagonMoveTHREE{0.3}{1} \& \hexagonMoveONE{0.3}{1} \\
	};	
	\path
	(m-1-1) edge[->, shorten <= 0em, shorten >= 0em] node[pos=0.5, above] {${\sss R^{-1}(e,c)}$} (m-1-2)
	(m-1-2) edge[->, shorten <= 0em, shorten >= 0em] node[pos=0.5, above] {${\sss \alpha(e,c,d)}$} (m-1-3)
	(m-1-3) edge[->, shorten <= 0em, shorten >= 0em] node[pos=0.5, above] {${\sss R(e,c+d)}$} (m-1-4)
	(m-1-1) edge[->, shorten <= -0.3em, shorten >= -0.3em, out=90, in=90, looseness=10, max distance = 3em] node[pos=0.5, above] {${\sss \mathfrak{c}(c,d,e)}$} (m-1-4)
	;
\end{tikzpicture}
\end{equation}
where we omitted the trivial maps. Note that this equation makes sense on its own, independently from fig.~\ref{fig:InterHalfHex}, according to \eqref{eq:effectBraid} and $\eqref{eq:effectFmove}$.

\bigskip \noindent
We will now repeat the previous analysis starting from a different cocycle condition in order to provide an alternative decomposition for the map $\mathfrak{c}$. Let us consider the cocycle condition
\begin{align}
	\nn
	\la d^{(4)}\omega, \snum{(012345)}\ra := d^{(4)}\omega(\zo,\zo,c,\zo|\zo,\zo, g|h,\zo|\zo) &= \frac{\omega(\zo,\zo,g|h,\zo|\zo)\,\omega(\zo,c,\zo|h,g|\zo)\, \omega(\zo,\zo,c|\zo,g|h)}{\omega(\zo, c,g|h,\zo|\zo) \, \omega(\zo,c,\zo|\zo,g|\zo) \, \omega(\zo,\zo,g|\zo, \zo|h)} \\
	\label{eq:secondHexSimp}
	&= \frac{R(h,g) \, \alpha(c,h,g)}{\mathfrak{c}(c,g,h) \, \alpha(c,g,h)} = 1
\end{align}
such that $\la \omega , \snum{(12345)}\ra = R(h,g)$, $\la \omega, \snum{(02345)}\ra =\mathfrak{c}(c,g,h)$, $\la \omega, \snum{(01345)}\ra = \alpha(c,h,g)$ and $\la \omega, \snum{(01235)}\ra =\alpha(c,g,h)^{-1}$. We made use between the first and the second line of the normalization conditions as well as \eqref{eq:3CocRelations}. As before, this cocycle condition can be represented graphically (see lower panel of fig.~\ref{fig:InterHalfHex}) by identifying all the 2-simplex variables and make a judicious choice of numbering of the vertices which dictates, among other things, which $\mathcal{P}_{2 \mapsto 3}$ move each term of \eqref{eq:secondHexSimp} corresponds to. In terms of string diagrams, this effectively boils down to:\footnote{The correspondence between fig.~\ref{fig:InterHalfHex} and the effective string diagrams is not as obvious as earlier where it was directly provided by the bold links of the dual complex. Indeed, because of the presence of additional links labeled by non-trivial group variables, the correspondence is not quite as transparent. However, by looking carefully at the value of each 2-simplex (or dual link) variables, the reader should be able to convince itself that it does reduce to the equation in terms of string diagrams. First of all, it should be clear that the links labeled by ${(014)}$ and ${(024)}$ are irrelevant from a combinatorial point of view in the top-left, top-right and right complexes (of the lower panel). Furthermore, from $\la \omega, {(01245)} \ra = \omega(g_{012},g_{014}-g_{012},g_{015}-g_{014}|g_{124},g_{125}-g_{124}|g_{245}) = \omega(\zo,c,\zo|\zo,g|\zo)$, we read off in particular that $g_{014}=g_{015}=c$ and $g_{045} = g_{145} =g$ so that we can effectively `forget' about the links labeled by ${(014)}$ and ${(015)}$ in the left and bottom-left complexes. }
\begin{equation}
	\begin{tikzpicture}[scale=1,baseline=0em]
		\matrix[matrix of math nodes,row sep =2em,column sep=2em, ampersand replacement=\&] (m) {
			\hexagonMoveTWO{0.3}{3}  \& \hexagonMoveSIX{0.3}{1} \& \hexagonMoveFIVE{0.3}{1} \& \hexagonMoveONE{0.3}{3} \\
		};	
		\path
		(m-1-1) edge[->, shorten <= 0em, shorten >= 0em] node[pos=0.5, above] {${\sss \alpha(c,h,g)}$} (m-1-2)
		(m-1-2) edge[->, shorten <= 0em, shorten >= 0em] node[pos=0.5, above] {${\sss R(h,g)}$} (m-1-3)
		(m-1-3) edge[->, shorten <= 0em, shorten >= 0em] node[pos=0.5, above] {${\sss \alpha^{-1}(c,g,h)}$} (m-1-4)
		(m-1-1) edge[->, shorten <= -0.3em, shorten >= -0.3em, out=90, in=90, looseness=10, max distance=3em] node[pos=0.5, above] {${\sss \mathfrak{c}(c,g,h)}$} (m-1-4)
		;
	\end{tikzpicture}
\end{equation}
where we omitted as before the trivial maps.

\bigskip \noindent
Let us summarize what we have shown so far: Using two special cases of the 2-form 4-cocycle condition, together with their geometrical interpretation that relies on the fact that the map performing a 2--3 Pachner move evaluates to a 2-form 4-cocycle, we have obtained two different decompositions for the map $\mathfrak{c}:(a,b,c)\mapsto \omega(\zo,a,b|c,\zo|\zo)$  in terms of the group 2-cochain $R$ and the group 3-cocycle $\alpha$, both algebraically and geometrically. Equating these two decompositions, we obtain
\begin{equation}
	\label{eq:Hex1}
	\boxed{
		\alpha(a,b,c) \, R(c,a+b) \, \alpha(c,a,b) = R(c, a) \, \alpha(a, c, b) \, R(c, b)
	}
\end{equation}
which is nothing else than one of the \emph{hexagon relations} appearing in the definition of a (abelian) braided monoidal category. This equation is the consistency condition for the braiding move whose amplitude is provided by the 2-cochain $R$, the same way \eqref{eq:Gr3Coc} is the one for the $\mathcal{P}_{2 \mapsto 2}$ move. We can deduce very easily the corresponding graphical interpretation in terms of string diagrams and it reads:
\begin{equation*}
	\begin{tikzpicture}[scale=1,baseline=0em]
	\matrix[matrix of math nodes,row sep =-1.5em,column sep=2em, ampersand replacement=\&] (m) {
		{} \& \hexagonMoveTHREE{0.3}{2}  \& \hexagonMoveFOUR{0.3}{2} \& {} \\
		\hexagonMoveONE{0.3}{2}  \& {} \& {} \& \hexagonMoveTWO{0.3}{2} \\
		{} \& \hexagonMoveFIVE{0.3}{2}  \& \hexagonMoveSIX{0.3}{2} \& {} \\
	};	
	\path
	(m-2-1) edge[<-, shorten <= 1em, shorten >= -2em] node[pos=1, right, yshift=-0.2em] {${\sss R(c,a+b)}$} (m-1-2)
	edge[->, shorten <= -2em, shorten >= 1em] node[pos=0, left, yshift=-0.2em] {${\sss \alpha(a,b,c)}$} (m-3-2)
	(m-1-2) edge[<-, shorten <= -0.5em, shorten >= -0.5em] node[pos=0.5, above] {${\sss \alpha(c,a,b)}$} (m-1-3)
	(m-3-2) edge[<-, shorten <= -0.5em, shorten >= -0.5em] node[pos=0.5, above] {${\sss R(c,b)}$} (m-3-3)
	(m-2-4) edge[<-, shorten <= 1em, shorten >= -2em] node[pos=1, left, yshift=-0.2em] {${\sss R(c,a)}$} (m-1-3)
	edge[->, shorten <= -2em, shorten >= 1em] node[pos=0, right, yshift=-0.2em] {${\sss \alpha(a,c,b)}$} (m-3-3)
	;
	\end{tikzpicture}
\end{equation*}
It turns out, there is another hexagon equation which can be obtained similarly starting from two others special cocycle conditions:
\begin{equation}
	\label{eq:Hex2}
	\boxed{
	R(c+a,b)\, \alpha(c,b,a) = \alpha(b,c,a) \, R(c,b) \, \alpha(c,a,b) \, R(a,b) 
	}
\end{equation}
whose graphical interpretation can be derived by proceeding as before. 

Thinking of $\omega$, $R$ and $\alpha$ as isomorphisms, what we have essentially shown in this part is that the input data of our model, namely $\{G,\omega\}$, reduces to the input data of an abelian braided monoidal category, namely $\{G,R,\alpha\}$, and that the consistency conditions of the constraints satisfied by the fixed point wave functions under $\mathcal{P}_{2 \mapsto 3}$ moves reduce to the consistency conditions of $R$ and $\alpha$. Note that a similar analysis has been carried out for instance in \cite{1998math.....11047Q, Stirling:2008bq, 2016arXiv160601414G} but using the bar construction of $B^2(G)$ instead of the $W$-construction as we did. The computations are in this case more straightforward. However, because the bar  presentation does not yield a geometrical interpretation in terms of 2-form flat connections defined on a triangulation, it is neither possible to show this correspondence from an intuitive simplicial point of view, nor to define the corresponding lattice Hamiltonian model as we are doing. We will provide more category theoretical details in sec.~\ref{sec:WW} and exploit this result to show to which extent our model is related to the Walker-Wang model.

\FloatBarrier

\subsection{Lattice Hamiltonian \label{sec:Ham}}
In this subsection, we introduce our lattice model which is an Hamiltonian realization of the 2-form topological invariant \eqref{eq:latticenform} whose ground states are described by the fixed point wave functions satisfying \eqref{eq:loc23} and \eqref{eq:loc41}. These ground states correspond to the physical states (as defined in app.~\ref{sec:app_quant}) that span the Hilbert space $\mathcal{H}_{\Sigma}$ obtained upon quantization of the 2-form TQFT on a manifold of the form  $\cM=\Sigma \times \mathbb{R}$.

\bigskip \noindent
We introduced in sec.~\ref{sec:LGT} a general formula \eqref{eq:latticenform} for the partition function of topological $q$-form lattice gauge theories. We reproduce below this formula for the case $q=2$ in a slightly different form:
\begin{equation}
	\label{eq:tentMove}
	\mathcal{Z}^{G_{[2]}}_\omega[\mathcal{M}] =\frac{1}{|G|^{|\triangle^1| - |\triangle^0|}} \sum_{g \in {\rm Col}(\mathcal{M},G_{[2]})}\prod_{\triangle^{4}}\la \omega, \triangle^4\ra^{\epsilon(\triangle^4)}
\end{equation} 
where $\omega \in Z^4(G_{[2]},\rU(1))$ is the 2-form 4-cocycle, $\la \omega , \triangle^4\ra^{\epsilon(\triangle^4)} \equiv {\mathcal S}_\omega[\triangle^4]$ is the topological action such that $\epsilon(\triangle^4) = \pm 1$ is determined according to conv.~\ref{conv:P23}, and ${\rm Col}(\mathcal{M},G_{[2]})$ is the set of $G_{[2]}$-colorings of $\mathcal{M}$, i.e. an assignment of group elements $g \in G$ to every 2-simplex which satisfy the 2-cocycle condition $dg=0$ where $\la dg, (wxyz)\ra  = g_{xyz} - g_{wyz} + g_{wxz} - g_{wxy}$. Note that we are now considering cocycles valued in $\rU(1)$ instead of cocycles valued in $\mathbb{R} / \mathbb{Z}$, this is merely a choice of convention. Furthermore, we now pick a cocycle in the cohomology $H^4(G_{[2]},\rU(1))$ instead of $H^4(B^2(G), \rU(1))$, which is the same by construction. The 2-form 4-cocycle condition ensures the topological invariance of the partition function. 

Let us now define an Hamiltonian realization of this topological field theory on a space-like three-dimensional hypersurface $\Sigma$ endowed with a triangulation $\triangle$ whose 2-simplices are labeled by group variables in the finite abelian group $G$. To every $3$-simplex of $\triangle$, we associate a projector $\mathbb{B}_{(\triangle^3)}$ which enforces the \emph{zero-flux condition}. To every $1$-simplex of $\triangle$, we associate a projector $\mathbb{A}_{(\triangle^1)}$ which enforces the \emph{twisted 1-form gauge invariance}. The zero-flux condition is particularly simple for our model since it boils down to the branching rules. In other words, it enforces the fact that the labeling of the 2-simplices define a $G_{[2]}$-coloring, i.e. a local description of a flat 2-form connection. Given a state $| \, \snum{(0123)} \, \ra$ that is the state of a labeled 3-simplex whose vertices are numbered $\snum{0}$, $\snum{1}$, $\snum{2}$ and $\snum{3}$, the action of the operator $\mathbb{B}_{(\triangle^3)}$ explicitly reads 
\begin{equation}
	\mathbb{B}_{\, {(0123)} \, } \, \triangleright | \, \snum{(0123)} \, \ra = \delta_{g_{123}-g_{023}+g_{013}-g_{012},0} | \, \snum{(0123)} \, \ra \; .
\end{equation}
The action of the operator $\mathbb{A}_{(\triangle^1)}$ is a little more subtle but there is a particularly convenient way of defining it via a so-called \emph{tent move} in terms of the state-sum invariant. Given a 2-simplex $(xy) \subset \triangle$, the operator $\mathbb{A}_{(xy)}$ acting on $(xy)$ can be written succinctly as\footnote{Instead of defining the operator $\mathbb A$ in terms of the topological action and an explicit sum, it would have been possible to define it directly in terms of the partition function so that only the group variables labeling faces in the bulk of the simplicial complex obtained via the join operation $\sqcup_{\rm j}$ are summed over.}
\begin{equation}
	\label{eq:defA}
	\mathbb{A}_{(xy)} = \frac{1}{|G|} \sum_{g_{xyz}}\mathcal{S}_\omega[(z) \sqcup_\mathsf{j} {\rm cl}(xy)] 
\end{equation}
where $\mathcal{S}_\omega[\triangle] = \prod_{\triangle^4 \subset \triangle}\la \omega, \triangle^4\ra^{\epsilon(\triangle^4)}$ is the topological action, ${\rm cl}(xy)$ is the minimal subcomplex of $\triangle$ that contains all the simplices such that $(xy)$ is one of their subsimplices, and $\sqcup_{\rm j}$ denotes the \emph{join} operation \cite{Williamson:2016evv}. 

\noindent
Let us illustrate the definitions of ${\rm cl}({\sss \bullet})$ and $\sqcup_{\rm j}$ with a simple example in one lower dimension. Let $\snum{(0)}$ be a $0$-simplex shared by only three 2-simplices, namely $\snum{(012)}$, $\snum{(023)}$ and $\snum{(013)}$. The operation $\snum{(0')} \sqcup_{\rm j} {\rm cl}(\snum{0})$ then reads
\begin{equation*}
	\snum{(0')} \sqcup_{\rm j} \actionA{0.8}{1}{1} = \actionA{0.8}{2}{2} \; .
\end{equation*}
Let us now consider the situation where one 1-simplex is shared by three 3-simplices.\footnote{We focus our definition of the Hamiltonian model on this special example in order to show later on how it is related to the Walker-Wang model. We postpone a more through study of this lattice Hamiltonian to another paper.} Let us write down the action of the operator $\mathbb{A}_{(xy)} =: \frac{1}{|G|}\sum_{g_{xyz}}\mathbb{A}_{(xy)}^{g_{xyz}}$ which enforces at this 1-simplex the twisted 1-form gauge invariance. Using \eqref{eq:tentMove}, one obtains
\begin{align}
	\label{eq:actionA}
	\mathbb{A}_{\rm (04)} \triangleright \Bigg| \HamONE{1}{1} \Bigg\rangle &= 
	\sum_{g_{045}} \mathcal{S}_\omega \Bigg[ \HamTWO{1}{1} \Bigg]
	\Bigg| \HamONE{1}{2} \Bigg\rangle \\[-0em]
	\label{eq:actionA2}
	& = \sum_{g_{045}} \frac{\la \omega , \snum{(02345)}\ra \, \la \omega , \snum{(01245)}\ra}{\la \omega, \snum{(01345)} \ra}
	\Bigg| \HamONE{1}{2} \Bigg\rangle
\end{align} 
where $g_{245} = g_{345} = g_{145} = \zo$.\footnote{The operator $\mathbb A$ acts on the 1-simplex ${(04)}$ only so that we must have $g_{025} = g_{024}+g_{045}$, $g_{035} = g_{034}+g_{045}$ and $g_{015} = g_{014}+g_{045}$. It then follows from the 2-cocycle condition $dg=0$ that $g_{245} = g_{345} = g_{145} =0$.} It is not obvious from the drawing but the complex $\snum{(5)} \sqcup_{\rm j} {\rm cl}\snum{(04)}$ does not contain the 2-simplex $\snum{(123)}$ and as such it only contains three 4-simplices, namely $\snum{(02345)}$, $\snum{(01245)}$ and $\snum{(01345)}$. 

Let us further suppose w.l.o.g that we have the following initial $G_{[2]}$-coloring: $g_{012} = a$, $g_{013} = a+b$, $g_{014} = a+b+c$, $g_{123} =e$, $g_{124} = e+f$ and $g_{234} = h$ and we denote the 1-form gauge parameter by $g_{045}=d$. 
Denoting the initial state in \eqref{eq:actionA} by $| \psi_{\rm init.} \ra$ and the final state in \eqref{eq:actionA2} by $| \psi_{\rm fin.} \ra$, the amplitude of the operator $\mathbb{A}^d_{(04)}$ for such coloring explicitly reads
\begin{equation*}
	\boxed{
	\la \psi_{\rm fin.} |\, \mathbb{A}_{(04)}^d \, | \psi_{\rm init.} \ra =  
	\frac{\la \omega , \snum{(02345)}\ra \, \la \omega , \snum{(01245)}\ra}{\la \omega, \snum{(01345)} \ra} = 	\frac{\omega(b+e,c+f,d|h,\zo|\zo)\, \omega(a,b+c,d|e+f,\zo|\zo)}{\omega(a+b,c,d|f+h,\zo|\zo)} }
\end{equation*}
where we used the correspondence \eqref{eq:pairing4Coc}. As it turns out, we could have anticipated this result. Indeed, if we embed the initial complex made of four 3-simplices meeting at $\snum{(4)}$ in a four-dimensional manifold, we can think of it as a 4-simplex $\snum{(01234)}$. But the topological action assigns to this 4-simplex an amplitude 
\begin{align}
	\nn
	\la \omega, \snum{(01234)} \ra &= \omega(g_{012},g_{013}-g_{012},g_{014}-g_{013}|g_{123},g_{124}-g_{123}|g_{234}) \\ &= \omega(a,b,c|e,f|h)	
\end{align}
together with the $G_{[2]}$-coloring defined above. Upon 1-form gauge transformation at the edge $\snum{(04)}$, one has $g_{014} \to g_{014} - \theta_{04}$, $g_{024} \to g_{024} - \theta_{04}$ and $g_{034} \to g_{034} - \theta_{04}$ so that the topological action transforms as
\begin{equation}
	\omega(a,b,c|e,f|h) \to \omega(a,b,c+d|e,f|h) 
\end{equation}
where $\theta_{04} = -d$.
We can then deduce from the 2-form 4-cocycle condition $d^{(4)}\omega(a,b,c,d|e,f,\zo|h,\zo|\zo)=1$ that the topological action is modified under this transformation by the following factor
\begin{equation}
	\omega(a,b,c|e,f|h) \to 
	\frac{\omega(b+e,c+f,d|h,\zo|\zo)\, \omega(a,b+c,d|e+f,\zo|\zo)}{\omega(a+b,c,d|f+h,\zo|\zo)} \cdot \omega(a,b,c|e,f|h)
\end{equation} 
which is exactly the amplitude of the operator as obtained above from the tent move. 

It follows from the 2-form 4-cocycle condition that the operators $\mathbb{A}_{(\triangle^1)}$ and $\mathbb{B}_{(\triangle^3)}$ as defined above commute\footnote{The only non-trivial case occurs when the operator $\mathbb{A}$ acts consecutively on two 1-simplices that bound the same 2-simplices. The amplitude of these consecutive actions is obtained as the partition function for the simplicial complex obtained via two consecutive join operations. Depending on the ordering of these consecutive actions, the corresponding simplicial complexes differ but they share the same boundary. It is therefore possible to go from one simplicial complex to another via a sequence of three-dimensional Pachner moves. The topological invariance of the partition function then ensures that the amplitude of the action of the operator $\mathbb A$ is the same for both cases, hence the commutativity.} and the lattice Hamiltonian projector finally reads
\begin{equation}
	\label{eq:Ham2Form}
	\mathbb{H} = - \sum_{\triangle^1} \mathbb{A}_{(\triangle^1)} - \sum_{\triangle^3} \mathbb{B}_{(\triangle^3)} \; .
\end{equation}
The fact that the ground states of this Hamiltonian satisfy equations \eqref{eq:loc23} and \eqref{eq:loc41} under local unitary transformations follows directly from the topological invariance of \eqref{eq:tentMove}, or more precisely from the 2-form 4-cocycle condition.

\subsection{Excitations\label{sec:tube}}

Given a \emph{closed} three-dimensional hypersurface $\Sigma$ endowed with a triangulation $\triangle$, the lattice Hamiltonian is provided by \eqref{eq:Ham2Form}, the states are defined as superpositions of labeled graph states, and the ground states of the Hamiltonian are defined as states $| \psi \ra$ satisfying $\mathbb{A}_{(\triangle^1)} \triangleright |\psi \ra = | \psi \ra$ and $\mathbb{B}_{(\triangle^3)} \triangleright |\psi \ra = | \psi \ra$ for each $\triangle^1, \triangle^3 \subset \triangle$. The Hilbert space of ground states on $\Sigma$ endowed with the triangulation $\triangle$ is denoted by $\mathcal{H}_\triangle$.

Recall that the two conditions enforced by the operators $\mathbb{B}_{(\triangle^3)}$ and $\mathbb{A}_{(\triangle^1)}$ are the 2-form flatness condition and the twisted 1-form gauge invariance, respectively. But, flat 2-form connections on $\Sigma$ can be defined as homomorphisms from the second homotopy group $\pi_2(\Sigma)$ to $G$, so that non-trivial 2-holonomies can only be found along non-contractible 2-paths. This means that by imposing the 2-form flatness condition at every $\triangle^3 \subset \triangle$, we make the implicit assumption that each $\triangle^3$ is associated to a contractible 2-path. We define an excitation as a local neighborhood of the triangulation where the energy density is higher than that of the ground state, i.e. a state for which the conditions $\mathbb{A}_{(\triangle^1)} \triangleright |\psi \ra = | \psi \ra$ and $\mathbb{B}_{(\triangle^3)} \triangleright |\psi \ra = | \psi \ra$ are violated in a local neighborhood. We refer to a state for which one constraint $\mathbb{A}_{(\triangle^1)} \triangleright |\psi \ra = | \psi \ra$ is violated as an \emph{electric charge} excitation and a state for which one constraint $\mathbb{B}_{(\triangle^3)} \triangleright |\psi \ra = | \psi \ra$ is violated as a \emph{magnetic flux} excitation.

Let us first focus on magnetic excitations. By definition, these excitations occur when a given state violates the 2-form flatness condition at one or several 3-simplices. But, if we want the 2-form connection interpretation to persist, this violation must be associated with a non-contractible closed 2-path. Given a closed three-dimensional manifold $\Sigma$, such a non-contractible 2-path can be produced by removing an appropriate three-manifold $\mathcal{B}$ from $\Sigma$, hence turning $\Sigma$ into an open manifold $\Sigma \backslash \mathcal{B}$ whose boundary is given by the boundary $\partial \mathcal{B}$ of the three-manifold $\mathcal{B}$. For instance, this can be done by removing a solid two-torus or a solid two-sphere from $\Sigma$. The resulting manifold would then have a torus boundary or a sphere boundary, respectively, that can support \emph{point-like} magnetic flux excitations. 

So we constrain the magnetic excitations to occur at boundary components of the three-manifold. Similarly, we restrict the electric charge excitations to occur at the boundary. More specifically, we endow each component of $\partial \Sigma$ that has at least one \emph{non-contractible 1-cycle} with a marked link for each {non-contractible 1-cycle} and allow for the 1-form twisted gauge invariance at a 1-simplex to be violated if and only if it coincides with a marked link. So the lattice Hamiltonian yields point-like magnetic flux excitations and string-like electric charge excitations, both located at the boundary of the manifold. Given a closed three-manifold $\Sigma$ equipped with a triangulation, we can remove three-manifolds from it and we think of the states defined on the resulting manifold as being excited with respect to the ground states defined on $\Sigma$.

\medskip \noindent
It is well-known that in such context manifolds of the form $\mathcal{N} \times [0,1]$ play a special role \cite{2010AnPhy.325.2707K, Lan:2013wia, Dittrich:2016typ, DDR1, Delcamp:2017pcw, Aasen:2017ubm}. For instance, in two dimensions, the Hamiltonian realization of the 3d Dijkgraaf-Witten TQFT yields point-like electric and magnetic excitations located at punctures. The twice-punctured two-sphere (or cylinder), i.e. $\mathfrak{T}[\mathbb{S}^1] \equiv \mathbb{S}^1 \times [0,1]$ is then the simplest topology supporting both type of excitations. Moreover, the gluing of two cylinders results in a manifold homeomorphic to a cylinder, hence defining an algebra on the Hilbert space of states that is referred to as \emph{Ocneanu's tube algebra} \cite{ocneanu1994chirality, ocneanu2001operator}. By defining specific excited states on the cylinder, we can confirm explicitly that this algebra is equivalent to the \emph{twisted Drinfel’d} double of the gauge group $G$ \cite{Drinfeld:1989st, Dijkgraaf1991, 1996q.alg.....5044K, Koornwinder:1998xg, Koornwinder:1999bg}. The irreducible representations of the twisted Drinfel'd double then classify the anyonic excitations of the theory. Similarly, in three dimensions, the Hamiltonian realization of the 4d Dijkgraaf-Witten TQFT yields point-like charge excitations and string-like flux excitations supported for instance by torus-boundaries. The manifold $\mathfrak{T}[\mathbb{T}^2] \equiv \mathbb{T}^2 \times [0,1]$ obtained by cutting open the three-torus supports states satisfying a higher-dimensional version of Ocneanu's tube algebra which yields an extension of the twisted Drinfel'd double referred to as the \emph{twisted quantum triple} \cite{Delcamp:2017pcw, Delcamp:2018efi, bullivant2019tube}. The irreducible representations of this algebraic structure then label the excitations of the theory. 

It turns out that the number of independent excited states on a manifold of the form $\mathcal N \times [0,1]$ corresponds to the ground state degeneracy on the manifold $\mathcal{N} \times \mathbb{S}^1$. Therefore, there is a systematic way to compute the ground state degeneracy of a given lattice Hamiltonian on a manifold of the form $\mathcal{N} \times \mathbb{S}^1$: Consider the tube algebra of $\mathfrak{T}[\mathcal N] \equiv \mathcal N \times [0,1]$, derive its irreducible representations, and find the ground states degeneracy as the number of such irreducible representations. 

\medskip \noindent
The strategy outlined above has been extensively employed to study gauge models of topological phases. But it can also be used in the context of 2-form topological models. We briefly sketch such strategy here and postpone to another paper a more thorough treatment. Let us first consider the manifold $\mathfrak{T}[\mathbb{S}^2] \equiv \mathbb{S}^2 \times [0,1]$. The two-sphere $\mathbb{S}^2$ has the following Betti numbers: $b_0=1$, $b_1=0$ and $b_2=1$. In other words, $\mathbb{S}^2$ has a single connected component, zero non-contractible 1-cycle and one non-contractible 2-cycle. In our context, this means that the tube $\mathfrak{T}[\mathbb{S}^2]$ can only support a single point-like magnetic excitation. 
In order to have a manifold that supports both electric and magnetic excitations, it is necessary to introduce non-contractible 1-cycles. The natural choice would be to consider the manifold $\mathfrak{T}[\mathbb{T}^2] \equiv \mathbb{T}^2 \times [0,1]$ obtained by cutting open the three-torus. This manifold would actually support one type of magnetic excitation and two types of electric excitations. However, the structure of the excitations associated with this manifold is rather involved. Therefore, as an intermediary step, we could focus instead on the three-pseudo-manifold $\mathfrak{T}[\mathbb{S}^2_{\rm no.}] \equiv \mathbb{S}^2_{\rm no.} \times [0,1]$ where $\mathbb{S}^2_{\rm no.}$ is the nodal sphere obtained after identifying two points of the two-sphere $\mathbb{S}^2$. This pseudo-manifold is homeomorphic to a pinched two-torus and can be graphically depicted as
\begin{align*}
\includegraphics[scale=1,valign=c]{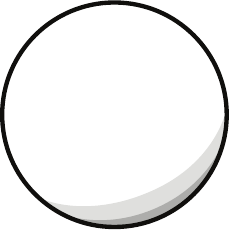}
\q \longrightarrow \q 
\includegraphics[scale=1,valign=c]{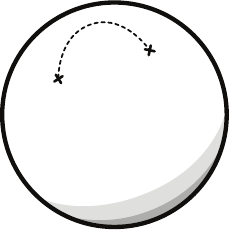}
\q \longleftrightarrow \q
\includegraphics[scale=1,valign=c]{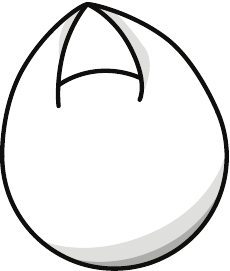}
\q \longleftrightarrow \q
\includegraphics[scale=1, valign =c]{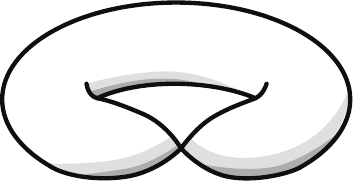} \; .
\end{align*}
The manifold $\mathbb{S}^2_{\rm no.}$ that possesses one non-contractible 2-cycle and one non-contractible 1-cycle (as opposed to the manifold $\mathfrak{T}[\mathbb{T}^2]$ that possesses two non-contractible 1-cycles) supports one point-like flux and string-like charge. As a matter of fact, we can think of $\mathfrak{T}[\mathbb{S}^2_{\rm no.}]$ as a special case of $\mathfrak{T}[\mathbb{T}^2]$ in the sense that some of the excitations supported by $\mathfrak{T}[\mathbb{T}^2]$ are condensed. As such, the study of the algebra associated to this tube cannot reveal as much information about the excitation content of the theory. It is however considerably simpler and constitutes an interesting intermediary case. We will compute in a follow-up work such tube algebras for several boundary manifolds and derive the ground state degeneracy on the corresponding closed manifolds. But let us conclude this section by computing explicitly the tube algebra for the nodal sphere in the case where the 2-form cocycle is \emph{trivial}.

So we are interested in the algebraic structure underlying the states defined on the manifold $\mathfrak{T}[\mathbb{S}^2_{\rm no.}] \equiv \mathbb{S}^2_{\rm no.} \times [0,1]$. Let us first find a basis for these states. To do so, we need to introduce a presentation of this pseudo-manifold. Since the two-sphere can be discretized by a 2-gon whose edges are identified, we can present the nodal sphere as a 2-gon whose edges and vertices are identified from which a discretization of $\mathfrak{T}[\mathbb{S}^2_{\rm no.}]$ can easily be obtained. Representing identified edges and identified vertices by an identical arrow and an identical dot, respectively, we have the following presentation:
\begin{align*}
\includegraphics[scale=1, valign=c]{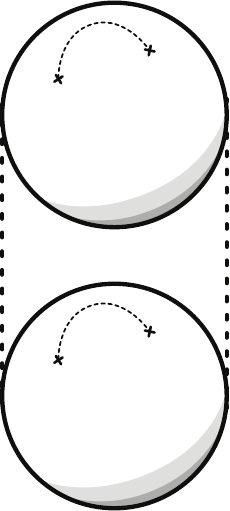} \q \longleftrightarrow \q \cylinderDis{0.3}\; .
\end{align*}
Each nodal two-sphere bounding $\mathfrak{T}[\mathbb{S}^2_{\rm no.}]$ corresponds to a non-contractible 2-cycle, hence supporting a flux excitation. Furthermore, each nodal sphere is equipped with a marked closed link which coincides with the edges of the discretization so that the 1-form gauge invariance is there relaxed. The electric excitation is captured by the face variable $b \in G$,\footnote{If we were to enforce the gauge invariance at the edges located at the boundary, we could `gauge fix away' the degree of freedom materialized by $b \in G$. This confirm that the variable $b \in G$ indeed captures an electric excitation, i.e. a violation of the 1-form gauge invariance.} while the magnetic excitation on the upper nodal sphere is captured by the face variable $a \in G$. It follows from the 2-flatness condition imposed on the 2-cycle `between the two spheres' that the face variable labeling the bottom sphere is also $a \in G$. This defines a basis of states associated with the manifold $\mathfrak{T}[\mathbb{S}^2_{\rm no.}]$ that is labeled by two group elements. 

Let us denote by $\mathcal{H}_{\mathfrak{T}[\mathbb{S}^2_{\rm no.}]}$ the Hilbert space spanned by these states. We can decompose this Hilbert space in terms of boundary colorings as follows
\begin{equation*}
\mathcal{H}_{\mathfrak{T}[\mathbb{S}^2_{\rm no.}]} =:  \bigoplus_{a \in {\rm Col}(\mathbb{S}^2_{\rm no.} \times \{0\},G_{[2]})} \mathcal{H}_{\mathfrak{T}[\mathbb{S}^2_{\rm no.}]}[ a ,a ]
\end{equation*}
where we implicitly made use of the fact that the 2-form flatness condition implies ${\rm Col}(\mathbb{S}^2_{\rm no.} \times \{0\},G_{[2]}) = {\rm Col}(\mathbb{S}^2_{\rm no.} \times \{1\},G_{[2]})$. The Hilbert space $\mathcal{H}_{\mathbb{S}^2_{\rm no.} \times [0,1]}[ a ,a ]$ is therefore spanned by states labeled by one group variable and is denoted by $(\snum{a} \! \xrightarrow{b} \! \snum{a}) \in \mathcal{H}_{\mathfrak{T}[\mathbb{S}^2_{\rm no.}]}[ a ,a ]$.

In general, the tube algebra for a manifold $\mathfrak{T}[\mathcal N]$ is defined by gluing two copies of $\mathfrak{T}[\mathcal N]$ along one of the boundary components. This gluing operation must be performed such that the marked links on the boundary are identified. The result is a manifold homeomorphic to the initial one. This yields an algebra product denoted by $\star$ which consists of two operations: $(i)$ A gluing map $\mathfrak{G}$ which identifies the boundary configurations, $(ii)$ the projection via $\mathbb{A}_{(\triangle^1)}$ onto the subspace of states satisfying the 1-form gauge invariance everywhere but at the marked links on the boundary.\footnote{A technicality we omitted is that for this gluing operation to be well-defined the two submanifolds which are identified must have opposite orientations and, correspondingly, the state spaces associated with these boundary submanifolds must be dual to each other.} The projection step $(ii)$ is required since after gluing there are new bulk 1-simplices, namely the ones that coincide with the marked links that are identified, which are not located at the boundary anymore and thus at which the 1-form gauge invariance must be enforced. Once the gauge invariance is enforced, the constraints are everywhere satisfied in the bulk of the resulting manifold so that the corresponding states satisfy the equations \eqref{eq:loc23} and \eqref{eq:loc41} under local unitary transformations. It is therefore possible to perform $\mathcal{P}_{2 \mapsto 3}$ moves in order to simplify the triangulation of the resulting manifold so as to obtain a state living in  $\mathfrak{T}[\mathcal N]$ again. Putting everything together, this definition in the case of $\mathfrak{T}[\mathbb{S}^2_{\rm no.}]$ reads
\begin{equation*}
\begin{array}{cccccc}
\star : &\mathcal{H}_{\mathfrak{T}[\mathbb{S}^2_{\rm no.}]} \otimes \mathcal{H}_{\mathfrak{T}[\mathbb{S}^2_{\rm no.}]}
&\xrightarrow{\;\; \mathfrak{G} \;\;}
& \mathcal{H}^{\rm ext.}_{\mathfrak{T}[\mathbb{S}^2_{\rm no.}]\; \cup_{\mathbb{S}^2_{\rm no.}} \mathfrak{T}[\mathbb{S}^2_{\rm no.}]} 
& \xrightarrow{\;\; \mathbb{A} \;\;} 
& \mathcal{H}_{\mathfrak{T}[\mathbb{S}^2_{\rm no.}] / \sim} \\
&
(\snum{a_1} \! \xrightarrow{b_1} \! \snum{a_1}) \smo (\snum{a_2} \! \xrightarrow{b_2} \! \snum{a_2}) 
& \longmapsto 
& \delta_{a_1,a_2} (\snum{a_1} \! \xrightarrow{b_1} \! \snum{a_1}) \otimes (\snum{a_1} \! \xrightarrow{b_2} \! \snum{a_1})
& \longmapsto 
& \delta_{a_1,a_2} \mathbb{A}\triangleright  (\snum{a_1} \! \xrightarrow{b_1} \! \snum{a_1}) \otimes (\snum{a_1} \! \xrightarrow{b_2} \! \snum{a_1})
\end{array}
\end{equation*}
where $ \mathcal{H}^{\rm ext.}_{\mathfrak{T}[\mathbb{S}^2_{\rm no.}]\; \cup_{\mathbb{S}^2_{\rm no.}} \mathfrak{T}[\mathbb{S}^2_{\rm no.}]} $ is the Hilbert space of states defined on the pseudo-manifold resulting from the gluing such that the 1-form gauge invariance is not yet enforced at the new bulk 1-simplex. Since the cocycle is taken to be trivial, both steps in the definition of the algebra product are particularly simple and the algebra simply reads
\begin{equation}
	  (\snum{a_1} \! \xrightarrow{b_1} \! \snum{a_1}) \star (\snum{a_2} \! \xrightarrow{b_2} \! \snum{a_2}) = \delta_{a_1,a_2} (\snum{a_1} \! \xrightarrow{b_1+b_2} \! \snum{a_1}) \; .
\end{equation}
Finding the irreducible representations is immediate and the ground state degeneracy on the manifold $\mathbb{S}^2_{\rm no.} \times \mathbb S^1$ is thus $|G|^2$ as expected. Naturally, the situation is considerably more complicated when the 2-form 4-cocycle is not trivial and this will be the subject of a follow-up work. 
\section{Correspondence with the Walker-Wang model \label{sec:WW}}
We exposed in the previous section how the local unitary transformations whose amplitudes are given in terms of a 2-form 4-cocycle can be reduced to a 2--2 Pachner move or a braiding move. Furthermore, we showed, algebraically and geometrically, how the 2-form 4-cocycle condition yields the so-called pentagon and hexagon relations which are the defining equations of a certain braided monoidal category. But it turns out that this braided monoidal category is the input data of another lattice Hamiltonian model, namely the Walker-Wang model \cite{Walker:2011mda}.\footnote{More precisely, the input data of the Walker-Wang model is a unitary fusion braided category. It is possible to endow the monoidal category we are interested in, namely the category of $G$-graded vector spaces, with the structures necessary to turn it into a unitary fusion braided category. However, for this specific example, it is not required to do so as far as the definition of the Hamiltonian is concerned.} In this section, we first provide further detail regarding the interplay between the 2-form cohomology group $H^4(G_{[2]},\rU(1))$ and abelian braided monoidal categories, then we study to which extent our 2-form gauge model is related to the Walker-Wang model.

\bigskip \noindent

\subsection{Braided monoidal categories}

 First let us provide some basic definitions of category theory. More details can be found for instance in \cite{etingof2016tensor}:

\begin{definition}[\emph{Monoidal category}] A \emph{monoidal category} is a sextuple  $(\mathcal{C}, \otimes, \unit,\ell,  r, \alpha)$ where:
	\begin{enumerate}[itemsep=0.4em,parsep=1pt,leftmargin=4em]
		\item[$\circ$] $\mathcal{C}$ is a \emph{category} whose collection of objects is denoted by ${\rm Ob}(\mathcal{C})$ and for $x,y \in {\rm Ob}(\mathcal{C})$, the collection of morphisms between them is denoted by ${\rm Hom}_{\mathcal{C}}(x,y)$.
		\item[$\circ$] $\otimes$ is a \emph{bifunctor} $\otimes: \mathcal{C} \times \mathcal{C} \rightarrow \mathcal{C}$ referred to as the \emph{tensor product}.
		\item[$\circ$] $\unit \in {\rm Ob}(\mathcal{C})$ is a unit object.
		\item[$\circ$] $\alpha$, $\ell$ and $r$ are natural isomorphisms:
		\begin{align*}
			\alpha_{x,y,z} : (x \otimes y) \otimes z &\xrightarrow{\sim} x \otimes (y \otimes z) \\
			\ell_x : \unit \otimes x &\xrightarrow{\sim} x \\
			r_x : x \otimes \unit &\xrightarrow{\sim} x		
		\end{align*}
		referred to as the \emph{associator}, the \emph{left unitor} and the \emph{right unitor}, respectively. These natural isomorphisms are subject to some coherence relations that we omit for now, namely the \emph{pentagon relation} and the \emph{triangle relation}.
	\end{enumerate}
\end{definition}
\noindent
In this article, we are only interested in a specific monoidal category, namely the category $\mathbb{C}$--${\rm Vec}_G$ of $G$-graded vector spaces over the field of complex numbers, where $G$ is a finite abelian group.\footnote{As mentioned earlier, the category $\mathbb{C}$--${\rm Vec}_G$ is actually an example of \emph{fusion category} but we do not need the corresponding additional structures in order to define the Walker-Wang model and show the correspondence with our 2-form gauge model.} We define a $G$-graded vector space as a vector space $V$ which satisfies $V = \bigoplus_{g \in G}V_g$ and the tensor product of two $G$-graded vector spaces reads
\begin{equation*}
	(V \otimes W)_g = \bigoplus_{h,k \in G \atop h+k=g}V_h \otimes W_k \; .
\end{equation*}
The category $\mathbb{C}$--${\rm Vec}_G$ has finitely many simple objects provided by the 1-dimensional $G$-graded vector spaces which are in one-to-one correspondence with group elements $g \in G$. We denote these simple objects by $\delta_{g \in G}$ and  they satisfy by definition ${\rm End}(\delta_g)= \mathbb{C}$, Naturally, the tensor product of simple objects boils down to the group multiplication: $\delta_{g} \otimes \delta_h \cong \delta_{g+h}$. Since it is enough to define the associator of the category on the simple objects, we are looking for an isomorphism determined by a function $\alpha: G^3 \rightarrow \mathbb{C}^\times$ such that
\begin{equation*}
	\alpha_{g,h,k} = \alpha(g,h,k) \cdot {\rm id}_{\delta_{g+h+k}} : (\delta_g \otimes \delta_h) \otimes \delta_k \xrightarrow{\sim} \delta_g \otimes (\delta_h \otimes \delta_k) \; .
\end{equation*}
The pentagon relation then implies that $\alpha$ is a group 3-cocyle in $H^3(G,\mathbb{C}^\times)$ (which is the same as $H^3(G,\rU(1))$\,). The triangle relation implies that if the left and the right unitors are trivial, then the 3-cocycle $\alpha$ is normalized, i.e. $\alpha(g, \unit ,h) = 1$, $\forall g,h \in G$. We will now turn the category $\mathbb{C}$--${\rm Vec}_G$ into a braided monoidal category:
\begin{definition}[\emph{Braided monoidal category}] 
	Given a monoidal category $\mathcal{C}$, a \emph{braiding} on $\mathcal{C}$ is a natural isomorphism $R_{x,y} : x \otimes y \xrightarrow{\sim} y \otimes x$ that is subject to the so-called \emph{hexagon relations}. A \emph{braided monoidal category} is then defined as a pair $\{\mathcal{C},R\}$.
\end{definition}
\noindent
In order to turn $\mathbb{C}$--${\rm Vec}_G$ into a braided monoidal category, we only need to add a braiding, i.e a group 2-cochain $R \in C^2(G,\mathbb{C}^\times)$, satisfying the hexagon equations which are exactly \eqref{eq:Hex1} and \eqref{eq:Hex2}. Interestingly, the set of associators and braidings as defined above enters the definition of the following cohomology:

\begin{definition}[\emph{Abelian cohomology group}]
	Pairs $\{\alpha, R\}$ satisfying \eqref{eq:Gr3Coc}, \eqref{eq:Hex1} and \eqref{eq:Hex2} are referred to as \emph{abelian cocycles} on $G$ and we denote the set of all abelian cocycles on $G$ by $Z^3_{\rm ab}(G,\mathbb{C}^\times)$. Let $\beta \in C^2(G,\mathbb{C}^\times)$, we call an \emph{abelian coboundary} a pair $\{\alpha, R\}$ such that
	\begin{align}
		\alpha(a,b,c) = \frac{\beta(b,c)\, \beta(a,b+c)}{\beta(a+b,c) \, \beta(a,b)} \q , \q R(a,b) = \frac{\beta(a,b)}{\beta(b,a)} \; .
	\end{align}
	The set of all abelian coboundaries is denoted by $B^3_{\rm ab}(G, \mathbb{C}^\times)$. We finally define the \emph{abelian cohomology  group} as the quotient space:
	\begin{equation}
		H^3_{\rm ab}(G,\mathbb{C}^\times) = \frac{Z^3_{\rm ab}(G,\mathbb{C}^\times)}{B^3_{\rm ab}(G, \mathbb{C}^\times)} \; .
	\end{equation}
\end{definition}
\noindent
It results from the definitions above that isomorphism classes of braided monoidal categories whose simple objects form an abelian group $G$ are classified by $H^3_{\rm ab}(G,\mathbb{C}^\times)$. 

\bigskip 
\noindent
Using these definitions, we can rephrase our previous result: Given a 2-form 4-cocycle $\omega$, the group cochains $\alpha:(a,b,c)\mapsto \omega(a,b,c|\zo,\zo|\zo)^{-1}$ and $R:(a,b)\mapsto \omega(\zo,\zo,b|a,\zo|\zo)$ form an abelian cocycle. Furthermore, it follows from \eqref{eq:cobounA} and \eqref{eq:cobounR} that the pair $\{d^{(3)}\alpha(a,b,c|\zo,\zo|\zo) ,d^{(3)}\alpha(\zo,\zo,a|b,\zo|\zo)\}$ forms an abelian coboundary. Putting everything together, it should not surprise the reader that \emph{there is a bijection between $H^4(G_{[2]},\rU(1))$ and $H^3_{\rm ab}(G,\rU(1))$}.\footnote{Here we are implicitly making use of the fact that when the group is finite, there is no difference between the cohomology of abelian cocycles valued in $\rU(1)$ and in $\mathbb{C}^\times$.} 

If the relationship between $H^4(G_{[2]},\rU(1))$ and $H^3_{\rm ab}(G,\rU(1))$ is natural in light of our derivations in the previous section, a complete proof of this bijection would require more care. Since our work does not strictly rely on this bijection, we refer the reader to \cite{1998math.....11047Q} instead. Nonetheless, let us assume this result until the end of this subsection and let us pursue our analysis. Recall that we defined earlier quadratic forms on a finite abelian group $G$ valued in $\mathbb{C}^\times$ as a function $q:G \to \mathbb{C}^\times$ such that $q(g) = q(-g)$ and 
	\begin{equation*}
		b:(g,h)\mapsto \frac{q(g)\, q(h)}{q(g+h)}
	\end{equation*}
is bilinear. We denote the group of quadratic forms on $G$ by ${\rm Quad}(G)$.
Given a braided monoidal category whose simple objects form the abelian group $G$ (such as  $\mathbb{C}$--${\rm Vec}_G$), we can construct easily a quadratic form $q: G \to \mathbb{C}^\times$ such that for all $g \in G$, $q(g) = R(g,g)\in {\rm Aut}_\mathcal{C}(g \otimes g) = \mathbb{C}^\times$. It follows directly that
\begin{align*}
	H^3_{\rm ab}(G, \mathbb{C}^\times) &\to {\rm Quad}(G) \\
	\{R,\alpha\} &\mapsto q(g)=R(g,g) 
\end{align*} 
is a homomorphism. But, and this is a result by Eilenberg and MacLane presented in a succinct way in \cite{etingof2016tensor}, this homomorphism turns out to be an \emph{isomorphism}. This means that abelian cocycles are classified by quadratic forms. Since we assumed that there was a bijection between $H^4(G_{[2]},\rU(1))$ and $H^3_{\rm ab}(G,\rU(1))$, this also proves that $H^4(G_{[2]},\rU(1))$ is classified by quadratic forms on $G$. Despite the numerous gaps we left, we hope this brief review provides some intuition as to why this is the case. This analysis thus completes the study initiated in sec.~\ref{sec:Sigma} where we made use of the same bijection in order to write down explicitly the action of a 2-form gauge theory in terms of a quadratic function $q$ and the Pontrjagin square $\mathfrak{P}$. In any case, we do not need this result to display how our 2-form gauge model is related to the Walker-Wang model for the braided monoidal category of $G$-graded vector spaces.

\subsection{Walker-Wang model for the category of $G$-graded vector spaces}
The Walker-Wang model was first introduced in \cite{Walker:2011mda} as a generalization of Levin-Wen models to 3+1 dimensions. In general, the input data for the Walker-Wang model is a \emph{unitary braided fusion category}. Crucially, the properties of the corresponding topological phase depends on whether the category is \emph{modular}. Indeed, if the category is modular, then the model is trivial in the sense that it displays neither ground state \emph{degeneracy} nor \emph{fractionalized excitations}. In this section, we are only interested in the Walker-Wang model based upon the braided (fusion) monoidal category of $G$-graded vector spaces whose input data is a finite abelian group $G$,  a group 3-cocycle $\alpha$ and a group 2-cochain $R$ which together satisfy the pentagon and the hexagon relations. In light of the correspondence between abelian braided monoidal categories and the cohomology class of 2-form 4-cocycles, we want to emphasize how our 2-form gauge model is related to this Walker-Wang model.

\bigskip \noindent
The lattice Hamiltonian introduced by Walker and Wang was originally defined on a cubic lattice such that all the nodes are six-valent. Crucially, in order to define the action of the commuting operators, it is necessary to split the six-valent nodes into three-valent ones so that the action of the plaquette operator can be expressed in terms of 2--2 Pachner moves (or \emph{F-moves}) and braiding moves. The Hilbert space of the model is then spanned by all graph states obtained by labeling the edges of the graph obtained after such splitting. Different splittings must lead to equivalent models as they all match in the continuum limit, but a specific choice needs to be made nonetheless and it is referred to as a choice of \emph{resolution} of the vertices. Note that this model can be generalized to richer input data such as $G$-crossed braided fusion categories, see \cite{Williamson:2016evv}.

In this section, we study the Walker-Wang model based upon the monoidal braided category of $G$-graded vector spaces. However, instead of working with a cubic discretization, we define the model on the one-skeleton of the 2-complex $\Upsilon$ dual to the triangulation $\triangle$. Naturally, since $\triangle$ is obtained as a gluing of $3$-simplices all the nodes of the one-skeleteon of $\Upsilon$ are four-valent. Therefore, it is still necessary to perform a (single) splitting of the nodes in order to obtain a graph whose nodes are all three-valent. The lattice Hamiltonian is given by 
\begin{equation}
	\mathbb{H}_{\rm WW} = - \sum_{\mathsf n} \mathbb{A}_{\mathsf n} - \sum_{\mathsf p}\mathbb{B}_{\mathsf p}
\end{equation}
such that to each three-valent node, we assign an operator $\mathbb{A}_{\mathsf n}$ which enforces the oriented product of the group variables labeling the edges meeting at the node ${\mathsf n}$ to vanish, and to each plaquette, we assign an operator $\mathbb{B}_{\mathsf p}$ which modifies the group configuration of the edges adjacent to ${\mathsf p}$ by `fusing' a loop of defect into the boundary of ${\mathsf p}$. We can define more precisely the action of $\mathbb{B}_{\mathsf p}$ using some graphical calculus in a way which is reminiscent of (2+1)d string net models.  To do so, we consider a special example, namely the triangular plaquette that is the one-skeleton of the dual graph of the union of the three 3-simplices $\snum{(0134)}$, $\snum{(0124)}$ and $\snum{(0234)}$ as depicted in \eqref{eq:actionA} so that we have the correspondence:
\begin{equation}
	\label{eq:corresWW2Form}
	\HamONE{1}{1}\q  \longleftrightarrow \q \WWHamONE{0.3}{2}
\end{equation}
where the 2-simplex $\snum{(123)}$ is not part of the 2-complex on the left-hand-side.
Without loss of generality, we make the following choice of splitting into three-valent nodes:
\begin{align}
	\label{eq:splittingWW}
	\WWHamONE{0.3}{1} \q \overset{\mathfrak{s}}{\longmapsto} \q  \WWHamTWO{0.3}
\end{align}
where in the second drawing we kept some labeling implicit as they can be deduced from the branching rules implemented at each node by the operator $\mathbb{A}_{\mathsf n}$. Furthermore, the orientation of the edges is also kept implicit, however it is always such that the group variable associated with an unlabeled link is obtained as the sum of the group variables labeling the other two links meeting at this node. We write the plaquette operator $\mathbb{B}_{\mathsf p} = \frac{1}{|G|}\sum_{d \in G}\mathbb{B}_{\mathsf p}^d$ where the action of $\mathbb{B}_{\mathsf p}^d$ is defined graphically via the insertion of a loop of defect $d$ as follows:
\begin{align}
	\mathbb{B}^d_{\mathsf p} \triangleright \Bigg| \; \WWHamTWO{0.3} \; \Bigg\rangle &= \Bigg| \; \WWHamTWObis{0.3} \; \Bigg\rangle \\ &=
	\Bigg| \; \WWHamTHREE{0.3} \; \Bigg\rangle 
\end{align}
where the last state is obtained by fusing the loop defect labeled by $d$ into the plaquette via trivial $\mathcal{P}_{2 \mapsto 2}$ moves. 

It now remains to use local unitary transformation so as to obtain a state whose underlying graph is identical to the initial one. To do so, we first perform three $R$-moves in order to move aside the links labeled by $e+f$, $f+h$ and $h$, so that $\mathcal{P}_{2 \mapsto 2}$ moves can be performed (as in 2d) without worrying about non-trivial braidings. Once all the $\mathcal{P}_{2 \mapsto 2}$ moves are performed, the links labeled by $e+f$, $f+h$ and $h$ are brought back to their original positions using three $R$-moves. Putting everything together, these transformations read
\begin{align} 
	\label{eq:WWinit}
	&\Bigg| \; \WWHamTHREE{0.3} \; \Bigg\rangle  \\ & \q =
	\frac{R(c,f+h)}{R(b+c,e+f)\,R(c+f,h)} \Bigg| \; \WWHamFOUR{0.3} \; \Bigg\rangle \\
	& \q =	\frac{R(c,f+h)}{R(b+c,e+f)\,R(c+f,h)} \\[-2em] & \q  \cdot \frac{\alpha(a,b+c,d)\,\alpha(f+h,c,d)\,\alpha(e+b,c+f,d)}{\alpha(e+f,b+c,d)\,\alpha(a+b,c,d)\,\alpha(h,c+f,d)}
	\Bigg| \; \WWHamFIVE{0.3} \; \Bigg\rangle \\[0.3em]
	& \q  =
	\frac{R(b+c+d,e+f)\,R(c+f+d,h)\,R(c,f+h)}{R(b+c,e+f)\,R(c+f,h)\,R(c+d,f+h)} \\[-2em] \label{eq:WWfin}
	& \q \cdot  \frac{\alpha(a,b+c,d)\,\alpha(f+h,c,d)\,\alpha(e+b,c+f,d)}{\alpha(e+f,b+c,d)\,\alpha(a+b,c,d)\,\alpha(h,c+f,d)}\Bigg| \; \WWHamSIX{0.3} \; \Bigg\rangle \; .
\end{align}
Denoting the initial state in \eqref{eq:WWinit} as $| \psi_{\rm init.} \ra$ and the final state in \eqref{eq:WWfin} as $| \psi_{\rm fin.} \ra$, the plaquette term for this configuration is $\la \psi_{\rm fin.} |\, \mathbb{B}_{\mathsf p}^d \, | \psi_{\rm init.} \ra$ and reads
\begin{empheq}[box=\fbox]{align}
	\nn
	& \la \psi_{\rm fin.} |\, \mathbb{B}_{\mathsf p}^d \, | \psi_{\rm init.} \ra 
	\\ \nn & \q = 
		\frac{R(b+c+d,e+f)\,R(c+f+d,h)\,R(c,f+h)}{R(b+c,e+f)\,R(c+f,h)\,R(c+d,f+h)} \cdot  \frac{\alpha(a,b+c,d)\,\alpha(f+h,c,d)\,\alpha(e+b,c+f,d)}{\alpha(e+f,b+c,d)\,\alpha(a+b,c,d)\,\alpha(h,c+f,d)} \; .
\end{empheq}
The example we chose in order to illustrate the definition of the plaquette operator is admittedly very special but it can be generalized easily to any other situation (in particular with a different distribution of legs pointing inward or outward the plaquette). But, since this special configuration is the one corresponding to the situation chosen to illustrate the definition of the operator $\mathbb{A}_{(\triangle^1)}$ of the 2-form gauge model in sec.~\ref{sec:Ham}, we are now able to draw a correspondence between the two Hamiltonian models. 

\subsection{From the 2-form gauge model to the Walker-Wang model}

In this section, we sketch the correspondence between the 2-form gauge model whose input data is $\{G, \omega \in Z^4(G_{[2]},\rU(1))\}$ and the Walker-Wang model for the braided monoidal category of $G$-graded vector spaces whose input data is $\{G, (\alpha,R) \in Z_{\rm ab}^3(G,\rU(1))\}$. We will not prove this correspondence in its full generality but merely focus on the specific example used above to define the two models.

\medskip \noindent
First of all, notice that the operator $\mathbb{A}_{\mathsf n}$ of the Walker-Wang model and the operator $\mathbb{B}_{(\triangle^3)}$ of the 2-form gauge model are essentially the same, both implement the branching rules. Therefore, our focus is on the action of the operator $\mathbb{B}_{\mathsf p}$ and $\mathbb{A}_{(\triangle^1)}$, More specifically, we want to compare their amplitude in the case of the configuration \eqref{eq:corresWW2Form}. It is clear that both of them enforce a twisted 1-form gauge invariance. Furthermore, it follows from the duality relation between $\triangle$ and $\Upsilon$ that a twisted 1-form gauge transformation at the 1-simplex $\snum{(04)}$ as performed by $\mathbb{A}_{(\triangle^1)}$ acts on the same 2-simplex variables as the plaquette operator $\mathbb{B}_{\mathsf p}$ via a loop of defect, namely $\snum{(014)}$, $\snum{(024)}$ and $\snum{(034)}$. However, it is not clear how the amplitudes of these two operators match, especially in light of the fact that the Walker-Wang model requires a splitting of nodes into three-valent ones. 

We reproduce below the amplitude of the operator $\mathbb{A}^d_{(04)}$:
\begin{equation}
	\label{eq:ampA}
	\la \psi_{\rm fin.} |\, \mathbb{A}_{(04)}^d \, | \psi_{\rm init.} \ra =  
	\frac{\omega(b+e,c+f,d|h,\zo|\zo)\, \omega(a,b+c,d|e+f,\zo|\zo)}{\omega(a+b,c,d|f+h,\zo|\zo)} 
\end{equation}
where we recognize that the same term appears three times, but for different variables. First, recall that the 2-form 4-cocycle $\omega \in Z^4(G_{[2]},\rU(1))$ reduces to the group 3-cocycle $\alpha \in Z^3(G,\rU(1))$ and the $R$-matrix $R \in C^2(G,\rU(1))$ such that $\alpha(a,b,c) = \omega^{-1}(a,b,c|\zo, \zo|\zo)$ and $R(a,b) = \omega(\zo,\zo,b|a,\zo|\zo)$, respectively. Now, let us consider the cocycle condition 
\begin{equation}
	d^{(4)}\omega(a,\zo,c,d|\zo,\zo,\zo|h,\zo|\zo) 
	= \frac{\omega(\zo,\zo,\zo|h,\zo|\zo)\, \omega(a,c,d|h,\zo|\zo)\, \omega(a,\zo,c+d|\zo,\zo|h)}{\omega(\zo,c,d|h,\zo|\zo)\, \omega(a,c,d|\zo,\zo|\zo)\, \omega(a,\zo,c|\zo,\zo|h)} = 1
\end{equation}
and let us rewrite it as follows 
\begin{equation}
	\label{eq:splittingONE}
	\boxed{
	\omega(a,c,d|h,\zo|\zo) = \mathfrak{s}(a,c,h) \, \alpha(a,c,d) \, \mathfrak{c}(c,d,h) \, \mathfrak{s}^{-1}(a,c+d,h)}
\end{equation}
where we defined $\mathfrak{s}(a,b,c):= \omega(a,\zo,b|\zo,\zo,|c)$. Moreover, we showed in \eqref{eq:firstHexSimp} that $\mathfrak{c}(c,d,h) = \omega(\zo,c,d|h,\zo|\zo)= R(c+d,h)\alpha^{-1}(h,c,d)R^{-1}(c,h)$ so that \eqref{eq:splittingONE} provides another expression for the terms appearing in the amplitude of the operator $\mathbb{A}^d_{(04)}$ in terms of $\alpha$, $R$ and a group 3-cochain $\mathfrak{s}$ that we have just defined. If we use equation \eqref{eq:splittingONE} in \eqref{eq:ampA}, we can rewrite the amplitude of the operator $\mathbb{A}_{(04)}^d$ as
\begin{figure}[]
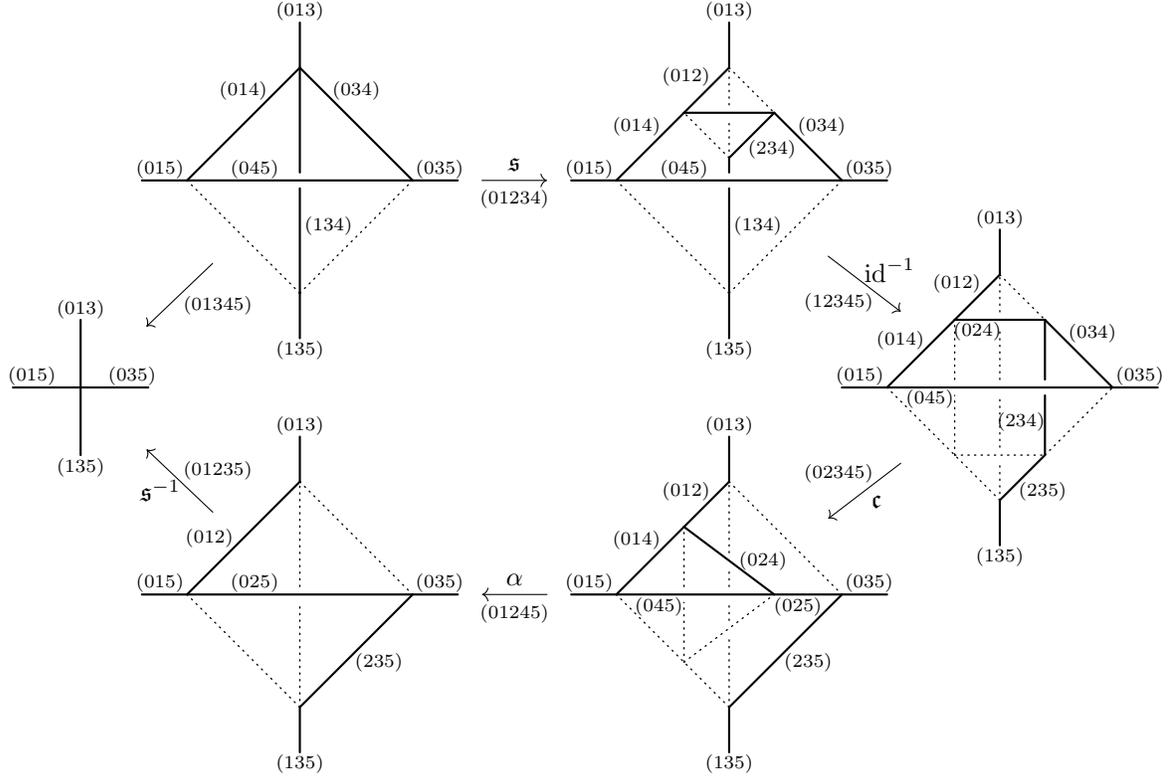

	\center
	\WWCocycle{}
	\caption{Graphical depiction of the 2-form cocycle condition $d^{(4)}\omega(a,\zo,c,d|\zo,\zo,\zo|h,\zo|\zo) \equiv \la d^{(4)}\omega ,(012345)\ra = 1$. This illustrates how the term  $\omega(a,c,d|h,\zo|\zo) = \la \omega ,(01345)\ra$ encodes all the defining steps of the plaquette operator in the Walker-Wang model: the splitting of the four-valent node into three valent ones, the combination of $\mathcal{P}_{2 \mapsto 2}$ moves and braiding moves as well as the recombination of three-valent nodes into a single four-valent one.}
	\label{fig:splittingWW}
\end{figure}
\begin{align}
	\nn
	&\la \psi_{\rm fin.} |\, \mathbb{A}_{(04)}^d \, | \psi_{\rm init.} \ra \\
	\nn
	& \q = 
	\frac{R(b+c+d,e+f)\,R(c+f+d,h)\,R(c,f+h)}{R(b+c,e+f)\,R(c+f,h)\,R(c+d,f+h)} \cdot  \frac{\alpha(a,b+c,d)\,\alpha(f+h,c,d)\,\alpha(e+b,c+f,d)}{\alpha(e+f,b+c,d)\,\alpha(a+b,c,d)\,\alpha(h,c+f,d)} \\
	\label{eq:ampAexp}
	& \q \cdot \frac{\mathfrak{s}(b+e,c+f,h)\, \mathfrak{s}(a,b+c,e+f)\, \mathfrak{s}(a+b,c+d,f+h)}{\mathfrak{s}(b+e,c+f+d,h)\, \mathfrak{s}(a,b+c+d,e+f)\, \mathfrak{s}(a+b,c,f+h)}
\end{align}
which reproduces exactly $ \la \psi_{\rm fin.} |\, \mathbb{B}_{\rm p}^d \, | \psi_{\rm init.} \ra $ up to the $\mathfrak{s}$-terms. 

So we are left to explain the role played by $\mathfrak{s}$. To do so, we use the same technique as in sec.~\ref{sec:consistency}, i.e. we identify $d^{(4)}\omega(a,\zo,c,d|\zo,\zo,\zo|h,\zo|\zo)$ with $\la d^{(4)}\omega ,\snum{(012345)}\ra$ and represent graphically the cocycle condition. However, this time the cocycle condition is not obtained as an equality between two sequences of $\mathcal{P}_{2 \mapsto 3}$ moves but by equating two sequences composed of two $\mathcal{P}_{1 \mapsto 4}$ moves and one $\mathcal{P}_{2 \mapsto 3}$ move so that each term appearing in $\la d^{(4)}\omega ,\snum{(012345)}\ra =1$ corresponds to a $\mathcal{P}_{2 \mapsto 3}$ move or a $\mathcal{P}_{1 \mapsto 4}$ move. The 2-simplex variables are identified using the correspondence \eqref{eq:pairing4Coc} and we represent by a dashed line vanishing variables. The result is represented fig.~\ref{fig:splittingWW}. We recognize that the different $\mathcal{P}_{2 \mapsto 3}$ and $\mathcal{P}_{1 \mapsto 4}$ moves reduce to: a move which splits the 4-valent node into two 3-valent ones whose amplitude is given by the function $\mathfrak{s}$, a trivial move which does not change the combinatorics of the graph built out of the bold links, a combination of braiding moves and $\mathcal{P}_{2 \mapsto 2}$ whose amplitude is given by the function $\mathfrak{c}$ as represented in \eqref{eq:decompC}, a $\mathcal{P}_{2 \mapsto 2}$ move, and finally a move which puts together two 3-valent nodes into a single 4-valent one whose amplitude is given by $\mathfrak{s}^{-1}$. We deduce that the effective action of $\mathfrak{s}$ can be graphically interpreted in terms of string diagrams as
\begin{equation}
	\Smove{0.3}{1} \q \overset{\mathfrak{s}(a,b,c)}{\longrightarrow} \q
	\Smove{0.3}{2} \; .
\end{equation}
The presence of the $\mathfrak{s}$-terms in \eqref{eq:ampAexp} is therefore explained by the fact that the definition of the operator $\mathbb{B}_{\mathsf p}$ in the Walker-Wang model requires an \emph{ad hoc} splitting of the nodes into 3-valent nodes while our model is defined directly in terms of the 4-valent initial ones.

So to summarize, the analysis carried out in this part confirms two things:  $(i)$ The correspondence between our model based on a 2-form 4-cocycle $\omega \in Z^4(G_{[2]}, \rU(1))$ and the Walker-Wang model for the category of $G$-graded vector spaces, $(ii)$ The fact that the \emph{ad hoc} resolution of the vertices required to define the plaquette term in the Walker-Wang model is directly included in the definition of the 2-form 4-cocycle. Furthermore, our approach makes transparent the fact that the plaquette operator of the Walker-Wang model  for the case of the category  $\mathbb{C}$--${\rm Vec}_G$ actually implements the invariance under twisted 1-form gauge transformations at the 1-simplex dual to the plaquette, as it is obvious from the definition of the operator $\mathbb{A}_{(\triangle^1)}$ of our 2-form gauge model.

\FloatBarrier

\section{Conclusion}
Gauge and higher gauge models of topological phases of matter have been under intense investigation in the past years, one reason being that they seem to encapsulate most of the known models displaying non-trivial topological order in (3+1)d. In this paper, we studied in detail models that have a 2-form gauge theory interpretation. The goal of this paper was two-fold: Study properties of 2-form topological gauge theories in the continuum and present an 3d exactly solvable model that is the Hamiltonian realization of a (3+1)d 2-form topological field theory.

Firstly, we defined 2-form topological theories as sigma models whose target space is provided by the second classifying space $B^2G$ of a finite abelian group $G$. These are classified by cohomology classes $[\omega] \in H^4(B^2G, \mathbb R / \mathbb Z)$. It turns out that this cohomology group is isomorphic to the group of (possibly degenerate) quadratic functions on $\mathbb R / \mathbb Z$ allowing for a more explicit expression of the partition function. Furthermore, such discrete 2-form gauge theories can be embedded into continuous strict 2-group gauge theories. We described the construction  of $\rU(1)$ strict 2-group connections using the language of Deligne-Beilinson cohomology. More specifically, we showed that the space of gauge inequivalent $q$-form $\rU(1)$ connections was isomorphic to the $q$-th Deligne-Beilinson cohomology and, within the same framework, how $\rU(1)$ 1-form and 2-form connections could be combined so as to obtain a strict 2-group connection. 

Secondly, we defined a lattice Hamiltonian realization of a 2-form gauge theory. To do so, we defined the 2-form cohomology of an abelian group that is isomorphic to the cohomology of its second classifying space as provided by the $W$-construction, and derived its properties. We showed in particular how a 2-form 4-cocycle reduces to an abelian 3-cocycle that is the input data of an abelian braided monoidal category. But these monoidal categories are classified by quadratic functions on the group, hence closing the loop with the results obtained in the first part. Correspondingly, we explained how our 2-form gauge model is related to the Walker-Wang model. Interestingly, we displayed how the \emph{ad hoc} splitting into three-valent vertices required for the definition of the Walker-Wang Hamiltonian is now directly encoded in the 2-form cocycle itself. 

The tools developed in this manuscript can be generalized and used for other purposes. For instance, the Deligne-Beilinson cohomology can be used to define \emph{weak} 2-group connections, while the strategy followed to define the 2-form cohomology can be extended to define a weak 2-group cohomology. As a matter of fact the study of weak 2-group gauge models of topological phases as initiated in \cite{delcamp2018gauge}
was one of the motivations for the present work and we believe that this work is useful to study more systematically these higher group gauge theories. Furthermore, in light of the correspondence between 2-form 4-cocycles and abelian braided monoidal categories, we believe that the tools developed in this manuscript could be used to study the braiding of higher-dimensional excitations from a cohomological point of view. More specifically, we anticipate 2-form 5-cocycles to be related to the braiding statistics of loop-like excitations \cite{Baez:2006un, Wang:2014xba, Wang:2014oya, putrov2016braiding, Bullivant:2018pju, Bullivant:2018djw}.

Apart from phases displaying \emph{intrinsic topological order} as studied in this manuscript, it is possible to define \emph{symmetry protected} topological phases of matter (SPTs). In general, SPTs are gapped phases of matter that are short-range entangled and have a global symmetry acting locally so that the phase can be adiabatically connected to the trivial one upon breaking the symmetry. It is possible for SPTs to contain operators that are localized on ($q$$-$1)-dimensional submanifolds (see e.g. \cite{Yoshida:2015cia, 2018arXiv180907325R}), in which case the global symmetry is referred to as a ($q$$-$1)-form global symmetry \cite{gaiotto2015generalized}. Gauging such a ($q$$-1$)-form global symmetry requires the introduction of $q$-form flat connection and the resulting theory is a $q$-form topological gauge theory. This gauging process was studied in \cite{delcamp2018gauge} both at the level of the action and in terms of its lattice realization, and could be reformulated in light of the constructions presented in this paper.

\begin{center}
	\textbf{Acknowledgments}
\end{center}
\noindent
CD would like to thank Alex Bullivant for several discussions on related topics and for reading an early version of this manuscript. AT would like to thank Faroogh Moosavian, Heidar Moradi and Shinsei Ryu for several useful discussions.
This project has received funding from the European Research Council (ERC) under the European Union’s Horizon 2020 research and innovation programme through the ERC Starting Grant WASCOSYS (No. 636201).
This research was supported in part by Perimeter Institute for Theoretical Physics.
Research at Perimeter Institute is supported by the Government of Canada through the Department of Innovation, Science and Economic Development Canada and by the Province of Ontario through the Ministry of Research, Innovation and Science.

\newpage
\appendix
\section{Postnikov towers and sigma models\label{app:Postnikov}}
In this appendix, we present further generalizations of the sigma models introduced in sec.~\ref{sec:Sigma} where the target space is provided by a $k$-stage Postnikov tower.

 As the name suggests, the $k$-stage Postnikov tower $E_k$ can be built in a sequence of $k$ steps. The \emph{first} stage is provided by a classifying space $E_1=B^{q_1}G_1$. The \emph{second} stage is provided by a fibration over $E_1$ with the fiber being isomorphic to $B^{q_2}G_2$. This step is captured in the sequence
\begin{align}
	0\to B^{q_2}G_2 \to E_2 \to E_1 \to 0
\end{align}
whose extension class is $[\alpha_2]\in H^{q_2+1}(E_1,G_2)$. At the \emph{third} stage, we build a space $E_{3}$ as a fibration over $E_2$ so that
\begin{align}
	0\to B^{q_2}G_3 \to E_3 \to E_2 \to 0
\end{align}
whose extension class is $[\alpha_3]\in H^{q_3+1}(E_2,G_3)$.\footnote{Throughout we assume that $q_1<q_2<n_3<\ldots < q_k$.} This sequence proceeds iteratively until
\begin{align}
	0\to B^{q_k}G_k \to E_k \to E_{k-1} \to 0
	\label{eq:k_stage}
\end{align}
whose extension class is $[\alpha_k]\in H^{q_k+1}(E_{k-1},G_k)$.  A homotopy class of map from $\cM$ to $E_k$ is provided by a $k$-tuple $\mathbb A_{k}$ 
\begin{align}
	\mathbb A_k=\left\{\left(A_1,A_2,\ldots, A_k\right) \in C^{q_1}(\cM,G_1)\times C^{q_2}(\cM,G_2)\times \cdots \times  
	C^{q_k}(\cM,G_k)
	\right\} \; .
\end{align}
Furthermore, we require that $\mathbb A_{k}\in \text{ker}(D_{E_k})$ which amounts to imposing the following cocycle conditions
\begin{align}
	dA_1=&\; 0 \nonumber \\
	dA_{2}=&\; \alpha_2(A_1) \nonumber \\
	dA_{3}=&\; \alpha_{3}(A_1,A_2) \nonumber \\
	\vdots&\; \nonumber \\
	dA_{k}=&\; \alpha_{k}(A_1,A_{2},\ldots, A_{k-1}) \; .
	\label{eq:cocycle_postnikov}
\end{align}
There is a gauge redundancy $\mathbb A_{k}\sim \mathbb A_k + D^{\flat}_{E_k}\mathbb{\Phi}_{k}$ generated by the null homotopy $D_{E_k}^{\flat}\mathbb{\Phi}_k$ where $\mathbb{\Phi}_k$ is the $k$-tuple
\begin{align}
	\mathbb{\Phi}_k=\left\{\left(\phi_1,\phi_2,\ldots, \phi_k\right) \in C^{q_1-1}(\cM,G_1)\times C^{q_2-1}(\cM,G_2)\times \cdots \times  
	C^{q_k-1}(\cM,G_k)
	\right\} \; .
\end{align}
The definition of $D^{\flat}_{E_k}$ is such that $\mathbb A_{k}\sim \mathbb A_k + D^{\flat}_{E_k}\mathbb{\Phi}_{k}$ implies
\begin{align}
	A_{1} \sim&\;  A_{1} + d\phi_1 \nonumber \\
	A_{2} \sim&\;  A_{2} + d\phi_2 +\zeta_2(A_1,\phi_1) \nonumber \\
	A_{3} \sim&\;  A_{3} + d\phi_3 +\zeta_3(A_1,\phi_1\, ; \, A_2, \phi_2) \nonumber \\
	&\; \vdots \nonumber \\
	A_{k} \sim&\;  A_{k} + d\phi_k +\zeta_k(A_1,\phi_1\, ; \, A_2, \phi_2 \, ; \, \ldots \, ; \, A_{k-1},\phi_{k-1}) 
\end{align}
where $\zeta_j$ is a descendent of the $j$-th Postnikov class $\alpha_j$, i.e
\begin{align*}
	d\zeta_j(A_1,\phi_1 \, ; \, \ldots \, ; \, A_{j-1},\phi_{j-1})= \alpha_{j}(A_1+d\phi_1 \, ; \, \ldots \, ; \, A_{j-1}+d\phi_{j-1})-\alpha_{j}(A_1 \, ; \, \ldots \, ; \, A_{j-1}) \; .
\end{align*}
One can check that $D_{E_k}\circ D^{\flat}_{E_k}=0$ so that we can define the cohomology 
\begin{align}
	H^{\vec{q}}_{E_k}(\cM):=\frac{\text{ker}(D_{E_k})}{\text{im}(D_{E_k}^{\flat})}
	\label{gen_coh}
\end{align}
where $\vec{q}=(q_1,q_2,\ldots, q_k)$, so that cohomology classes label isomorphism classes of data $\mathbb A_k$.
Finally, we may build a generalized topological gauge theory by constructing a topological action from a cohomology class $[\omega]\in H^{d+1}(E_k,\mathbb R/\mathbb Z)$ whose partition function reads
\begin{align}
	\mathcal Z_\omega^{E_k}[\cM]=\frac{1}{\prod_{j=1}^{k}|G_j|^{b_{0\to q_j-1}}}\sum_{[\mathbb A_k]\in H^{\vec{q}}_{E_k}(\cM)} e^{2\pi i\la\omega(\mathbb A_k),[\cM ]\ra } \; .
\end{align}

\medskip \noindent
Following the examples provided in sec.~\ref{sec:Sigma}, we know that we can obtain a lattice realization of a topological model whose target space is given by a $k$-stage Postnikov tower by reproducing the construction above, except that we now work with a triangulation $\triangle$ of $\cM$ and that instead of summing over cohomology classes in a generalized cohomology group $H^{\vec{q}}_{E_k}(\cM)$, we sum over colorings ${g}\in {\rm Col}(\cM, {E_k})$. An element ${g}=\left\{g_1, g_{2}, \dots, g_k \right\} \in {\rm Col}(\cM, {E_k})$ is such that $g_{i}$ is a coloring of the $q_i$-simplices of $\triangle$. These colorings  are such that the corresponding group variables satisfy local constraints which are the analogue of the (twisted) cocycle conditions presented earlier that depend on the cohomology classes $[\alpha_{p}]\in H^{q_p+1}(E_{p-1},G_p)$. Using the differential on cochains, these local constraints read  
	\begin{align}
		\la dg_1, \triangle^{q_1+1}\ra =&\; 0 \nonumber \\
		\la dg_2-\alpha_2(g_1), \triangle^{q_2+1}\ra =&\; 0 \nonumber \\
		\vdots&\; \nonumber \\
		\la dg_k-\alpha_k(g_1,g_2,\dots,g_k), \triangle^{q_k+1}\ra =&\; 0 \; .
	\end{align} 
	Finally, the partition function is provided by
	\begin{align}
		\mathcal Z_\omega^{E_k}[\cM]=\frac{1}{\prod_{j=1}^{k}|G_j|^{|\triangle^{0\to q_j-1}|}}
		\sum_{g\in {\rm Col}(\cM , {E_k})} \prod_{\triangle^{d+1}} e^{2 \pi i \mathcal{S}_\omega[g,\triangle^{d+1}]} \; .
	\end{align}

\section{Pontrjagin square\label{sec:app_Pontr}}
In this appendix, we collect some important properties of the Pontrjagin square $\mathfrak P$ \cite{whitehead1949simply}: 
\begin{property}[]
	If $f\in Z^{2}(\cM,\mathbb Z_{n})$, then $\mathfrak P(f) \in Z^{4}(\cM,\mathbb Z_{2n})$ with $n$ even, while with $n$ odd we have $\mathfrak P(f) \in Z^{4}(\cM,\mathbb Z_{n})$. An explicit expression for $\mathfrak P(f)$ can then be written in terms of \emph{Steenrod's higher cup products}\footnote{Given $f\in C^{p}(\cM,\mathcal A)$ and $g\in C^{q}(\cM,\mathcal A)$, we write $f\smilo_{\! i}g \in C^{p+q-i}(\cM,\mathcal A)$ to denote Steenrod's higher generalization of the cup product \cite{steenrod1947products} that satisfies in particular the property
		\begin{align}
			f\smilo_{\! i} g- (-1)^{pq-i}g \smilo_{\! i} f =&\; (-1)^{p+q-i-1}\left[d(f \smilo_{\! i+1} g)-df \smilo_{\! i+1} g - (-1)^{p}f \smilo_{\! i+1} dg\right] \; .
		\end{align}} 
	as  
	\begin{align}
		\mathfrak P(f)=
		\begin{cases}
			\tilde{f} \smilo \tilde{ f}+\tilde{f} \smilo_{\! 1} d \tilde{f}, & \text{if}\ n  \ \text{is even}  \\
			f \smilo f, & \text{if} \ n \ \text{is odd}
		\end{cases}
	\end{align}
	where $\tilde{f}$ is the \emph{integer lift} of $f$, i.e $\tilde{f} \in C^{2}(\cM,\mathbb Z)$ such that $d \tilde{f}=nu$ for $u\in B^{3}(\cM,\mathbb Z)$ and $f=\tilde{f} \ \text{mod} \ n$. 
\end{property}
\noindent
We can check that $\mathfrak P(f)$ as defined above is indeed closed. We consider the two cases separately. When $n$ is odd, $d \mathfrak P(f)= d f \smilo f + f \smilo d f =0$, and when $n$  is even
\begin{align}
	d \mathfrak P(f)=&\; d \left [ \tilde{f} \smilo \tilde{f} + \tilde{f} \smilo_{\! 1} d\tilde{f} \right] \nn \\
	=&\; \cancel{d \tilde{f} \smilo \tilde{f}} + \tilde{f} \smilo d \tilde{f} + d \tilde{f} \smilo_{\! 1} d \tilde{f} + \tilde{f} \smilo d \tilde{f} - \cancel{d \tilde{f} \smilo \tilde{f}} \nn \\
	=&\; 2n \tilde{f} \smilo u + n^2 u \smilo_{1} u =  0 \q (\text{mod} \ 2n) \; .
\end{align}
\begin{property}[]
	The Pontrjagin square refines the bilinear form $2f\smilo g$. Indeed, let $f,g \in Z^{2}(\cM,\mathbb Z_n)$. If $n$ is odd, one has
	\begin{align}
		\mathfrak P(f+g)-\mathfrak P(f)-\mathfrak P(g)=&\; f\smilo g + g\smilo f \nn \\
		=&\; 2 f\smilo g+d(f \smilo_{\! 1} g) \nn \\
		\myeq&\; 2 f \smilo g
	\end{align}
	where $\myeq$ is an equality up to exact terms. If $n$ is even, we write $d \tilde{f}=nu$ and $d \tilde{g}=nv$, and we get
	\begin{align}
		\mathfrak P(f+g)-\mathfrak P(f)-\mathfrak P(g)=&\; \tilde{f}\smilo \tilde{g} + \tilde{g}\smilo \tilde{f}  + n(\tilde{f} \smilo_{\! 1} v + \tilde{g} \smilo_{\! 1} u ) \nn \\
		=&\; 2\tilde{f} \smilo \tilde{g} + d(\tilde{f} \smilo_{\! 1} \tilde{g}) +n\left[-u \smilo_{\! 1} g - \cancel{f \smilo_{\! 1} v } + \cancel{f\smilo_{\! 1} v} + g \smilo_{\! 1} u \right] \nn \\
		=&\; 2\tilde{f} \smilo \tilde{g} + d(\tilde{f} \smilo_{\! 1} \tilde{g}) -2nu \smilo_{\! 1} \tilde{g} -n d \left(u \smilo_{\! 2} g\right) -n^2 u \smilo_{\! 2} v \nn \\
		\myeq &\; 2\tilde{f} \smilo \tilde{g} \q (\text{mod} \ 2n) \; .
	\end{align}
\end{property}
\begin{property}For a group $G=\bigoplus_{I}\mathbb Z_{n_I}$, we write $f^{I}\in Z^{2}(\cM,\mathbb Z_{n_I}) $ and the Pontrjagin square satisfies 	
	\begin{align}
		\mathfrak P(\sum_{I}f^{I})= \sum_{I}\mathfrak P(f^I)+ \sum_{I<J}f^{I}\smilo f^J \; .
		\label{copies_pontrjagin}
	\end{align}
\end{property}

\section{Operators, quantization and invertibility of  2-form topological theories \label{sec:app_quant}}

In this appendix, we review some of the properties of the 2-form topological theory introduced in sec.~\ref{sec:Sigma}. More precisely, we consider the partition function of the 2-form gauge theory formulated as a continuous topological field theory, construct its gauge invariant operators, quantize it, and study its invertibility. We follow closely the analysis of \cite{kapustin2014coupling, gaiotto2015generalized}. 

\medskip \noindent In order to keep the notations lighter and focus on the physical aspects, we consider the simpler case of a 2-form gauge theory with gauge group $\mathbb Z_{n}$. Let us consider the topological action 
\begin{align}
\mathcal{S}_p[A,B,\cM]=2\pi i\int_{\cM}\left[nB\wedge d_{\dr}A +\frac{pn}{2}B\wedge B \right] \; ,
\label{eq:single_BFBB}
\end{align}
which is exactly \eqref{eq:action_BwedgeB} for $G=\mathbb Z_n$, where $p$ prescribes a choice of homomorphism in $ \text{Hom}(\Gamma(\mathbb Z_n),\mathbb R/\mathbb Z)$. Instead of differentiating the cases for $n$ being an odd or even integer, we work with general $n$ and restrict the values of $p$. More precisely, one takes $p\in \mathbb Z$ when $n$ is even, and $p\in 2\mathbb Z$ when $n$ is odd. Since $p\sim p+2n$,\footnote{This can be seen for example by integrating over the 1-form gauge field $A$ in the path integral which reduces $B$ to a 2-form $\mathbb Z_{n}$ gauge field. Then it is apparent that the topological action for the 2-form $\mathbb Z_{n}$ theory evaluates to the same number in $\mathbb R/2\pi \mathbb Z$ for the theories labeled by $p$ and $p+2n$.} there are $n$ distinct topological gauge theories for $n$ odd and $2n$ distinct topological gauge theories for $n$ even. This agrees with the order of the universal quadratic group for $\mathbb Z_{n}$. Usually gauge invariant operators in (3+1)d topological gauge theories are defined on closed lines and surfaces. Such operators assign topological data in the form of correlation functions to certain linked configurations for the corresponding lines and surfaces. For example, the (3+1)d BF theory assigns a non-trivial phase to a linked line and surface embedded in the (3+1)-dimensional spacetime manifold. More interestingly, non-trivial discrete gauge theories, namely Dijkgraaf-Witten theories, may have topological correlation functions associated to linked configurations of three or four surface operators (cf. for example \cite{Wang:2014xba, Chen:2015gma, Tiwari:2016zru, putrov2016braiding}). Gauge invariance \eqref{eq:strict_gt} dictates that the Wilson operators of the 2-form continuous topological gauge theory \eqref{eq:single_BFBB} be defined on closed surfaces $\mathfrak L^{(2)}$ and closed lines $\mathfrak L^{(1)}$, which have open surfaces $\partial^{-1}\mathfrak L^{(1)}$ attached to them:
\begin{align}
U^{\mc}(\mathfrak L^{(2)}):=&\; \exp\left\{2\pi i\mc\oint_{\mathfrak L^{(2)}}B\right\} 
\label{eq:Wilson_BB1} \\
W^{\ec}(\mathfrak L^{(1)},\partial^{-1}\mathfrak L^{(1)}):=&\; \exp\left\{2\pi i \ec \oint_{\mathfrak L^{(1)}}A+2\pi ip\ec\int_{\partial^{-1}\mathfrak L^{(1)}}B\right\} \;.
\label{eq:Wilson_BB2}
\end{align}
It was pointed out in \cite{gaiotto2015generalized} that operators with support on open manifolds are topologically trivial since open manifolds cannot link with other manifolds embedded in the spacetime manifold $\cM$. Therefore, in a topological field theory, correlation functions of such operators with all other observables in the theory are trivial. But, for a given choice of parameter $p$ in \eqref{eq:single_BFBB}, all Wilson line operators are not trivialized. In fact $W^{\ec}(\mathfrak L^{(1)}, \partial^{-1}\mathfrak L^{(1)})$ is an inherent line operator if $p \cdot \ec \in n\mathbb Z$. The reason for this is that $\exp\left\{2\pi i n \int_{\partial^{-1}\mathfrak L^{(1)}}B \right\}$ is the identity operator so that the operator $W^{\ec}$ does not have a surface attached to it (or equivalently has a transparent surface attached to it) and is therefore a genuine line operator. From the above constraint on genuine line operators, one may read off that $\widetilde{W}:=W^{n/\text{gcd}(n,p)}$ is the simplest non-trivial line operator and since $W^n$ is trivial, there are $\text{gcd}(n,p)$ such non-trivial operators. Similarly, some surface operators can end on closed lines and are therefore topologically trivial. The number of surface operators that cannot end on lines match the number of line operators, namely $\text{gcd}(n,p)$. As a quick illustration of this last point, let us have a look at two examples:

\begin{example}[{$n=12$ and $p=4$}] 
	Naively, one would say that the surface operators are $U^{\mc}(\mathfrak{L}^{(2)})$ with $\mc=0,\dots,12$, however, when $\mc/4\in \mathbb Z$, such a surface operator can end on a line. For instance, if $\mc=4$, one could have the operator $\exp\left\{2\pi i\oint_{\partial \mathfrak L^{(2)}}A+8\pi i\int_{ \mathfrak L^{(2)}}B\right\}$. Therefore, the number of surface operators modulo the number of trivial surface operators is ${\text{gcd}(n,p)}$.
\end{example}

\begin{example}[{$n=12$ and $p=5$}] 
	Following the above argument, we expect to get no genuine line of surface operators in this case since $n$ and $p$ are coprime. This can be explicitly checked. Any surface operator $U^{\mc}(\mathfrak L^{(2)})$ can be trivialized by attaching a line with charge $\ec=5\mc  \, (\text{mod} \ 12)$ to it. Equivalently, a line with charge $\ec$ can be trivialized by adding an open surface with flux $\mc=5\ec \; (\text{mod} \ 12)$ to it. Hence, there are no non-trivial operators in the theory when $\text{gcd}(n,p)=1$. 
\end{example}
\noindent Open non-trivial line operators create \emph{magnetic point-like excitations} whereas open non-trivial surface operators create \emph{string or loop-like electric excitations}. These operators as well as the states they generate can be constructed explicitly within the lattice Hamiltonian formalism. This lattice formalism can then be used to study the braiding statistics of the corresponding excitations. Within the partition function approach, the non-trivial operators in the 2-form theory have correlation functions that are identical to an ordinary BF theory. These correlation function are in turn related to the braiding statistics of the corresponding excitations as seen from the lattice picture. This is a consequence of the fact that the cohomological twist ($\propto B\wedge B$) does not alter the canonical commutation relations of the theory. Therefore, the correlation functions take the form 
\begin{align}
\la \, U^{\mc}(\mathfrak L^{(2)})\widetilde{W}^{\ec}(\mathfrak L^{(1)}) \, \ra =\exp\bigg\{\frac{2\pi i \mc 
	\cdot \ec  \, \text{link}(\mathfrak L^{(2)},\mathfrak L^{(1)})}{\text{gcd}(n,p)}\bigg\} 
\end{align} 
where $\text{link}(\mathfrak L^{(2)},\mathfrak L^{(1)})$ is the linking number of the 2-cycle $\mathfrak L^{(2)}$ and the 1-cycle $\mathfrak L^{(1)}$ embedded in the 4-manifold $\cM$, and $\widetilde{W}^{\ec}(\mathfrak L^{(1)}) = W^{\ec \cdot n / {\rm gcd}(n,p)}$. The partition function whose action is \eqref{eq:single_BFBB} then takes the form \cite{gaiotto2015generalized}
\begin{align}
	\mathcal Z_p^{B^2G}[\mathcal M]=\bigg(\frac{n}{\text{gcd}(n,p)}\bigg)^{\chi(\cM)/2} \cdot e^{i\sigma(\cM)/8} \cdot \text{gcd}(n,p)^{\chi(\cM)} \cdot\frac{|H^{1}(\cM,\mathbb Z_{\text{gcd}(n,p)})|}{ |H^{0}(\cM,\mathbb Z_{\text{gcd}(n,p)})|}
	\label{eq:2form_pf}
\end{align}
where $\sigma(\cM)$ is the signature of the manifold $\cM$ and $\chi(\cM)=\sum_{i=0}^{4}(-1)^{i}b_i$ is the Euler characteristic of $\cM$. It is important to note that $\chi(\cM)$ can be written as an integral over purely geometric data and thus it is not topological in the strict sense. It is illustrative to split this partition function into the product of two terms: The first term $\mathcal Z^{\text{inv}}$ incurs contributions in the partition sum only from the trivial (transparent operators), whereas the second term $\mathcal Z^{\text{non-inv}}$ incurs contributions from the non-trivial operators:
\begin{align}
	\mathcal Z_{p}^{\text{inv}}[\cM]=&\;\bigg(\frac{n}{\text{gcd}(n,p)}\bigg)^{\chi(\cM)/2} \cdot e^{i\sigma(\cM)/8}
	\nonumber \\
	\mathcal Z_{p}^{\text{non-inv}}[\cM]=&\;  \text{gcd}(n,p)^{\chi(\cM)} \cdot \frac{|H^{1}(\cM,\mathbb Z_{\text{gcd}(n,p)})|}{ |H^{0}(\cM,\mathbb Z_{\text{gcd}(n,p)})|}=\text{gcd}(n,p)^{b_{2}(\cM)-b_{1}(\cM)+b_0(\cM)} \; ,
\end{align}
where the last equality follows from the Poincar\'{e} which implies that $b_k = b_{4-k}$ in 4d. The first term $\mathcal Z_{p}^{\text{inv}}[\cM]$ is the partition function for an \emph{invertible} topological theory.\footnote{A ($d$+1)-dimensional invertible topological field theory is a TQFT that simply assigns a $\rU(1)$ phase to any closed ($d$+1)-manifold $\mathcal M$ so that it assigns a unique state on any $d$-manifold $\Sigma$. From a physical standpoint these TQFTs describe invertible topological (gapped) phases of matter (see for example \cite{Freed:2014eja, Freed:2016rqq}) that are short-range entangled phases of matter, i.e. they can be smoothly connected to a reference trivial phase upon stacking with another invertible phase of matter. The TQFT corresponding to the trivial reference phase assigns the number 1 to every ($d$+1)-manifold $\mathcal M$. A necessary and sufficient condition for a once-extended TQFT to be invertible is that the partition function assigned to a tori $\mathbb{T}^{d+1}$ is unity \cite{Schommer-Pries:2015lnx}.} This partition function can be thought of as a pure $\rU(1)$ phase since the term $(n/\text{gcd}(n,p))^{\chi(\cM)/2}$ can be absorbed into a geometric counterterm $\frac{\chi(\cM)}{2}\ln(n/\text{gcd}(n,p))$.  The second term $\mathcal Z_{p}^{\text{non-inv}}[\cM]$ is essentially the partition function for an untwisted Dijkgraaf-Witten theory with gauge group $G=\mathbb Z_{\text{gcd}(n,p)}$ (equivalently a $\mathbb Z_{\text{gcd}(n,p)}$ BF theory) up to a geometric counterterm $\chi(\cM)\ln(\text{gcd}(n,p))$, or alternatively is the partition function of a 2-form topological gauge theory with gauge group $G=\mathbb Z_{\text{gcd}(n,p)}$ and trivial cohomology class.

\bigskip \noindent
By quantizing the theory on manifolds of the form $\cM = \Sigma \times \mathbb R$, we obtain Hilbert spaces $\mathcal{H}_\Sigma$ of physical states. The dimension of these Hilbert spaces is an interesting class of objects. Indeed, given a surface $\Sigma$, the dimensions of $\mathcal{H}_\Sigma$ corresponds to the ground state degeneracy of the lattice Hamiltonian realization of the theory on $\Sigma$. By computing explicitly the ground state degeneracy, we can then confirm for which values of $n$ and $p$ the theory is invertible.

By definition, an invertible (3+1)d TQFT assigns a single physical state to any 3-manifold so that the dimension of the Hilbert space $\mathcal H_\Sigma$ obtained upon quantization of the theory on $\Sigma \times \mathbb R$ incurs a contribution only from the non-invertible part of the theory. Since the non-invertible part of the partition function can be mapped (dualized) to an untwisted $\mathbb Z_{\text{gcd}(n,p)}$ Dijkgraaf-Witten theory, one has
\begin{align}
	{\rm dim}\, \mathcal H_\Sigma=\mathcal Z[\Sigma \times \mathbb{S}^{1}]= \text{gcd}(n,p)^{b_{1}(\Sigma)=b_2(\Sigma)} \; .
\end{align}
A basis for the Hilbert space $\mathcal H_\Sigma$ can be labeled by non-trivial line or surface operators on $\Sigma$. Let $[\mathfrak L^{(1)}]_i$ be a basis in $H_{1}(\cM,\mathbb Z)$ and $[\mathfrak L^{(2)}]_i$ the dual basis in $H_{2}(\cM,\mathbb Z)$ such that the intersection pairing $\mathbb I([\mathfrak L^{(1)}]_i,[\mathfrak L^{(2)}]_j)=\delta_{ij}$. A convenient basis for the states on $\Sigma$ is labeled by the vector $\vec{\mc}=({\mc_1},\ldots , {\mc_{b_1(\Sigma)}})$ such that 
\begin{equation}
	U^{\mc}([\mathfrak L^{(2)}]_i)|\vec{\mc}\ra= \exp\bigg\{\frac{2\pi i {\mc_i}}{\text{gcd}(n,p)}\bigg\}|\vec{\mc}\ra 
\end{equation}
where the surface operators $U$ are defined according to \eqref{eq:Wilson_BB1}.
Such a basis can be explicitly constructed as
\begin{align}
	|\vec{\mc}\ra = \prod_{i=1}^{b_1(\Sigma)}\widetilde{W}^{{\mc_i}}([\mathfrak L^{(1)}]_i)| \varnothing \ra
\end{align} 
where the vacuum is normalized to have unit eigenvalue for all the non-trivial surface operators. A similar basis can be constructed that diagonalizes the line operators $\widetilde{W}$ defined in \eqref{eq:Wilson_BB2}. Denoting this basis by $| \vec{\ec} \ra$, one has the following overlap
\begin{align}
	\la \, \vec{\ec} \, | \, \vec{\mc} \, \ra = \exp\left\{\frac{2\pi i \vec{\ec} \cdot \vec{\mc}}{\text{gcd}(n,p)}\right\}
	\label{eq:BF_overlap}
\end{align} 
that can be viewed as the partition function on a four-sphere $\mathbb{S}^4$ with the line operators $\widetilde{W}^{{\ec_i}}([\mathfrak L^{(1)}]_i) $ and $U^{{\mc_j}}([\mathfrak L^{(2)}]_j)$ inserted such that the linking number $\text{link}([\mathfrak L^{(1)}]_{i},[\mathfrak L^{(2)}]_j)=\delta_{ij}$. In order to visualize this, it is possible to start with a four-sphere and hollow out a four manifold $\mathcal B_{\Sigma}$ whose boundary is ${\Sigma}$. Then using standard \emph{surgery}
\begin{align}
	\mathbb C\xrightarrow{\mathcal Z[\mathbb{S}^{4} \backslash \mathcal B_X]} \mathcal H_{X} \xrightarrow{\mathcal Z[\overline{\mathbb{S}^{4}\backslash \mathcal B_{X}}]} \mathbb C 
\end{align}
where the surgery involves carving out $\mathcal B_{\Sigma}$ from $\mathbb{S}^4$ and inserting line operators in the carved out $\mathcal B_{\Sigma}$ so as to create the state $|\vec{\ec}\ra$ on $\Sigma$. Similarly inserting surface operators in $\mathcal Z \backslash \mathcal B_{\Sigma}$ such as to create the state $|\vec{\mc}\ra$ on $\partial (\mathcal Z \backslash \mathcal B_{\Sigma})=\Sigma$, then filling in $\mathcal B_{\Sigma}$ into $\mathcal Z \backslash \mathcal B_{\Sigma}$, which amounts to the overlap $\la \, \vec{\ec} \, | \, \vec{\mc} \, \ra$. This is nothing but the $\mathbb{S}^4$ partition function with linked configuration of lines and surfaces as described above. 

\medskip \noindent
Since it is known that invertible topological theories are short-range entangled it is illustrative to compute the \emph{topological entanglement entropy} \cite{Kitaev:2005dm, Levin:2006zz, Grover:2011fa, Zheng:2017yta, Wen:2017xwk} to confirm this. The computation is rather straightforward.  We are interested in the situation where $\Sigma = \mathbb{S}^{3}$ and want to compute the topological piece in the topological entanglement entropy. We bipartition $\mathbb{S}^3$ into subregions $\Sigma_1$ and $\Sigma_2$ such that for $\partial \Sigma_1 =\overline{\partial \Sigma_2} =\mathbb D^{2}$. Following a well-known recipe \cite{Dong:2008ft,
	Wen:2016snr, Wen:2017xwk} to compute the topological entanglement entropy, we first need to compute the $n$-th Renyi entropy $S_{\Sigma_1}^{(n)}$: 
\begin{align}
	S_{\Sigma_1}^{(n)}=\frac{1}{1-n}\ln\frac{\text{tr}\rho_{\Sigma_1}^{n}}{(\text{tr}\rho_{\Sigma_1})^n} =\ln \mathcal Z [\mathbb{S}^{4}] 
\end{align}
where we have used the result from \cite{Dong:2008ft} $\text{tr}(\rho_{\Sigma_1}^n)=\text{tr}(\rho_{\Sigma_1})= \mathcal Z[\mathbb S^{4}]$. Then, the topological entanglement entropy is defined as $S^{\text{topo}}_{A}:=\lim_{n\to 1}S_{A}^{(n)}$ which indeed only captures the topological piece in the entanglement entropy. This suffices for our current purpose, however computing the geometrical piece in the entanglement entropy requires more careful considerations. Using the above expression and \eqref{eq:2form_pf} one gets
\begin{align}
	S^{\text{topo}}_A=-\ln \text{gcd}(n,p)
\end{align}  
where we have implicitly absorbed the terms that depends on the Euler characteristic $\chi(\cM)$ into local geometric counterterms. So as expected the 2-form TQFT is short-range entangled (or invertible) when $\text{gcd}(n,p)=1$, and not otherwise. 

 It is known \cite{Walker:2011mda, vonKeyserlingk:2012zn, kapustin2014coupling} that when $\text{gcd}(n,p)=1$, i.e when the quadratic form defining the topological action is non-degenerate, the theory admits a gapped boundary condition with non-trivial line operators which form a modular tensor category. On the other hand when $\text{gcd}(n,p)\neq 1$, there also exist gapped boundaries with non-trivial operators however these do not form a modular category anymore but a premodular one.

\section{Topological actions in terms of Deligne-Beilinson cocycles\label{sec:app_DB}}

In this appendix we derive expressions for various topological actions built from Deligne-Beilinson cohomological data. 

\addtocontents{toc}{\protect\setcounter{tocdepth}{-10}}
\subsection{(2+1)d BF theory}

First, let  us consider BF theory in (2+1)d. The BF topological action is commonly written as 
\begin{align}
	\mathcal{S}[A,B,\cM]=2\pi n i \int_{\cM} B\wedge d_{\dr}A
	\label{eq:BF_2+1}
\end{align}
where $n\in \mathbb Z$ is a parameter of the theory, $B$ and $A$ are 1-form $\rU(1)$ connections and $\cM$ is an oriented 3-manifold. But this expression does not make sense when we include topological sectors of $A$ and $B$. To give a more precise definition of the BF topological action, we consider $\mathbb A,\mathbb B \in H^{1}_{\rm DB}(\cM,\mathbb Z)$ together with the following pairing:
\begin{align}
	H^{1}_{\rm DB}(\cM,\mathbb Z)\times H^{1}_{\rm DB}(\cM,\mathbb Z)\to H^{3}(\cM, \mathbb R/\mathbb Z) \sim \mathbb R/\mathbb Z \; .
\end{align}  
Let the local data that defines $\mathbb A$ and $\mathbb B$ as DB 1-cochains be denoted by $\mathbb A=\left\{\mu_0^1,\mu_1^0,n_2^{\mathbb A}\right\}$ and $\mathbb B=\left\{\nu_0^1,\nu_1^0, n_{2}^{\mathbb B} \right\}$, respectively,  and the corresponding gauge transformations be parametrized by DB 0-cochains $\Lambda=\left\{\lambda_0^0,m_1^{\mathbb A}\right\}$ and $\Theta=\left\{\theta_0^0,m_1^{\mathbb B}\right\}$. Then, the BF topological action can be derived term by term. The first term is the usual BF expression on 3-chains that are contained within open sets as per usual for a polyhedral decomposition of a 3-manifold:
\begin{align}
	\mathcal T_1=\sum_{i_0}\int_{\mathfrak{l}^{(3)}_{i_0}}\left(\nu_0^1\wedge d_{\dr}^1 \mu^{1}_0\right)_{i_0} \; .
\end{align} 
Under gauge transformations $\mu_0^1\to \mu_0^1 + d_{\dr}^0 \lambda_0^0$ and $\nu_0^1 \to \nu_0^1 +d_{\dr}^{0}\theta_0^0$, the term $\mathcal T_1$ transforms as
\begin{align}
	\mathcal T_1\to&\;  \mathcal T_1 + \sum_{i_0}\int_{\partial \mathfrak{l}^{(3)}_{i_0}}\left(\theta_0^0 \wedge d_{\dr}^1 \mu_0^1\right)_{i_0} =\mathcal T_1 + \sum_{i_0,i_1}\int_{ \mathfrak{l}^{(2)}_{i_0i_1}}\left(d_0 \theta_0^0 \wedge d_{\dr}^1 \mu_0^1\right)_{i_0i_1} \; . 
\end{align}
To compensate for the variational term we need to add the term 
\begin{align}
	\mathcal T_2=-\sum_{i_0,i_1}\int_{\mathfrak{l}^{(2)}_{i_0i_1}}\left(\nu_1^0 d_{\dr}^1 \mu^1_0\right)_{i_0i_1} \; ,
\end{align}
however, $\mathcal T_1+ \mathcal T_{2}$ together is not yet gauge invariant, indeed it transforms as
\begin{align}
	\mathcal T_1 +\mathcal T_{2}\to &\; \mathcal T_1 +\mathcal T_{2}+ \sum_{i_0,i_1}\int_{\mathfrak{l}^{(2)}_{i_0i_1}}\left( m_1^{\mathbb B}d_{\dr}^1\mu^1_0\right)_{i_0i_1} \nn \\
	= &\; \mathcal T_1 +\mathcal T_{2}+ \sum_{i_0,i_1}\int_{\partial \mathfrak{l}^{(2)}_{i_0i_1}}\left( m_1^{\mathbb B}\mu^1_0\right)_{i_0i_1} \nn \\
	= &\; \mathcal T_1 +\mathcal T_{2}+ \sum_{i_0,i_1,i_2}\int_{\mathfrak{l}^{(1)}_{i_0i_1i_2}}d_{1}\big( m_1^{\mathbb B}\mu^1_0\big)_{i_0i_1i_2} \nn \\
	=&\;  \mathcal T_1 +\mathcal T_{2}+ \sum_{i_0,i_1,i_2}\int_{\mathfrak{l}^{(1)}_{i_0i_1i_2}}\left(d_{1} m_1^{\mathbb B}\mu^1_0-m_{1}^{\mathbb B}d_0\mu_0^1\right)_{i_0i_1i_2} \; .
	\label{eq:T_1T_2}
\end{align}
Using the gluing cocycle conditions \eqref{eq:1form_gluing_rel} for DB 1-cocycles, the last term is integer-valued and can therefore be dropped as the action in \eqref{eq:BF_2+1} is valued in $\mathbb R/2\pi\mathbb Z$. Then, in order to cancel the gauge non-invariant contribution in \eqref{eq:T_1T_2}, we add a third term
\begin{align}
	\mathcal T_{3}=\sum_{i_0,i_1,i_2}\int_{\mathfrak{l}^{(1)}_{i_0i_1i_2}}\left(n_2^{\mathbb B}\mu_0^1\right)_{i_0i_1i_2} 
\end{align}
which itself transforms as 
\begin{equation}
	\mathcal T_3 \to \mathcal T_3 + \sum_{i_0,i_1,i_2}\int_{l^{i_0i_1i_2}}\left(n_2^{\mathbb B}d_{\dr}^0\lambda_0^0\right)_{i_0i_1i_2} \; .
\end{equation}
And finally, in order to cancel the term which prevents the gauge invariance of $\mathcal T_3$, we add 
\begin{equation}
	\mathcal T_{4}=-\sum_{i_0,i_1,i_2,i_3}\int_{\mathfrak{l}^{(0)}_{i_0i_1i_2i_3}}\left(n_2^{\mathbb B}\mu_1^0\right)_{i_0i_1i_2i_3} \; .
\end{equation}
To summarize, the BF topological action \eqref{eq:BF_2+1} written in terms of Deligne-Beilinson cocycles takes the form 
\begin{align}
	\nn
	 \mathcal{S}[\mathbb A, \mathbb B,\cM] &=2\pi in \big(\mathcal T_1 +\mathcal T_2 +\mathcal T_3 +\mathcal T_4 \big) \\ \nn
	&  = 2\pi i n \bigg(\sum_{i_0}\int_{\mathfrak{l}^{(3)}_{i_0}}\left(\nu_0^1\wedge d_{\dr}^1 \mu^{1}_0\right)_{i_0}
	-\sum_{i_0,i_1}\int_{\mathfrak{l}^{(2)}_{i_0i_1}}\left(\nu_1^0 \wedge d_{\dr}^1 \mu^1_0\right)_{i_0i_1} 
	\\	\label{eq:BF_DB_2+1} 
	&\hspace{2.5em}
	+\sum_{i_0,i_1,i_2}\int_{\mathfrak{l}^{(1)}_{i_0i_1i_2}}\left(n_2^{\mathbb B}\mu_0^1\right)_{i_0i_1i_2}-\sum_{i_0,i_1,i_2,i_3}\int_{\mathfrak{l}^{(0)}_{i_0i_1i_2i_3}}\left(n_2^{\mathbb B}\mu_1^0\right)_{i_0i_1i_2i_3}\bigg) \; .
\end{align}  
The quantum partition function for a theory defined via the action \eqref{eq:BF_DB_2+1} is the same as that for an untwisted $\mathbb Z_{n}$ gauge theory. In order to see this, we integrate over $\mathbb B$ in the path integral. Firstly, integrating over $\nu_0^1$ imposes $d^1_{\dr}\mu_0^1=0$, and since $\mu_0^1$ is defined on a simply connected open set, we can always write $\mu_0^1 =d^0_{\dr} \alpha_0^0$ that can be gauged away by choosing $\lambda_{0}^{0}=-\alpha_0^0$. Hence, one obtains $\mu_0^1=0$. This sets the first three terms in \eqref{eq:BF_DB_2+1} to zero. Secondly, upon performing a sum over $n_{2}^{\mathbb B}\in \mathbb Z$, we obtain a delta function which imposes that $\mu_1^0$ is an integral multiple of $1/n$. Putting everything together, one obtains a $\mathbb Z_{n}$ connection from the BF theory. This should be viewed as living on a triangulation that is dual to the \v{C}ech complex, i.e the open sets are vertices of the triangulation, overlaps are 1-simplices, and so on and so forth. 

Note finally that the $\mathbb Z_{n}$ gauge theory has an \emph{electromagnetic-duality}, which is manifest in the BF theory formulation, under the exchange $\mathbb A\leftrightarrow \mathbb B$. This duality may be understood as an embedding of the quantum double $\mathcal D(\mathbb Z_{n})$ into a $\rU(1)\times \rU(1)$ gauge theory which is the gauge group of the `level' $n$ BF theory. As a corollary, one may integrate over $\mathbb A$ instead of $\mathbb B$ and obtain a $\mathbb Z_n$ gauge theory for the Pontrjagin dual group $\widetilde{\mathbb Z}_{n}\simeq \mathbb Z_n$. Performing an integration by parts together with the gluing relations \eqref{eq:1form_gluing_rel}, we can rewrite \eqref{eq:BF_DB_2+1} as
\begin{align}
	\nn
	\mathcal{S}[\mathbb A, \mathbb B,\cM] &= 2\pi in \bigg(\sum_{i_0}\int_{\mathfrak{l}_{i_0}^{(3)}}\left(d_{\dr}^1\nu_{0}^{1}\wedge \mu_{0}^{1} \right)_{i_0}
	-\sum_{i_0,i_1}\int_{\mathfrak{l}_{i_0i_1}^{(1)}}\left(\nu_{0}^{1}\wedge d\mu _{1}^{0} \right)_{i_0i_1} \\
	&\hspace{2.5em}+\sum_{i_0,i_1,i_2}\int_{\mathfrak{l}^{(1)}_{i_0i_1i_2}}\left(d_{\dr}^0\nu_1^0 \wedge\mu_{1}^0 \right)_{i_0i_1i_2}
	-\sum_{i_0,i_1,i_2,i_3} \int_{\mathfrak{l}^{(0)}_{i_0i_1i_2i_3}} \left(n_2^{\mathbb A}\nu_1^{0}\right)_{i_0i_1i_2i_3}\bigg) \; .
\end{align}
The integral over $\mu_{0}^{1}$ imposes $d^{1}_{\dr}\nu_0^1=0$, hence $\nu_0^1$ can be set to zero by making gauge choice. Then, the integral over $\mu_1^0$ imposes that $d_{1}\nu_1^{0}=0$. Finally,  the sum over $n_{2}^{\mathbb A}$ imposes that $\nu_1^{0}\in \frac{1}{n}\mathbb Z$. Together this makes $\mathbb B$ a $\mathbb Z_n$-valued \v{C}ech 1-cocycle.  	

\subsection{(3+1)d BF theory}
Let us now consider $\rU(1)$ (3+1)d BF theory. The theory is built from a 1-form $\rU(1)$ connection $A$ and a 2-form $\rU(1)$ connection B 
\begin{align}
	\mathcal{S}[A,B, \cM]=2\pi in\int_{\cM}B\wedge d_{\dr}A \; .
	\label{eq:BF_3+1}
\end{align}
However, when evaluated on topological sectors of the $\rU(1)$ bundles, the above integral does not make sense. This calls for a more rigorous definition of the BF topological action using a DB 2-cocycle $\mathbb B=\left\{\nu_0^2,\nu_1^1,\nu_2^0, n_{3}^{\mathbb B}\right\}$ and DB 1-cocycle $\mathbb A=\left\{\mu_0^1,\mu_1^0,n_2^{\mathbb A}\right\}$ as defined for the (2+1)d case. The gauge transformations of $\mathbb A$ and $\mathbb B$ are labeled by DB $0,1$-cochains  $\Lambda$ and $\Theta$, respectively, as described in sec.~\ref{sec:DB}. The topological action can be defined term by term as before. On 4-chains contained within open sets, we define the term
\begin{align}
	\mathcal T_{1}=\sum_{i_0}\int_{\mathfrak{l}^{(4)}_{i_0}}(\nu_0^2 \wedge d_{\dr}^1 \mu_0^1)_{i_0} \; .
\end{align}
Under the gauge transformations $\mu_0^1\to \mu_0^1 + d_{\dr}^0 \lambda_0^0$ and $\nu_0^2 \to \nu_0^2 +d_{\dr}^{1}\theta_0^1$, the term $\mathcal T_1$ transforms as
\begin{align}
	\mathcal T_1 \to&\;  \mathcal T_1 + \sum_{i_0}\int_{\partial \mathfrak{l}^{(4)}_{i_0}}\left(\theta_0^1 \wedge d_{\dr}^1  \mu_0^1\right)_{i_0} =\mathcal T_1 + \sum_{i_0,i_1}\int_{ \mathfrak{l}^{(3)}_{i_0i_1}}\left(d_0\theta_0^1 \wedge d_{\dr}^1 \mu_0^1\right)_{i_0i_1} \nn \; .
\end{align}
To compensate for the variational term, we need to add the term
\begin{align}
	\mathcal T_2=-\sum_{i_0,i_1}\int_{\mathfrak{l}^{(3)}_{i_0i_1}}\left(\nu_1^1\wedge  d_{\dr}^1 \mu^1_0\right)_{i_0i_1} \; ,
\end{align}
however, $\mathcal T_1+ \mathcal T_{2}$ is not yet gauge invariant, indeed it transforms as 
\begin{align}
	\mathcal T_1 +\mathcal T_2 \to&\; \mathcal T_1 +\mathcal T_2 + \sum_{i_0,i_1}\int_{\mathfrak{l}^{(3)}_{i_0i_1}}\left(d_{\dr}^0\theta_1^0 \wedge  d_{\dr}^1 \mu^1_0\right)_{i_0i_1} \nn \\
	=&\; \mathcal T_1 +\mathcal T_2 + \sum_{i_0,i_1}\int_{\partial \mathfrak{l}^{(3)}_{i_0i_1}}\left(\theta_1^0 \wedge d_{\dr}^1 \mu^1_0\right)_{i_0i_1} \nn \\
	=&\; \mathcal T_1 +\mathcal T_2 + \sum_{i_0,i_1,i_2}\int_{\mathfrak{l}^{(2)}_{i_0i_1i_2}}\left(d_1\theta_1^0\wedge d_{\dr}^1 \mu^1_0\right)_{i_0i_1i_2} 
\end{align}
where we have used $d^2_{\dr} \circ d_{\dr}^1\mu_0^1 =0$. In order to cancel the gauge non-invariant contribution, we add a third term
\begin{equation} 
	\mathcal T_{3}= \sum_{i_0,i_1,i_2}\int_{\mathfrak{l}^{(2)}_{i_0i_1i_2}}\left( \nu_2^0 \wedge d_{\dr}^1 \mu_0^1\right)_{i_0i_1i_2}
\end{equation}	
which itself transforms as
\begin{align}
	\mathcal T_{3}\to&\; \mathcal T_{3} + \sum_{i_0,i_1,i_2}\int_{\mathfrak{l}^{(2)}_{i_0i_1i_2}}d^{1}_{\dr}\big( m_2^{\mathbb B}  \mu_0^1\big)_{i_0i_1i_2} \nn  \\
	=&\; \mathcal T_{3} + \sum_{i_0,i_1,i_2}\int_{\partial \mathfrak{l}^{(2)}_{i_0i_1i_2}}\left( m_2^{\mathbb B}  \mu_0^1\right)_{i_0i_1i_2} \nn  \\
	=&\; \mathcal T_{3} + \sum_{i_0,i_1,i_2,i_3}\int_{\mathfrak{l}^{(1)}_{i_0i_1i_2i_3}}\left( d_2 (m_2^{\mathbb B})  \mu_0^1+ m_2^{\mathbb B}  d_0 \mu_0^1\right)_{i_0i_1i_2i_3} \; .
\end{align} 
And finally, in order to cancel the term which prevents the gauge invariance of $\mathcal T_3$, we add  
\begin{align}
	\mathcal T_4=&\;-\sum_{i_0,i_1,i_2,i_3}\int_{\mathfrak{l}^{(1)}_{i_0i_1i_2i_3}}\left(n_{3}^{\mathbb B} \mu_0^1\right)_{i_0i_1i_2i_3}\nn \\
	\mathcal T_{5}=&\; \sum_{i_0,i_1,i_2,i_3,i_4}\int_{\mathfrak{l}^{(0)}_{i_0i_1i_2i_3i_4}}\left(n_{3}^{\mathbb B}\mu_1^{0}\right)_{i_0i_1i_2i_3i_4} \; .
\end{align}
Eventually, the topological action \eqref{eq:BF_3+1} takes the form $\mathcal{S}[A,B,\cM]=2\pi in \sum_{j=1}^{5}\mathcal T_j$. Similar to the case of (2+1)d BF theory, $\mathbb B$ can be readily integrated out in the partition function in order to obtain a 2-form $\mathbb Z_{n}$ gauge theory. This can be implemented by first integrating over $\nu_{0}^{2}$ that sets $\mu_{0}^1\sim 0$ (by fixing a gauge). This sets the first four terms $\mathcal T_{1,2,3,4}$ to zero. In the last term, $n_3^{\mathbb B}$ can be summed over which enforces $\mu_1^0\in \frac{1}{n}\mathbb Z$. 

 Similarly we may first perform an integration by parts and then impose the gluing relations. Doing so $\mathbb A$ can be integrated out instead of $\mathbb B$. This reduces $\mathbb B$ to a $\mathbb Z_n$ valued \v{C}ech 2-cocycle. Thus establishing the duality between 1-form and 2-form gauge fields within the (3+1)d BF theory.

\bibliographystyle{JHEP}
\bibliography{ref_cat}

\end{document}

%% file: tikzsb.tex
\tikzset{->1-/.style={decoration={
			markings,
			mark=at position #1 with {\arrow{Triangle[fill=black, width=5pt, length=5pt]}}},postaction={decorate}}}
\tikzset{->2-/.style={decoration={
					markings,
					mark=at position #1 with {\arrow{Triangle[fill=white, width=5pt, length=5pt]}}},postaction={decorate}}}
\tikzset{->3-/.style={decoration={
			markings,
			mark=at position #1 with {\arrow{Triangle[fill=lightgray, width=5pt, length=5pt]}}},postaction={decorate}}}

\tikzset{
	every picture/.append style={
		line join=round,
		line cap=round,
	}
}

\tikzstyle{dot}=[dotted, line width=0.5pt]

\newcommand{\cylinderDis}[1]{
	\begin{tikzpicture}[scale=#1,baseline=2em, line width=0.8pt]
	\coordinate (0) at (0,0);
	\coordinate (1) at (0,6);
	\coordinate (6) at (-6,0);
	\coordinate (7) at (-6,6);
	
	\draw[->3-=.55] (0) -- (1);
	\draw[] (0) edge [->1-=.53, in=30, out=150, looseness=1] (6);
	\draw[] (0) edge [->1-=.53, in=-30, out=-150, looseness=1] (6);
	\draw[] (1) edge [->2-=.53, in=30, out=150, looseness=1] (7);
	\draw[] (1) edge [->2-=.53, in=-30, out=-150, looseness=1] (7);
	\draw[->3-=.55] (6) -- (7);
	
	\node[draw,circle,scale=0.4,fill] at (0) {};
	\node[draw,circle,scale=0.4,fill] at (6) {};
	\node[draw,circle,scale=0.4,fill=white] at (1) {};
	\node[draw,circle,scale=0.4,fill=white] at (7) {};
	
	\node[] at (-3,0) {$\sss{a}$};
	\node[] at (-3,6) {$\sss{a}$};
	\node[] at (-3,3) {$\sss{b}$};
	\end{tikzpicture}
}

\newcommand{\prismoSINGLE}[2]{
	\begin{tikzpicture}[scale=#1,baseline=2em, line width=0.8pt]
	\coordinate (0) at (0,0);
	\coordinate (1) at (0,6);
	\coordinate (3) at (-3,-1.5);
	\coordinate (4) at (-3,4.5);
	\coordinate (6) at (-6,0);
	\coordinate (7) at (-6,6);
	
	\coordinate (0634) at (intersection of 0--6 and 3--4);
	\coordinate (0637) at (intersection of 0--6 and 3--7);
	\coordinate (0734) at (intersection of 0--7 and 3--4);
	\draw[] (0) -- (1);
	\draw[] (0) -- (3);
	\draw[] (1) -- (4);
	\draw[] (6) -- (7);
	\draw[] (3) -- (6);
	\draw[] (4) -- (7);
	\draw[] (3) -- (4);
	\draw[] (1) -- (7);
	\draw[shorten >= 0.3em] (0) -- (0634);
	\draw[shorten <= 0.3em, shorten >= 0.3em] (0634) -- (0637);
	\draw[shorten <= 0.3em] (0637) -- (6);
	\draw[] (3) -- (7);
	\draw[] (0) -- (4);
	\draw[shorten >= 0.3em] (0) -- (0734);
	\draw[shorten <= 0.3em] (0734) -- (7);
	
	\ifthenelse{#2 = 2}{
		\node[below] at (0) {$\sss{0}$};
		\node[above] at (1) {$\sss{1}$};
		\node[below] at (3) {$\sss{3}$};
		\node[above] at (4) {$\sss{4}$};
		\node[below] at (6) {$\sss{6}$};
		\node[above] at (7) {$\sss{7}$};
	}{}

	\ifthenelse{#2 = 3}{
		\node[below] at (0) {$\sss{1'}$};
		\node[above] at (1) {$\sss{2 }$};
		\node[below] at (3) {$\sss{4'}$};
		\node[above] at (4) {$\sss{5}$};
		\node[below] at (6) {$\sss{7'}$};
		\node[above] at (7) {$\sss{8}$};
	}{}
	\end{tikzpicture}
}

\newcommand{\convONE}[4]{
	\begin{tikzpicture}[scale=#1,baseline=0.4em, line width=0.8pt]
	\coordinate (0) at (0,{sin(60)});
	\coordinate (1) at ({cos(60)},0);
	\coordinate (2) at (-{cos(60)},0);
	\coordinate (3) at (0,0.3);
	
	\draw[] (0) to (1);
	\draw[] (1) to (2);
	\draw[] (2) to (0);
	
	\node[] at (3) {{$\otimes$}};
	
	\node[above=-0.2em] at (0) {{$\scriptstyle{#2}$}};
	\node[right=-0.2em] at (1) {{$\scriptstyle{#3}$}};	
	\node[left=-0.2em] at (2) {{$\scriptstyle{#4}$}};
	\end{tikzpicture}
}

\newcommand{\IsoTrans}[3]{
	\begin{tikzpicture}[scale=#1,baseline=#3 em, line width=0.8pt]
	\coordinate (2) at (0,0);
	\coordinate (1) at (1.4,0);
	\coordinate (0) at (0.7,1.2);

	\node[above] at (0) {{$\scriptstyle{0}$}};	
	\node[below] at (1) {{$\scriptstyle{1}$}};
	\node[below] at (2) {{$\scriptstyle{2}$}};
	
	\ifthenelse{#2 = 1}{	
		\draw[] (0) -- (1);
		\draw[] (1) -- (2);
		\draw[] (0) -- (2);
	}{}
		
	\ifthenelse{#2 = 2}{
		\draw[] (0) edge [out=290, in=95, looseness=1.2](1);
		\draw[] (1) edge [out=170, in=340, looseness=1.2] (2);
		\draw[] (0) edge [out=210, in=60, looseness=1.2] (2);	
	}{}
	\end{tikzpicture}
}

\newcommand{\PFourOne}[3]{
	\begin{tikzpicture}[scale=#1, baseline=#3 em, line width=0.8pt]
	\coordinate (0) at (0,1);
	\coordinate (1) at (1,-0.4);
	\coordinate (2) at (0.55,-1);
	\coordinate (3) at (-1.2,-0.55);
	\coordinate (4) at (0,-0.1);	
	
	\node[above] at (0) {{$\scriptstyle{0}$}};	
	\node[right] at (1) {{$\scriptstyle{1}$}};
	\node[below] at (2) {{$\scriptstyle{2}$}};
	\node[left] at (3) {{$\scriptstyle{3}$}};
	
	\draw[] (0) -- (1);
	\draw[] (1) -- (2);
	\draw[] (2) -- (3);
	\draw[] (0) -- (3);
	\draw[] (0) -- (2);

	\ifthenelse{#2 = 1}{	
		\node[left, yshift=1] at (4) {{$\scriptstyle{4}$}};
		\draw[] (3) -- (4);
		\draw[] (0) -- (4);
		\draw[] (2) -- (4);
		\draw[shorten >= 0.3em] (1) -- (intersection of 1--4 and 0--2);
		\draw[shorten <= 0.3em] (intersection of 1--4 and 0--2) -- (4);
		\draw[shorten >= 0.3em] (1) -- (intersection of 1--3 and 0--2);
		\draw[shorten <= 0.3em, shorten >=0.3em] (intersection of 1--3 and 0--2) -- (intersection of 1--3 and 2--4);
		\draw[shorten <= 0.3em] (intersection of 1--3 and 2--4) -- (3);
	}{}
	\ifthenelse{#2 = 2}{
		\draw[shorten >= 0.3em] (1) -- (intersection of 1--3 and 0--2);
		\draw[shorten <= 0.3em] (intersection of 1--3 and 0--2) -- (3); 
	}{}
	
	\end{tikzpicture}
}

\newcommand{\PPONE}[3]{
	\begin{tikzpicture}[scale=#1,baseline=#3 em, line width=0.8pt]
	\coordinate (0) at(0,1);
	\coordinate (1) at ({sin(72)} ,{cos(72)} );
	\coordinate (2) at ({sin(144)} ,{-cos(36)} );
	\coordinate (3) at ({-sin(144)} ,{-cos(36)} );
	\coordinate (4) at ({-sin(72)} ,{cos(72)} );
	
	\draw[] (0) to (1);
	\draw[] (1) to (2);
	\draw[] (2) to (3);
	\draw[] (3) to (4);
	\draw[] (4) to (0);
	
	\draw[] (0) to (3);
	\draw[] (1) to (3);
	\coordinate (a) at (intersection of 3--1 and 0--2);
	\draw[shorten >= 0.3em] (0) to (a);
	\draw[shorten <= 0.3em] (a) to (2);
	\coordinate (b) at (intersection of 0--3 and 4--2);
	\coordinate (c) at (intersection of 3--1 and 4--2);
	\draw[shorten >= 0.3em] (4) to (b);
	\draw[shorten <= 0.3em, shorten >= 0.3em] (b) to (c);
	\draw[shorten <= 0.3em] (c) to (2);
	
	\ifthenelse{#2 = 1}{
		\node[above] at (0) {{$\scriptstyle{0}$}};	
		\node[right] at (1) {{$\scriptstyle{1}$}};
		\node[below] at (2) {{$\scriptstyle{2}$}};
		\node[below] at (3) {{$\scriptstyle{3}$}};
		\node[left] at (4) {{$\scriptstyle{4}$}};}{}	
	\end{tikzpicture}
}

\newcommand{\PPTWO}[3]{
	\begin{tikzpicture}[scale=#1,baseline=#3 em, line width=0.8pt]
	\coordinate (0) at(0,1);
	\coordinate (1) at ({sin(72)} ,{cos(72)} );
	\coordinate (2) at ({sin(144)} ,{-cos(36)} );
	\coordinate (3) at ({-sin(144)} ,{-cos(36)} );
	\coordinate (4) at ({-sin(72)} ,{cos(72)} );
	
	\draw[] (0) to (1);
	\draw[] (1) to (2);
	\draw[] (2) to (3);
	\draw[] (3) to (4);
	\draw[] (4) to (0);
	
	\draw[] (0) to (3);
	\draw[] (1) to (3);
	\coordinate (b) at (intersection of 3--1 and 0--2);
	\coordinate (a) at (intersection of 4--1 and 0--2);
	\draw[shorten >= 0.3em] (0) to (a);
	\draw[shorten <= 0.3em, shorten >= 0.3em] (a) to (b);
	\draw[shorten <= 0.3em] (b) to (2);
	\coordinate (c) at (intersection of 0--3 and 4--2);
	\coordinate (d) at (intersection of 3--1 and 4--2);
	\draw[shorten >= 0.3em] (4) to (c);
	\draw[shorten <= 0.3em, shorten >= 0.3em] (c) to (d);
	\draw[shorten <= 0.3em] (d) to (2);
	\coordinate (e) at (intersection of 0--3 and 4--1);
	\draw[shorten >= 0.3em] (4) to (e);
	\draw[shorten <= 0.3em] (e) to (1);
	
	\ifthenelse{#2 = 1}{	
		\node[above] at (0) {{$\scriptstyle{0}$}};	
		\node[right] at (1) {{$\scriptstyle{1}$}};
		\node[below] at (2) {{$\scriptstyle{2}$}};
		\node[below] at (3) {{$\scriptstyle{3}$}};
		\node[left] at (4) {{$\scriptstyle{4}$}};}{}	
	\end{tikzpicture}
}

\newcommand{\PPPONE}[1]{
	\begin{tikzpicture}[scale=1,baseline=0em, line width=0.8pt]
	\coordinate (0) at (0,1);
	\coordinate (1) at ({sin(60)} ,{cos(60)} );
	\coordinate (2) at ({sin(60)} ,{-cos(60)} );
	\coordinate (3) at (0,-1);
	\coordinate (4) at ({-sin(60)} ,{-cos(60)} );
	\coordinate (5) at ({-sin(60)} ,{cos(60)} );
	
	\draw[] (0) to (1);
	\draw[] (1) to (2);
	\draw[] (2) to (3);
	\draw[] (3) to (4);
	\draw[] (4) to (5);
	\draw[] (5) to (0);
	
	\draw[] (1) to (3);
	\draw[] (0) to (3);
	\draw[] (5) to (3);
	\coordinate (a) at (intersection of 3--1 and 0--2);
	\draw[shorten >= 0.3em] (0) to (a);
	\draw[shorten <= 0.3em, shorten >= 0em] (a) to (2);
	\coordinate (b) at (intersection of 5--3 and 4--2);
	\coordinate (c) at (intersection of 0--3 and 4--2);
	\coordinate (d) at (intersection of 3--1 and 4--2);
	\draw[shorten >= 0.3em] (4) to (b);
	\draw[shorten <= 0.3em, shorten >= 0.3em] (b) to (c);
	\draw[shorten <= 0.3em, shorten >= 0.3em] (c) to (d);
	\draw[shorten <= 0.3em] (d) to (2);
	\coordinate (e) at (intersection of 0--3 and 5--2);
	\coordinate (f) at (intersection of 3--1 and 5--2);
	\draw[shorten >= 0.3em] (5) to (e);
	\draw[shorten <= 0.3em, shorten >= 0.3em] (e) to (f);
	\draw[shorten <= 0.3em] (f) to (2);
	
	\ifthenelse{#1 = 1}{
		\node[above] at (0) {{$\scriptstyle{0}$}};	
		\node[right] at (1) {{$\scriptstyle{1}$}};
		\node[right] at (2) {{$\scriptstyle{2}$}};
		\node[below] at (3) {{$\scriptstyle{3}$}};
		\node[left] at (4) {{$\scriptstyle{4}$}};
		\node[left] at (5) {{$\scriptstyle{5}$}};}{}
	\end{tikzpicture}
}

\newcommand{\PPPTWO}[1]{
	\begin{tikzpicture}[scale=1,baseline=0em, line width=0.8pt]
	\coordinate (0) at (0,1);
	\coordinate (1) at ({sin(60)} ,{cos(60)} );
	\coordinate (2) at ({sin(60)} ,{-cos(60)} );
	\coordinate (3) at (0,-1);
	\coordinate (4) at ({-sin(60)} ,{-cos(60)} );
	\coordinate (5) at ({-sin(60)} ,{cos(60)} );
	
	\draw[] (0) to (1);
	\draw[] (1) to (2);
	\draw[] (2) to (3);
	\draw[] (3) to (4);
	\draw[] (4) to (5);
	\draw[] (5) to (0);
	
	\draw[] (1) to (3);
	\draw[] (0) to (3);
	\draw[] (5) to (3);
	\coordinate (a) at (intersection of 3--1 and 0--2);
	\draw[shorten >= 0.3em] (0) to (a);
	\draw[shorten <= 0.3em, shorten >= 0em] (a) to (2);
	\coordinate (b) at (intersection of 5--3 and 4--2);
	\coordinate (c) at (intersection of 0--3 and 4--2);
	\coordinate (d) at (intersection of 3--1 and 4--2);
	\draw[shorten >= 0.3em] (4) to (b);
	\draw[shorten <= 0.3em, shorten >= 0.3em] (b) to (c);
	\draw[shorten <= 0.3em, shorten >= 0.3em] (c) to (d);
	\draw[shorten <= 0.3em] (d) to (2);
	\coordinate (h) at (intersection of 0--4 and 5--2);
	\coordinate (e) at (intersection of 0--3 and 5--2);
	\coordinate (f) at (intersection of 3--1 and 5--2);
	\draw[shorten >= 0.3em] (5) to (h);
	\draw[shorten >= 0.3em, shorten <= 0.3em] (h) to (e);
	\draw[shorten <= 0.3em, shorten >= 0.3em] (e) to (f);
	\draw[shorten <= 0.3em] (f) to (2);
	\coordinate (g) at (intersection of 0--4 and 5--3);
	\draw[shorten >= 0.3em] (0) to (g);
	\draw[shorten <= 0.3em, shorten >= 0.3em] (g) to (4);
	
	\ifthenelse{#1 = 1}{
		\node[above] at (0) {{$\scriptstyle{0}$}};	
		\node[right] at (1) {{$\scriptstyle{1}$}};
		\node[right] at (2) {{$\scriptstyle{2}$}};
		\node[below] at (3) {{$\scriptstyle{3}$}};
		\node[left] at (4) {{$\scriptstyle{4}$}};
		\node[left] at (5) {{$\scriptstyle{5}$}};}{}
	\end{tikzpicture}
}

\newcommand{\PPPTHREE}[1]{
	\begin{tikzpicture}[scale=1,baseline=0em, line width=0.8pt]
	\coordinate (0) at (0,1);
	\coordinate (1) at ({sin(60)} ,{cos(60)} );
	\coordinate (2) at ({sin(60)} ,{-cos(60)} );
	\coordinate (3) at (0,-1);
	\coordinate (4) at ({-sin(60)} ,{-cos(60)} );
	\coordinate (5) at ({-sin(60)} ,{cos(60)} );
	
	\draw[] (0) to (1);
	\draw[] (1) to (2);
	\draw[] (2) to (3);
	\draw[] (3) to (4);
	\draw[] (4) to (5);
	\draw[] (5) to (0);
	
	\draw[] (1) to (3);
	\draw[] (0) to (3);
	\draw[] (5) to (3);
	\coordinate (a) at (intersection of 3--1 and 0--2);
	\coordinate (k) at (intersection of 1--4 and 0--2);
	\draw[shorten >= 0.3em] (0) to (k);
	\draw[shorten <= 0.3em, shorten >= 0.3em] (k) to (a);
	\draw[shorten <= 0.3em] (a) to (2);
	\coordinate (b) at (intersection of 5--3 and 4--2);
	\coordinate (c) at (intersection of 0--3 and 4--2);
	\coordinate (d) at (intersection of 3--1 and 4--2);
	\draw[shorten >= 0.3em] (4) to (b);
	\draw[shorten <= 0.3em, shorten >= 0.3em] (b) to (c);
	\draw[shorten <= 0.3em, shorten >= 0.3em] (c) to (d);
	\draw[shorten <= 0.3em] (d) to (2);
	\coordinate (h) at (intersection of 0--4 and 5--2);
	\coordinate (e) at (intersection of 0--3 and 5--2);
	\coordinate (f) at (intersection of 3--1 and 5--2);
	\draw[shorten >= 0.3em] (5) to (h);
	\draw[shorten >= 0.3em, shorten <= 0.3em] (h) to (e);
	\draw[shorten <= 0.3em, shorten >= 0.3em] (e) to (f);
	\draw[shorten <= 0.3em] (f) to (2);
	\coordinate (g) at (intersection of 0--4 and 5--3);
	\draw[shorten >= 0.3em] (0) to (g);
	\draw[shorten <= 0.3em, shorten >= 0.3em] (g) to (4);
	\coordinate (i) at (intersection of 0--3 and 1--4);
	\coordinate (j) at (intersection of 5--3 and 1--4);
	\draw[shorten >= 0.3em] (1) to (i);
	\draw[shorten <= 0.3em, shorten >= 0.3em] (i) to (j);
	\draw[shorten <= 0.3em] (j) to (4);
	
	\ifthenelse{#1 = 1}{	
		\node[above] at (0) {{$\scriptstyle{0}$}};	
		\node[right] at (1) {{$\scriptstyle{1}$}};
		\node[right] at (2) {{$\scriptstyle{2}$}};
		\node[below] at (3) {{$\scriptstyle{3}$}};
		\node[left] at (4) {{$\scriptstyle{4}$}};
		\node[left] at (5) {{$\scriptstyle{5}$}};}{}
	\end{tikzpicture}
}

\newcommand{\PPPFOUR}[1]{
	\begin{tikzpicture}[scale=1,baseline=0em, line width=0.8pt]
	\coordinate (0) at (0,1);
	\coordinate (1) at ({sin(60)} ,{cos(60)} );
	\coordinate (2) at ({sin(60)} ,{-cos(60)} );
	\coordinate (3) at (0,-1);
	\coordinate (4) at ({-sin(60)} ,{-cos(60)} );
	\coordinate (5) at ({-sin(60)} ,{cos(60)} );
	
	\draw[] (0) to (1);
	\draw[] (1) to (2);
	\draw[] (2) to (3);
	\draw[] (3) to (4);
	\draw[] (4) to (5);
	\draw[] (5) to (0);
	
	\draw[] (1) to (3);
	\draw[] (0) to (3);
	\draw[] (5) to (3);
	\coordinate (a) at (intersection of 3--1 and 0--2);
	\coordinate (k) at (intersection of 1--4 and 0--2);
	\coordinate (m) at (intersection of 1--5 and 0--2);
	\draw[shorten >= 0.3em] (0) to (m);
	\draw[shorten <= 0.3em, shorten >= 0.3em] (m) to (k);
	\draw[shorten <= 0.3em, shorten >= 0.3em] (k) to (a);
	\draw[shorten <= 0.3em] (a) to (2);
	\coordinate (b) at (intersection of 5--3 and 4--2);
	\coordinate (c) at (intersection of 0--3 and 4--2);
	\coordinate (d) at (intersection of 3--1 and 4--2);
	\draw[shorten >= 0.3em] (4) to (b);
	\draw[shorten <= 0.3em, shorten >= 0.3em] (b) to (c);
	\draw[shorten <= 0.3em, shorten >= 0.3em] (c) to (d);
	\draw[shorten <= 0.3em] (d) to (2);
	\coordinate (h) at (intersection of 0--4 and 5--2);
	\coordinate (e) at (intersection of 0--3 and 5--2);
	\coordinate (f) at (intersection of 3--1 and 5--2);
	\draw[shorten >= 0.3em] (5) to (h);
	\draw[shorten >= 0.3em, shorten <= 0.3em] (h) to (e);
	\draw[shorten <= 0.3em, shorten >= 0.3em] (e) to (f);
	\draw[shorten <= 0.3em] (f) to (2);
	\coordinate (g) at (intersection of 0--4 and 5--3);
	\draw[shorten >= 0.3em] (0) to (g);
	\draw[shorten <= 0.3em, shorten >= 0.3em] (g) to (4);
	\coordinate (i) at (intersection of 0--3 and 1--4);
	\coordinate (j) at (intersection of 5--3 and 1--4);
	\draw[shorten >= 0.3em] (1) to (i);
	\draw[shorten <= 0.3em, shorten >= 0.3em] (i) to (j);
	\draw[shorten <= 0.3em] (j) to (4);
	\coordinate (l) at (intersection of 0--4 and 5--1);
	\coordinate (m) at (intersection of 0--3 and 5--1);
	\draw[shorten >= 0.3em] (5) to (l);
	\draw[shorten <= 0.3em, shorten >= 0.3em] (l) to (m);
	\draw[shorten <= 0.3em] (m) to (1);
	
	\ifthenelse{#1 = 1}{	
		\node[above] at (0) {{$\scriptstyle{0}$}};	
		\node[right] at (1) {{$\scriptstyle{1}$}};
		\node[right] at (2) {{$\scriptstyle{2}$}};
		\node[below] at (3) {{$\scriptstyle{3}$}};
		\node[left] at (4) {{$\scriptstyle{4}$}};
		\node[left] at (5) {{$\scriptstyle{5}$}};}{}
	\end{tikzpicture}
}

\newcommand{\PPPFIVE}[1]{
	\begin{tikzpicture}[scale=1,baseline=0em, line width=0.8pt]
	\coordinate (0) at (0,1);
	\coordinate (1) at ({sin(60)} ,{cos(60)} );
	\coordinate (2) at ({sin(60)} ,{-cos(60)} );
	\coordinate (3) at (0,-1);
	\coordinate (4) at ({-sin(60)} ,{-cos(60)} );
	\coordinate (5) at ({-sin(60)} ,{cos(60)} );
	
	\draw[] (0) to (1);
	\draw[] (1) to (2);
	\draw[] (2) to (3);
	\draw[] (3) to (4);
	\draw[] (4) to (5);
	\draw[] (5) to (0);
	
	\draw[] (1) to (3);
	\draw[] (0) to (3);
	\draw[] (5) to (3);
	\coordinate (a) at (intersection of 3--1 and 0--2);
	\coordinate (k) at (intersection of 1--4 and 0--2);
	\coordinate (m) at (intersection of 1--5 and 0--2);
	\draw[shorten >= 0.3em] (0) to (m);
	\draw[shorten <= 0.3em, shorten >= 0.3em] (m) to (k);
	\draw[shorten <= 0.3em, shorten >= 0.3em] (k) to (a);
	\draw[shorten <= 0.3em] (a) to (2);
	\coordinate (b) at (intersection of 5--3 and 4--2);
	\coordinate (c) at (intersection of 0--3 and 4--2);
	\coordinate (d) at (intersection of 3--1 and 4--2);
	\draw[shorten >= 0.3em] (4) to (b);
	\draw[shorten <= 0.3em, shorten >= 0.3em] (b) to (c);
	\draw[shorten <= 0.3em, shorten >= 0.3em] (c) to (d);
	\draw[shorten <= 0.3em] (d) to (2);
	\coordinate (e) at (intersection of 0--3 and 5--2);
	\coordinate (f) at (intersection of 3--1 and 5--2);
	\draw[shorten >= 0.3em] (5) to (e);
	\draw[shorten <= 0.3em, shorten >= 0.3em] (e) to (f);
	\draw[shorten <= 0.3em] (f) to (2);
	\coordinate (m) at (intersection of 0--3 and 5--1);
	\draw[shorten >= 0.3em] (5) to (m);
	\draw[shorten <= 0.3em] (m) to (1);
	\coordinate (i) at (intersection of 0--3 and 1--4);
	\coordinate (j) at (intersection of 5--3 and 1--4);
	\draw[shorten >= 0.3em] (1) to (i);
	\draw[shorten <= 0.3em, shorten >= 0.3em] (i) to (j);
	\draw[shorten <= 0.3em] (j) to (4);
	
	\ifthenelse{#1 = 1}{	
		\node[above] at (0) {{$\scriptstyle{0}$}};	
		\node[right] at (1) {{$\scriptstyle{1}$}};
		\node[right] at (2) {{$\scriptstyle{2}$}};
		\node[below] at (3) {{$\scriptstyle{3}$}};
		\node[left] at (4) {{$\scriptstyle{4}$}};
		\node[left] at (5) {{$\scriptstyle{5}$}};}{}
	\end{tikzpicture}
}

\newcommand{\PPPSIX}[1]{
	\begin{tikzpicture}[scale=1,baseline=0em, line width=0.8pt]
	\coordinate (0) at (0,1);
	\coordinate (1) at ({sin(60)} ,{cos(60)} );
	\coordinate (2) at ({sin(60)} ,{-cos(60)} );
	\coordinate (3) at (0,-1);
	\coordinate (4) at ({-sin(60)} ,{-cos(60)} );
	\coordinate (5) at ({-sin(60)} ,{cos(60)} );
	
	\draw[] (0) to (1);
	\draw[] (1) to (2);
	\draw[] (2) to (3);
	\draw[] (3) to (4);
	\draw[] (4) to (5);
	\draw[] (5) to (0);
	
	\draw[] (1) to (3);
	\draw[] (0) to (3);
	\draw[] (5) to (3);
	\coordinate (a) at (intersection of 3--1 and 0--2);
	\coordinate (m) at (intersection of 1--5 and 0--2);
	\draw[shorten >= 0.3em] (0) to (m);
	\draw[shorten <= 0.3em, shorten >= 0.3em] (m) to (a);
	\draw[shorten <= 0.3em] (a) to (2);
	\coordinate (b) at (intersection of 5--3 and 4--2);
	\coordinate (c) at (intersection of 0--3 and 4--2);
	\coordinate (d) at (intersection of 3--1 and 4--2);
	\draw[shorten >= 0.3em] (4) to (b);
	\draw[shorten <= 0.3em, shorten >= 0.3em] (b) to (c);
	\draw[shorten <= 0.3em, shorten >= 0.3em] (c) to (d);
	\draw[shorten <= 0.3em] (d) to (2);
	\coordinate (e) at (intersection of 0--3 and 5--2);
	\coordinate (f) at (intersection of 3--1 and 5--2);
	\draw[shorten >= 0.3em] (5) to (e);
	\draw[shorten <= 0.3em, shorten >= 0.3em] (e) to (f);
	\draw[shorten <= 0.3em] (f) to (2);
	\coordinate (m) at (intersection of 0--3 and 5--1);
	\draw[shorten >= 0.3em] (5) to (m);
	\draw[shorten <= 0.3em] (m) to (1);
	
	\ifthenelse{#1 = 1}{	
		\node[above] at (0) {{$\scriptstyle{0}$}};	
		\node[right] at (1) {{$\scriptstyle{1}$}};
		\node[right] at (2) {{$\scriptstyle{2}$}};
		\node[below] at (3) {{$\scriptstyle{3}$}};
		\node[left] at (4) {{$\scriptstyle{4}$}};
		\node[left] at (5) {{$\scriptstyle{5}$}};}{}
	\end{tikzpicture}
}

\newcommand{\assocTWO}[0]{
	\begin{tikzpicture}[scale=1,baseline=0em]
	\matrix[matrix of math nodes,row sep =-2.5em,column sep=2.6em, ampersand replacement=\&] (m) {
		{}  \&[-1em] \PPPTWO{1} \& \PPPTHREE{1} \&[-1em] {}
		\\
		\PPPONE{1} \& {} \& {} \& \PPPFOUR{1}
		\\
		{} \& \PPPSIX{1}  \& \PPPFIVE{1} \& {} \\
	};	
	\path
	(m-2-1) edge[->, shorten <= -0.3em, shorten >= -0.3em] node[pos=0.4, above] {$\omega$} node[pos=0.15, right, xshift=0.2em] {$\sss{(02345)}$} (m-1-2)
	edge[->, shorten <= -0.3em, shorten >= -0.3em] node[pos=0.4, below] {$\omega$} node[pos=0.15, right, xshift=0.2em] {$\sss{(01235)}$} (m-3-2)
	(m-1-2) edge[->, shorten <= 0em, shorten >= 0em] node[pos=0.4, above] {$\omega$} node[pos=0.5, below] {$\sss{(01234)}$}(m-1-3)
	(m-3-2) edge[->, shorten <= 0em, shorten >= 0em] node[pos=0.4, below] {$\omega$} node[pos=0.5, above] {$\sss{(12345)}$} (m-3-3)
	(m-2-4) edge[<-, shorten <= -0.3em, shorten >= -0.3em] node[pos=0.4, above] {$\omega$} node[pos=0.15, left, xshift=-0.2em] {$\sss{(01245)}$}(m-1-3)
	edge[<-, shorten <= -0.3em, shorten >= -0.3em] node[pos=0.4, below] {$\omega$} node[pos=0.15, left, xshift=-0.2em] {$\sss{(01345)}$} (m-3-3)
	;
	\end{tikzpicture}
}

\newcommand{\hexoo}[6]{
	\begin{tikzpicture}[scale=#4,baseline=0em, line width=0.8pt]
	\coordinate (0) at (0,1);
	\coordinate (1) at ({sin(60)} ,{cos(60)} );
	\coordinate (2) at ({sin(60)} ,{-cos(60)} );
	\coordinate (3) at (0,-1);
	\coordinate (4) at ({-sin(60)} ,{-cos(60)} );
	\coordinate (5) at ({-sin(60)} ,{cos(60)} );
	
	\draw[opacity = #6] (0) to (1);
	\draw[opacity = #6] (1) to (2);
	\draw[opacity = #6] (2) to (3);
	\draw[opacity = #6] (3) to (4);
	\draw[opacity = #6] (4) to (5);
	\draw[opacity = #6] (5) to (0);
	
	\foreach \x in {#1, #2, #3}
	{
		\ifthenelse{\x = 20 \OR \x = 02}{\draw[opacity = #6] (0) to (2);}{}
		\ifthenelse{\x = 30 \OR \x = 03}{\draw[opacity = #6] (0) to (3);}{}
		\ifthenelse{\x = 40 \OR \x = 04}{\draw[opacity = #6] (0) to (4);}{}
		\ifthenelse{\x = 13 \OR \x = 31}{\draw[opacity = #6] (1) to (3);}{}
		\ifthenelse{\x = 14 \OR \x = 41}{\draw[opacity = #6] (1) to (4);}{}
		\ifthenelse{\x = 15 \OR \x = 51}{\draw[opacity = #6] (1) to (5);}{}
		\ifthenelse{\x = 24 \OR \x = 42}{\draw[opacity = #6] (2) to (4);}{}
		\ifthenelse{\x = 25 \OR \x = 52}{\draw[opacity = #6] (2) to (5);}{}
		\ifthenelse{\x = 35 \OR \x = 53}{\draw[opacity = #6] (3) to (5);}{}
	}
	
	\ifthenelse{#5 = 1}{
		\node[above] at (0) {{$\scriptstyle{0}$}};	
		\node[right] at (1) {{$\scriptstyle{1}$}};
		\node[right] at (2) {{$\scriptstyle{2}$}};
		\node[below] at (3) {{$\scriptstyle{3}$}};
		\node[left] at (4) {{$\scriptstyle{4}$}};
		\node[left] at (5) {{$\scriptstyle{5}$}};}{}	
	\end{tikzpicture}
}

\newcommand{\dualONE}[2]{
	\begin{tikzpicture}[scale=#1,baseline=0em, line width=0.8pt]
	\coordinate (0) at(0, 0);
	\coordinate (1) at (2, 0);
	\coordinate (2) at (-2 , 0);
	\coordinate (3) at (0, 4);
	\coordinate (4) at (2, 4);
	\coordinate (5) at (-2, 4);
	\coordinate (6) at (0, -4);
	\coordinate (7) at (2, -4);
	\coordinate (8) at (-2, -4);
	\coordinate (9) at (0, 2);
	\coordinate (10) at (0, -2);
	
	\ifthenelse{#2 = 1}{
		\draw[dot] (0) -- (1);
		\draw[dot] (0) -- (2) node[anchor=center, pos=1.5] {{$\phantom{\sss{(024)}}$}};
		\draw[] (0) -- (9) node[right, pos=0.5, xshift=-0.2em] {{$\sss{(035)}$}};
		\draw[] (9) -- (3) node[anchor=center, pos=1.25] {{$\sss{(045)}$}};
		\draw[dot] (9) -- (4);
		\draw[] (9) -- (5) node[anchor=center, pos=1.25] {{$\sss{(034)}$}};
		\draw[] (0) -- (10) node[left, pos=0.5, xshift=0.2em] {{$\sss{(015)}$}};  
		\draw[] (10) -- (6) node[anchor=center, pos=1.25] {{$\sss{(025)}$}};
		\draw[] (10) -- (7) node[anchor=center, pos=1.25] {{$\sss{(125)}$}};
		\draw[dot] (10) -- (8);		
	}{}
	\ifthenelse{#2 = 2}{
		\draw[dot] (0) -- (1);
		\draw[] (0) -- (2) node[anchor=center, pos=1.5] {{$\sss{(024)}$}};
		\draw[] (0) -- (9) node[right, pos=0.5, xshift=-0.2em] {{$\sss{(045)}$}};
		\draw[] (9) -- (3) node[anchor=center, pos=1.25] {{$\sss{(015)}$}};
		\draw[] (9) -- (4) node[anchor=center, pos=1.25] {{$\sss{(145)}$}};
		\draw[] (9) -- (5) node[anchor=center, pos=1.25] {{$\sss{(014)}$}};
		\draw[] (0) -- (10) node[left, pos=0.5, xshift=0.2em] {{$\sss{(025)}$}};  
		\draw[] (10) -- (6) node[anchor=center, pos=1.25] {{$\sss{(035)}$}};
		\draw[] (10) -- (7) node[anchor=center, pos=1.25] {{$\sss{(235)}$}};
		\draw[dot] (10) -- (8);			
	}{}
	\end{tikzpicture}
}

\newcommand{\dualTWO}[2]{
	\begin{tikzpicture}[scale=#1,baseline=0em, line width=0.8pt]
	\coordinate (0) at(0, 0);
	\coordinate (1) at (4, 2);
	\coordinate (2) at (-4, 2);
	\coordinate (3) at (0, 4);
	\coordinate (4) at (4, 4);
	\coordinate (5) at (-4, 4);
	\coordinate (6) at (0, -4);
	\coordinate (7) at (2, -4);
	\coordinate (8) at (-2, -4);
	\coordinate (9) at (0, 2);
	\coordinate (10) at (0, -2);
	\coordinate (11) at (-2, 2);
	\coordinate (12) at (2,2);
	
	\ifthenelse{#2 = 1}{
		\draw[dot] (0) -- (4);
		\draw[] (0) -- (11) node[left, pos=0.2] {{$\sss{(014)}$}};
		\draw[] (11) -- (5) node[anchor=center, pos=1.25] {{$\sss{(034)}$}};
		\draw[dot] (2) -- (1);
		\draw[] (0) -- (10) node[left, pos=0.5, xshift=0.2em] {{$\sss{(015)}$}}; 
		\draw[] (10) -- (6) node[anchor=center, pos=1.25] {{$\sss{(025)}$}};
		\draw[] (10) -- (7) node[anchor=center, pos=1.25] {{$\sss{(125)}$}};
		\draw[dot] (10) to (8);
		\draw[shorten >= 0.3em] (0) -- (9);
		\draw[shorten <= 0.3em] (9) -- (3) node[anchor=center, pos=1.22] {{$\sss{(045)}$}};
	}{}
	\ifthenelse{#2 = 2}{
		\draw[] (0) -- (12) node[right, pos=0.2] {{$\sss{(125)}$}};
		\draw[] (12) -- (4) node[anchor=center, pos=1.25] {{$\sss{(145)}$}};
		\draw[dot] (0) -- (11);
		\draw[] (11) -- (5) node[anchor=center, pos=1.25] {{$\sss{(014)}$}};
		\draw[] (11) -- (2) node[anchor=center, pos=1.5] {{$\sss{(024)}$}};
		\draw[dot] (11) -- (1);
		\draw[] (0) -- (10) node[left, pos=0.5, xshift=0.2em] {{$\sss{(025)}$}}; 
		\draw[] (10) -- (6) node[anchor=center, pos=1.25] {{$\sss{(035)}$}};
		\draw[] (10) -- (7) node[anchor=center, pos=1.25] {{$\sss{(235)}$}};
		\draw[dot] (10) to (8);
		\draw[shorten >= 0.3em] (0) -- (9);
		\draw[shorten <= 0.3em] (9) -- (3) node[anchor=center, pos=1.22] {{$\sss{(015)}$}};
	}{}	
	\end{tikzpicture}
}

\newcommand{\dualTHREE}[2]{
	\begin{tikzpicture}[scale=#1,baseline=0em, line width=0.8pt]
	\coordinate (0) at(0, 0);
	\coordinate (1) at (6, 2);
	\coordinate (2) at (-6, 2);
	\coordinate (3) at (0, 4);
	\coordinate (4) at (6, 4);
	\coordinate (5) at (-6, 4);
	\coordinate (6) at (0, -4);
	\coordinate (7) at (4, -2);
	\coordinate (8) at (-4, -2);
	\coordinate (9) at (0, 2);
	\coordinate (10) at (0, -2);
	\coordinate (11) at (2, 0);
	\coordinate (12) at (-2, 0);
	\coordinate (13) at (-4, 2);
	\coordinate (14) at (4, 2);
	
	\ifthenelse{#2 = 1}{
		\draw[] (12) -- (11) node[above, pos=0.21, yshift=-0.28em] {{$\sss{(124)}$}};
		\draw[] (11) -- (7)  node[anchor=center, pos=1.25] {{$\sss{(125)}$}};
		\draw[dot] (12) -- (8);
		\draw[dot] (10) -- (4);
		\draw[] (10) -- (12) node[left, pos=0.2] {{$\sss{(024)}$}};
		\draw[] (12) -- (13) node[left, pos=0.2] {{$\sss{(014)}$}};
		\draw[] (13) -- (5) node[anchor=center, pos=1.25] {{$\sss{(034)}$}};
		\draw[dot] (1) -- (2);
		\draw[] (10) -- (6) node[anchor=center, pos=1.25] {{$\sss{(025)}$}};
		\draw[shorten >= 0.3em, shorten <= 0.3em] (0) -- (9);
		\draw[shorten <= 0.3em] (9) -- (3) node[anchor=center, pos=1.22] {{$\sss{(045)}$}};
		\draw[shorten >= 0.3em] (10) -- (0);
	}{}
	\ifthenelse{#2 = 2}{
		\draw[dot] (12) -- (11);
		\draw[] (11) -- (7) node[anchor=center, pos=1.25] {{$\sss{(235)}$}};
		\draw[dot] (12) -- (8);
		\draw[] (10) -- (11) node[right, pos=0.2] {{$\sss{(135)}$}};
		\draw[] (11) -- (14) node[right, pos=0.2] {{$\sss{(125)}$}};
		\draw[] (14) -- (4) node[anchor=center, pos=1.25] {{$\sss{(145)}$}};
		\draw[dot] (10) -- (12);
		\draw[dot] (12) -- (13);
		\draw[] (13) -- (5) node[anchor=center, pos=1.25] {{$\sss{(014)}$}};
		\draw[] (13) -- (2) node[anchor=center, pos=1.5] {{$\sss{(024)}$}};
		\draw[dot] (13) -- (1);
		\draw[] (10) -- (6) node[anchor=center, pos=1.25] {{$\sss{(035)}$}};
		\draw[shorten >= 0.3em, shorten <= 0.3em] (0) -- (9);
		\draw[shorten <= 0.3em] (9) -- (3) node[anchor=center, pos=1.22] {{$\sss{(015)}$}};
		\draw[shorten >= 0.3em] (10) -- (0);
	}{}	
	\end{tikzpicture}
}

\newcommand{\dualFOUR}[2]{
	\begin{tikzpicture}[scale=#1,baseline=0em, line width=0.8pt]
	\coordinate (10) at(0, -2);
	\coordinate (1) at (6, 0);
	\coordinate (2) at (-6, 2);
	\coordinate (3) at (0, 4);
	\coordinate (4) at (6, 4);
	\coordinate (5) at (-6, 4);
	\coordinate (6) at (0, -4);
	\coordinate (7) at (6, -2);
	\coordinate (8) at (-4, -2);
	\coordinate (9) at (0, 2);
	\coordinate (11) at (2, 0);
	\coordinate (12) at (-2, 0);
	\coordinate (13) at (2, 2);
	\coordinate (14) at (3, 1);
	\coordinate (15) at (-4, 2);
	\coordinate (16) at (4,0);
	
	\ifthenelse{#2 = 1}{
		\draw[] (10) -- (12) node[left, pos=0.2] {{$\sss{(024)}$}};
		\draw[] (12) -- (15) node[left, pos=0.2] {{$\sss{(014)}$}};
		\draw[] (15) -- (5) node[anchor=center, pos=1.25] {{$\sss{(034)}$}};
		\draw[dot] (2) -- (13);
		\draw[dot] (12) -- (8);
		\draw[] (16) -- (13) node[right, pos=0.4] {{$\sss{(123)}$}};
		\draw[] (16) -- (7)  node[anchor=center, pos=1.25] {{$\sss{(125)}$}};
		\draw[dot] (11) -- (1);
		\draw[dot] (11) -- (13);
		\draw[] (12) -- (13) node[above, pos=0.2] {{$\sss{(124)}$}};;
		\draw[shorten >= 0.3em, dot] (10) -- (14);
		\draw[shorten <= 0.3em, dot] (14) -- (4);
		\coordinate (a) at (intersection of 12--13 and 10--3);
		\draw[] (10) -- (6) node[anchor=center, pos=1.25] {{$\sss{(025)}$}};
		\draw[shorten >= 0.3em] (10) -- (a);
		\draw[shorten <= 0.3em, shorten >= 0.3em] (a) -- (9);
		\draw[shorten <= 0.3em] (9) -- (3)  node[anchor=center, pos=1.22] {{$\sss{(045)}$}};
	}{}
	\ifthenelse{#2 = 2}{
		\draw[dot] (10) -- (12);
		\draw[dot] (12) -- (15);
		\draw[] (15) -- (5) node[anchor=center, pos=1.25] {{$\sss{(014)}$}};
		\draw[] (15) -- (2) node[anchor=center, pos=1.5] {{$\sss{(024)}$}};
		\draw[dot] (15) -- (13);
		\draw[dot] (12) -- (8);
		\draw[] (16) -- (13) node[right, pos=0.4] {{$\sss{(234)}$}};
		\draw[] (16) -- (7)  node[anchor=center, pos=1.25] {{$\sss{(235)}$}};
		\draw[dot] (11) -- (1);
		\draw[] (11) -- (13) node[left, pos=0.3, xshift=0.43em] {{$\sss{(134)}$}};
		\draw[dot] (12) -- (13);
		\draw[] (10) -- (11)  node[right, pos=0.2] {{$\sss{(135)}$}};
		\draw[shorten >= 0.3em] (11) -- (14);
		\draw[shorten <= 0.3em] (14) -- (4) node[anchor=center, pos=1.18] {{$\sss{(145)}$}};
		\coordinate (a) at (intersection of 12--13 and 10--3);
		\draw[] (10) -- (6) node[anchor=center, pos=1.25] {{$\sss{(035)}$}};
		\draw[shorten >= 0.3em] (10) -- (a);
		\draw[shorten <= 0.3em, shorten >= 0.3em] (a) -- (9);
		\draw[shorten <= 0.3em] (9) -- (3)  node[anchor=center, pos=1.22] {{$\sss{(015)}$}};
	}{}	
	\end{tikzpicture}
}

\newcommand{\dualFIVE}[2]{
	\begin{tikzpicture}[scale=#1,baseline=0em, line width=0.8pt]
	\coordinate (0) at(0, 0);
	\coordinate (1) at (6, 2);
	\coordinate (2) at (-6, 2);
	\coordinate (3) at (0, 4);
	\coordinate (4) at (4, 4);
	\coordinate (5) at (-4, 4);
	\coordinate (6) at (0, -4);
	\coordinate (7) at (6, 0);
	\coordinate (8) at (-6, 0);
	\coordinate (9) at (0, 2);
	\coordinate (10) at (0, -2);
	\coordinate (11) at (2, 0);
	\coordinate (12) at (-2, 0);
	\coordinate (13) at (-2, 2);
	\coordinate (14) at (2, 2);
	\coordinate (15) at (-4, 2);
	\coordinate (16) at (4, 2);
	
	\ifthenelse{#2 = 1}{
		\draw[dot] (12) -- (11);
		\draw[dot] (10) -- (16);
		\draw[] (10) -- (12) node[left, pos=0.2] {{$\sss{(024)}$}};
		\draw[] (12) -- (15) node[left, pos=0.2] {{$\sss{(023)}$}};
		\draw[dot] (2) -- (15);
		\draw[] (15) -- (13);
		\draw[] (13) -- (14) node[above, pos=0.21, yshift=-0.28em] {{$\sss{(123)}$}};
		\draw[]	(14) -- (16);	
		\draw[dot] (16) -- (1);
		\draw[dot] (8) -- (15);
		\draw[] (16) -- (7) node[anchor=center, pos=1.25] {{$\sss{(125)}$}};
		\draw[] (10) -- (6) node[anchor=center, pos=1.25] {{$\sss{(025)}$}};
		\draw[shorten >= 0.3em] (10) -- (0);
		\draw[shorten >= 0.3em, shorten <= 0.3em] (0) -- (9);
		\draw[shorten <= 0.3em] (9) -- (3)  node[anchor=center, pos=1.22] {{$\sss{(045)}$}};
		\draw[shorten >= 0.3em] (12) -- (13);
		\draw[shorten <= 0.3em] (13) -- (5) node[anchor=center, pos=1.22] {{$\sss{(034)}$}};
		\draw[shorten >= 0.3em, dot] (11) -- (14);
		\draw[shorten >= 0.3em, shorten <= 0.3em, dot] (14) -- (4);
	}{}
	\ifthenelse{#2 = 2}{
		\draw[] (12) -- (11) node[above, pos=0.245, yshift=-0.28em] {{$\sss{(134)}$}};
		\draw[] (10) -- (11) node[right, pos=0.2] {{$\sss{(135)}$}};
		\draw[dot] (11) -- (16);
		\draw[dot] (10) -- (12);
		\draw[] (12) -- (15) node[left, pos=0.2] {{$\sss{(034)}$}};
		\draw[] (15) -- (2)  node[anchor=center, pos=1.5] {{$\sss{(024)}$}};
		\draw[] (15) -- (13);
		\draw[] (13) -- (14) node[above, pos=0.21, yshift=-0.28em] {{$\sss{(234)}$}};
		\draw[]	(14) -- (16);	
		\draw[dot] (16) -- (1);
		\draw[dot] (8) -- (15);
		\draw[] (16) -- (7) node[anchor=center, pos=1.25] {{$\sss{(235)}$}};
		\draw[] (10) -- (6) node[anchor=center, pos=1.25] {{$\sss{(035)}$}};
		\draw[shorten >= 0.3em] (10) -- (0);
		\draw[shorten >= 0.3em, shorten <= 0.3em] (0) -- (9);
		\draw[shorten <= 0.3em] (9) -- (3)  node[anchor=center, pos=1.22] {{$\sss{(015)}$}};
		\draw[shorten >= 0.3em] (12) -- (13);
		\draw[shorten <= 0.3em] (13) -- (5) node[anchor=center, pos=1.22] {{$\sss{(014)}$}};
		\draw[shorten >= 0.3em] (11) -- (14);
		\draw[shorten <= 0.3em] (14) -- (4) node[anchor=center, pos=1.22] {{$\sss{(145)}$}};
	}{}	
	\end{tikzpicture}
}

\newcommand{\dualSIX}[2]{
	\begin{tikzpicture}[scale=#1,baseline=0em, line width=0.8pt]
	\coordinate (0) at(0, 0);
	\coordinate (1) at (4, 0);
	\coordinate (2) at (-4, 0);
	\coordinate (3) at (0, 4);
	\coordinate (4) at (2, 4);
	\coordinate (5) at (-2, 4);
	\coordinate (6) at (0, -4);
	\coordinate (7) at (4, -2);
	\coordinate (8) at (-4, -2);
	\coordinate (9) at (0, 2);
	\coordinate (10) at (0, -2);
	\coordinate (11) at (2, 0);
	\coordinate (12) at (-2, 0);
	
	\ifthenelse{#2 = 1}{
		\draw[dot] (2) -- (12);
		\draw[] (12) -- (11) node[above, pos=0.21, yshift=-0.28em] {{$\sss{(123)}$}};
		\draw[dot] (8) -- (12);
		\draw[] (11) -- (7) node[anchor=center, pos=1.25] {{$\sss{(125)}$}};
		\draw[dot] (11) -- (1);
		\draw[] (10) -- (12) node[left, pos=0.2] {{$\sss{(023)}$}};
		\draw[dot] (10) -- (11);
		\draw[] (9) -- (5) node[anchor=center, pos=1.25] {{$\sss{(034)}$}};
		\draw[dot] (9) -- (4);
		\draw[] (10) -- (6) node[anchor=center, pos=1.25] {{$\sss{(025)}$}};  
		\draw[shorten >= 0.3em] (10) -- (0);
		\draw[shorten <= 0.3em] (0) -- (9) node[right, pos=0.6, xshift=-0.2em] {{$\sss{(035)}$}};
		\draw[] (9) -- (3) node[anchor=center, pos=1.25] {{$\sss{(045)}$}};
	}{}
	\ifthenelse{#2 = 2}{
		\draw[] (12) -- (2)  node[anchor=center, pos=1.5] {{$\sss{(024)}$}};
		\draw[] (12) -- (11) node[above, pos=0.21, yshift=-0.28em] {{$\sss{(234)}$}};
		\draw[dot] (8) -- (12);
		\draw[] (11) -- (7)  node[anchor=center, pos=1.25] {{$\sss{(235)}$}};
		\draw[dot] (11) -- (1);
		\draw[] (10) -- (12) node[left, pos=0.2] {{$\sss{(034)}$}};
		\draw[dot] (10) -- (11);
		\draw[] (9) -- (5) node[anchor=center, pos=1.25] {{$\sss{(014)}$}};
		\draw[] (9) -- (4) node[anchor=center, pos=1.25] {{$\sss{(145)}$}};;
		\draw[] (10) -- (6) node[anchor=center, pos=1.25] {{$\sss{(035)}$}};  
		\draw[shorten >= 0.3em] (10) -- (0);
		\draw[shorten <= 0.3em] (0) -- (9) node[right, pos=0.6, xshift=-0.2em] {{$\sss{(045)}$}};
		\draw[] (9) -- (3) node[anchor=center, pos=1.25] {{$\sss{(015)}$}};
	}{}	
	\end{tikzpicture}
}

\newcommand{\fourCocycleONE}[1]{
	\begin{tikzpicture}[scale=1,baseline=0em]
	\matrix[matrix of math nodes,row sep =-2em,column sep=2.4em, ampersand replacement=\&] (m) {
		{}  \&[-1.5em] \hspace{-3em} \dualTWO{0.3}{#1} \& \hspace{-3.2em} \dualTHREE{0.3}{#1} \&[-5em] {}
		\\
		\dualONE{0.3}{#1} \& {} \& {} \& \dualFOUR{0.3}{#1}
		\\
		{} \& \dualSIX{0.3}{#1}  \& \dualFIVE{0.3}{#1} \& {} \\
	};	
	\path
	(m-2-1) edge[->, shorten <= 0.6em, shorten >= 2.5em, out=85, in=185, looseness=1] node[pos=0.5, left, yshift=0.6em] {$\sss{(01345)}$} node[pos=0.3, right] {${\rm id}$}(m-1-2)
	edge[<-, shorten <= 0.5em, shorten >= 1em, out=275, in=175, looseness=1.1] node[pos=0.6, left, yshift=-0.5em] {$\sss{(01235)}$} node[pos=0.5, above] {$R$} (m-3-2)
	(m-1-2) edge[<-, shorten <= 0.5em, shorten >= 2em] node[pos=0.4, above] {$\mathfrak{c}$} node[pos=0.4, below] {$\sss{(01245)}$} (m-1-3)
	(m-3-2) edge[->, shorten <= -1.2em, shorten >= 0.8em] node[pos=0.1, below] {$\alpha^{-1}$} node[pos=0.1, above] {$\sss{(02345)}$} (m-3-3)
	(m-2-4) edge[<-, shorten <= 0.6em, shorten >= 2.5em, out=95, in=355, looseness=1] node[pos=0.5, right, yshift=0.6em] {$\sss{(12345)}$} node[pos=0.3, left] {${\rm id}$} (m-1-3)
	edge[<-, shorten <= 0.5em, shorten >= 1em, out=265, in=5, looseness=1.1] node[pos=0.6, right, yshift=-0.5em] {$\sss{(01234)}$} node[pos=0.5, above] {$R$} (m-3-3)
	;
	\end{tikzpicture}
}

\newcommand{\fourCocycleTWO}[1]{
	\begin{tikzpicture}[scale=1,baseline=0em]
	\matrix[matrix of math nodes,row sep =-2em,column sep=2.4em, ampersand replacement=\&] (m) {
		{}  \&[-3.5em] \hspace{-0em} \dualTWO{0.3}{#1} \&[-2.1em] \hspace{-0em} \dualTHREE{0.3}{#1} \&[-6em] {}
		\\
		\dualONE{0.3}{#1} \& {} \& {} \& \dualFOUR{0.3}{#1}
		\\
		{} \&  \dualSIX{0.3}{#1}  \& \dualFIVE{0.3}{#1} \& {} \\
	};	
	\path
	(m-2-1) edge[->, shorten <= 0.6em, shorten >= -0.8em, out=85, in=185, looseness=1.1] node[pos=0.9, left, yshift=0.4em] {$\sss{(01245)}$} node[pos=0.5, right, yshift=-0.2em] {${\rm id}$}(m-1-2)
	edge[<-, shorten <=0.5em, shorten >= -0.5em, out=275, in=175, looseness=1.1] node[pos=0.9, left, yshift=-0.5em] {$\sss{(02345)}$} node[pos=0.7, above] {$\mathfrak{c}$} (m-3-2)
	(m-1-2) edge[<-, shorten <= -1.5em, shorten >= -1em] node[pos=0.4, above] {$\alpha^{-1}$} node[pos=0.4, below] {$\sss{(01235)}$} (m-1-3)
	(m-3-2) edge[->, shorten <= -1.1em, shorten >= -1.1em] node[pos=0.1, below] {$\alpha$} node[pos=0.1, above] {$\sss{(01345)}$} (m-3-3)
	(m-2-4) edge[->, shorten <= 0.6em, shorten >= -0em, out=95, in=355, looseness=1.1] node[pos=0.9, right, yshift=0.4em] {$\sss{(12345)}$} node[pos=0.5, left, yshift= -0.3em] {$R$} (m-1-3)
	edge[->, shorten <= 0.5em, shorten >= 0em, out=265, in=5, looseness=1.1] node[pos=0.9, right, yshift=-0.5em] {$\sss{(01234)}$} node[pos=0.7, above] {${\rm id}$} (m-3-3)
	;
	\end{tikzpicture}
}

\newcommand{\PachnerDualONE}[2]{
	\begin{tikzpicture}[scale=#1,baseline=0em, line width=0.8pt]
	\coordinate (0) at(0, 0);
	\coordinate (1) at (-2, 3);
	\coordinate (2) at (0 , 3);
	\coordinate (3) at (2, 3);
	\coordinate (4) at (0, 1);
	\coordinate (5) at (0, -1);
	\coordinate (6) at (-2, -3);
	\coordinate (7) at (0, -3);
	\coordinate (8) at (2, -3);	
	
	\ifthenelse{#2 = 1}{
		\node[right, xshift=-0.2em] at (0) {{$\sss{(023)}$}};	
		\draw[] (0) -- (2) node[anchor=center, pos=1.16] {{$\sss{(234)}$}};
		\draw[] (0) -- (7) node[anchor=center, pos=1.16] {{$\sss{(123)}$}};
		\draw[] (4) -- (1) node[anchor=center, pos=1.25] {{$\sss{(034)}$}};;
		\draw[] (4) -- (3) node[anchor=center, pos=1.25] {{$\sss{(024)}$}};
		\draw[] (5) -- (6) node[anchor=center, pos=1.25] {{$\sss{(013)}$}};
		\draw[] (5) -- (8) node[anchor=center, pos=1.25] {{$\sss{(012)}$}};	
	}{}	
	\ifthenelse{#2 = 2}{
		\node[right, xshift=-0.2em] at (0) {{$\sss{(034)}$}};	
		\draw[] (0) -- (2) node[anchor=center, pos=1.16] {{$\sss{(014)}$}};
		\draw[] (0) -- (7) node[anchor=center, pos=1.16] {{$\sss{(024)}$}};
		\draw[dot] (4) -- (1);
		\draw[dot] (4) -- (3);
		\draw[dot] (5) -- (6);
		\draw[] (5) -- (8)  node[anchor=center, pos=1.25] {{$\sss{(023)}$}};	
	}{}
	\ifthenelse{#2 = 3}{
		\node[right, xshift=-0.2em] at (0) {{$\sss{(024)}$}};	
		\draw[] (4) -- (5);
		\draw[] (4) -- (2)  node[anchor=center, pos=1.25] {{$\sss{(034)}$}};
		\draw[] (4) -- (5);
		\draw[] (5) -- (7) node[anchor=center, pos=1.25] {{$\sss{(014)}$}};
		\draw[] (4) -- (1)   node[anchor=center, pos=1.25] {{$\sss{(023)}$}};
		\draw[dot] (4) -- (3);
		\draw[] (5) -- (6)   node[anchor=center, pos=1.25] {{$\sss{(012)}$}};
		\draw[dot] (5) -- (8);	
	}{}		
	\end{tikzpicture}
}

\newcommand{\PachnerDualTWO}[2]{
	\begin{tikzpicture}[scale=#1,baseline=0em, line width=0.8pt]
	\coordinate (0) at(0, 0);
	\coordinate (1) at (-4, 3);
	\coordinate (2) at (0 , 3);
	\coordinate (3) at (4, 3);
	\coordinate (4) at (0, 1);
	\coordinate (5) at (0, -1);
	\coordinate (6) at (-4, -1);
	\coordinate (7) at (0, -3);
	\coordinate (8) at (4, -1);
	\coordinate (9) at (-2, 1);
	\coordinate (10) at (2, 1);
	
	\ifthenelse{#2 = 1}{
		\draw[] (9) -- (1) node[anchor=center, pos=1.25] {{$\sss{(034)}$}};
		\draw[] (9) -- (6) node[anchor=center, pos=1.25] {{$\sss{(013)}$}};
		\draw[] (9) -- (10) node[above, pos=0.21, yshift=-0.28em] {{$\sss{(014)}$}};
		\draw[] (5) -- (9) node[left, pos=0.2] {{$\sss{(134)}$}};
		\draw[] (5) -- (10) node[right, pos=0.2] {{$\sss{(124)}$}};
		\draw[] (10) -- (3) node[anchor=center, pos=1.25] {{$\sss{(024)}$}};
		\draw[] (10) -- (8) node[anchor=center, pos=1.25] {{$\sss{(012)}$}};
		\draw[] (5) -- (7) node[anchor=center, pos=1.24] {{$\sss{(123)}$}};;
		\draw[shorten >= 0.3em] (5) -- (4);
		\draw[shorten <= 0.3em] (4) -- (2) node[anchor=center, pos=1.21] {{$\sss{(234)}$}};;
	}{}
	\ifthenelse{#2 = 2}{
		\draw[dot] (9) -- (1);
		\draw[dot] (9) -- (6);
		\draw[] (9) -- (10) node[above, pos=0.21, yshift=-0.28em] {{$\sss{(123)}$}};
		\draw[] (5) -- (9) node[left, pos=0.2] {{$\sss{(124)}$}};
		\draw[dot] (5) -- (10);
		\draw[dot] (10) -- (3);
		\draw[] (10) -- (8)  node[anchor=center, pos=1.25] {{$\sss{(023)}$}};
		\draw[] (5) -- (7) node[anchor=center, pos=1.24] {{$\sss{(024)}$}};;
		\draw[shorten >= 0.3em] (5) -- (4);
		\draw[shorten <= 0.3em] (4) -- (2) node[anchor=center, pos=1.21] {{$\sss{(014)}$}};
	}{}
		\ifthenelse{#2 = 3}{
		\draw[] (9) -- (1) node[anchor=center, pos=1.25] {{$\sss{(023)}$}};
		\draw[] (9) -- (6) node[anchor=center, pos=1.25] {{$\sss{(012)}$}};
		\draw[dot] (9) -- (10);
		\draw[] (5) -- (9) node[left, pos=0.2] {{$\sss{(013)}$}};
		\draw[dot] (5) -- (10);
		\draw[dot] (10) -- (3);
		\draw[dot] (10) -- (8);
		\draw[] (5) -- (7) node[anchor=center, pos=1.25] {{$\sss{(014)}$}};
		\draw[shorten >= 0.3em] (5) -- (4);
		\draw[shorten <= 0.3em] (4) -- (2) node[anchor=center, pos=1.25] {{$\sss{(034)}$}};
	}{}
	\end{tikzpicture}
}

\newcommand{\PachnerDualFOUR}[2]{
	\begin{tikzpicture}[scale=#1,baseline=0em, line width=0.8pt]
	\coordinate (0) at(0, 0);
	\coordinate (1) at (0, -2);
	\coordinate (2) at (-2 , 2);
	\coordinate (3) at (0, 2);
	\coordinate (4) at (2, 2);	
	
	\ifthenelse{#2 = 1}{
		\draw[] (0) -- (1) node[anchor=center, pos=1.25] {{$\sss{(023)}$}};
		\draw[] (0) -- (3) node[anchor=center, pos=1.25] {{$\sss{(234)}$}};
		\draw[] (0) -- (2) node[anchor=center, pos=1.25] {{$\sss{(034)}$}};;
		\draw[] (0) -- (4) node[anchor=center, pos=1.25] {{$\sss{(024)}$}};
	}{}		
	\end{tikzpicture}
}

\newcommand{\PachnerDualTHREE}[2]{
	\begin{tikzpicture}[scale=#1,baseline=0em, line width=0.8pt]
	\coordinate (0) at(0, 0);
	\coordinate (1) at (0, -2);
	\coordinate (2) at (-2 , 0);
	\coordinate (3) at (2, 0);
	\coordinate (4) at (0, 2);
	\coordinate (5) at (-4, 2);
	\coordinate (6) at (0, 4);
	\coordinate (7) at (4, 2);
	\coordinate (8) at (0, -4);
	
	\ifthenelse{#2 = 1}{
		\draw[] (2) -- (5) node[anchor=center, pos=1.25] {{$\sss{(034)}$}};
		\draw[] (4) -- (6) node[anchor=center, pos=1.25] {{$\sss{(234)}$}};
		\draw[] (3) -- (7) node[anchor=center, pos=1.25] {{$\sss{(024)}$}};
		\draw[] (2) -- (3);
		\draw[] (4) -- (2) node[left, pos=0.2] {{$\sss{(134)}$}};
		\draw[] (4) -- (3) node[right, pos=0.2] {{$\sss{(124)}$}};
		\draw[] (1) -- (2) node[left, pos=0.2] {{$\sss{(013)}$}};
		\draw[] (1) -- (3) node[right, pos=0.2] {{$\sss{(012)}$}};
		\draw[] (1) -- (8) node[anchor=center, pos=1.25] {{$\sss{(023)}$}};
		\draw[shorten >= 0.3em] (1) -- (0);
		\draw[shorten <= 0.3em] (0) -- (4);
	}{}
	\end{tikzpicture}
}

\newcommand{\actionA}[3]{
	\begin{tikzpicture}[scale=#1,baseline=#3 em, line width=0.8pt]
	\coordinate (0) at (0,0);
	\coordinate (1) at (1,1);
	\coordinate (2) at (-2,0.8);
	\coordinate (3) at (-1, 3);
	\coordinate (4) at (intersection of 1--2 and 0--3);	
	\coordinate (5) at (-0.6,0.6);
	\coordinate (a) at (intersection of 3--5 and 1--2);
	\coordinate (b) at (intersection of 1--5 and 0--3);
	
	\ifthenelse{#2 = 1}{
		\draw[] (0)--(1); 
		\draw[] (1)--(2);
		\draw[] (0)--(2); 
		\draw[] (5)--(0);
		\draw[] (5)--(1);
		\draw[] (5)--(2);
		\node[below] at (0) {{$\scriptstyle{2}$}};
		\node[right] at (1) {{$\scriptstyle{1}$}};
		\node[left] at (2) {{$\scriptstyle{3}$}};
		\node[below=-0.1em] at (5) {{$\scriptstyle{0}$}};	
	}{}	
	
	\ifthenelse{#2 = 2}{
		\draw[] (0)--(1);
		\draw[shorten >= 0.3em] (1)--(4);
		\draw[shorten >= 0.3em, shorten <= 0.3em] (4)--(a);
		\draw[shorten <= 0.3em] (a)--(2) node[pos=0.5, above=-0.2em]{};
		\draw[] (2)--(3) node[pos=0.5, left]{};
		\draw[] (0)--(3) node[pos=0.5, right]{};
		\draw[] (1)--(3);
		\draw[] (0)--(2);
		\draw[] (5)--(0);
		\draw[shorten >= 0.3em] (1)--(b);
		\draw[shorten <= 0.3em] (b)--(5);
		\draw[] (5)--(2);
		\draw[] (3)--(5);
		\node[below] at (0) {{$\scriptstyle{2}$}};
		\node[right] at (1) {{$\scriptstyle{1}$}};
		\node[left] at (2) {{$\scriptstyle{3}$}};
		\node[below=-0.1em] at (5) {{$\scriptstyle{0}$}};
		\node[above] at (3) {{$\scriptstyle{0'}$}};	
	}{}
	
	\end{tikzpicture}
}

\newcommand{\WWONE}[1]{
	\begin{tikzpicture}[scale=#1,baseline=0em, line width=0.8pt]
	\coordinate (0) at(0, 0);
	\coordinate (1) at (0, -5);
	\coordinate (2) at (-5 , 0);
	\coordinate (3) at (5, 0);
	\coordinate (4) at (0, 5);
	\coordinate (5) at (-7, 0);
	\coordinate (6) at (0, 7);
	\coordinate (7) at (7, 0);
	\coordinate (8) at (0, -7);

	\draw[] (2) -- (5) node[above, pos=0.6, yshift=-0.28em] {{$\sss{(015)}$}};
	\draw[] (4) -- (6) node[anchor=center, pos=1.25] {{$\sss{(013)}$}};
	\draw[] (3) -- (7) node[above, pos=0.6, yshift=-0.28em] {{$\sss{(035)}$}};
	\draw[] (2) -- (3) node[above, pos=0.3, yshift=-0.28em] {{$\sss{(045)}$}};
	\draw[] (4) -- (2) node[left, pos=0.2] {{$\sss{(014)}$}};
	\draw[] (4) -- (3) node[right, pos=0.2] {{$\sss{(034)}$}};
	\draw[dot] (1) -- (2);
	\draw[dot] (1) -- (3);
	\draw[] (1) -- (8) node[anchor=center, pos=1.25] {{$\sss{(135)}$}};
	\draw[shorten >= 0.3em] (1) -- (0)  node[right, pos=0.6, xshift=-0.2em] {{$\sss{(134)}$}};
	\draw[shorten <= 0.3em] (0) -- (4);
	\end{tikzpicture}
}

\newcommand{\WWTWO}[1]{
	\begin{tikzpicture}[scale=#1,baseline=0em, line width=0.8pt]
	\coordinate (0) at(0, 0);
	\coordinate (1) at (0, -5);
	\coordinate (2) at (-5 , 0);
	\coordinate (3) at (5, 0);
	\coordinate (4) at (0, 5);
	\coordinate (5) at (-7, 0);
	\coordinate (6) at (0, 7);
	\coordinate (7) at (7, 0);
	\coordinate (8) at (0, -7);
	\coordinate (9) at (-2, 3);
	\coordinate (10) at (2, 3);
	\coordinate (11) at (0, 1);
	\coordinate (12) at (0,3);
	
	\draw[] (2) -- (5) node[above, pos=0.6, yshift=-0.28em] {{$\sss{(015)}$}};
	\draw[] (4) -- (6) node[anchor=center, pos=1.25] {{$\sss{(013)}$}};
	\draw[] (3) -- (7) node[above, pos=0.6, yshift=-0.28em] {{$\sss{(035)}$}};
	\draw[] (2) -- (3) node[above, pos=0.3, yshift=-0.28em] {{$\sss{(045)}$}};
	\draw[] (4) -- (9) node[left, pos=0.2] {{$\sss{(012)}$}};
	\draw[dot] (4) -- (10);
	\draw[dot] (1) -- (2);
	\draw[dot] (1) -- (3);
	\draw[] (1) -- (8) node[anchor=center, pos=1.25] {{$\sss{(135)}$}};
	\draw[shorten >= 0.3em] (1) -- (0) node[right, pos=0.6, xshift=-0.2em] {{$\sss{(134)}$}};
	\draw[shorten <= 0.3em] (0) -- (11);
	\draw[dot, shorten >= 0.3em] (11) -- (12);
	\draw[dot, shorten <= 0.3em] (12) -- (4);
	\draw[] (11) -- (10) node[right, pos=0.2] {{$\sss{(234)}$}};
	\draw[] (10) -- (3) node[right, pos=0.2] {{$\sss{(034)}$}};
	\draw[dot] (11) -- (9);
	\draw[] (9) -- (2) node[left, pos=0.2] {{$\sss{(014)}$}};
	\draw[] (9) -- (10);
	\end{tikzpicture}
}

\newcommand{\WWTHREE}[1]{
	\begin{tikzpicture}[scale=#1,baseline=0em, line width=0.8pt]
	\coordinate (0) at(0, 0);
	\coordinate (1) at (0, -5);
	\coordinate (2) at (-5 , 0);
	\coordinate (3) at (5, 0);
	\coordinate (4) at (0, 5);
	\coordinate (5) at (-7, 0);
	\coordinate (6) at (0, 7);
	\coordinate (7) at (7, 0);
	\coordinate (8) at (0, -7);
	\coordinate (9) at (-2, 3);
	\coordinate (10) at (2, 3);
	\coordinate (11) at (-2, -3);
	\coordinate (12) at (2,-3);
	\coordinate (13) at (0,3);
	\coordinate (14) at (0,-3);
	\coordinate (15) at (-2,0);
	\coordinate (16) at (2,0);
	
	\draw[] (2) -- (5) node[above, pos=0.6, yshift=-0.28em] {{$\sss{(015)}$}};
	\draw[] (4) -- (6) node[anchor=center, pos=1.25] {{$\sss{(013)}$}};
	\draw[] (3) -- (7) node[above, pos=0.6, yshift=-0.28em] {{$\sss{(035)}$}};
	\draw[] (2) -- (3) node[below, pos=0.19, yshift=+0.28em] {{$\sss{(045)}$}};
	\draw[] (4) -- (9) node[left, pos=0.2] {{$\sss{(012)}$}};
	\draw[dot] (4) -- (10);
	\draw[dot] (1) -- (2);
	\draw[] (1) -- (12) node[right, pos=0.2] {{$\sss{(235)}$}};
	\draw[dot] (12) -- (3);
	\draw[] (1) -- (8) node[anchor=center, pos=1.25] {{$\sss{(135)}$}};
	\draw[dot, shorten >= 0.3em] (1) -- (14);
	\draw[dot, shorten <= 0.3em, shorten >= 0.3em] (14) -- (0);
	\draw[dot, shorten <= 0.3em, shorten >= 0.3em] (0) -- (13);
	\draw[dot, shorten <= 0.3em] (13) -- (4);
	\draw[dot] (11) -- (12);
	\draw[] (10) -- (3) node[right, pos=0.2] {{$\sss{(034)}$}};
	\draw[] (9) -- (2) node[left, pos=0.3] {{$\sss{(014)}$}};
	\draw[] (9) -- (10) node[below, pos=0.25, yshift=0.28em] {{$\sss{(024)}$}};
	\draw[dot, shorten >= 0.3em] (11) -- (15);
	\draw[dot, shorten <= 0.3em] (15) -- (9);
	\draw[shorten >= 0.3em] (12) -- (16) node[left, pos=0.5, xshift=0.38em] {{$\sss{(234)}$}};
	\draw[shorten <= 0.3em] (16) -- (10);
	\end{tikzpicture}
}

\newcommand{\WWFOUR}[1]{
	\begin{tikzpicture}[scale=#1,baseline=0em, line width=0.8pt]
	\coordinate (0) at(0, 0);
	\coordinate (1) at (0, -5);
	\coordinate (2) at (-5 , 0);
	\coordinate (3) at (5, 0);
	\coordinate (4) at (0, 5);
	\coordinate (5) at (-7, 0);
	\coordinate (6) at (0, 7);
	\coordinate (7) at (7, 0);
	\coordinate (8) at (0, -7);
	\coordinate (9) at (-2, 3);
	\coordinate (11) at (-2, -3);
	\coordinate (15) at (-2,0);
	\coordinate (16) at (2,0);
	\coordinate (13) at (intersection of 9--16 and 0--4);
	\coordinate (14) at (intersection of 11--16 and 0--1);
	
	\draw[] (2) -- (5) node[above, pos=0.6, yshift=-0.28em] {{$\sss{(015)}$}};
	\draw[] (4) -- (6) node[anchor=center, pos=1.25] {{$\sss{(013)}$}};
	\draw[] (3) -- (7) node[above, pos=0.6, yshift=-0.28em] {{$\sss{(035)}$}};
	\draw[] (2) -- (16) node[below, pos=0.27, yshift=+0.28em] {{$\sss{(045)}$}};
	\draw[] (16) -- (3) node[below, pos=0.35, yshift=0.28em] {{$\sss{(025)}$}};
	\draw[] (4) -- (9) node[left, pos=0.2] {{$\sss{(012)}$}};
	\draw[dot] (4) -- (3);
	\draw[dot] (1) -- (2);
	\draw[] (1) -- (3) node[right, pos=0.4] {{$\sss{(235)}$}};
	\draw[] (1) -- (8) node[anchor=center, pos=1.25] {{$\sss{(135)}$}};
	\draw[dot, shorten >= 0.3em] (1) -- (14);
	\draw[dot, shorten <= 0.3em, shorten >= 0.3em] (14) -- (0);
	\draw[dot, shorten <= 0.3em, shorten >= 0.3em] (0) -- (13);
	\draw[dot, shorten <= 0.3em] (13) -- (4);
	\draw[dot] (11) -- (16);
	\draw[] (9) -- (2)  node[left, pos=0.2] {{$\sss{(014)}$}};
	\draw[] (9) -- (16) node[right, pos=0.5] {{$\sss{(024)}$}};
	\draw[dot, shorten >= 0.3em] (11) -- (15);
	\draw[dot, shorten <= 0.3em] (15) -- (9);
	\end{tikzpicture}
}

\newcommand{\WWFIVE}[1]{
	\begin{tikzpicture}[scale=#1,baseline=0em, line width=0.8pt]
	\coordinate (0) at(0, 0);
	\coordinate (1) at (0, -5);
	\coordinate (2) at (-5 , 0);
	\coordinate (3) at (5, 0);
	\coordinate (4) at (0, 5);
	\coordinate (5) at (-7, 0);
	\coordinate (6) at (0, 7);
	\coordinate (7) at (7, 0);
	\coordinate (8) at (0, -7);
	
	\draw[] (2) -- (5) node[above, pos=0.6, yshift=-0.28em] {{$\sss{(015)}$}};
	\draw[] (4) -- (6) node[anchor=center, pos=1.25] {{$\sss{(013)}$}};
	\draw[] (3) -- (7) node[above, pos=0.6, yshift=-0.28em] {{$\sss{(035)}$}};
	\draw[] (2) -- (3) node[above, pos=0.3, yshift=-0.28em] {{$\sss{(025)}$}};
	\draw[] (4) -- (2) node[left, pos=0.5] {{$\sss{(012)}$}};
	\draw[dot] (4) -- (3);
	\draw[dot] (1) -- (2);
	\draw[] (1) -- (3) node[right, pos=0.4] {{$\sss{(235)}$}};
	\draw[] (1) -- (8) node[anchor=center, pos=1.25] {{$\sss{(135)}$}};
	\draw[dot, shorten >= 0.3em] (1) -- (0);
	\draw[dot, shorten <= 0.3em] (0) -- (4);
	\end{tikzpicture}
}

\newcommand{\WWSIX}[1]{
	\begin{tikzpicture}[scale=#1,baseline=0em, line width=0.8pt]
	\coordinate (5) at (-3, 0);
	\coordinate (6) at (0, 3);
	\coordinate (7) at (3, 0);
	\coordinate (8) at (0, -3);
	
	\draw[] (5) -- (7) node[above, pos=0.14, yshift=-0.28em] {{$\sss{(015)}$}} node[above, pos=0.88, yshift=-0.28em] {{$\sss{(035)}$}};
	\draw[] (6) -- (8) node[anchor=center, pos=-0.08] {{$\sss{(013)}$}} node[anchor=center, pos=1.09] {{$\sss{(135)}$}};
	\end{tikzpicture}
}

\newcommand{\WWCocycle}[0]{
	\begin{tikzpicture}[scale=1,baseline=0em]
	\matrix[matrix of math nodes,row sep =-7em,column sep=2.5em, ampersand replacement=\&] (m) {
	{}  \&[-4.6em]  \WWONE{0.3} \&  \WWTWO{0.3} \&[-6em] {}
	\\
	\WWSIX{0.3} \& {} \& {} \& \WWTHREE{0.3}
	\\
	{} \&  \WWFIVE{0.3}  \&\WWFOUR{0.3} \& {} \\
};		
	\path
	(m-2-1) edge[->, shorten <= -2em, shorten >= 1.5em] node[pos=0.1, right, xshift=0.2em] {$\sss{(01345)}$} (m-1-2)
	edge[->, shorten <= -2em, shorten >= 1.5em] node[pos=0.1, right, xshift=0.2em] {$\sss{(01235)}$} node[pos=0.1, below, xshift=-0.3em] {$\mathfrak{s}^{-1}$} (m-3-2)
	(m-1-2) edge[->, shorten <= 0em, shorten >= 0em] node[pos=0.5, above] {$\mathfrak{s}$} node[pos=0.5, below] {$\sss{(01234)}$} (m-1-3)
	(m-3-2) edge[<-, shorten <= 0em, shorten >= 0em] node[pos=0.5, above] {$\alpha$} node[pos=0.5, below] {$\sss{(01245)}$} (m-3-3)
	(m-2-4) edge[->, shorten <= 0.5em, shorten >= 0.5em] node[pos=0.75, left, xshift=-0.2em] {$\sss{(12345)}$} node[pos=0.6, above, xshift=0.6em] {$\rm{id}^{-1}$} (m-1-3)
	edge[<-, shorten <= 0.5em, shorten >= 0.5em] node[pos=0.75, left, xshift=-0.2em] {$\sss{(02345)}$} node[pos=0.57, below, xshift=0.2em] {$\mathfrak{c}$} (m-3-3)
	;
	\end{tikzpicture}
}

\newcommand{\WWHamONE}[2]{
	\begin{tikzpicture}[scale=#1,baseline=-2.5em, line width=0.8pt]
	\coordinate (0) at (0, 0);
	\coordinate (1) at (0, 2);
	\coordinate (2) at (0, -2);
	\coordinate (3) at (5,-5);
	\coordinate (4) at (7, -3);
	\coordinate (5) at (2.6, -4.2);
	\coordinate (6) at (-5,-5);
	\coordinate (7) at (-7, -3);
	\coordinate (8) at (-2.6, -4.2);

	\ifthenelse{#2 = 1}{
		\draw[] (0) -- (1) node[anchor=center, pos=1.25] {{$\sss{a}$}};
		\draw[] (0) -- (2) node[anchor=center, pos=1.25] {{$\sss{e+f}$}};
		\draw[] (0) -- (3) node[right,pos=0.4] {{$\sss{b+c+e+f}$}};
		\draw[] (3) -- (4) node[anchor=center, pos=1.25] {{$\sss{e+b}$}};
		\draw[] (3) -- (5) node[anchor=center, pos=1.25] {{$\sss{h}$}};	
		\draw[] (0) -- (6) node[left, pos=0.4] {{$\sss{a+b+c}$}};
		\draw[] (6) -- (7) node[anchor=center, pos=1.25] {{$\sss{a+b}$}};
		\draw[] (6) -- (8) node[anchor=center, pos=1.25, xshift=0.4em] {{$\sss{f+h}$}};		
		\draw[] (3) -- (6) node[below, pos=0.5, yshift=0.28em] {{$\sss{c+f+h}$}};
	}{}
	\ifthenelse{#2 = 2}{
		\draw[] (0) -- (1) node[anchor=center, pos=1.25] {{$\sss{(012)}$}};
		\draw[] (0) -- (2) node[anchor=center, pos=1.25] {{$\sss{(124)}$}};
		\draw[] (0) -- (3) node[right,pos=0.4] {{$\sss{(024)}$}};
		\draw[] (3) -- (4) node[anchor=center, pos=1.25] {{$\sss{(023)}$}};
		\draw[] (3) -- (5) node[anchor=center, pos=1.25, xshift=-0.35em] {{$\sss{(234)}$}};	
		\draw[] (0) -- (6) node[left, pos=0.4] {{$\sss{(014)}$}};
		\draw[] (6) -- (7) node[anchor=center, pos=1.25] {{$\sss{(013)}$}};
		\draw[] (6) -- (8) node[anchor=center, pos=1.25, xshift=0.35em] {{$\sss{(134)}$}};		
		\draw[] (3) -- (6) node[below, pos=0.5, yshift=0.28em] {{$\sss{(034)}$}};	
	}{}
	\end{tikzpicture}
}

\newcommand{\WWHamTWO}[1]{
	\begin{tikzpicture}[scale=#1,baseline=-2.5em, line width=0.8pt]
	\coordinate (1) at (-1, 1);
	\coordinate (2) at (1, -3);
	\coordinate (4) at (6, -2);
	\coordinate (5) at (0.6, -4.2);
	\coordinate (7) at (-6, -2);
	\coordinate (8) at (-0.6, -4.2);	
	\coordinate (20) at (-1, -1);
	\coordinate (21) at (1, -1);
	\coordinate (22) at (4, -4);
	\coordinate (23) at (3, -5);	
	\coordinate (24) at (-4, -4);
	\coordinate (25) at (-3, -5);
	
	\draw[] (20) -- (1) node[anchor=center, pos=1.25] {{$\sss{a}$}};
	\draw[] (21) -- (2);
	\draw[] (20) -- (24);
	\draw[] (21) -- (22);
	\draw[] (25) -- (23);
	\draw[] (20) -- (21) node[above, pos=0.5, yshift=-0.28em] {{$\sss{b+c}$}};
	\draw[] (22) -- (23) node[right, pos=0.7] {{$\sss{c+f}$}};
	\draw[] (24) -- (25) node[left, pos=0.7] {{$\sss{c}$}};			
	\draw[] (22) -- (4) node[anchor=center, pos=1.25] {{$\sss{e+b}$}};
	\draw[] (23) -- (5) node[above, pos=0.8, yshift=-0.08em] {{$\sss{h}$}};	
	\draw[] (24) -- (7) node[anchor=center, pos=1.25] {{$\sss{a+b}$}};
	\draw[] (25) -- (8) node[above, pos=0.8, yshift=-0.2em] {{$\sss{f+h}$}};			
	\end{tikzpicture}
}

\newcommand{\WWHamTWObis}[1]{
	\begin{tikzpicture}[scale=#1,baseline=-2.5em, line width=0.8pt]
	\coordinate (1) at (-1, 1);
	\coordinate (2) at (1, -3);
	\coordinate (4) at (6, -2);
	\coordinate (5) at (0.6, -4.2);
	\coordinate (7) at (-6, -2);
	\coordinate (8) at (-0.6, -4.2);	
	\coordinate (14) at (-1,-5);
	\coordinate (20) at (-1, -1);
	\coordinate (21) at (1, -1);
	\coordinate (22) at (4, -4);
	\coordinate (23) at (3, -5);	
	\coordinate (24) at (-4, -4);
	\coordinate (25) at (-3, -5);
	
	\coordinate (a) at (0,-1.1);
	\coordinate (b) at (-3.2,-4.5);
	\coordinate (c) at (3.2,-4.5);
	
	\draw[] (20) -- (1) node[anchor=center, pos=1.25] {{$\sss{a}$}};
	\draw[shorten >= 0.3em] (21) -- (intersection of 21--2 and a--c);
	\draw[shorten <= 0.3em] (intersection of 21--2 and a--c) -- (2);
	\draw[] (20) -- (24);
	\draw[] (21) -- (22);
	\draw[] (25) -- (23);
	\draw[] (20) -- (21) node[above, pos=0.5, yshift=-0.28em] {{$\sss{b+c}$}};
	\draw[] (22) -- (23) node[right, pos=0.7] {{$\sss{c+f}$}};
	\draw[] (24) -- (25) node[left, pos=0.7] {{$\sss{c}$}};			
	\draw[] (22) -- (4) node[anchor=center, pos=1.25] {{$\sss{e+b}$}};
	\path (23) -- (5) node[above, pos=0.8, yshift=-0.08em] {{$\sss{h}$}};
	\draw[shorten >= 0.4em] (23) -- (intersection of 23--5 and b--c);
	\draw[shorten <= 0.4em] (intersection of 23--5 and b--c) -- (5);	
	\draw[] (24) -- (7) node[anchor=center, pos=1.25] {{$\sss{a+b}$}};
	\path (25) -- (8) node[above, pos=0.8, yshift=-0.2em] {{$\sss{f+h}$}};
	\draw[shorten >= 0.4em] (25) -- (intersection of 25--8 and b--c);
	\draw[shorten <= 0.4em] (intersection of 25--8 and b--c) -- (8);
	
	\draw[rounded corners=0.6em] (a) -- (b) -- (c) -- cycle;
	\draw[] node[above, yshift=-1.5em, xshift=-0.3em] at (a) {{$\sss{d}$}}; 	
	\end{tikzpicture}
}

\newcommand{\WWHamTHREE}[1]{
	\begin{tikzpicture}[scale=#1,baseline=-2.5em, line width=0.8pt]
	\coordinate (1) at (-1, 1);
	\coordinate (2) at (1, -3);
	\coordinate (4) at (6, -2);
	\coordinate (5) at (0.6, -4.2);
	\coordinate (7) at (-6, -2);
	\coordinate (8) at (-0.6, -4.2);
	\coordinate (9) at (2,-2);
	\coordinate (10) at (3,-3);	
	\coordinate (11) at (1,-5);
	\coordinate (12) at (-2,-2);
	\coordinate (13) at (-3,-3);	
	\coordinate (14) at (-1,-5);
	\coordinate (20) at (-1, -1);
	\coordinate (21) at (1, -1);
	\coordinate (22) at (4, -4);
	\coordinate (23) at (3, -5);	
	\coordinate (24) at (-4, -4);
	\coordinate (25) at (-3, -5);
		
	\draw[] (20) -- (1) node[anchor=center, pos=1.25] {{$\sss{a}$}};
	\draw[shorten >= 0.3em] (21) -- (intersection of 9--12 and 21--2);
	\draw[shorten <= 0.3em] (intersection of 9--12 and 21--2) -- (2);
	\draw[] (20) -- (24);
	\draw[] (21) -- (22);
	\draw[] (25) -- (23);
	\draw[] (20) -- (21) node[above, pos=0.5, yshift=-0.28em] {{$\sss{b+c}$}};
	\draw[] (22) -- (23) node[right, pos=0.7] {{$\sss{c+f}$}};
	\draw[] (24) -- (25) node[left, pos=0.7] {{$\sss{c}$}};			
	\draw[] (22) -- (4) node[anchor=center, pos=1.25] {{$\sss{e+b}$}};
	\path (23) -- (5) node[above, pos=0.8, yshift=-0.08em] {{$\sss{h}$}};
	\draw[shorten >= 0.3em] (23) -- (intersection of 23--5 and 10--11);
	\draw[shorten <= 0.3em] (intersection of 23--5 and 10--11) -- (5);	
	\draw[] (24) -- (7) node[anchor=center, pos=1.25] {{$\sss{a+b}$}};
	\path (25) -- (8) node[above, pos=0.8, yshift=-0.2em] {{$\sss{f+h}$}};
	\draw[shorten >= 0.3em] (25) -- (intersection of 25--8 and 13--14);
	\draw[shorten <= 0.3em] (intersection of 25--8 and 13--14) -- (8);		
	\draw[] (9) -- (12) node[below, pos=0.5, yshift=0.28em] {{$\sss{d}$}};
	\draw[] (10) -- (11);
	\draw[] (13) -- (14);	
	\end{tikzpicture}
}

\newcommand{\WWHamFOUR}[1]{
	\begin{tikzpicture}[scale=#1,baseline=-2.5em, line width=0.8pt]
	\coordinate (1) at (-1, 1);
	\coordinate (2) at (0, -3);
	\coordinate (4) at (6, -2);
	\coordinate (5) at (1.3, -3.5);
	\coordinate (7) at (-6, -2);
	\coordinate (8) at (-1.3, -3.5);
	\coordinate (9) at (2,-2);
	\coordinate (10) at (3,-3);	
	\coordinate (11) at (1,-5);
	\coordinate (12) at (-2,-2);
	\coordinate (13) at (-3,-3);	
	\coordinate (14) at (-1,-5);
	\coordinate (20) at (-1, -1);
	\coordinate (21) at (1, -1);
	\coordinate (22) at (4, -4);
	\coordinate (23) at (3, -5);	
	\coordinate (24) at (-4, -4);
	\coordinate (25) at (-3, -5);
	\coordinate (26) at (1, 0);
	\coordinate (27) at (0, 0);
	\coordinate (28) at (-4.2, -5.4);
	\coordinate (29) at (4.2, -5.4);
	\coordinate (30) at (-4.3, -4.5);
	\coordinate (31) at (4.3, -4.5);

	\draw[] (20) -- (1) node[anchor=center, pos=1.25] {{$\sss{a}$}};
	\draw[] (20) -- (24);
	\draw[] (21) -- (22);
	\draw[] (25) -- (23) node[below, pos=0.5] {$\sss{f+h+c+d}$};
	\draw[] (20) -- (21);
	\draw[] (22) -- (23);
	\draw[] (24) -- (25);			
	\draw[] (22) -- (4) node[anchor=center, pos=1.25] {{$\sss{e+b}$}};
	\draw[] (24) -- (7) node[anchor=center, pos=1.25] {{$\sss{a+b}$}};
	\draw[] (9) -- (12) node[below, pos=0.75, yshift=0.28em] {{$\sss{d}$}};
	\draw[] (10) -- (11) node[left, pos=0.6] {$\sss{d}$};
	\draw[] (13) -- (14) node[right, pos=0.6] {$\sss{d}$};

	\draw[] (23) -- (29);	
	\draw[] (29) arc(-120:60:0.45) -- (31) node[right, pos=1, xshift=0.2em] {$\sss{h}$};
	\draw[shorten >= 0.3em] (31) -- (intersection of 31--5 and 22--23);
	\draw[shorten >= 0.3em, shorten <= 0.3em] (intersection of 31--5 and 22--23) -- (intersection of 31--5 and 10--11);
	\draw[shorten <= 0.3em] (intersection of 31--5 and 10--11) -- (5);

	\draw[] (25) -- (28);	
	\draw[] (28) arc(300:120:0.45) -- (30) node[left, pos=1, xshift=-0.2em] {$\sss{f+h}$};
	\draw[shorten >= 0.3em] (30) -- (intersection of 30--8 and 24--25);
	\draw[shorten >= 0.3em, shorten <= 0.3em] (intersection of 30--8 and 24--25) -- (intersection of 30--8 and 13--14);
	\draw[shorten <= 0.3em] (intersection of 30--8 and 13--14) -- (8);

	\draw[] (21) -- (26) node[right, pos=1] {{$\sss{e+f}$}};
	\draw[] (26) arc(0:180:0.5) -- (27);
	\draw[shorten >= 0.3em] (27) -- (intersection of 20--21 and 27--2);
	\draw[shorten <= 0.3em, shorten >= 0.3em] (intersection of 20--21 and 27--2) -- (intersection of 9--12 and 27--2);
	\draw[shorten <= 0.3em] (intersection of 9--12 and 27--2) -- (2);	
	\end{tikzpicture}
}

\newcommand{\WWHamFIVE}[1]{
	\begin{tikzpicture}[scale=#1,baseline=-2.5em, line width=0.8pt]
	\coordinate (1) at (-1, 1);
	\coordinate (2) at (0, -3);
	\coordinate (4) at (6, -2);
	\coordinate (5) at (1.3, -3.5);
	\coordinate (7) at (-6, -2);
	\coordinate (8) at (-1.3, -3.5);
	\coordinate (9) at (2,-2);
	\coordinate (10) at (3,-3);	
	\coordinate (11) at (1,-5);
	\coordinate (12) at (-2,-2);
	\coordinate (13) at (-3,-3);	
	\coordinate (14) at (-1,-5);
	\coordinate (20) at (-1, -1);
	\coordinate (21) at (1, -1);
	\coordinate (22) at (4, -4);
	\coordinate (23) at (3, -5);	
	\coordinate (24) at (-4, -4);
	\coordinate (25) at (-3, -5);
	\coordinate (26) at (1, 0);
	\coordinate (27) at (0, 0);
	\coordinate (28) at (-4.2, -5.4);
	\coordinate (29) at (4.2, -5.4);
	\coordinate (30) at (-4.3, -4.5);
	\coordinate (31) at (4.3, -4.5);

	\draw[] (20) -- (1) node[anchor=center, pos=1.25] {{$\sss{a}$}};
	\draw[] (20) -- (24);
	\draw[] (21) -- (22);
	\draw[] (25) -- (23) node[below, pos=0.5] {$\sss{f+h+c+d}$};
	\draw[] (20) -- (21);
	\draw[] (22) -- (23);
	\draw[] (24) -- (25);			
	\draw[] (22) -- (4) node[anchor=center, pos=1.25] {{$\sss{e+b}$}};
	\draw[] (24) -- (7) node[anchor=center, pos=1.25] {{$\sss{a+b}$}};

	\draw[] (23) -- (29);	
	\draw[] (29) arc(-120:60:0.45) -- (31) node[right, pos=1, xshift=0.2em] {$\sss{h}$};
	\draw[shorten >= 0.3em] (31) -- (intersection of 31--5 and 22--23);
	\draw[shorten <= 0.3em] (intersection of 31--5 and 22--23) -- (5);

	\draw[] (25) -- (28);	
	\draw[] (28) arc(300:120:0.45) -- (30) node[left, pos=1, xshift=-0.2em] {$\sss{f+h}$};
	\draw[shorten >= 0.3em] (30) -- (intersection of 30--8 and 24--25);
	\draw[shorten <= 0.3em] (intersection of 30--8 and 24--25) -- (8);

	\draw[] (21) -- (26) node[right, pos=1] {{$\sss{e+f}$}};
	\draw[] (26) arc(0:180:0.5) -- (27);
	\draw[shorten >= 0.3em] (27) -- (intersection of 20--21 and 27--2);
	\draw[shorten <= 0.3em] (intersection of 20--21 and 27--2) -- (2);				
	\end{tikzpicture}
}

\newcommand{\WWHamSIX}[1]{
	\begin{tikzpicture}[scale=#1,baseline=-2.5em, line width=0.8pt]
	\coordinate (1) at (-1, 1);
	\coordinate (2) at (1, -3);
	\coordinate (4) at (6, -2);
	\coordinate (5) at (0.6, -4.2);
	\coordinate (7) at (-6, -2);
	\coordinate (8) at (-0.6, -4.2);	
	\coordinate (20) at (-1, -1);
	\coordinate (21) at (1, -1);
	\coordinate (22) at (4, -4);
	\coordinate (23) at (3, -5);	
	\coordinate (24) at (-4, -4);
	\coordinate (25) at (-3, -5);
	
	\draw[] (20) -- (1) node[anchor=center, pos=1.25] {{$\sss{a}$}};
	\draw[] (21) -- (2);
	\draw[] (20) -- (24);
	\draw[] (21) -- (22);
	\draw[] (25) -- (23);
	\draw[] (20) -- (21) node[above, pos=0.8, yshift=-0.28em] {{$\sss{b+c+d}$}};
	\draw[] (22) -- (23) node[right, pos=0.7] {{$\sss{c+f+d}$}};
	\draw[] (24) -- (25) node[left, pos=0.7] {{$\sss{c+d}$}};			
	\draw[] (22) -- (4) node[anchor=center, pos=1.25] {{$\sss{e+b}$}};
	\draw[] (23) -- (5) node[above, pos=0.8, yshift=-0.08em] {{$\sss{h}$}};	
	\draw[] (24) -- (7) node[anchor=center, pos=1.25] {{$\sss{a+b}$}};
	\draw[] (25) -- (8) node[above, pos=0.8, yshift=-0.2em] {{$\sss{f+h}$}};			
	\end{tikzpicture}
}

\newcommand{\HamONE}[2]{
	\begin{tikzpicture}[scale=#1,baseline=-0.5em, line width=0.8pt]
		\coordinate (0) at (0,1);
		\coordinate (1) at (1,-0.4);
		\coordinate (2) at (0.55,-1);
		\coordinate (3) at (-1.2,-0.55);
		\coordinate (4) at (0,-0.1);	
	
		\draw[] (0) -- (1);
		\draw[] (1) -- (2);
		\draw[] (2) -- (3);
		\draw[] (0) -- (3);
		\draw[] (0) -- (2);
		\draw[] (3) -- (4);
		\draw[] (0) -- (4);
		\draw[] (2) -- (4);
		\draw[shorten >= 0.3em] (1) -- (intersection of 1--4 and 0--2);
		\draw[shorten <= 0.3em] (intersection of 1--4 and 0--2) -- (4);
		\draw[shorten >= 0.3em] (1) -- (intersection of 1--3 and 0--2);
		\draw[shorten <= 0.3em, shorten >=0.3em] (intersection of 1--3 and 0--2) -- (intersection of 1--3 and 2--4);
		\draw[shorten <= 0.3em] (intersection of 1--3 and 2--4) -- (3);

	\ifthenelse{#2 = 1}{	
		\node[above] at (0) {{$\scriptstyle{0}$}};	
		\node[right] at (1) {{$\scriptstyle{2}$}};
		\node[below] at (2) {{$\scriptstyle{3}$}};
		\node[left] at (3) {{$\scriptstyle{1}$}};	
		\node[left, yshift=1.5] at (4) {{$\scriptstyle{4}$}};}{}
	
	\ifthenelse{#2 = 2}{	
		\node[above] at (0) {{$\scriptstyle{0}$}};	
		\node[right] at (1) {{$\scriptstyle{2}$}};
		\node[below] at (2) {{$\scriptstyle{3}$}};
		\node[left] at (3) {{$\scriptstyle{1}$}};	
		\node[left, yshift=1.5] at (4) {{$\scriptstyle{5}$}};}{}
	\end{tikzpicture}
}

\newcommand{\HamTWO}[2]{
	\begin{tikzpicture}[scale=#1,baseline=-0.5em, line width=0.8pt]
		\coordinate (0) at (0,1);
		\coordinate (1) at (1,-0.4);
		\coordinate (2) at (0.55,-1);
		\coordinate (3) at (-1.2,-0.55);
		\coordinate (4) at (0,-0.1);
		\coordinate (5) at (-1.5,1.1);	
		
		\draw[] (0) -- (1);
		\draw[] (1) -- (2);
		\draw[] (2) -- (3);
		\draw[shorten >= 0.3em] (0) -- (intersection of 0--3 and 2--5);
		\draw[shorten <= 0.3em] (intersection of 0--3 and 2--5) -- (3);
		\draw[] (0) -- (2);
		\draw[shorten >= 0.3em] (3) -- (intersection of 3--4 and 2--5);
		\draw[shorten <= 0.3em] (intersection of 3--4 and 2--5) -- (4);
		\draw[] (0) -- (4);
		\draw[] (2) -- (4);
		\draw[] (0) -- (5);
		\draw[] (3) -- (5);
		\draw[shorten >= 0.3em] (1) -- (intersection of 1--4 and 0--2);
		\draw[shorten <= 0.3em] (intersection of 1--4 and 0--2) -- (4);
		\draw[shorten >= 0.3em] (1) -- (intersection of 1--3 and 0--2);
		\draw[shorten <= 0.3em, shorten >=0.3em] (intersection of 1--3 and 0--2) -- (intersection of 1--3 and 2--4);
		\draw[shorten <= 0.3em, shorten >= 0.3em] (intersection of 1--3 and 2--4) -- (intersection of 1--3 and 2--5);
		\draw[shorten <= 0.3em] (intersection of 1--3 and 2--5) -- (3);
		\draw[shorten >= 0.3em] (4) -- (intersection of 0--3 and 4--5);
		\draw[shorten <= 0.3em] (intersection of 0--3 and 4--5) -- (5);
		\draw[] (2) -- (5);
		\draw[shorten >= 0.3em] (1) -- (intersection of 1--5 and 0--2);
		\draw[shorten >= 0.3em, shorten <= 0.3em] (intersection of 1--5 and 0--2) -- (intersection of 1--5 and 0--4);
		\draw[shorten >= 0.3em, shorten <= 0.3em] (intersection of 1--5 and 0--4) -- (intersection of 1--5 and 0--3);
		\draw[shorten <= 0.3em] (intersection of 1--5 and 0--3) -- (5);

	\ifthenelse{#2 = 1}{	
		\node[above] at (0) {{$\scriptstyle{0}$}};	
		\node[right] at (1) {{$\scriptstyle{2}$}};
		\node[below] at (2) {{$\scriptstyle{3}$}};
		\node[left] at (3) {{$\scriptstyle{1}$}};
		\node[above] at (5) {{$\scriptstyle{5}$}};		
		\node[left, yshift=1.4] at (4) {{$\scriptstyle{4}$}};}{}
	\end{tikzpicture}
}

\newcommand{\Rmove}[2]{
	\begin{tikzpicture}[scale=#1,baseline=1.5em, line width=0.8pt]
	\coordinate (0) at(0, 0);
	\coordinate (1) at (-2, 4);
	\coordinate (2) at (0 , 4);
	\coordinate (3) at (2, 4);
	\coordinate (4) at (0, 2);
	
	\ifthenelse{#2 = 1}{
		\draw[] (0) -- (4);	
		\draw[] (4) -- (1) node[anchor=center, pos=1.25] {{$\sss{b}$}};
		\draw[] (4) -- (3) node[anchor=center, pos=1.25] {{$\sss{a}$}};
	}{}

	\coordinate (5) at (0, 3);
	\ifthenelse{#2 = 2}{
		\draw[] (0) -- (4);	
		\draw[shorten >= 0.3em] (4) edge [out=10, in=-30, looseness=1.5] (5);
		\draw[shorten <= 0.3em] (5) -- (1) node[anchor=center, pos=1.25, yshift=0.1em] {{$\sss{b}$}};
		\draw[] (4) edge [out=170, in=210, looseness=1.5] (5);
		\draw[] (5) -- (3) node[anchor=center, pos=1.25, yshift=0.1em] {{$\sss{a}$}};
	}{}
	\ifthenelse{#2 = 3}{
		\draw[] (0) -- (4);	
		\draw[] (4) edge [out=10, in=-30, looseness=1.5] (5);
		\draw[] (5) -- (1) node[anchor=center, pos=1.25, yshift=0.1em] {{$\sss{b}$}};
		\draw[shorten >= 0.3em] (4) edge [out=170, in=210, looseness=1.5] (5);
		\draw[shorten <= 0.3em] (5) -- (3) node[anchor=center, pos=1.25, yshift=0.1em] {{$\sss{a}$}};
	}{}
	\end{tikzpicture}
}

\newcommand{\hexagonMoveONE}[2]{
	\begin{tikzpicture}[scale=#1,baseline=1.5em, line width=0.8pt]
	\coordinate (0) at(0, 0);
	\coordinate (1) at (0, 2);
	\coordinate (2) at (-2, 4);
	\coordinate (3) at (-4, 6);
	\coordinate (4) at (0, 6);
	\coordinate (5) at (4, 6);
	
	\ifthenelse{#2 = 1}{
		\draw[] (0) -- (1);
		\draw[] (1) -- (2);
		\draw[] (2) -- (3) node[anchor=center, pos=1.25] {{$\sss{c}$}};
		\draw[] (2) -- (4) node[anchor=center, pos=1.25] {{$\sss{d}$}};			
		\draw[] (1) -- (2, 4);
		\draw[] (2, 4) -- (5) node[anchor=center, pos=1.25] {{$\sss{e}$}};
	}{}	

	\ifthenelse{#2 = 2}{
		\draw[] (0) -- (1);
		\draw[] (1) -- (2);
		\draw[] (2) -- (3) node[anchor=center, pos=1.25] {{$\sss{a}$}};
		\draw[] (2) -- (4) node[anchor=center, pos=1.25] {{$\sss{b}$}};			
		\draw[] (1) -- (2, 4);
		\draw[] (2, 4) -- (5) node[anchor=center, pos=1.25] {{$\sss{c}$}};
	}{}	

	\ifthenelse{#2 = 3}{
		\draw[] (0) -- (1);
		\draw[] (1) -- (2);
		\draw[] (2) -- (3) node[anchor=center, pos=1.25] {{$\sss{c}$}};
		\draw[] (2) -- (4) node[anchor=center, pos=1.25] {{$\sss{g}$}};			
		\draw[] (1) -- (2, 4);
		\draw[] (2, 4) -- (5) node[anchor=center, pos=1.25] {{$\sss{h}$}};
	}{}	
	\end{tikzpicture}
}

\newcommand{\hexagonMoveTWO}[2]{
	\begin{tikzpicture}[scale=#1,baseline=1.5em, line width=0.8pt]
	\coordinate (0) at(0, 0);
	\coordinate (1) at (0, 2);
	\coordinate (2) at (-2, 4);
	\coordinate (3) at (-4, 6);
	\coordinate (4) at (0, 6);
	\coordinate (5) at (4, 6);
	\coordinate (6) at (0, 4);
	\coordinate (7) at (1, 2.5);
	\coordinate (8) at (3, 3.5);
	\coordinate (9) at (0, 4);
	\coordinate (10) at (2, 5);
	
	\coordinate (11) at (intersection of 10--4 and 2--5);
	
	\draw[] (0) -- (1);
	\draw[] (1) -- (8);
	\draw[] (8) edge [out=35, in=-30, looseness=1.5] (10);
	\draw[shorten >= 0.3em] (10) -- (11);
	\draw[shorten <= 0.3em] (11) -- (4);
	\draw[] (1) -- (3);
	\draw[] (2)  -- (11);
	\draw[] (11) -- (5);
	
	\ifthenelse{#2 = 1}{
		\path (2, 4) -- (5) node[anchor=center, pos=1.25] {{$\sss{e}$}};
		\path (2) -- (3) node[anchor=center, pos=1.25] {{$\sss{c}$}};
		\path (2, 4) -- (4) node[anchor=center, pos=1.25] {{$\sss{d}$}};
	}{}

	\ifthenelse{#2 = 2}{
		\path (2, 4) -- (5) node[anchor=center, pos=1.25] {{$\sss{c}$}};
		\path (2) -- (3) node[anchor=center, pos=1.25] {{$\sss{a}$}};
		\path (2, 4) -- (4) node[anchor=center, pos=1.25] {{$\sss{b}$}};
	}{}

	\ifthenelse{#2 = 3}{
		\path (2, 4) -- (5) node[anchor=center, pos=1.25] {{$\sss{h}$}};
		\path (2) -- (3) node[anchor=center, pos=1.25] {{$\sss{c}$}};
		\path (2, 4) -- (4) node[anchor=center, pos=1.25] {{$\sss{g}$}};
	}{}
	\end{tikzpicture}
}

\newcommand{\hexagonMoveTHREE}[2]{
	\begin{tikzpicture}[scale=#1,baseline=1.5em, line width=0.8pt]
	\coordinate (0) at(0, 0);
	\coordinate (1) at (0, 2);
	\coordinate (2) at (-2, 4);
	\coordinate (3) at (-4, 6);
	\coordinate (4) at (0, 6);
	\coordinate (5) at (4, 6);
	\coordinate (6) at (0, 3);
	
	\ifthenelse{#2 = 1}{
		\draw[] (0) -- (1);
		\draw[] (2) -- (3) node[anchor=center, pos=1.25] {{$\sss{c}$}};
		\draw[] (2) -- (4) node[anchor=center, pos=1.25] {{$\sss{d}$}};				
		
		\draw[shorten >= 0.3em] (1) edge [out=10, in=-30, looseness=1.5] (6);
		\draw[shorten <= 0.3em] (6) -- (2);
		\draw[] (1) edge [out=170, in=210, looseness=1.5] (6);
		\draw[] (6) -- (5);
		
		\path (2, 4) -- (5) node[anchor=center, pos=1.25] {{$\sss{e}$}};
	}{}

	\ifthenelse{#2 = 2}{
		\draw[] (0) -- (1);
		\draw[] (2) -- (3) node[anchor=center, pos=1.25] {{$\sss{a}$}};
		\draw[] (2) -- (4) node[anchor=center, pos=1.25] {{$\sss{b}$}};				
		
		\draw[shorten >= 0.3em] (1) edge [out=10, in=-30, looseness=1.5] (6);
		\draw[shorten <= 0.3em] (6) -- (2);
		\draw[] (1) edge [out=170, in=210, looseness=1.5] (6);
		\draw[] (6) -- (5);
		
		\path (2, 4) -- (5) node[anchor=center, pos=1.25] {{$\sss{c}$}};
	}{}
	\end{tikzpicture}
}

\newcommand{\hexagonMoveFOUR}[2]{
	\begin{tikzpicture}[scale=#1,baseline=1.5em, line width=0.8pt]
	\coordinate (0) at(0, 0);
	\coordinate (1) at (0, 2);
	\coordinate (2) at (-2, 4);
	\coordinate (3) at (-4, 6);
	\coordinate (4) at (0, 6);
	\coordinate (5) at (4, 6);
	\coordinate (6) at (0, 4);
	\coordinate (7) at (-1, 3);
	\coordinate (8) at (3, 3.5);
	\coordinate (10) at (2, 5);
	
	\coordinate (11) at (intersection of 10--4 and 2--5);
	\coordinate (12) at (intersection of 3--6 and 2--5);
	
	\draw[] (0) -- (1);
	\draw[] (1) -- (8);
	\draw[] (8) edge [out=35, in=-30, looseness=1.5] (10);
	\draw[shorten >= 0.3em] (10) -- (11);
	\draw[shorten <= 0.3em] (11) -- (4);
	\draw[] (1) -- (7);
	\draw[] (7) edge [out=135, in=205, looseness=1.5] (12);
	\draw[shorten >= 0.3em] (7) edge [out=10, in=-30, looseness=1.5] (12);
	\draw[shorten <= 0.3em] (12) -- (3);
	\draw[] (12) -- (5);
	
	\ifthenelse{#2 = 1}{
		\path (2, 4) -- (5) node[anchor=center, pos=1.25] {{$\sss{e}$}};
		\path (2) -- (3) node[anchor=center, pos=1.25] {{$\sss{c}$}};
		\path (2, 4) -- (4) node[anchor=center, pos=1.25] {{$\sss{d}$}};
	}{}

	\ifthenelse{#2 = 2}{
		\path (2, 4) -- (5) node[anchor=center, pos=1.25] {{$\sss{c}$}};
		\path (2) -- (3) node[anchor=center, pos=1.25] {{$\sss{a}$}};
		\path (2, 4) -- (4) node[anchor=center, pos=1.25] {{$\sss{b}$}};
	}{}
	\end{tikzpicture}
}

\newcommand{\hexagonMoveFIVE}[2]{
	\begin{tikzpicture}[scale=#1,baseline=1.5em, line width=0.8pt]
	\coordinate (0) at(0, 0);
	\coordinate (1) at (0, 2);
	\coordinate (2) at (2, 4);
	\coordinate (3) at (-4, 6);
	\coordinate (4) at (0, 6);
	\coordinate (5) at (4, 6);
	
	\ifthenelse{#2 = 2}{
		\draw[] (0) -- (1);
		\draw[] (1) -- (3);
		\draw[] (2) -- (4) node[anchor=center, pos=1.25] {{$\sss{b}$}};			
		\draw[] (1) -- (2, 4);
		\draw[] (2, 4) -- (5) node[anchor=center, pos=1.25] {{$\sss{c}$}};
		
		\path (-2, 4) -- (3) node[anchor=center, pos=1.25] {{$\sss{a}$}};
	}{}	

	\ifthenelse{#2 = 1}{
		\draw[] (0) -- (1);
		\draw[] (1) -- (3);
		\draw[] (2) -- (4) node[anchor=center, pos=1.25] {{$\sss{g}$}};			
		\draw[] (1) -- (2, 4);
		\draw[] (2, 4) -- (5) node[anchor=center, pos=1.25] {{$\sss{h}$}};
		
		\path (-2, 4) -- (3) node[anchor=center, pos=1.25] {{$\sss{c}$}};
	}{}	
	\end{tikzpicture}
}

\newcommand{\hexagonMoveSIX}[2]{
	\begin{tikzpicture}[scale=#1,baseline=1.5em, line width=0.8pt]
	\coordinate (0) at(0, 0);
	\coordinate (1) at (0, 2);
	\coordinate (2) at (2, 4);
	\coordinate (3) at (-4, 6);
	\coordinate (4) at (0, 6);
	\coordinate (5) at (4, 6);
	\coordinate (7) at (2, 5);
	
	\draw[] (0) -- (1);
	\draw[] (1) -- (3);
	\draw[] (2) edge [out=170, in=210, looseness=1.5] (7);	
	\draw[shorten >= 0.3em] (2) edge [out=45, in=-30, looseness=1.5] (7);
	\draw[shorten <= 0.3em] (7) -- (4);
	\draw[] (7) -- (5);	
	\draw[] (1) -- (2);
	
	\ifthenelse{#2 = 2}{	
		\path (-2, 4) -- (3) node[anchor=center, pos=1.25] {{$\sss{a}$}};
		\path (2) -- (4) node[anchor=center, pos=1.25] {{$\sss{b}$}};
		\path (2) -- (5) node[anchor=center, pos=1.25] {{$\sss{c}$}};
	}{}

	\ifthenelse{#2 = 1}{	
		\path (-2, 4) -- (3) node[anchor=center, pos=1.25] {{$\sss{c}$}};
		\path (2) -- (4) node[anchor=center, pos=1.25] {{$\sss{g}$}};
		\path (2) -- (5) node[anchor=center, pos=1.25] {{$\sss{h}$}};
	}{}
	\end{tikzpicture}
}

\newcommand{\Smove}[2]{
	\begin{tikzpicture}[scale=#1,baseline=0em, line width=0.8pt]
	\coordinate (0) at(0, 0);
	\coordinate (1) at (0, 2);
	\coordinate (2) at (2 , 0);
	\coordinate (3) at (0, -2);
	\coordinate (4) at (-2, 0);
	
	\ifthenelse{#2 = 1}{
		\draw[] (0) -- (1) node[anchor=center, pos=1.25] {{$\sss{a}$}};
		\draw[] (0) -- (2) node[anchor=center, pos=1.5] {{$\sss{b+c}$}};	
		\draw[] (0) -- (3) node[anchor=center, pos=1.25] {{$\sss{c}$}};	
		\draw[] (0) -- (4) node[anchor=center, pos=1.5] {{$\sss{a+b}$}};								

	}{}
	
	\coordinate (5) at (2, -2);
	\coordinate (6) at (4, 0);
	\ifthenelse{#2 = 2}{
		\draw[] (0) -- (1) node[anchor=center, pos=1.25] {{$\sss{a}$}};	
		\draw[] (0) -- (4) node[anchor=center, pos=1.5] {{$\sss{a+b}$}};
		\draw[] (0) -- (2) node[above, pos=0.5, yshift=-0.28em] {{$\sss{b}$}};	
		\draw[] (2) -- (5) node[anchor=center, pos=1.25] {{$\sss{c}$}};
		\draw[] (2) -- (6) node[anchor=center, pos=1.5] {{$\sss{b+c}$}};	
	}{}
	\end{tikzpicture}
}

%% file: QFormExcitations.bbl
\providecommand{\href}[2]{#2}\begingroup\raggedright\begin{thebibliography}{100}

\bibitem{Atiyah:1989vu}
M.~Atiyah, \emph{{Topological quantum field theories}},
  \href{http://dx.doi.org/10.1007/BF02698547}{\emph{Inst. Hautes Etudes Sci.
  Publ. Math.} {\bfseries 68} (1989) 175--186}.

\bibitem{Baez:1995xq}
J.~C. Baez and J.~Dolan, \emph{{Higher dimensional algebra and topological
  quantum field theory}}, \href{http://dx.doi.org/10.1063/1.531236}{\emph{J.
  Math. Phys.} {\bfseries 36} (1995) 6073--6105},
  [\href{https://arxiv.org/abs/q-alg/9503002}{{\ttfamily q-alg/9503002}}].

\bibitem{lurie2009higher}
J.~Lurie, \emph{Higher Topos Theory (AM-170)}.
\newblock Princeton University Press, 2009.

\bibitem{Lurie:2009keu}
J.~Lurie, \emph{{On the Classification of Topological Field Theories}},
  \href{https://arxiv.org/abs/0905.0465}{{\ttfamily 0905.0465}}.

\bibitem{Freed:2012hx}
D.~S. Freed, \emph{{The cobordism hypothesis}},
  \href{https://arxiv.org/abs/1210.5100}{{\ttfamily 1210.5100}}.

\bibitem{Gaiotto:2017zba}
D.~Gaiotto and T.~Johnson-Freyd, \emph{{Symmetry Protected Topological phases
  and Generalized Cohomology}},
  \href{https://arxiv.org/abs/1712.07950}{{\ttfamily 1712.07950}}.

\bibitem{dijkgraaf1990topological}
{Dijkgraaf, Robbert and Witten, Edward}, \emph{{Topological gauge theories and
  group cohomology}},
  \href{http://dx.doi.org/10.1007/BF02096988}{\emph{{Communications in
  Mathematical Physics}} {\bfseries 129} (Apr, 1990) 393--429}.

\bibitem{Hu:2012wx}
Y.~Hu, Y.~Wan and Y.-S. Wu, \emph{{Twisted quantum double model of topological
  phases in two dimensions}},
  \href{http://dx.doi.org/10.1103/PhysRevB.87.125114}{\emph{Phys. Rev.}
  {\bfseries B87} (2013) 125114},
  [\href{https://arxiv.org/abs/1211.3695}{{\ttfamily 1211.3695}}].

\bibitem{Wan:2014woa}
Y.~Wan, J.~C. Wang and H.~He, \emph{{Twisted Gauge Theory Model of Topological
  Phases in Three Dimensions}},
  \href{http://dx.doi.org/10.1103/PhysRevB.92.045101}{\emph{Phys. Rev.}
  {\bfseries B92} (2015) 045101},
  [\href{https://arxiv.org/abs/1409.3216}{{\ttfamily 1409.3216}}].

\bibitem{eilenberg1953groups}
S.~Eilenberg and S.~M. Lane, \emph{{On the groups H ($\pi$, n), I}},
  {\emph{Annals of Mathematics} (1953) 55--106}.

\bibitem{eilenberg1954groups}
S.~Eilenberg and S.~MacLane, \emph{{On the groups H ($\pi$, n), II: Methods of
  computation}}, {\emph{Annals of Mathematics} (1954) 49--139}.

\bibitem{kapustin2014topological}
A.~Kapustin and R.~Thorngren, \emph{{Topological Field Theory on a Lattice,
  Discrete Theta-Angles and Confinement}},
  \href{http://dx.doi.org/10.4310/ATMP.2014.v18.n5.a4}{\emph{Adv. Theor. Math.
  Phys.} {\bfseries 18} (2014) 1233--1247},
  [\href{https://arxiv.org/abs/1308.2926}{{\ttfamily 1308.2926}}].

\bibitem{Aharony:2013hda}
O.~Aharony, N.~Seiberg and Y.~Tachikawa, \emph{{Reading between the lines of
  four-dimensional gauge theories}},
  \href{http://dx.doi.org/10.1007/JHEP08(2013)115}{\emph{JHEP} {\bfseries 08}
  (2013) 115}, [\href{https://arxiv.org/abs/1305.0318}{{\ttfamily 1305.0318}}].

\bibitem{gaiotto2015generalized}
D.~Gaiotto, A.~Kapustin, N.~Seiberg and B.~Willett, \emph{{Generalized Global
  Symmetries}}, \href{http://dx.doi.org/10.1007/JHEP02(2015)172}{\emph{JHEP}
  {\bfseries 02} (2015) 172},
  [\href{https://arxiv.org/abs/1412.5148}{{\ttfamily 1412.5148}}].

\bibitem{Pfeiffer:2003je}
H.~Pfeiffer, \emph{{Higher gauge theory and a nonAbelian generalization of
  2-form electrodynamics}},
  \href{http://dx.doi.org/10.1016/S0003-4916(03)00147-7}{\emph{Annals Phys.}
  {\bfseries 308} (2003) 447--477},
  [\href{https://arxiv.org/abs/hep-th/0304074}{{\ttfamily hep-th/0304074}}].

\bibitem{Freed:2009qp}
D.~S. Freed, M.~J. Hopkins, J.~Lurie and C.~Teleman, \emph{{Topological Quantum
  Field Theories from Compact Lie Groups}},  in \emph{{A Celebration of Raoul
  Bott's Legacy in Mathematics Montreal, Canada, June 9-13, 2008}}, 2009,
  \href{https://arxiv.org/abs/0905.0731}{{\ttfamily 0905.0731}}.

\bibitem{Walker:2011mda}
K.~Walker and Z.~Wang, \emph{{(3+1)-TQFTs and Topological Insulators}},
  \href{https://arxiv.org/abs/1104.2632}{{\ttfamily 1104.2632}}.

\bibitem{vonKeyserlingk:2012zn}
C.~W. von Keyserlingk, F.~J. Burnell and S.~H. Simon, \emph{{Three-dimensional
  topological lattice models with surface anyons}},
  \href{http://dx.doi.org/10.1103/PhysRevB.87.045107}{\emph{Phys. Rev.}
  {\bfseries B87} (2013) 045107},
  [\href{https://arxiv.org/abs/1208.5128}{{\ttfamily 1208.5128}}].

\bibitem{gaiotto2017theta}
D.~Gaiotto, A.~Kapustin, Z.~Komargodski and N.~Seiberg, \emph{{Theta, Time
  Reversal, and Temperature}},
  \href{http://dx.doi.org/10.1007/JHEP05(2017)091}{\emph{JHEP} {\bfseries 05}
  (2017) 091}, [\href{https://arxiv.org/abs/1703.00501}{{\ttfamily
  1703.00501}}].

\bibitem{delcamp2018gauge}
C.~Delcamp and A.~Tiwari, \emph{{From gauge to higher gauge models of
  topological phases}},
  \href{http://dx.doi.org/10.1007/JHEP10(2018)049}{\emph{JHEP} {\bfseries 10}
  (2018) 049}, [\href{https://arxiv.org/abs/1802.10104}{{\ttfamily
  1802.10104}}].

\bibitem{Benini:2018reh}
F.~Benini, C.~Cordova and P.-S. Hsin, \emph{{On 2-Group Global Symmetries and
  their Anomalies}},  \href{https://arxiv.org/abs/1803.09336}{{\ttfamily
  1803.09336}}.

\bibitem{Hsin:2018vcg}
P.-S. Hsin, H.~T. Lam and N.~Seiberg, \emph{{Comments on One-Form Global
  Symmetries and Their Gauging in 3d and 4d}},
  \href{https://arxiv.org/abs/1812.04716}{{\ttfamily 1812.04716}}.

\bibitem{Wan:2018bns}
Z.~Wan and J.~Wang, \emph{{Non-Abelian Gauge Theories, Sigma Models, Higher
  Anomalies, Symmetries, and Cobordisms}},
  \href{https://arxiv.org/abs/1812.11967}{{\ttfamily 1812.11967}}.

\bibitem{Wan:2018zql}
Z.~Wan and J.~Wang, \emph{{New Higher Anomalies, SU(N) Yang-Mills Gauge Theory
  and $\mathbb{CP}^{\mathrm{N}-1}$ Sigma Model}},
  \href{https://arxiv.org/abs/1812.11968}{{\ttfamily 1812.11968}}.

\bibitem{Guo:2018vij}
M.~Guo, K.~Ohmori, P.~Putrov, Z.~Wan and J.~Wang, \emph{{Fermionic Finite-Group
  Gauge Theories and Interacting Symmetric/Crystalline Orders via Cobordisms}},
   \href{https://arxiv.org/abs/1812.11959}{{\ttfamily 1812.11959}}.

\bibitem{Wan:2018djl}
Z.~Wan and J.~Wang, \emph{{Adjoint QCD$_4$, Deconfined Critical Phenomena,
  Symmetry-Enriched Topological Quantum Field Theory, and Higher
  Symmetry-Extension}},
  \href{http://dx.doi.org/10.1103/PhysRevD.99.065013}{\emph{Phys. Rev.}
  {\bfseries D99} (2019) 065013},
  [\href{https://arxiv.org/abs/1812.11955}{{\ttfamily 1812.11955}}].

\bibitem{Wan:2019oyr}
Z.~Wan, J.~Wang and Y.~Zheng, \emph{{Quantum Yang-Mills 4d Theory and
  Time-Reversal Symmetric 5d Higher-Gauge TQFT: Anyonic-String/Brane Braiding
  Statistics to Topological Link Invariants}},
  \href{https://arxiv.org/abs/1904.00994}{{\ttfamily 1904.00994}}.

\bibitem{kapustin2014anomalies}
A.~Kapustin and R.~Thorngren, \emph{{Anomalies of discrete symmetries in
  various dimensions and group cohomology}},
  \href{https://arxiv.org/abs/1404.3230}{{\ttfamily 1404.3230}}.

\bibitem{Turaev:1994xb}
V.~G. Turaev, \emph{{Quantum invariants of knots and three manifolds}},
  {\emph{De Gruyter Stud. Math.} {\bfseries 18} (1994) 1--588}.

\bibitem{Turaev:1992hq}
V.~G. Turaev and O.~Y. Viro, \emph{{State sum invariants of 3 manifolds and
  quantum 6j symbols}},
  \href{http://dx.doi.org/10.1016/0040-9383(92)90015-A}{\emph{Topology}
  {\bfseries 31} (1992) 865--902}.

\bibitem{Barrett:1993ab}
J.~W. Barrett and B.~W. Westbury, \emph{{Invariants of piecewise linear three
  manifolds}},
  \href{http://dx.doi.org/10.1090/S0002-9947-96-01660-1}{\emph{Trans. Am. Math.
  Soc.} {\bfseries 348} (1996) 3997--4022},
  [\href{https://arxiv.org/abs/hep-th/9311155}{{\ttfamily hep-th/9311155}}].

\bibitem{Levin:2004mi}
M.~A. Levin and X.-G. Wen, \emph{{String net condensation: A Physical mechanism
  for topological phases}},
  \href{http://dx.doi.org/10.1103/PhysRevB.71.045110}{\emph{Phys. Rev.}
  {\bfseries B71} (2005) 045110},
  [\href{https://arxiv.org/abs/cond-mat/0404617}{{\ttfamily
  cond-mat/0404617}}].

\bibitem{Crane:1993cm}
L.~Crane, L.~H. Kauffman and D.~Yetter, \emph{{Evaluating the Crane-Yetter
  invariant}},  \href{https://arxiv.org/abs/hep-th/9309063}{{\ttfamily
  hep-th/9309063}}.

\bibitem{Crane:1993if}
L.~Crane and D.~Yetter, \emph{{A Categorical construction of 4-D topological
  quantum field theories}},  in \emph{{In *Dayton 1992, Proceedings, Quantum
  topology* 120-130}}, 1993,
  \href{https://arxiv.org/abs/hep-th/9301062}{{\ttfamily hep-th/9301062}}.

\bibitem{Crane:1994ji}
L.~Crane, L.~H. Kauffman and D.~N. Yetter, \emph{{State sum invariants of four
  manifolds. 1.}},  \href{https://arxiv.org/abs/hep-th/9409167}{{\ttfamily
  hep-th/9409167}}.

\bibitem{baez2004higher}
J.~C. {Baez} and A.~D. {Lauda}, \emph{{Higher-Dimensional Algebra V:
  2-Groups}}, {\emph{ArXiv Mathematics e-prints} (July, 2003) },
  [\href{https://arxiv.org/abs/math/0307200}{{\ttfamily math/0307200}}].

\bibitem{Baez:2003fs}
J.~C. Baez and A.~S. Crans, \emph{{Higher-Dimensional Algebra VI: Lie
  2-Algebras}}, {\emph{Theor. Appl. Categor.} {\bfseries 12} (2004) 492--528},
  [\href{https://arxiv.org/abs/math/0307263}{{\ttfamily math/0307263}}].

\bibitem{Baez:2004in}
J.~Baez and U.~Schreiber, \emph{{Higher gauge theory: 2-connections on
  2-bundles}},  \href{https://arxiv.org/abs/hep-th/0412325}{{\ttfamily
  hep-th/0412325}}.

\bibitem{Delcamp:2017pcw}
C.~Delcamp, \emph{{Excitation basis for (3+1)d topological phases}},
  \href{http://dx.doi.org/10.1007/JHEP12(2017)128}{\emph{JHEP} {\bfseries 12}
  (2017) 128}, [\href{https://arxiv.org/abs/1709.04924}{{\ttfamily
  1709.04924}}].

\bibitem{deligne1971theorie}
P.~Deligne, \emph{{Th{\'e}orie de Hodge, II}}, {\emph{Publications
  Math{\'e}matiques de l'Institut des Hautes {\'E}tudes Scientifiques}
  {\bfseries 40} (1971) 5--57}.

\bibitem{Thuillier:2015vma}
F.~Thuillier, \emph{{Deligne-Beilinson Cohomology in U(1) Chern-Simons
  Theories}},  in \emph{{Proceedings, Winter School in Mathematical Physics:
  Mathematical Aspects of Quantum Field Theory: Les Houches, France, January
  29-February 3, 2012}}, pp.~233--271, Springer, Springer, 2015,
  \href{http://dx.doi.org/10.1007/978-3-319-09949-1_8}{DOI}.

\bibitem{Mathieu:2015mda}
P.~Mathieu and F.~Thuillier, \emph{{Abelian BF theory and Turaev-Viro
  invariant}}, \href{http://dx.doi.org/10.1063/1.4942046}{\emph{J. Math. Phys.}
  {\bfseries 57} (2016) 022306},
  [\href{https://arxiv.org/abs/1509.04236}{{\ttfamily 1509.04236}}].

\bibitem{PACHNER1991129}
{Udo Pachner}, \emph{{P.L. Homeomophic Manifolds are Equivalent by Elementary
  Shellings}},
  \href{http://dx.doi.org/https://doi.org/10.1016/S0195-6698(13)80080-7}{\emph{{European
  Journal of Combinatorics}} {\bfseries 12} (1991) 129 -- 145}.

\bibitem{Freed:1991bn}
D.~S. Freed and F.~Quinn, \emph{{Chern-Simons theory with finite gauge group}},
  \href{http://dx.doi.org/10.1007/BF02096860}{\emph{Commun. Math. Phys.}
  {\bfseries 156} (1993) 435--472},
  [\href{https://arxiv.org/abs/hep-th/9111004}{{\ttfamily hep-th/9111004}}].

\bibitem{hatcher2002algebraic}
A.~Hatcher, \emph{Algebraic topology. 2002}, {\emph{Cambridge UP, Cambridge}
  {\bfseries 606} (2002) }.

\bibitem{Kapustin:2013uxa}
A.~Kapustin and R.~Thorngren, \emph{{Higher symmetry and gapped phases of gauge
  theories}},  \href{https://arxiv.org/abs/1309.4721}{{\ttfamily 1309.4721}}.

\bibitem{Cordova:2018cvg}
C.~Cordova, T.~T. Dumitrescu and K.~Intriligator, \emph{{Exploring 2-Group
  Global Symmetries}},  \href{https://arxiv.org/abs/1802.04790}{{\ttfamily
  1802.04790}}.

\bibitem{Bullivant:2016clk}
A.~Bullivant, M.~Calçada, Z.~Kádár, P.~Martin and J.~F. Martins,
  \emph{{Topological phases from higher gauge symmetry in 3+1 dimensions}},
  \href{http://dx.doi.org/10.1103/PhysRevB.95.155118}{\emph{Phys. Rev.}
  {\bfseries B95} (2017) 155118},
  [\href{https://arxiv.org/abs/1606.06639}{{\ttfamily 1606.06639}}].

\bibitem{Cui:2016bmd}
S.~X. Cui, \emph{{Higher Categories and Topological Quantum Field Theories}},
  \href{https://arxiv.org/abs/1610.07628}{{\ttfamily 1610.07628}}.

\bibitem{Bullivant:2017sjz}
A.~Bullivant, M.~Calcada, Z.~Kádár, J.~F. Martins and P.~Martin,
  \emph{{Higher lattices, discrete two-dimensional holonomy and topological
  phases in (3+1) D with higher gauge symmetry}},
  \href{https://arxiv.org/abs/1702.00868}{{\ttfamily 1702.00868}}.

\bibitem{2018arXiv180809394Z}
C.~{Zhu}, T.~{Lan} and X.-G. {Wen}, \emph{{Topological non-linear
  $\sigma$-model, higher gauge theory, and a realization of all 3+1D
  topological orders for boson systems}}, {\emph{ArXiv e-prints} (Aug., 2018)
  }, [\href{https://arxiv.org/abs/1808.09394}{{\ttfamily 1808.09394}}].

\bibitem{freed2007heisenberg}
D.~S. Freed, G.~W. Moore and G.~Segal, \emph{{Heisenberg groups and
  noncommutative fluxes}}, {\emph{Annals of Physics} {\bfseries 322} (2007)
  236--285}.

\bibitem{Kitaev:1997wr}
A.~{\relax Yu}. Kitaev, \emph{{Fault tolerant quantum computation by anyons}},
  \href{http://dx.doi.org/10.1016/S0003-4916(02)00018-0}{\emph{Annals Phys.}
  {\bfseries 303} (2003) 2--30},
  [\href{https://arxiv.org/abs/quant-ph/9707021}{{\ttfamily
  quant-ph/9707021}}].

\bibitem{Wang:2014pma}
J.~C. Wang, Z.-C. Gu and X.-G. Wen, \emph{{Field theory representation of
  gauge-gravity symmetry-protected topological invariants, group cohomology and
  beyond}}, \href{http://dx.doi.org/10.1103/PhysRevLett.114.031601}{\emph{Phys.
  Rev. Lett.} {\bfseries 114} (2015) 031601},
  [\href{https://arxiv.org/abs/1405.7689}{{\ttfamily 1405.7689}}].

\bibitem{Tiwari:2016zru}
A.~Tiwari, X.~Chen and S.~Ryu, \emph{{Wilson operator algebras and ground
  states of coupled BF theories}},
  \href{http://dx.doi.org/10.1103/PhysRevB.95.245124}{\emph{Phys. Rev.}
  {\bfseries B95} (2017) 245124},
  [\href{https://arxiv.org/abs/1603.08429}{{\ttfamily 1603.08429}}].

\bibitem{Tiwari:2017wqf}
A.~Tiwari, X.~Chen, K.~Shiozaki and S.~Ryu, \emph{{Bosonic topological phases
  of matter: bulk-boundary correspondence, SPT invariants and gauging}},
  \href{https://arxiv.org/abs/1710.04730}{{\ttfamily 1710.04730}}.

\bibitem{putrov2016braiding}
P.~Putrov, J.~Wang and S.-T. Yau, \emph{{Braiding Statistics and Link
  Invariants of Bosonic/Fermionic Topological Quantum Matter in 2+1 and 3+1
  dimensions}}, \href{http://dx.doi.org/10.1016/j.aop.2017.06.019}{\emph{Annals
  Phys.} {\bfseries 384} (2017) 254--287},
  [\href{https://arxiv.org/abs/1612.09298}{{\ttfamily 1612.09298}}].

\bibitem{wang2018tunneling}
J.~Wang, K.~Ohmori, P.~Putrov, Y.~Zheng, H.~Lin, P.~Gao et~al.,
  \emph{{Tunneling Topological Vacua via Extended Operators: TQFT Spectra and
  Boundary Deconfinement in Various Dimensions}},
  \href{https://arxiv.org/abs/1801.05416}{{\ttfamily 1801.05416}}.

\bibitem{kapustin2014coupling}
A.~Kapustin and N.~Seiberg, \emph{{Coupling a QFT to a TQFT and Duality}},
  \href{http://dx.doi.org/10.1007/JHEP04(2014)001}{\emph{JHEP} {\bfseries 04}
  (2014) 001}, [\href{https://arxiv.org/abs/1401.0740}{{\ttfamily 1401.0740}}].

\bibitem{cheeger1985differential}
{Cheeger, Jeff and Simons, James}, \emph{{Differential characters and geometric
  invariants}},  in \emph{Geometry and Topology}, (Berlin, Heidelberg),
  pp.~50--80, Springer Berlin Heidelberg, 1985.

\bibitem{hopkins2002quadratic}
M.~J. Hopkins and I.~M. Singer, \emph{{Quadratic functions in geometry,
  topology, and M theory}}, {\emph{J. Diff. Geom.} {\bfseries 70} (2005)
  329--452}, [\href{https://arxiv.org/abs/math/0211216}{{\ttfamily
  math/0211216}}].

\bibitem{simons2008axiomatic}
J.~{Simons} and D.~{Sullivan}, \emph{{Axiomatic Characterization of Ordinary
  Differential Cohomology}}, {\emph{ArXiv Mathematics e-prints} (Jan., 2007) },
  [\href{https://arxiv.org/abs/math/0701077}{{\ttfamily math/0701077}}].

\bibitem{eilenberg1943relations}
S.~Eilenberg and S.~MacLane, \emph{{Relations between homology and homotopy
  groups}}, {\emph{Proceedings of the National Academy of Sciences} {\bfseries
  29} (1943) 155--158}.

\bibitem{eilenberg1950relations}
S.~Eilenberg and S.~MacLane, \emph{{Relations between homology and homotopy
  groups of spaces. II}}, {\emph{Annals of mathematics} (1950) 514--533}.

\bibitem{may1992simplicial}
J.~P. May, \emph{Simplicial objects in algebraic topology}, vol.~11.
\newblock University of Chicago Press, 1992.

\bibitem{may1999concise}
J.~P. May, \emph{A concise course in algebraic topology}.
\newblock University of Chicago press, 1999.

\bibitem{Chen:2010gda}
X.~Chen, Z.~C. Gu and X.~G. Wen, \emph{{Local unitary transformation,
  long-range quantum entanglement, wave function renormalization, and
  topological order}},
  \href{http://dx.doi.org/10.1103/PhysRevB.82.155138}{\emph{Phys. Rev.}
  {\bfseries B82} (2010) 155138},
  [\href{https://arxiv.org/abs/1004.3835}{{\ttfamily 1004.3835}}].

\bibitem{1998math.....11047Q}
F.~{Quinn}, \emph{{Group categories and their field theories}}, {\emph{ArXiv
  Mathematics e-prints} (Nov., 1998) },
  [\href{https://arxiv.org/abs/math/9811047}{{\ttfamily math/9811047}}].

\bibitem{Stirling:2008bq}
S.~D. Stirling, \emph{{Abelian Chern-Simons theory with toral gauge group,
  modular tensor categories, and group categories}}, Ph.D. thesis, Texas U.,
  Math Dept., 2008.
\newblock \href{https://arxiv.org/abs/0807.2857}{{\ttfamily 0807.2857}}.

\bibitem{2016arXiv160601414G}
C.~{Galindo} and N.~{Jaramillo Torres}, \emph{{Solutions of the hexagon
  equation for abelian anyons}}, {\emph{ArXiv e-prints} (June, 2016) },
  [\href{https://arxiv.org/abs/1606.01414}{{\ttfamily 1606.01414}}].

\bibitem{Williamson:2016evv}
D.~J. Williamson and Z.~Wang, \emph{{Hamiltonian models for topological phases
  of matter in three spatial dimensions}},
  \href{http://dx.doi.org/10.1016/j.aop.2016.12.018}{\emph{Annals Phys.}
  {\bfseries 377} (2017) 311--344},
  [\href{https://arxiv.org/abs/1606.07144}{{\ttfamily 1606.07144}}].

\bibitem{2010AnPhy.325.2707K}
R.~{Koenig}, G.~{Kuperberg} and B.~W. {Reichardt}, \emph{{Quantum computation
  with Turaev-Viro codes}},
  \href{http://dx.doi.org/10.1016/j.aop.2010.08.001}{\emph{Annals of Physics}
  {\bfseries 325} (Dec., 2010) 2707--2749},
  [\href{https://arxiv.org/abs/1002.2816}{{\ttfamily 1002.2816}}].

\bibitem{Lan:2013wia}
T.~Lan and X.-G. Wen, \emph{{Topological quasiparticles and the holographic
  bulk-edge relation in (2+1) -dimensional string-net models}},
  \href{http://dx.doi.org/10.1103/PhysRevB.90.115119}{\emph{Phys. Rev.}
  {\bfseries B90} (2014) 115119},
  [\href{https://arxiv.org/abs/1311.1784}{{\ttfamily 1311.1784}}].

\bibitem{Dittrich:2016typ}
B.~Dittrich and M.~Geiller, \emph{{Quantum gravity kinematics from extended
  TQFTs}}, \href{http://dx.doi.org/10.1088/1367-2630/aa54e2}{\emph{New J.
  Phys.} {\bfseries 19} (2017) 013003},
  [\href{https://arxiv.org/abs/1604.05195}{{\ttfamily 1604.05195}}].

\bibitem{DDR1}
C.~Delcamp, B.~Dittrich and A.~Riello, \emph{{Fusion basis for lattice gauge
  theory and loop quantum gravity}},
  \href{http://dx.doi.org/10.1007/JHEP02(2017)061}{\emph{JHEP} {\bfseries 02}
  (2017) 061}, [\href{https://arxiv.org/abs/1607.08881}{{\ttfamily
  1607.08881}}].

\bibitem{Aasen:2017ubm}
D.~Aasen, E.~Lake and K.~Walker, \emph{{Fermion condensation and super pivotal
  categories}},  \href{https://arxiv.org/abs/1709.01941}{{\ttfamily
  1709.01941}}.

\bibitem{ocneanu1994chirality}
A.~Ocneanu, \emph{Chirality for operator algebras}, {\emph{Subfactors (Kyuzeso,
  1993)} (1994) 39--63}.

\bibitem{ocneanu2001operator}
A.~Ocneanu, \emph{Operator algebras, topology and subgroups of quantum
  symmetry--construction of subgroups of quantum groups},  in \emph{Taniguchi
  Conference on Mathematics Nara}, vol.~98, pp.~235--263, 2001.

\bibitem{Drinfeld:1989st}
V.~G. Drinfeld, \emph{{Quasi Hopf algebras}}, {\emph{Alg. Anal.} {\bfseries
  1N6} (1989) 114--148}.

\bibitem{Dijkgraaf1991}
R.~{Dijkgraaf}, V.~{Pasquier} and P.~{Roche}, \emph{{Quasi hopf algebras, group
  cohomology and orbifold models}},
  \href{http://dx.doi.org/10.1016/0920-5632(91)90123-V}{\emph{Nuclear Physics B
  Proceedings Supplements} {\bfseries 18} (Jan., 1991) 60--72}.

\bibitem{1996q.alg.....5044K}
T.~H. {Koornwinder} and N.~M. {Muller}, \emph{{Quantum double of a (locally)
  compact group}},  in \emph{eprint arXiv:q-alg/9605044}, May, 1996.

\bibitem{Koornwinder:1998xg}
T.~H. Koornwinder, F.~A. Bais and N.~M. Muller, \emph{{Tensor product
  representations of the quantum double of a compact group}},
  \href{http://dx.doi.org/10.1007/s002200050475}{\emph{Commun. Math. Phys.}
  {\bfseries 198} (1998) 157--186},
  [\href{https://arxiv.org/abs/q-alg/9712042}{{\ttfamily q-alg/9712042}}].

\bibitem{Koornwinder:1999bg}
T.~H. Koornwinder, B.~J. Schroers, J.~K. Slingerland and F.~A. Bais,
  \emph{{Fourier transform and the Verlinde formula for the quantum double of a
  finite group}}, \href{http://dx.doi.org/10.1088/0305-4470/32/48/313}{\emph{J.
  Phys.} {\bfseries A32} (1999) 8539--8549},
  [\href{https://arxiv.org/abs/math/9904029}{{\ttfamily math/9904029}}].

\bibitem{Delcamp:2018efi}
C.~Delcamp and B.~Dittrich, \emph{{Towards a dual spin network basis for (3+1)d
  lattice gauge theories and topological phases}},
  \href{http://dx.doi.org/10.1007/JHEP10(2018)023}{\emph{JHEP} {\bfseries 10}
  (2018) 023}, [\href{https://arxiv.org/abs/1806.00456}{{\ttfamily
  1806.00456}}].

\bibitem{bullivant2019tube}
A.~Bullivant and C.~Delcamp, \emph{Tube algebras, excitations statistics and
  compactification in gauge models of topological phases}, {\emph{arXiv
  preprint arXiv:1905.08673} (2019) }.

\bibitem{etingof2016tensor}
P.~Etingof, S.~Gelaki, D.~Nikshych and V.~Ostrik, \emph{Tensor categories},
  vol.~205.
\newblock American Mathematical Soc., 2016.

\bibitem{Baez:2006un}
J.~C. Baez, D.~K. Wise and A.~S. Crans, \emph{{Exotic statistics for strings in
  4d BF theory}},
  \href{http://dx.doi.org/10.4310/ATMP.2007.v11.n5.a1}{\emph{Adv. Theor. Math.
  Phys.} {\bfseries 11} (2007) 707--749},
  [\href{https://arxiv.org/abs/gr-qc/0603085}{{\ttfamily gr-qc/0603085}}].

\bibitem{Wang:2014xba}
C.~Wang and M.~Levin, \emph{{Braiding statistics of loop excitations in three
  dimensions}},
  \href{http://dx.doi.org/10.1103/PhysRevLett.113.080403}{\emph{Phys. Rev.
  Lett.} {\bfseries 113} (2014) 080403},
  [\href{https://arxiv.org/abs/1403.7437}{{\ttfamily 1403.7437}}].

\bibitem{Wang:2014oya}
J.~Wang and X.-G. Wen, \emph{{Non-Abelian string and particle braiding in
  topological order: Modular SL(3,$\mathbb{Z}$) representation and (3+1)
  -dimensional twisted gauge theory}},
  \href{http://dx.doi.org/10.1103/PhysRevB.91.035134}{\emph{Phys. Rev.}
  {\bfseries B91} (2015) 035134},
  [\href{https://arxiv.org/abs/1404.7854}{{\ttfamily 1404.7854}}].

\bibitem{Bullivant:2018pju}
A.~Bullivant, J.~F. Martins and P.~Martin, \emph{{From Aharonov-Bohm type
  effects in discrete (3+1)-dimensional higher gauge theory to representations
  of the Loop Braid Group}},
  \href{https://arxiv.org/abs/1807.09551}{{\ttfamily 1807.09551}}.

\bibitem{Bullivant:2018djw}
A.~Bullivant, A.~Kimball, P.~Martin and E.~C. Rowell, \emph{{Representations of
  the Necklace Braid Group: Topological and Combinatorial Approaches}},
  \href{https://arxiv.org/abs/1810.05152}{{\ttfamily 1810.05152}}.

\bibitem{Yoshida:2015cia}
B.~Yoshida, \emph{{Topological phases with generalized global symmetries}},
  \href{http://dx.doi.org/10.1103/PhysRevB.93.155131}{\emph{Phys. Rev.}
  {\bfseries B93} (2016) 155131},
  [\href{https://arxiv.org/abs/1508.03468}{{\ttfamily 1508.03468}}].

\bibitem{2018arXiv180907325R}
A.~{Rasmussen} and Y.-M. {Lu}, \emph{{Classification and construction of
  higher-order symmetry protected topological phases of interacting bosons}},
  {\emph{ArXiv e-prints} (Sept., 2018) },
  [\href{https://arxiv.org/abs/1809.07325}{{\ttfamily 1809.07325}}].

\bibitem{whitehead1949simply}
J.~Whitehead, \emph{{On simply connected, 4-dimensional polyhedra}},
  {\emph{Commentarii Mathematici Helvetici} {\bfseries 22} (1949) 48--92}.

\bibitem{steenrod1947products}
N.~E. Steenrod, \emph{Products of cocycles and extensions of mappings},
  {\emph{Annals of Mathematics} (1947) 290--320}.

\bibitem{Chen:2015gma}
X.~Chen, A.~Tiwari and S.~Ryu, \emph{{Bulk-boundary correspondence in
  (3+1)-dimensional topological phases}},
  \href{http://dx.doi.org/10.1103/PhysRevB.94.045113,
  10.1103/PhysRevB.94.079903}{\emph{Phys. Rev.} {\bfseries B94} (2016) 045113},
  [\href{https://arxiv.org/abs/1509.04266}{{\ttfamily 1509.04266}}].

\bibitem{Freed:2014eja}
D.~S. Freed, \emph{{Short-range entanglement and invertible field theories}},
  \href{https://arxiv.org/abs/1406.7278}{{\ttfamily 1406.7278}}.

\bibitem{Freed:2016rqq}
D.~S. Freed and M.~J. Hopkins, \emph{{Reflection positivity and invertible
  topological phases}},  \href{https://arxiv.org/abs/1604.06527}{{\ttfamily
  1604.06527}}.

\bibitem{Schommer-Pries:2015lnx}
C.~Schommer-Pries, \emph{{Tori Detect Invertibility of Topological Field
  Theories}},  \href{https://arxiv.org/abs/1511.01772}{{\ttfamily 1511.01772}}.

\bibitem{Kitaev:2005dm}
A.~Kitaev and J.~Preskill, \emph{{Topological entanglement entropy}},
  \href{http://dx.doi.org/10.1103/PhysRevLett.96.110404}{\emph{Phys. Rev.
  Lett.} {\bfseries 96} (2006) 110404},
  [\href{https://arxiv.org/abs/hep-th/0510092}{{\ttfamily hep-th/0510092}}].

\bibitem{Levin:2006zz}
M.~Levin and X.-G. Wen, \emph{{Detecting Topological Order in a Ground State
  Wave Function}},
  \href{http://dx.doi.org/10.1103/PhysRevLett.96.110405}{\emph{Phys. Rev.
  Lett.} {\bfseries 96} (2006) 110405},
  [\href{https://arxiv.org/abs/cond-mat/0510613}{{\ttfamily
  cond-mat/0510613}}].

\bibitem{Grover:2011fa}
T.~Grover, A.~M. Turner and A.~Vishwanath, \emph{{Entanglement Entropy of
  Gapped Phases and Topological Order in Three dimensions}},
  \href{http://dx.doi.org/10.1103/PhysRevB.84.195120}{\emph{Phys. Rev.}
  {\bfseries B84} (2011) 195120},
  [\href{https://arxiv.org/abs/1108.4038}{{\ttfamily 1108.4038}}].

\bibitem{Zheng:2017yta}
Y.~Zheng, H.~He, B.~Bradlyn, J.~Cano, T.~Neupert and B.~A. Bernevig,
  \emph{{Structure of the entanglement entropy of (3+1)-dimensional gapped
  phases of matter}},
  \href{http://dx.doi.org/10.1103/PhysRevB.97.195118}{\emph{Phys. Rev.}
  {\bfseries B97} (2018) 195118},
  [\href{https://arxiv.org/abs/1710.01747}{{\ttfamily 1710.01747}}].

\bibitem{Wen:2017xwk}
X.~Wen, H.~He, A.~Tiwari, Y.~Zheng and P.~Ye, \emph{{Entanglement entropy of
  (3+1)D topological orders with excitations}},
  \href{https://arxiv.org/abs/1710.11168}{{\ttfamily 1710.11168}}.

\bibitem{Dong:2008ft}
S.~Dong, E.~Fradkin, R.~G. Leigh and S.~Nowling, \emph{{Topological
  Entanglement Entropy in Chern-Simons Theories and Quantum Hall Fluids}},
  \href{http://dx.doi.org/10.1088/1126-6708/2008/05/016}{\emph{JHEP} {\bfseries
  05} (2008) 016}, [\href{https://arxiv.org/abs/0802.3231}{{\ttfamily
  0802.3231}}].

\bibitem{Wen:2016snr}
X.~Wen, S.~Matsuura and S.~Ryu, \emph{{Edge theory approach to topological
  entanglement entropy, mutual information and entanglement negativity in
  Chern-Simons theories}},
  \href{http://dx.doi.org/10.1103/PhysRevB.93.245140}{\emph{Phys. Rev.}
  {\bfseries B93} (2016) 245140},
  [\href{https://arxiv.org/abs/1603.08534}{{\ttfamily 1603.08534}}].

\end{thebibliography}\endgroup
